\documentclass[a4paper,11pt]{article}
\usepackage[utf8]{inputenc}
\usepackage{verbatim,graphics,graphicx,color,slashed,amsmath}
\usepackage[T1]{fontenc}
\usepackage{subfigure}
\usepackage[table,svgnames]{xcolor}
\usepackage{color}
\usepackage{jheppub}
\usepackage{amsfonts}
\usepackage{amssymb}
\usepackage{bbding}
\usepackage{appendix}
\usepackage{mhchem}
\usepackage{stmaryrd}
\usepackage{blkarray}
\usepackage[export]{adjustbox}
\graphicspath{ {./images/} }
\usepackage{bbold}
\usepackage{CJKutf8}
\usepackage[T1]{fontenc}

\allowdisplaybreaks[3]

\newcommand{\be}{\begin{equation}}
\newcommand{\ee}{\end{equation}}
\newcommand{\bea}{\begin{eqnarray}}
\newcommand{\eea}{\end{eqnarray}}

\usepackage[normalem]{ulem}

\title{SMEFTs living on the edge: determining the UV theories from positivity and extremality}

\author[a,b,c]{Cen Zhang}

\affiliation[a]{Theoretical Physics Division, Institute of High Energy Physics, Chinese Academy of Sciences, Beijing 100049, China}
\affiliation[b]{School of Physical Sciences, University of Chinese Academy of Sciences, Beijing 100049, China}
\affiliation[c]{Center for High Energy Physics, Peking University, Beijing 100871, China}

\abstract{We study the ``inverse problem'' in the context of the Standard Model Effective Field Theory (SMEFT): how and to what extend can one reconstruct the UV theory, given the measured values of the operator coefficients in the IR? The main obstacle of this problem is the degeneracies in the space of coefficients: a given SMEFT truncated at a finite dimension can be mapped to infinitely many UV theories. We discuss these degeneracies at the dimension-8 level, and show that positivity bounds play a crucial role in the inverse problem. In particular, the degeneracies either vanish or become significantly limited for SMEFTs that live on or close to the positivity bounds. The UV particles of these SMEFTs, and their properties such as spin, charge, other quantum numbers, and interactions with the SM particles, can often be uniquely determined, assuming dimension-8 coefficients are measured. The allowed region for SMEFTs, which forms a convex cone, can be systematically constructed by enumerating its generators. We show that a geometric notion, extremality, conveniently connects the positivity problem with the inverse problem. We discuss the implications of a SMEFT living on an extremal ray, on a $k$-face, and on the vertex of the positive cone. We also show that the information of the dimension-8 coefficients can be used to set exclusion limits on all individual UV states that interact with the SM, independent of specific model assumptions. Our results indicate that the dimension-8 operators encode much more information about the UV than one would naively expect, which can be used to reverse engineer the UV physics from the SMEFT.}

\begin{document}
\maketitle
\flushbottom


\section{Introduction}
\label{sec1}
In the Standard Model Effective Field Theory (SMEFT) approach to new physics~\cite{Weinberg:1979sa,Buchmuller:1985jz,Grzadkowski:2010es, Brivio:2017vri, Lehman:2014jma,Henning:2015alf, Li:2020gnx,Murphy:2020rsh,Li:2020xlh,Liao:2020jmn}, coefficients of operators are to be determined by experimental data via global fits~\cite{Ethier:2021bye, Almeida:2021asy, Ellis:2020unq, Dawson:2020oco, DeBlas:2019qco, deBlas:2019rxi, Hartland:2019bjb, Durieux:2019rbz, Falkowski:2019hvp, Durieux:2018ggn, Durieux:2018tev, Ellis:2018gqa, Barklow:2017suo, Durieux:2017rsg, Falkowski:2017pss, Falkowski:2015jaa, Efrati:2015eaa}.
Writing the \((B\)- and \(L\)-number conserving) SMEFT Lagrangian as
\begin{equation}
\mathcal{L}=\mathcal{L}_{\mathrm{SM}}+\sum_{n=3}^{\infty} \frac{1}{\Lambda^{(2 n)}} \sum_{i} C_{i}^{(2 n)} O^{(2 n)}
\end{equation}
where \(C^{(2 n)}\) and \(O^{(2 n)}\) are the coefficients and operators of dimension \(2 n\) respectively, we hope that, with gradually increasing data coming from LHC and future colliders, we will eventually determine as many coefficients \(C^{(2 n)}\) as possible, at least for the lower \(n\)'s. These coefficients contain vital information that can be used to reconstruct the UV completion. It is, therefore, natural to ask the following question: once the coefficients \(C^{(2 n)}\) are known up to a certain dimension, how and to what extend can we extract the UV physics from this information? This question needs to be answered in order for SMEFT to be a useful bottom-up approach to new physics.

This question is often referred to as the ``inverse problem'', in different contexts, such as SUSY~\cite{Arkani-Hamed:2005qjb}, Higgs~\cite{peskintalk}, and SMEFT~\cite{Dawson:2020oco, Gu:2020thj}. This paper focuses on the context of SMEFT. The problem can be viewed as the inverse of the EFT matching: the calculation of the operator coefficients from a known UV theory. While the latter is a well-studied and systematized procedure~\cite{Cohen:2020fcu, Fuentes-Martin:2016uol, Henning:2016lyp, Henning:2014wua}, 
the inverse problem, however, goes in the opposite direction, and has been rarely discussed in the literature. The main difficulty is that each SMEFT\footnote{In this work, ``a SMEFT'' means the SMEFT with its coefficients taking a given set of values, i.e. a single point in the parameter space.}, truncated at a finite dimension, can be mapped to infinitely many UV theories. We will refer to this situation as ``degeneracy''.

There are two sources of degeneracies. The more obvious one is due to the uncertainties in real measurements. They prevent us from resolving the two SMEFTs that are close to each other, so that their corresponding UV completions cannot be distinguished. Studies of this kind, for example those in Ref.~\cite{Ethier:2021bye, Almeida:2021asy, Ellis:2020unq, Dawson:2020oco, DeBlas:2019qco, deBlas:2019rxi, Hartland:2019bjb, Durieux:2019rbz, Falkowski:2019hvp, Durieux:2018ggn, Durieux:2018tev, Ellis:2018gqa, Barklow:2017suo, Durieux:2017rsg, Falkowski:2015krw, Falkowski:2015jaa, Efrati:2015eaa}, 
allow us to quantify the potential of an experiment in probing and discriminating between different scenarios beyond the SM (BSM), and thus provide valuable inputs for motivating the building of future colliders.

However, even if one could determine the SMEFT without any uncertainty, an intrinsic degeneracy still exists in the problem: each SMEFT, truncated at a finite mass dimension, can be UV completed by an infinite number of BSM theories. This is a purely theoretical problem, and is what we will discuss in this paper.

As a simple example of the degeneracy at dim-6, after integrating out a heavy vector which couples to the right-handed SM electron current will generate an operator \(O_{e e}^{(6)}=\) \(\left(\bar{e}_{R} \gamma_{\mu} e_{R}\right)\left(\bar{e}_{R} \gamma^{\mu} e_{R}\right)\) with coefficient \(-g_{1}^{2} / 2 M_{1}^{2}\), where \(M_{1}\) is the vector mass and \(g_{1}\) its coupling strength. The same procedure for a scalar with mass \(M_{2}\) and coupling strength \(g_{2}\) to the \(\bar{e}_{R}^{c} e_{R}\) term will generate instead a coefficient \(g_{2}^{2} / 2 M_{2}^{2}\) with an opposite sign. A measured coefficient \(C_{e e}^{(6)}\) admits an infinite number of solutions for \(g_{1} / M_{1}\) and \(g_{2} / M_{2}\), with the only constraint being
\begin{equation}
\frac{C_{e e}^{(6)}}{\Lambda^{2}}=-\frac{g_{1}^{2}}{2 M_{1}^{2}}+\frac{g_{2}^{2}}{2 M_{2}^{2}}\label{eq:1.2}
\end{equation}
As such, it cannot resolve the flat direction \(g_{1}^{2} / M_{1}^{2}-g_{2}^{2} / M_{2}^{2}=\) const. This prevents us from determining even just the ratio \(g / M\) for each particle type. Note that this ``flat direction'' is different from what is often discussed in the literature: it is not a flat direction in the space of coefficients due to real measurements not being able to probe certain directions, but rather, it is one in the space of UV models, which cannot be lifted at dim-6, even if all coefficients are precisely measured.

Naively, including even higher-dimensional coefficients, which carry additional information, seems to be the only solution. While this is in general true, increasing the dimension does not fully resolve the degeneracy, as there can always be an infinite number of particles in the UV spectrum. At any finite dimension, an infinite number of UV theories remain to be degenerate. Furthermore, experimentally measuring operators beyond dimension-8 (dim-8) is challenging, as in reality the lower-dimensional operators dominate. For this reason, including more and more coefficients at increasingly higher dimensions does not seem to be a promising solution to resolve the degeneracy. In fact, in the literature, SMEFTs beyond a dim-6 truncation are rarely discussed, except in certain problems, such as the classification and counting of higher dimensional operators~\cite{Lehman:2014jma, Henning:2015alf, Li:2020gnx, Murphy:2020rsh, Li:2020xlh, Liao:2020jmn}, where dim-6 operators are known to be unimportant (see, {\it e.g.}, Ref.~\cite{Degrande:2013kka,Eboli:2016kko,Ellis:2020ljj, Gu:2020ldn}),  
or studies of the impacts of (ignoring) dim-8 effects in dim-6 analyses~\cite{Hays:2018zze, Hays:2020scx, Corbett:2021eux, Alioli:2020kez, Boughezal:2021tih, Dawson:2021xei}.

In this paper, we will present a different point of view: studying a subset of dim-8 operators can provide us vital information about the UV theory. In certain regions of the dim-8 coefficient space, degeneracies drastically reduce, sometimes completely vanish, allowing us to uniquely pin down the particle contents of the UV theory. The reason is the so called ``positivity bounds'' arising at dim-8 \cite{Zhang:2020jyn, Li:2021cjv, Zhang:2018shp}. The positivity bounds \cite{Adams:2006sv, Pham:1985cr, Ananthanarayan:1994hf} have received increasing attention in the recent years (see \cite{Zhang:2020jyn, Li:2021cjv, Tolley:2020gtv, Caron-Huot:2020cmc, Chiang:2021ziz, Sinha:2020win, Raman:2021pkf, deRham:2017avq, deRham:2017zjm, Arkani-Hamed:2020blm, Bellazzini:2020cot, Guerrieri:2020bto, Grall:2021xxm, Caron-Huot:2021rmr, Caron-Huot:2021enk, Bern:2021ppb, Du:2021byy} for the recent rapid progress in extending the scope and strength of the bounds, and see, {\it e.g.}, \cite{Zhang:2018shp, Zhang:2020jyn, Li:2021cjv, Bi:2019phv, Yamashita:2020gtt, Fuks:2020ujk, Gu:2020ldn, bellazzini_symmetries_2014, Bellazzini:2017bkb, Bellazzini:2018paj, Remmen:2019cyz, Remmen:2020vts, Bonnefoy:2020yee, Trott:2020ebl,  Chala:2021wpj,
Distler:2006if, Manohar:2008tc, Cheung:2016yqr, Bonifacio:2016wcb, deRham:2017imi, deRham:2018qqo, Bonifacio:2018vzv, Melville:2019wyy, Herrero-Valea:2019hde, deRham:2019ctd, Alberte:2019xfh, Alberte:2019zhd, Chen:2019qvr, Wang:2020jxr, Wang:2020xlt, Huang:2020nqy, Tokuda:2020mlf, Herrero-Valea:2020wxz, Henriksson:2021ymi, Aoki:2021ffc} for applications of the positivity bounds in SMEFT and other scenarios), and as we will show, they are related to the inverse problem in an interesting way. They come from the assumption that the EFT admits a UV completion that is consistent with the fundamental principles of Quantum Field Theory (QFT). (The positivity bounds are of a similar nature of the swampland idea \cite{Vafa:2005ui}, but only conservatively rely on well-established QFT principles.) The dim-6 operators in SMEFT are not subject to these bounds (for amplitudes with only single insertions of them), whereas a subset of dim-8 coefficients (more precisely, those that induce \(E^{4}\) dependence in four-point amplitudes) are confined by a set of homogeneous polynomial bounds. The latter carve out a convex cone in the parameter space, which we dub the positivity cone. It is perhaps not surprising that, being aware of which SMEFT cannot be UV completed at all, these bounds are related to the inverse problem in a specific way. Another hint of the connection is a well-known fact: positivity implies that the leading BSM effects may show up at dim-6 or dim-8, but not higher than dim-8 (see for example \cite{Zhang:2018shp}). This is equivalent to the following statement: the origin of the dim-8 coefficient space has no degeneracy, because the only possible UV completion there is the SM itself. As a very simple example, the analogue of Eq.~(\ref{eq:1.2}) at dim-8 has a plus sign between the two terms, as required by positivity
\begin{equation}
-\frac{C_{e e}^{(8)}}{\Lambda^{4}}=\frac{g_{1}^{2}}{2 M_{1}^{4}}+\frac{g_{2}^{2}}{M_{2}^{4}}
\end{equation}
where $C_{e e}^{(8)}$ is the coefficient of $O_{e e}^{(8)}=\partial^{\nu}\left(\bar{e}_{R} \gamma^{\mu} e_{R}\right) \partial_{\nu}\left(\bar{e}_{R} \gamma_{\mu} e_{R}\right)$. The flat direction now does not exist anymore, thanks to the positiveness of both terms. If $C_{e e}^{(8)}=0$, we immediately conclude that both the vector and the scalar cannot exist in the UV.

The main purpose of this paper is to explore the pattern of degeneracy in the dim-8 coefficient space, and study its relation with positivity bounds. The main finding will be that the SMEFTs on or near the boundary of the positive cone are special, in that they have limited or no degeneracies. In particular, a geometric notion called ``extremality'' \cite{Zhang:2020jyn, bellazzini_symmetries_2014} can be used to study what exactly we can say about the UV completions of these theories. The boundary of the positivity cone consists of its vertex (the origin), the extremal rays, and the \(k\)-faces. A \(k\)-face is a \(k\)-dimensional face of the cone, and the origin and the extremal rays are simply the 0- and 1-faces. Geometrically, these objects are defined by extremality. The latter requires that if any element on a \(k\)-face of a convex cone is a sum of several other elements of the same cone, the latter must all live on the same face. To see the implication of extremality in physics, consider the possible tree-level UV completions of some SMEFT on a \(k\)-face. A particle \(i\) in their UV spectrum, after being integrated out, will generate a coefficient vector \(\vec{C}_{i}^{(8)}\) at dim-8. Positivity requires that this vector lives inside the cone, whereas extremality requires that it lives on exactly the same \(k\)-face. The latter sets a clear restriction on the quantum numbers of particle \(i\) and how it is allowed to interact with the SM particles. We will see that this interpretation can be extended even beyond tree-level UV completions.

Another finding of this work is that even though degeneracies do exist for SMEFTs in the interior of the cone, the SMEFTs close to the boundary have less degeneracies, or equivalently less arbitrariness in finding their UV completions. In particular, exclusion limits on all types of BSM particles can be set, without having to first assume a specific UV theory. Being model-independent, these bounds are of great help for reconstructing the BSM scenario, and serve as guidance for further experimental studies. Exclusion limits of this type, unfortunately, cannot be set by studying the SMEFT truncated only at dim-6, unless very specific assumptions are made about the UV models. We shall emphasize that, in this work, when we say ``determine the UV theory'', we are only interested in the interaction aspects the theories, while the mass spectrum of the UV particles will not be considered, for which information beyond dim-8 will be required. This will be clarified with examples.

All these intriguing features of dim-8 coefficients suggest that studying the SMEFT at the dim-8 level is of special interest. It not only brings forth information in addition to the normally considered dim-6 ones, but more importantly, depending on what the actual UV theory is, it potentially provides the opportunity to completely and uniquely determine the particle content of the UV theory. While there is of course no guarantee that the nature prefers a SMEFT that lives on the boundary, evidence for the opposite is also absent, and this fact alone is already a good motivation to study the phenomenology aspects of dim-8 SMEFT~\cite{Li:2020gnx,Murphy:2020rsh}. Furthermore, even if the nature lives in the interior of the cone, model-independent limits on individual UV particles are of great value by themselves.

In practice, however, learning from dim-8 is based on two requirements: 1) one needs to know where $exactly$ the boundary is, which requires a technique to derive the complete and most constraining positivity bounds at dim-8; and 2) one needs to be able to actually measure the dim-8 coefficients to a reasonable accuracy level, without being affected by the possible existence of dim-6 ones.

The first issue is relatively better studied. Recent progresses in extending the scope of positivity bounds can be categorized in three directions: the inclusion of higher-dimensional operators (or higher powers of \(s\) dependence) \cite{Arkani-Hamed:2020blm, Bellazzini:2020cot}, the inclusion of higher-angular momenta in the scattering (or higher powers of \(t\) dependence) \cite{Tolley:2020gtv, Caron-Huot:2020cmc} (see also \cite{deRham:2017avq, deRham:2017zjm, Sinha:2020win, Chiang:2021ziz, Raman:2021pkf}), 
and the inclusion of multiple particle species \cite{Zhang:2020jyn, Li:2021cjv} (see also \cite{Zhang:2018shp, Bi:2019phv, Yamashita:2020gtt, Fuks:2020ujk, Gu:2020ldn, bellazzini_symmetries_2014, Bellazzini:2017bkb, Bellazzini:2018paj, Remmen:2019cyz, Remmen:2020vts, Bonnefoy:2020yee, Trott:2020ebl}). 
Progress in the 3rd direction is the most relevant in the inverse problem, as it allows us to discuss the boundary of SMEFTs in a large-dimensional parameter space, and therefore to infer how a UV particle interacts with multiple SM species. Progresses in the first two directions do not improve bounds at the dim-8 level, and are thus less relevant in this specific context, as precise measurements of coefficients beyond dim-8 seem unpromising.

Focusing on SMEFTs truncated at dim-8, the standard way to derive bounds was to use a 2-to-2 scattering amplitudes, $elastic$ and $forward$, \(\mathbf{A}_{i j \rightarrow i j}\), where \(i, j\) are some particle states. Positivity requires, roughly,
\begin{equation}
\frac{d^{2}}{d s^{2}} \mathbf{A}_{i j \rightarrow i j}(s) \geq 0
\end{equation}
Here, the incoming states \(i\) and \(j\) can each be a superposition of different basis states. Enumerating all possible superposed states leads to the generalized elastic bounds. Ref.~\cite{Zhang:2020jyn, Li:2021cjv}, however, pointed out that even these generalized elastic bounds fail to capture the precise boundary of UV-completable SMEFTs at dim-8. Additional bounds arise from amplitudes in which the two incoming particles are entangled \cite{Zhang:2020jyn}. One way to capture the full bounds, if the particles being studied are charged under some symmetry group(s), is to construct the allowed amplitude as a convex hull of the projective operators. This was first proposed in Ref.~\cite{bellazzini_symmetries_2014}, in which the positivity region is identified as a polyhedral cone, whose edge vectors are the projectors. More recently, this approach is reformulated using extremal rays and generalized to cases where the positivity cones have curved boundaries ~\cite{Zhang:2020jyn}, see also Refs.~\cite{Yamashita:2020gtt} and \cite{Fuks:2020ujk} for further developments. Ref.~\cite{Zhang:2020jyn} also pointed out the connection between the extremal rays and the inverse problem, on which this work is based. In the first half of this paper, we will further explore this approach in details using a ``generator'' point of view, which makes manifest the relation between bounds and UV theories. Alternatively, a different approach proposed by Ref.~\cite{Li:2021cjv} studies the dual cone of the positivity region. It turns the positivity problem into a semidefinite programming, which is numerically efficient, in particular if a large number of particles are involved. Its drawback, however, is that the relation between positivity and the UV completions becomes obscured in the dual space. We therefore refrain from using this approach in the discussion of the inverse problem, keeping in mind that it could always serve as an efficient alternative to determine the precise boundary.

The second problem arises from a more realistic consideration: will we be able to actually measure precisely the dim-8 coefficients, to the extend that the picture described above can be practically relevant? A complete answer remains unclear, especially because most SMEFT studies in the literature focused on dim-6 operators. However, several works have studied the phenomenological aspects of certain dim-8 operators, and demonstrated that reasonable sensitivities can in general be achieved at HL-LHC or future colliders, either by global fitting or by constructing novel observables~\cite{Hays:2020scx, Ellis:2020ljj, Fuks:2020ujk, Gu:2020ldn, Alioli:2020kez, Boughezal:2021tih}. 
In particular, Ref.~\cite{Fuks:2020ujk} actually showed that positivity bounds, when combined with realistic measurements, do provide useful model-independent exclusion limits to all types of UV particles.

Our take is that more studies on dim-8 coefficients are needed to fully understand our potential reach in reality, but to this end, a motivation is needed. Why should one study dim-8 operators, given that in most cases the dominant effects of a BSM theory are described by the dim-6 ones? The goal of this work is exactly to provide such a motivation: rather than just fixing more operator coefficients, a measurement of the dim-8 coefficients could provide, depending on where the SMEFT lives in the positive cone, much more crucial information about UV particles. In this paper we shall, therefore, first concentrate on establishing this motivation, and defer the phenomenological studies of certain dim-8 coefficients to future works. For this same reason, we will also avoid using dim-6 coefficients in the inverse problem, so as to have a clear understanding of what exactly we can learn from dim-8 coefficients in the ideal case. We shall keep in mind that dim-6 coefficients could always add additional information in realistic problems.

The paper is organized as follows. In Section 2, we consider a simple EFT with two scalars. The purpose is to provide a heuristic description of the main findings of this paper. In Section 3, we explain the positivity approach of Refs.~\cite{Zhang:2020jyn} in more details. In particular, we define the ``generators'' of the positivity cone, which naturally serves as a connection between positivity bounds and the inverse problem. Section 4 is devoted to a more detailed discussion of positivity bounds, in which we illustrate various aspects of the cone construction, with a series of examples. We continue to discuss the inverse problem in Section 5, assuming tree-level UV completions, with a focus on the implication of SMEFTs saturating positivity bounds. In Section 6, we generalize our results to several loop-level UV completions. The main findings of this work is summarized and discussed in Section 7.\\

\section{A toy example}
\label{sec2}
In this section, we consider a toy EFT with two scalar fields, and discuss the implications of this EFT saturating certain positivity bounds. The purpose is to give a flavor of the main conclusions of this work.

\subsection{Bounds for 2-scalar EFT}
\label{sec2.1}
Consider an EFT with two scalar fields, \(\phi_{1}\) and \(\phi_{2}\), with two discrete symmetries imposed:

\begin{enumerate}
  \item the permutation symmetry under \(\phi_{1} \leftrightarrow \phi_{2}\);

  \item a \(Z_{2}\) symmetry \(\phi_{1} \rightarrow-\phi_{1}\) (or equivalently \(\phi_{2} \rightarrow-\phi_{2}\) ).

\end{enumerate}
We are interested in the operators that enter the 4-point amplitudes and give rise to the \(E^{2}\) and \(E^{4}\) dependence. The independent ones at dim-6 and dim-8 can be easily enumerated. At dim-6, we have only one operator,
\begin{equation}
O^{(6)}=\partial_{\mu} \phi_{1}^{2} \partial^{\mu} \phi_{2}^{2}
\end{equation}
and at dim-8
\begin{align}
&O_{1}^{(8)}=\partial_{\mu} \phi_{1} \partial^{\mu} \phi_{1} \partial_{\nu} \phi_{1} \partial^{\nu} \phi_{1}+\partial_{\mu} \phi_{2} \partial^{\mu} \phi_{2} \partial_{\nu} \phi_{2} \partial^{\nu} \phi_{2} \\
&O_{2}^{(8)}=\partial_{\mu} \phi_{1} \partial^{\mu} \phi_{1} \partial_{\nu} \phi_{2} \partial^{\nu} \phi_{2} \\
&O_{3}^{(8)}=\partial_{\mu} \phi_{1} \partial^{\mu} \phi_{2} \partial_{\nu} \phi_{1} \partial^{\nu} \phi_{2}
\end{align}
Their dimensionless Wilson coefficients are denoted as \(C_{1}^{8}, C_{2}^{8}\) and \(C_{3}^{8}\) respectively.

\begin{figure}[h]
	\begin{center}
		\includegraphics[width=.5\linewidth]{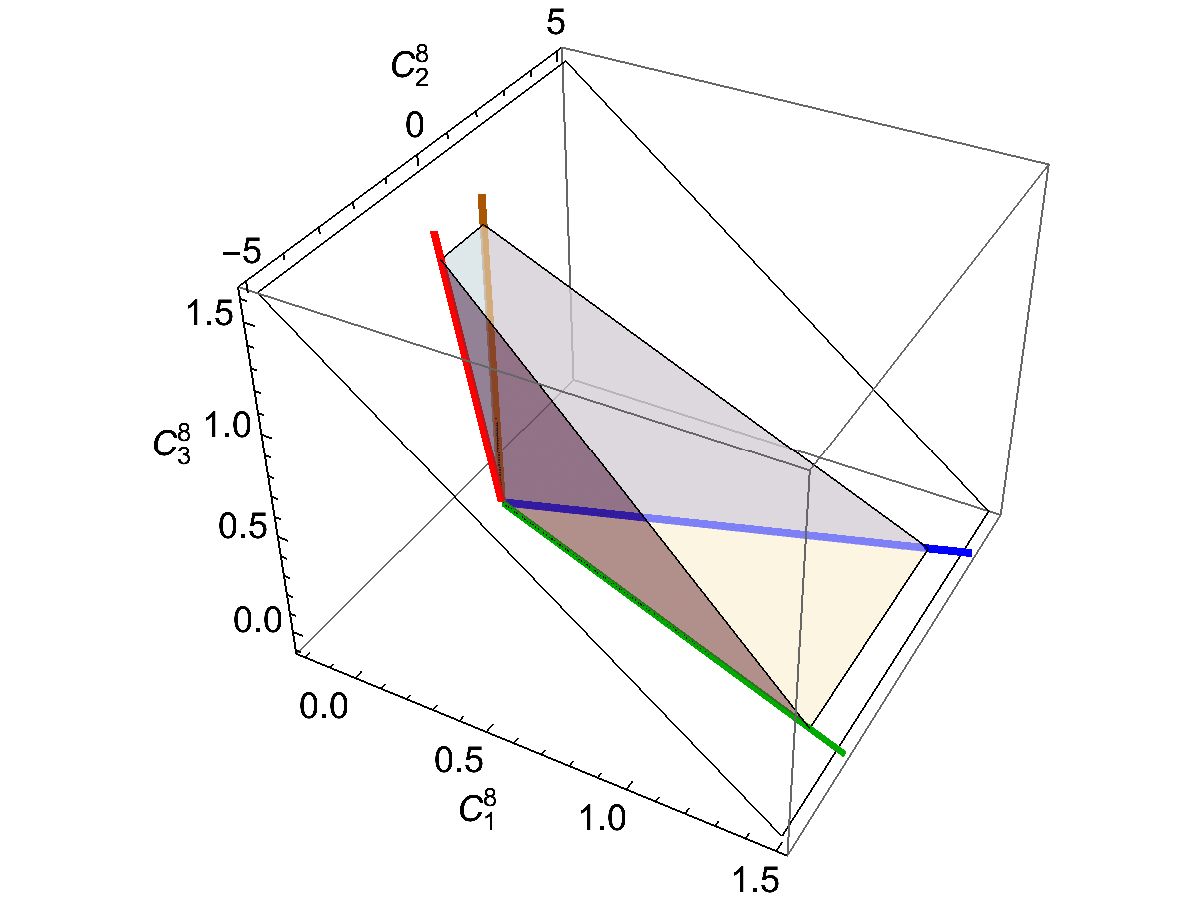}
        \includegraphics[width=.4\linewidth]{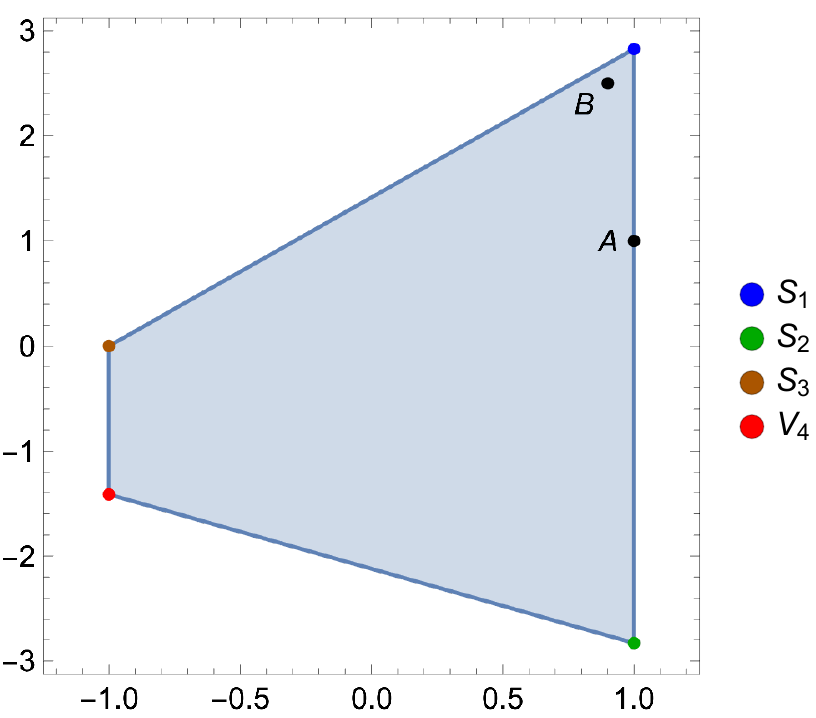}
	\end{center}
	\caption{Left: the positivity cone in the \(\left(C_{1}^{8}, C_{2}^{8}, C_{3}^{8}\right)\) space, which is a pyramid. The edge vectors are labeled in different colors. Right: a cross section of this cone, which is a quadrilateral. The cross section is obtained with a hyperplane \(C_{1}+C_{3}=\sqrt{2}\) (also shown in the left figure) intersecting the cone.}
	\label{fig1}
\end{figure}

Let us first investigate the boundary of dim-8 parameter space. The easiest way to derive positivity bounds is to use the forward and elastic scattering amplitudes \cite{Adams:2006sv}:
\begin{equation}
\left.\frac{d^{2}}{d s^{2}} \mathcal{M}_{\phi_{i} \phi_{j} \rightarrow \phi_{i} \phi_{j}}(s)\right|_{s \rightarrow 0} \geq 0
\end{equation}
where \(\phi_{i, j}\) can be an arbitrary superposition of \(\phi_{1}\) and \(\phi_{2}\). While we are going to justify these bounds in Section 3, for now let us consider their implications on the Wilson coefficients. Defining the states \(\left|\phi_{\pm}\right\rangle \equiv \frac{1}{\sqrt{2}}\left|\phi_{1}\right\rangle \pm \frac{1}{\sqrt{2}}\left|\phi_{2}\right\rangle\), and consider the following four elastic channels:
\begin{align}
&\frac{1}{2} \frac{d^{2}}{d s^{2}} \mathcal{M}_{\phi_{1} \phi_{1} \rightarrow \phi_{1} \phi_{1}}=4 C_{1}^{8} \geq 0 \label{eq:2.6}\\
&\frac{1}{2} \frac{d^{2}}{d s^{2}} \mathcal{M}_{\phi_{+} \phi_{+} \rightarrow \phi_{+} \phi_{+}}=4\left(2 C_{1}^{8}+C_{2}^{8}+C_{3}^{8}\right) \geq 0 \\
&\frac{1}{2} \frac{d^{2}}{d s^{2}} \mathcal{M}_{\phi_{+} \phi_{-} \rightarrow \phi_{+} \phi_{-}}=4\left(2 C_{1}^{8}-C_{2}^{8}\right) \geq 0 \\
&\frac{1}{2} \frac{d^{2}}{d s^{2}} \mathcal{M}_{\phi_{1} \phi_{2} \rightarrow \phi_{1} \phi_{2}}=C_{3}^{8} \geq 0\label{eq:2.9}
\end{align}
These four bounds carve out a pyramid in the dim-8 parameter space. Its vertex is the origin, and each face corresponds to one of the inequalities above, see Figure \ref{fig1} left. This pyramid turns out to be the tightest possible positivity bounds at dim-8: other elastic channels with differently superposed fields contain no new information. One may ask why these four channels are special. A general explanation is provided in Ref.~\cite{Zhang:2020jyn}, based on the duality of convex cones.\\

\subsection{Mapping the extremal rays with UV particles}
\label{sec2.2}
Instead of the four bounds, let us take a different point of view: a pyramid can also be determined by its four edge vectors. Any ray inside a pyramid is a positive combination of these vectors. We now ask the following question: which UV completions lead to EFTs that live on these edge vectors?

Consider all possible UV completions at the tree level. The operators listed above can be generated by integrating out heavy particles that couple to the two scalar fields. They are classified by the parity under the two discrete symmetries. There are four possible states with spin less than or equal to one. We list them below.
\[
\begin{array}{c|ccc}
\text { Particle } & \text { Spin } & \begin{array}{c}
\text { Parities } \\
\left(\phi_{1} \rightarrow-\phi_{1}, \phi_{1} \leftrightarrow \phi_{2}\right)
\end{array} & \text { Interaction } \\
\hline S_{1} & 0 & +, \quad+ & g_{1} M_{1}\left(\phi_{1}^{2}+\phi_{2}^{2}\right) S_{1} \\
S_{2} & 0 & +,\quad- & g_{2} M_{2}\left(\phi_{1}^{2}-\phi_{2}^{2}\right) S_{2} \\
S_{3} & 0 & -,\quad+ & g_{3} M_{3} \phi_{1} \phi_{2} S_{3} \\
V_{4} & 1 & -, \quad- & g_{4}(\phi_{1} \stackrel{\leftrightarrow}{D}_{\mu} \phi_{2}) V_{4}^{\mu}
\end{array}
\]
where \(g_{i}\) and \(M_{i}\) are the coupling and mass of the corresponding particle \(S_{i}\) or \(V_{i}\) respectively. Now, integrating out each particle will generate a set of dim-8 coefficients, and we will denote them by a vector, \(\vec{C}_{X}^{8}=\left(C_{1}^{8}, C_{2}^{8}, C_{3}^{8}\right)\), where \(X\) labels the heavy particle integrated out. These vectors can be easily computed, and written as:
\begin{equation}
\vec{C}_{S_{i}}^{8}=w_{i} \vec{c}_{i}, \text { for } i=1,2,3 ; \quad \vec{C}_{V_{4}}^{8}=w_{4} \vec{c}_{4}
\end{equation}
where the prefactor \(w_{i}\) is always non-negative:
\begin{equation}
w_{i} \equiv g_{i}^{2} / M_{i}^{4} \geq 0
\end{equation}
and the \(\vec{c}_{i}\) represents the direction of the vector:
\begin{align}
\vec{c}_{1}^{\,8}=2 \times(1,2,0), &\quad \vec{c}_{2}^{\,8}=2 \times(1,-2,0), \\
\vec{c}_{3}^{\,8}=2 \times(0,0,1), &\quad \vec{c}_{4}^{\,8}=2 \times(0,-1,1) .
\end{align}
We show all \(\vec{c}_{X}^{\,8}\)'s in Figure \ref{fig1} in different colors. Interestingly, these four vector are exactly the four edge vectors of the pyramid, carved out by positivity.

We now have a simple, but incomplete answer to the aforementioned question: the ``one-particle UV completions'' lead to EFTs that live on these edge vectors. By one-particle UV completion, we mean a UV completion that contains only one heavy particle in the UV spectrum. Depending on which particle it is, the corresponding EFT falls on one of the four edge vectors. The answer is incomplete because we have not yet ruled out the possibility of other UV completions mapping also to the same edge vectors.

The mapping from one-particle UV completions to edge vectors is not surprising. After all, the dim-8 coefficients generated by integrating out \(S_{1}, S_{2}, S_{3}\) and \(V_{4}\) at the tree level are as follows:
\begin{equation}
\vec{C}^{8}=\sum_{i=1}^{4} w_{i} \vec{c}^{\,8}_{i}, \quad w_{i} \geq 0 \label{eq:2.14}
\end{equation}
The positiveness of the \(w_{i}\) implies that the coefficients of all tree-level UV completions are positively generated by the \(\vec{c}_{i}^{\,8}\) vectors, and therefore they fill the convex hull of these vectors, which is a pyramid. Geometrically, the generators of this pyramid are its edge vectors, \(\vec{c}_{i}^{\,8}\), just like physically the generators of all tree-level UV-completions are all the one-particle UV completions. Therefore the correspondence between the edge vectors of the positivity pyramid and the one-particle UV completions is expected. In fact, Eq.~(\ref{eq:2.14}) gives the edge-representation of the pyramid, while Eqs.~(\ref{eq:2.6})-(\ref{eq:2.9}) give its face-representation. What is nontrivial is that Eqs.~(\ref{eq:2.6})-(\ref{eq:2.9}) are actually derived without assuming a tree-level or even a weakly-coupled UV completion, and therefore this picture remains valid beyond the tree level.

So far, this mapping is established only in the top-down direction: a one-particle UV completion, after matching, falls onto one of the edge vectors. One of the main observations of Ref.~\cite{Zhang:2020jyn}, however, is that this mapping actually goes in both directions. In other words, SMEFTs on the edge vectors have no degeneracy, as the only possible UV completions are the one-particle extensions. The implication is that if data tells us that \(\vec{C}^{8}\) is proportional to, say, \(\vec{c}_{1}\), we can immediately conclude that only \(S_{1}\) exists in the UV theory. This then completes the answer to the aforementioned question: only the one-particle UV completions can lead to EFTs that live on these edge vectors. As a result, the inverse problem are solved for these edge vectors.

There is an intuitive way to see why it is so: an EFT generated by the \(S_{1}\) scalar stays on the top right corner of the quadrilateral in Figure \ref{fig1} right, and obviously the existence of any particle of a different type will ``drag'' the total coefficient vector towards inside of the pyramid. To take it back to the top-right corner, contributions outside of the pyramid is needed, which then violates positivity bounds. This is exactly how ``extremality'' plays a role in the inverse problem. The edge vectors are the extremal rays of the pyramid, and so they cannot be written as a sum of two other vectors, which are linearly independent and both contained in the same pyramid. Physically, it implies that the UV theory cannot have multiple (different kinds of) heavy particles, because integrating out each of them will generate some non-vanishing \(w_{i} \vec{c}_{i}^{\,8}\), and with more than one heavy particles, \(\vec{C}^{8}\) is a sum of different \(w_{i} \vec{c}_{i}^{\,8}\), which cannot be extremal. Note that it is important that the positivity bounds need to exist in the first place, carving out a convex cone in which these extremal rays can be defined. We will show that this is always the case at dim-8, but in general not true at dim-6.

The same conclusion can be obtained from a different point of view, by exploiting the following fact: a positivity bound, when saturated, rules out the possible existence of certain heavy states. In fact, in terms of \(w_{i}=g_{i}^{2} / M_{i}^{4}\), these bounds can be written as:
\begin{align}
&C_{1}^{8}=2\left(\frac{g_{1}^{2}}{M_{1}^{4}}+\frac{g_{2}^{2}}{M_{2}^{4}}\right) \geq 0 \label{eq:2.15}\\
&2 C_{1}^{8}+C_{2}^{8}+C_{3}^{8}=2\left(4 \frac{g_{1}^{2}}{M_{1}^{4}}+\frac{g_{3}^{2}}{M_{3}^{4}}\right) \geq 0 \\
&2 C_{1}^{8}-C_{2}^{8}=2\left(4 \frac{g_{2}^{2}}{M_{2}^{4}}+\frac{g_{4}^{2}}{M_{4}^{4}}\right) \geq 0 \\
&C_{3}^{8}=2\left(\frac{g_{3}^{2}}{M_{3}^{4}}+\frac{g_{4}^{2}}{M_{4}^{4}}\right) \geq 0 \label{eq:2.18}
\end{align}
Obviously, each saturated bound can rule out the possible existence of two heavy particles (in this example). If the observed \(\vec{C}^{8}\) is \(\propto \vec{c}_{1} \propto(1,2,0)\), it saturates the last two bounds, and so \(S_{2}, S_{3}, V_{4}\) cannot exist. The only allowed particle in the UV spectrum is \(S_{1}\). Note that there may be multiple particles of the same type, and in this case we should replace \(g_{i}^{2} / M_{i}^{4}\) by \(\sum_{j} g_{i, j}^{2} / M_{i, j}^{4}\) where \(j\) labels different particles of the same type. Since each term in the summation is individually positive, they will all be ruled out by a saturated bound, and therefore the above argument remains valid. Also note that new contributions on the r.h.s.~may arise, if loop-level UV completions incorporated, but they are also individually positive and do not spoil this argument.

\subsection{Degeneracies in the dim-8 coefficient space}
\label{sec2.3}
The vanishing degeneracy at the extremal rays suggests that the distribution of degeneracies inside the pyramid may exhibit a nontrivial pattern. To quantify the degeneracy, let us be more specific about the inverse problem. At dim-8, we are mostly interested in the following question: given the measured values of \(\vec{C}^{8}\), to what extend can we solve Eq.~(\ref{eq:2.14}) and determine the \(w_{i}\) factors? These factors depend on the couplings and masses of particles of each type: \(w_{i}=\sum_{j} g_{i, j}^{2} / M_{i, j}^{4}\). Of course, the \(w_{i}\)'s do not tell us all details of what the UV theory is, but they do tell us which kinds of heavy particles exist in the UV completion, and how large their contributions are, which is crucial for understanding the UV theory. Limited at dim-8, knowing the values of \(\vec{w} \equiv\left(w_{1}, w_{2}, \cdots\right)\) is already a satisfying result for the inverse problem, and so we do not attempt to further extract more information inside the summation, for which even higher dimensional operators need to be studied. We consider the determination of the feasible solutions of Eq.~(\ref{eq:2.14}) for \(\vec{w}\) as a weaker version of the inverse problem, and it is this problem that we will focus on for the rest of the paper.

Obviously, even this weaker version cannot have a definite answer. The reason is that Eq.~(\ref{eq:2.14}) gives three constraints (as there are three operator coefficients), but we have four \(w_{i}\)'s to be determined. The situation is even worse in more realistic problems, where the number of possible UV particles can be much larger than the number of coefficients, which some times can be even infinity. As a result, the solution space for \(\vec{w}\) can have a very large dimension, which means large uncertainties are expected in the determination of each \(w_{i}\).

However, if a positivity cone exists, the picture is completely different. Let us denote the set of feasible solutions for \(\vec{w}\) by \(\mathcal{W}\). Outside the positivity cone, \(\mathcal{W}\) must be empty since UV completions cannot exist. If the distribution of \(\mathcal{W}\) is continuous, we should expect \(\mathcal{W}\) to be ``small'' for \(\vec{C}^{8}\) near the boundary. If data tells us \(\vec{C}^{8}\) is indeed near the boundary, we expect that certain concrete information about the UV theory can be extracted. We have seen that the extremal rays, or the edge vectors, are examples where a unique solution exists: only one of the \(w_{i}\)'s can be nonzero. This fully determines the UV particle content.

There are other points that admit a unique solution for \(w_{i}\). Take a point that lives on one of the four faces, say the one represented by the bound of Eq.~(\ref{eq:2.9}). In the right plot of Figure \ref{fig1}, its projection would stay on the line segment connecting \(S_{1}\) and \(S_{2}\), and we label it by point ``A''. What can we say about the UV theory? First, \(S_{3}\) and \(V_{4}\) cannot exist. Intuitively, their existence will ``drag'' this point towards inside the pyramid, and therefore it cannot stay on the boundary, unless additional contributions violating this bound exist. Indeed, according to Eq.~(\ref{eq:2.18}), the bound \(C_{3}^{8} \geq 0\) being saturated excludes exactly \(S_{3}\) and \(V_{4}\). Now, \(w_{1}\) and \(w_{2}\), being the only nonzero factors, can be uniquely determined, because a given point on a 2-dimensional face fixes exactly two degrees of freedom. In fact, for point ``A'' we find
\begin{equation}
w_{1}=\frac{1}{4}+\frac{1}{\sqrt{2}}, \quad w_{2}=-\frac{1}{4}+\frac{1}{\sqrt{2}}
\end{equation}
Together, we can conclude that the UV theory consists of two types of heavy scalars, \(S_{1}\) and \(S_{2}\), and \(g_{1}^{2} / M_{1}^{4}\) and \(g_{2}^{2} / M_{2}^{4}\) are uniquely determined.

In general, the boundary of a \(d\)-dimensional positivity cone is a collection of \(k\)-faces, where \(k=0,1,2, \cdots, d-1\). In this example, we have one 0-face --- the origin, four 1-faces --- the edge vectors, and four 2-faces. The degeneracy vanishes at the origin, because Eqs.~(\ref{eq:2.15})-(\ref{eq:2.18}) are all saturated. It also vanishes at the 1-faces, because extremal rays cannot be split. For the 2-faces, we can similarly use their extremality: for a theory that lives on a 2-face, if one decomposes its UV spectrum, all individual particles must live on the same face. This is exactly how we excluded \(S_{3}\) and \(V_{4}\) for the point ``A''. More generally, if a \(k\)-face is spanned by \(l\) ``one-particle UV completions'', all EFTs on that face can only have these \(l\) particles in their UV spectrum. If \(l=k\), the \(w_{i}\) for these \(l\) particles can be completely fixed. Extremality plays a central role in this kind of arguments.

On the other hand, the EFTs more inside the pyramid do not in general have a unique UV completion. However, we can still quantify and constrain the arbitrariness in finding the UV completion of a given EFT. Taking the point ``B'' in Figure \ref{fig1} as an example, we have \(C_{1}=1.34, C_{2}=2.5\), and \(C_{3}=0.07\) in some unit. The feasible values for the \(w_{i}\)'s are all constrained in small intervals:
\begin{equation}
w_{1} \in[1.297,1.314], \quad w_{2} \in[0.029,0.047], \quad w_{3}, w_{4} \in[0,0.071]
\end{equation}
simply because this point is near the boundary. We then conclude that the dominant contribution comes from \(S_{1}\), and one can set limits on the existence of \(S_{2}, S_{3}\) and \(V_{4}\).

\begin{figure}[h]
	\begin{center}
		\includegraphics[width=.5\linewidth]{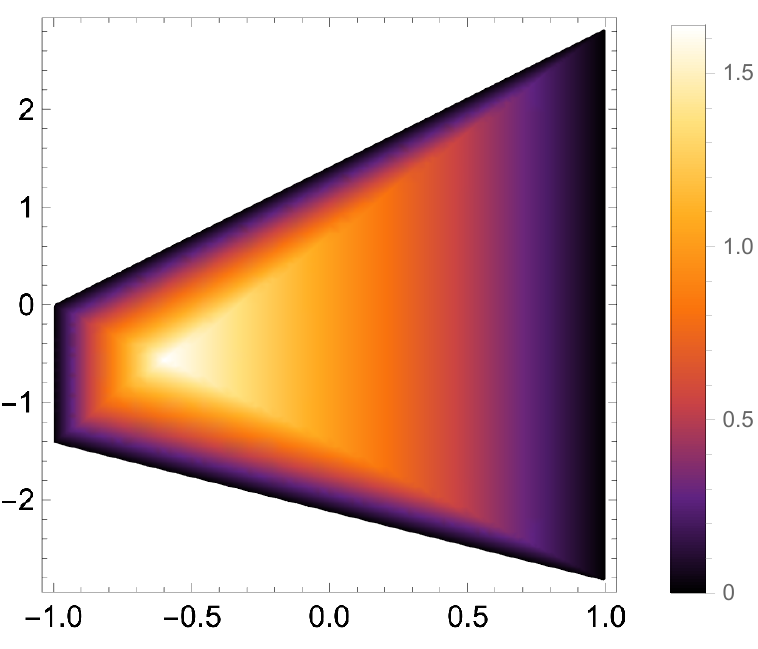}
	\end{center}
	\caption{The degeneracy \(\Delta\) on a cross section of the positivity pyramid, defined by \(C_{1}+C_{3}=\sqrt{2}\). See Eq.~(\ref{eq:2.21}) for definition of \(\Delta\).}
	\label{fig2}
\end{figure}

More generally, the range of possible values for the feasible solution \(\vec{w}\) can be used to quantify the degeneracy. Define
\begin{equation}
\Delta \equiv \max _{\vec{w}_{(1)}, \vec{w}_{(2)} \in \mathcal{W}}\left|\vec{w}_{(1)}-\vec{w}_{(2)}\right| \label{eq:2.21}
\end{equation}
{\it i.e.}~the largest ``distance'' between two feasible \(\vec{w}\) solutions. We plot \(\Delta\) in Figure \ref{fig2}. As expected, the EFTs on the boundary of the pyramid have zero uncertainty on \(w_{i}\), which implies that the contribution of each type of heavy particles can be uniquely determined. On the other hand, the EFTs more inside the pyramid have a larger degeneracy, and thus more arbitrariness in their UV completions, while those more close to the boundary have a smaller degeneracy.

One last nontrivial point is that \(\Delta\) is always finite over the entire cross section of the pyramid. This implies that with any dim-8 measurement it is possible to set exclusion limits on all UV particles, in a model-independent way. Later we will see that this is related to the fact that the positivity cone at dim-8 is always salient, {\it i.e.}~it does not contain any straight line. This fact, unfortunately, does not hold at dim-6. In fact, in this toy example, there is only one dim-6 operator, but its coefficient can be either positive (for \(\left.S_{1}, V_{4}\right)\) or negative (for \(\left.S_{2}, S_{3}\right)\). The allowed coefficient space is then the entire real axis. In this case the degeneracy of the \(w_{i}\)'s at dim-6 is always infinity: \(w_{1}\) and \(w_{4}\) will cancel against \(w_{2}\) and \(w_{3}\), and therefore each of them is allowed to be arbitrarily large. A simple consequence is that the observation of vanishing dim-6 operators cannot completely rule out the potential existence of heavy new physics, and in fact it is not even possible to exclude any single heavy particle up to any mass scale. In contrast, the observation of vanishing dim-8 operators would confidently rule out all kinds of heavy states, independent of the UV theory assumptions. This difference between dim-6 and dim-8 coefficients illustrates one of the reasons why dim-8 operators is special in the context of the inverse problem, and deserve more attention in particle phenomenology.

Let us summarize some interesting features of the dim-8 space we found in this example:

\begin{itemize}
  \item Positivity bounds carve out the boundary of UV-completable EFTs, which is a pyramid with four edge vectors.

  \item The EFTs outside the pyramid cannot have UV completions.

  \item The EFTs on the edge vectors uniquely correspond to UV completions with a single type of heavy particles.

  \item The EFTs on the faces uniquely correspond to UV completions consisting of two types of heavy particles.

  \item The EFTs near the edge vectors or the boundary have limited arbitrariness in their possible particle contents.

  \item The EFTs more inside the pyramid have more uncertainties in determining their UV completions, but the degeneracy is always finite.

\end{itemize}
These features are all absent at dim-6, as positivity bounds do not exist there.

In more practical problems, the parameter space can be of much larger dimension, and the positivity bounds may carve out a cone with much more edge vectors, but the overall picture does not change a lot. The points summarized above are in general valid, but one needs be aware of some additional complications. Let us list some of them:

\begin{itemize}
  \item Integrating out a heavy particle could generate a coefficient vector that is not necessarily an edge vector. Instead, it could stay on the faces, or even inside the cone. This changes the distribution of the degeneracy.

  \item Integrating out a heavy loop contribution does not change the positivity bound, but could also generate a coefficient vector on the faces or inside the cone. Similarly, this also modifies the distribution of the degeneracy.

  \item Positivity region may have curved boundaries. In this case, the number of extremal rays is infinity. This could happen if the symmetries of the problem does not allow the intermediate states to be classified into a finite number of categories, each with a fixed coupling to light particles. Nevertheless, the fact that an extremal ray must correspond to a one-particle UV completion, remains valid.

\end{itemize}
All these points will be illustrated with examples in Sections 4 and 6 .

\section{Theory framework}
\label{sec3}
In this section we set up the main formalism of this work. Our goals are: 1) to identify the exact boundary of the SMEFTs at dim-8, and 2) to connect the bounds to UV completions, which then allows the discussion of the inverse problem.

More specifically, we consider the 2-to-2 scattering amplitude at the tree level in an EFT:
\begin{equation}
\overline{\mathbf{M}}_{i j \rightarrow k l}=\sum_{m, n} a_{m, n}^{i j k l} s^{m} t^{n}
\end{equation}
where \(\overline{\mathbf{M}}_{i j \rightarrow k l}\) is the amplitude with poles subtracted, which is a polynomial of \(s\) and \(t\), at the tree level. The SM particle masses are neglected. We only focus on the \(s^{2}\) term, \(a_{2,0}^{i j k l}\) which represents the forward \(t \rightarrow 0\) limit. We do not consider higher energy dependence with \(m+n>2\), because this inevitably requires a knowledge of dim-10 operators or higher, which seems challenging given the reach at the LHC and even the future colliders. We do not consider the \(n>0\) terms where the \(t\)-dependence arises, because at dim-8 they are not independent of \(a_{2,0}^{i j k l}\), thanks to crossing. In fact, \(k, l\) crossing and \(j, k\) crossing
\begin{align}
&\overline{\mathbf{M}}_{i j \rightarrow l k}(s, t)=\overline{\mathbf{M}}_{i j \rightarrow k l}(s,-s-t) \\
&\overline{\mathbf{M}}_{i k \rightarrow j l}(s, t)=\overline{\mathbf{M}}_{i j \rightarrow k l}(t, s)
\end{align}
lead to the following relations between \(m+n=2\) coefficients:
\begin{align}
&a_{2,0}^{i j l k}=a_{2,0}^{i j k l}-a_{1,1}^{i j k l}+a_{0,2}^{i j k l} \\
&a_{2,0}^{i k j l}=a_{0,2}^{i j k l}
\end{align}
Therefore at the \(m+n=2\) level, we have
\begin{equation}
\overline{\mathbf{M}}_{i j \rightarrow k l}(s, t)=a_{2,0}^{i j k l} s^{2}+\left(a_{2,0}^{i j k l}-a_{2,0}^{i j l k}+a_{2,0}^{i k j l}\right) s t+a_{2,0}^{i k j l} t^{2} \ldots
\end{equation}
which means that the kinematic dependence in \(\overline{\mathbf{M}}_{i j \rightarrow k l}\) are fully encoded in the \(a_{2,0}\) coefficients, which can be accessed with forward amplitudes of different channels. We will base our approach on a study of the full space spanned by \(a_{2,0}^{i j k l}\) for all \(i, j, k, l\) combinations. Once we find the precise bounds in this space, going non-forward does not bring additional results, as the \(t\)-dependence has no independent degree of freedom.

The \(a_{2,0}^{i j k l}\) coefficient, computed at the tree level, is a linear combination of dim-8 coefficients, plus a quadratic form of dim-6 coefficients. The dim-8 operators are those involving exactly four particles with an \(s^{2}\) dependence, while the dim-6 operators are those involving exactly three (vector) particles. In this paper, when we say dim-8 coefficient/SMEFT space, we refer to the space spanned by the coefficients of these operators. In particular, at dim-8, the relevant operators are the type \(1,4,7,11,14,16\), and 21 operators of the Table 1 of Ref.~\cite{Li:2020gnx,Murphy:2020rsh}. The total number is 250 for one generation (including \(B\) number violating operators) and 6076 for three generations.

The situation can be different if loop corrections within the SMEFT enter. \(\overline{\mathbf{M}}_{i j \rightarrow k l}\) is no longer a polynomial of \(s, t .\) We instead define the \(\mathcal{M}^{i j k l}\) (see next sections) as a substitute of \(a_{2,0}^{i j k l}\). Its mapping to the coefficient space is possible but can be more complicated. We will, however, still ambiguously use the word ``dim-8 SMEFT space'' to mean the space spanned by \(\mathcal{M}^{i j k l}\). As we will see, both positivity bounds and the inverse problem can be discussed at the level of \(\mathcal{M}^{i j k l}\), but mapping them to actual operators is the only way to link \(\mathcal{M}^{i j k l}\) to real measurements.

\subsection{Notations and basic concepts}
Let us clarify some notations that will be useful in this paper. We will use \(\mathbf{M}_{i \rightarrow f}\) to denote an amplitude with initial state \(i\) and final state \(f\). In particular, for a \(2 \rightarrow 2\) amplitude \(i j \rightarrow k l\), we define the following rank-4 tensor:
\begin{equation}
\mathcal{M}^{i j k l}=\left.\frac{d^{2}}{d s^{2}} \overline{\mathbf{M}}_{i j \rightarrow k l}(s, t=0)\right|_{s \rightarrow 0}-\int_{-(\epsilon \Lambda)^{2}}^{(\epsilon \Lambda)^{2}} \frac{\mathrm{d} \mu \operatorname{Disc} \overline{\mathbf{M}}_{i j \rightarrow k l}(\mu)}{2 i \pi \mu^{3}}\label{eq:3.7}
\end{equation}
where \(\overline{\mathbf{M}}\) is the amplitude with poles subtracted, and \(i, j, k, l\) run through all low energy degrees of freedom, including those of different particle species, polarization and other quantum numbers. The second term subtracts the low-energy dispersive integral, which will be clarified in Section 3.2. We will simply call this tensor, \(\mathcal{M}^{i j k l}\), the ``amplitude''.

For any tensor, \(T^{i(j|k| l)} \equiv T^{i j k l}+T^{i l k j}\). We use \(\left(\mathbf{r}_{3}, \mathbf{r}_{2}\right)_{Y}\) to label a state in the irreducible representation (irrep) \(\mathbf{r}_{3}\) under \(\mathrm{SU}(3), \mathbf{r}_{2}\) under \(\mathrm{SU}(2)\), and has a hypercharge \(Y\). Similarly, \(\mathbf{r}_{Y}\) indicates just the \(\mathrm{SU}(2)\) irrep \(\mathbf{r}\) and the hypercharge \(Y\). We frequently use a vector \(\vec{C}\) to represent a set of operator coefficients, which corresponds to a single point in the coefficient space.

Some basic concepts from convex geometry can be useful:

\begin{itemize}
  \item A \(convex \ cone\) (or cone) is a subset of a vector space, closed under additions and positive scalar multiplications. A \(salient\) cone is a cone which contains no straight lines. If \(\mathbf{C}\) is salient cone, then \(\pm x \in \mathbf{C}\) implies \(x=0\).

  \item An $extremal\ ray$ (ER) of a cone \(\mathbf{C}\) is an element \(x \in \mathbf{C}\) that cannot be a sum of two other elements in \(\mathbf{C}\). If an ER \(x\) can be written as \(x=y_{1}+y_{2}\) with \(y_{1}, y_{2} \in \mathbf{C}\), we must have \(x=\lambda y_{1}\) or \(x=\lambda y_{2}\), with \(\lambda\) a real constant. The ERs of a polyhedral cone are its edges.

  \item A subset \(\mathbf{F}\) of \(\mathbf{C}\) is called a {\it face}, if for every \(x \in \mathbf{F}\) and every \(y, z \in \mathbf{C}\) such that \(x \in(y, z)\), we have \(y, z \in \mathbf{F}\). A face of dimension \(k\) is called a \(k\)-face. An ER is a 1-face. The origin of a salient cone is a 0-face. A facet of a \(d\)-dimensional cone is a \((d-1)\)-face. The boundary of some cone \(\mathbf{C}\) consists of its 0-, 1-, \(\ldots,(d-1)\)-faces. A polyhedral cone has a finite number of faces.

  \item A \(PSD\ cone\) of dimension \(n\) is the set of \(n \times n\) positive semidefinite (PSD) matrices, which is a convex cone. Its ERs are the rank-1 PSD matrices. Its faces are the subsets whose elements have the same null space.

  \item The \(conical\ hull\) of a set \(\mathbf{X}\), is the ensemble of all positive linear combinations of elements in \(\mathbf{X}\). We denote it by cone \((\mathbf{X})\). The ERs of cone \((\mathbf{X})\) belong to \(\mathbf{X}\).

\end{itemize}
\subsection{Dispersion relation}
\label{sec3.2}
In the forward limit, a twice-subtracted dispersion relation can be derived for \(\mathcal{M}^{i j k l}\), assuming that a UV completion exists and is consistent with the fundamental principles of QFT:
\begin{equation}
\mathcal{M}^{i j k l}\left((\epsilon \Lambda)^{2}\right)=\int_{(\epsilon \Lambda)^{2}}^{\infty} \frac{\mathrm{d} \mu \operatorname{Disc} \mathbf{M}_{i j \rightarrow k l}(\mu)}{2 i \pi \mu^{3}}+(j \leftrightarrow l)
\end{equation}
where we have assumed that the SM particle masses are negligible compared to \(\Lambda\). The derivation can be found in, e.g., Refs.~\cite{Zhang:2020jyn}. The dispersive integration on the r.h.s.~normally starts from the lowest branch point, but we have subtracted the dispersive contribution below a properly chosen scale, \(\epsilon \Lambda\), in the definition of \(\mathcal{M}^{i j k l}\), such that the r.h.s.~starts from \(\epsilon \Lambda\), see Eq.~(\ref{eq:3.7}). \(\epsilon\) is chosen to be less than one, so that the l.h.s. is still calculable in the EFT. This trick is following the ``improved positivity bounds'' of Refs.~\cite{deRham:2017imi}, and can also be thought of as the ``arc'' defined in Ref.~\cite{Bellazzini:2020cot}, with a radius \((\epsilon \Lambda)^{2}\). If \(\mathcal{M}^{i j k l}\) is computed at the tree level, this subtraction term is a higher order contribution (as the discontinuity arises from loops), and in this case \(\mathcal{M}^{i j k l}\) is simply the \(a_{2,0}^{i j k l}\) coefficient in the previous section.

If \(\mathcal{M}^{i j k l}\) is computed at the loop level, the subtraction term cannot be ignored. It leads to a dependence of \(\mathcal{M}^{i j k l}\) on \(\epsilon \Lambda .\) As we will see in Section 6, choosing a large \(\epsilon\) without breaking the EFT validity always leads to better bounds. It also allows better information from the UV theory to be extracted. Before Section 6, however, we will fix \(\epsilon \Lambda\) and aim at deriving the boundary. We thus drop this scale dependence until Section 6.

Upon using the generalized optical theorem, we can rewrite the r.h.s.:
\begin{equation}
\mathcal{M}^{i j k l}=\frac{1}{2 \pi} \int_{(\epsilon \Lambda)^{2}}^{\infty} \frac{d s}{s^{3}} \sum_{X}\left[\mathbf{M}_{i j \rightarrow X} \mathbf{M}_{k l \rightarrow X}^{*}+(j \leftrightarrow l)\right] \label{eq:3.9}
\end{equation}
where the sum is over all intermediate state, denoted by \(X\), which may be infinite and continuous. The r.h.s.~is not calculable without knowing the UV theory, but certain bounds can be extracted. The most obvious one is \(\mathcal{M}^{i j i j} \geq 0 .\) More generally, the following bounds have an interpretation of positiveness in an elastic amplitude:
\begin{equation}
\left.\frac{d^{2}}{d s^{2}} \overline{\mathbf{M}}_{u v \rightarrow u v}(s, t=0)\right|_{s \rightarrow 0}=u_{i} v_{j} u_{k}^{*} v_{l}^{*} \mathcal{M}^{i j k l} \geq 0
\end{equation}
where \(u, v\) are superpositions of basis particle states, \(|u\rangle=u_{i}|i\rangle\) and \(|v\rangle=v_{i}|i\rangle .\) The last inequality simply follows from Eq.~(\ref{eq:3.9}). This is the origin of the bounds, Eqs.~(\ref{eq:2.6})-(\ref{eq:2.9}), that we have used in Section 2.

Elasticity is a notion that depends on the basis of particle states, which is why superposed states lead to additional bounds. However, Refs.~\cite{Zhang:2020jyn, Li:2021cjv} have shown that even the superposed elastic bounds, after enumerating all \(u, v\) vectors, may not be sufficient. For the discussion of the inverse problem, we need an approach that guarantees the exact boundary of all UV-completable SMEFTs.

Without knowing the size of possible UV contributions, the relevant information from Eq.~(\ref{eq:3.9}) is:
\begin{align}
&\mathcal{M} \in \mathbf{C}^{n^{4}} \\
&\mathbf{C}^{n^{4}} \equiv \operatorname{cone}\left(\left\{m^{i j} m^{* k l}+m^{i \bar{l}} m^{* k \bar{j}}\right\}\right)
\end{align}
namely the allowed \(\mathcal{M}\) must be contained in the set \(\mathbf{C}\), which is a conical hull of all rank-4 tensors that have the form \(m^{i j} m^{* k l}+m^{i \bar{l}} m^{* k \bar{j}}\), where \(m\) is an arbitrary \(n \times n\) matrix, and \(n\) is the number of independent particle modes involved in the problem. This is simply because in Eq.~(\ref{eq:3.9}) we can take \(\mathbf{M}_{i j \rightarrow X} \rightarrow m^{i j}\), and all other factors apart from \(m^{i j} m^{* k l}+m^{i \bar{l}} m^{* k \bar{j}}\) is positive. Our goal is to determine the boundary of \(\mathbf{C}^{n^{4}}\).

At this point, we should make a choice of the basis for particle states. While all bases give the same physics result, it is sometimes convenient to work with self-conjugate states, so that \(\bar{l}=l\) and \(\bar{j}=j\). This has the advantage that one essentially works with real quantities. On the other hand, when fermion states are present, or if particles live in complex representations of some internal symmetries, it is more natural to work with complex fields. In this work, for scalars and gauge bosons we will work with self-conjugate states, while for fermions we will work with helicity basis.

\subsection{Scalars and vectors}
\label{sec3.3}
For self-conjugate fields, we further split the real and imaginary part of \(\mathbf{M}_{i j \rightarrow X}\):
\begin{equation}
\mathbf{M}_{i j \rightarrow X} \rightarrow m_{R}^{i j}+i m_{I}^{i j}
\end{equation}
with this, we write the amplitude \(\mathcal{M}^{i j k l}\) as
\begin{align}
\mathcal{M}^{i j k l} &=\sum_{\alpha}\left(m_{\alpha, R}^{i j}+i m_{\alpha, I}^{i j}\right)\left(m_{\alpha, R}^{k l}-i m_{\alpha, I}^{k l}\right)  \\
&=\sum_{\alpha}\left(m_{\alpha, R}^{i j} m_{\alpha, R}^{k l}+m_{\alpha, I}^{i j} m_{\alpha, I}^{k l}\right)-i \sum_{\alpha}\left(m_{\alpha, R}^{i j} m_{\alpha, I}^{k l}-m_{\alpha, I}^{i j} m_{\alpha, R}^{k l}\right)
\end{align}
where \(\alpha\) labels all intermediate states, and the positive factors from Eq.~(\ref{eq:3.9}) are absorbed to \(m_{\alpha}^{i j}\). If the amplitude is time reversal invariant, the second term actually vanishes by invoking the \(j \leftrightarrow l\) crossing symmetry (i.e., adding the \(j \leftrightarrow l\) term in Eq.~(\ref{eq:3.9})). With the first term, we can define the set of allowed values of \(\mathcal{M}^{i j k l}\) by
\begin{equation}
\mathbf{C}^{n^{4}}=\operatorname{cone}\left(\left\{m^{i j} m^{k l}+m^{i l} m^{k j}\right\}\right)\label{eq:3.16}
\end{equation}
This is similar to Eq.~(\ref{eq:3.16}), but has the advantage that one essentially only deals with real quantities.

The possible values of \(m^{i j}\) are further restricted by symmetries of the system. For example, discrete symmetries, such as parity, could directly impose constraints on certain elements of \(m\), depending on the parity of the intermediate state \(X\). Continuous symmetries, such as gauge symmetries, could further group several \(X\) states to form a multiplet, and in this case the \(m^{i j} m^{k l}+m^{i l} m^{k j}\) term should be understood as an inner product:
\begin{equation}
\sum_{\alpha}\left(m_{\alpha}^{i j} m_{\alpha}^{k l}+m_{\alpha}^{i l} m_{\alpha}^{k j}\right) \Rightarrow \mathbf{m}^{i j} \cdot \mathbf{m}^{k l}+\mathbf{m}^{i l} \cdot \mathbf{m}^{k j}
\end{equation}
where \(\alpha\) here labels different states in the multiplet. In this case \(\mathbf{m}\) can often be fixed as the CG coefficients. These will be illustrated in Section 4.

For vector bosons, instead of helicity states, in this work we work with linearly polarized states, so that each vector \(V\) is described by two real fields, \(V_{x}\) and \(V_{y}\), which are connected by an \(S O(2)\) rotational symmetry around the ``beam direction'' (because we only consider forward scattering). These fields can then be treated as two scalars charged under some internal \(S O(2)\) group, with a small difference related to parity violation \cite{Li:2021cjv}, which will be discussed in Section 4.2.1. Vector bosons could also be dealt with in the helicity basis, see discussions in Ref.~\cite{Trott:2020ebl}.

Another important symmetry we shall consider is the simultaneous exchange \(i \leftrightarrow j\). It carries the information from the spin of the intermediate particle \(X\), which we did not use so far. For the scalar case, a spin \(J\) state \(X\) couples to two scalars in the following form~\cite{Arkani-Hamed:2017jhn}
\begin{equation}
m^{i j} \propto g_{i j}[12]^{J}\langle 1 \mathbf{X}\rangle^{J}\langle 2 \mathbf{X}\rangle^{J}
\end{equation}
where \(g_{i j}\) is the coupling constant. In the forward limit, this amplitude simply reduces to a scalar function of energy, and the only information we would need is \(m^{i j} \propto g_{i j}\). However, the above amplitude must be symmetric under \(p_{1} \leftrightarrow p_{2}\) and \(i \leftrightarrow j\), which means
\begin{equation}
g_{i j}=(-)^{J} g_{j i}
\end{equation}
i.e. \(m\) needs to be either symmetric or anti-symmetric.

This symmetry, at the level of the 2-to-2 amplitude \(i j \rightarrow k l\), is reflected by the fact that \(i \leftrightarrow j\) and \(k \leftrightarrow l\) is a symmetry for \(\mathcal{M}^{i j k l}\). For scalar particles, this is equivalent to a rotation of \(\pi\) around the \(y\) axis (perpendicular to the beam axis). The situation can be slightly different for vectors: with linearly polarized states, this double exchange corresponds to parity transformation, as one has to flip the polarization along the \(x\) direction after the rotation around \(y\)-axis. If parity is conserved, we have the same requirement as the scalar case, i.e. \(m^{i j}\) is either symmetric or anti-symmetric, and \(\mathcal{M}\) is invariant under this double exchange; if parity is violated, transitions between symmetric and anti-symmetric states are allowed, and the double exchange is not a symmetry anymore. We will illustrate this point in Section 4.2.1, taking photon-photon scattering as an example.

To sum up, for self-conjugate fields, we construct the positivity cone for the allowed \(\mathcal{M}^{i j k l}\) by
\begin{equation}
\mathbf{C}=\operatorname{cone}\left(\left\{m^{i j} m^{k l}+m^{i l} m^{k j} \mid m^{i j}=\pm m^{j i}, m \in \mathbb{R}^{n \times n}\right\}\right) \label{eq:3.20}
\end{equation}
where \(n\) is the number of particle modes consider in the problem. The \(m^{i j}=\pm m^{j i}\) requirement may be dropped for parity violating vector interactions. When continuous symmetries are present, \(m^{i j} m^{k l}+m^{i l} m^{k j}\) should be interpreted as \(\mathbf{m}^{i j} \cdot \mathbf{m}^{k l}+\mathbf{m}^{i l} \cdot \mathbf{m}^{k j}\).

\subsection{Fermions}
\label{sec3.4}
For SM fermions, we are going to work with the helicity basis, where states are not self-conjugate. Consider \(m^{i j}\) where \(i, j\) are both right handed. We have
\begin{equation}
m^{i j} \propto g_{i j}[12]^{J+1}\langle 1 \mathbf{X}\rangle^{J}\langle 2 \mathbf{X}\rangle^{J}\label{eq:3.21}
\end{equation}
This leads to \(m^{i j}=(-)^{J} m^{j i}\) similar to the scalar case. Similar conclusion holds for left-handed fermions, or simply \(\bar{i}, \bar{j}\).

If an \(m^{i j}\) is generated by an intermediate state \(X\) that couples to \(i\) and \(j\), we expect an additional contribution generated by the CP conjugate of this coupling, \(\bar{X}\) to \(\bar{i}\) and \(\bar{j}\). To take this into account, We find it convenient to simply invoke the crossing symmetry under \(i \leftrightarrow k\) and write
\begin{equation}
\begin{aligned}
\mathcal{M}^{i j k l} &=\frac{1}{2}\left(\mathcal{M}^{i j k l}+\mathcal{M}^{\bar{k} j \bar{i} l}\right) \\
&=\frac{1}{2} \sum_{\alpha}\left(m_{\alpha}^{i j} m_{\alpha}^{* k l}+m_{\alpha}^{i \bar{l}} m_{\alpha}^{* k \bar{j}}+m_{\alpha}^{\bar{k} j} m_{\alpha}^{* \bar{i} l}+m_{\alpha}^{\bar{k} \bar{l}} m_{\alpha}^{*} {}^{\bar{i} \bar{j}}\right)
\end{aligned}
\end{equation}
This essentially combines the contributions of \(X\) and \(\bar{X}\), and has the advantage of making crossing symmetry manifest. CP-violation may occur if \(X\) couples to \(i j\) and \(\overline{i j}\) simultaneously. If \(\mathrm{CP}\) is conserved, \(\mathcal{M}\) is real-analytic and therefore \(\mathcal{M}^{i j k l}=\mathcal{M}^{k l i j} .\) We may conversely use this condition to construct the \(\mathcal{M}\) for CP-conserving theories, and this is often more convenient than imposing CP-conservation for each \(m_{\alpha}\). Examples will be given in Section 4.3.

Consider now \(m^{i \bar{j}}, m^{\bar{i} j}\) where \(i, j\) and \(\bar{i}, \bar{j}\) are right- and left-handed respectively. We have
\begin{align}
&m^{i \bar{j}}=\mathbf{M}_{i j \rightarrow X^{+}} \propto g_{i j}[12]^{J}\langle 1 \mathbf{X}\rangle^{J-1}\langle 2 \mathbf{X}\rangle^{J+1} \\
&m^{\bar{i} j}=\mathbf{M}_{i j \rightarrow X^{-}} \propto g_{i j}^{\prime}[12]^{J}\langle 1 \mathbf{X}\rangle^{J+1}\langle 2 \mathbf{X}\rangle^{J-1}
\end{align}
These two amplitudes are connected by a rotational symmetry around the \(y\)-axis, under which
\begin{equation}
m^{i \bar{j}} \Rightarrow-m^{\overline{j} i}=\eta m^{i \bar{j}}
\end{equation}
where \(\eta\) must be a pure phase. This gives \(g_{i j}=\eta g_{j i}^{\prime}.\) Alternatively, it is more convenient to take into account the contribution from the \(m^{i j}\) term by imposing the double exchange symmetry, \(i \leftrightarrow j, k \leftrightarrow l\), at the \(\mathcal{M}^{i j k l}\) level. Similar to the scalar case, the symmetry corresponds to a rotation around the \(y\)-axis. For the intermediate state with \(J_{3}=0\) angular momentum along the \(z\)-axis, this forces \(m^{i j}\) to be either symmetric or anti-symmetric; for the \(J_{3}=+1\) state, it automatically combines the \(J_{3}=-1\) contribution. We will give more details in Section 4.3.

To sum up, in the helicity basis we take advantage of full crossing symmetries of \(\mathcal{M}^{i j k l}\):
\begin{equation}
\mathbf{C}=\text{cone}\left(\left\{m^{i j} m^{* k l}+m^{i \bar{l}} m^{* k \bar{j}}+m^{\bar{k} j} m^{* \bar{i} l}+m^{\bar{k} \bar{l}} m^{* \bar{i}\bar{j}}+(i \leftrightarrow j, k \leftrightarrow l) \mid m \in \mathbb{C}^{2 n \times 2 n}\right\}\right) \label{eq:3.26}
\end{equation}
Again, when continuous symmetries are present, \(m^{i j} m^{* k l}+m^{i \bar{l}} m^{* k \bar{j}}+m^{\bar{k} j} m^{* \overline{i }l}+m^{\bar{k}\bar{l}} m^{*\bar {i} \bar{j} }\) should be interpreted as \(\mathbf{m}^{i j} \cdot \mathbf{m}^{* k l}+\mathbf{m}^{i \bar{l}} \cdot \mathbf{m}^{* k \bar{j}}+\mathbf{m}^{\bar{k} j} \cdot \mathbf{m}^{* \overline{i}l}+\mathbf{m}^{\bar{k} \bar{l}} \cdot \mathbf{m}^{* \bar{i} \bar{j}}\).

\subsection{Generating the coefficient space}
\label{sec3.5}
We are now ready to determine positivity cone from the generation point of view. For this purpose, it is convenient to define the ``generators'' of the positivity cone. Later we will see that they play a crucial role in connecting positivity bounds with the inverse problem.

A generator is any rank-4 tensor structure that could potentially appear in the integrand of the dispersion relation and is allowed by the symmetries of the theory. We define them following our master equations. For Eq.~(\ref{eq:3.20}), we define
\begin{equation}
\mathcal{G}^{i j k l} \equiv m^{i j} m^{k l}+m^{i l} m^{k j}, \quad m^{i j}=\pm m^{j i}, m \in \mathbb{R}^{n \times n} \label{eq:3.27}
\end{equation}
while for Eq.~(\ref{eq:3.26}), we define
\begin{equation}
\mathcal{G}^{i j k l} \equiv m^{i j} m^{* k l}+m^{i \bar{l}} m^{* k \bar{j}}+m^{\bar{k} j} m^{* \bar{i }l}+m^{\bar{k} \bar{l}} m^{* \bar{i} \bar{j}}+(i \leftrightarrow j, k \leftrightarrow l), \quad m \in \mathbb{C}^{2 n \times 2 n} \label{eq:3.28}
\end{equation}
The \(m\) matrices are restricted by the symmetries of the theory. Note that our definition is up to an arbitrary overall factor, which plays no role in the generation of the cone \(\mathbf{C}\). This reflects the fact that the scale of the BSM physics is unknown and unrestricted. In the rest of the paper, equations for \(\mathcal{G}\) or \(m\) are to be interpreted as valid only up to an overall factor, unless otherwise specified.

With this definition, the positivity cone is positively generated from \(\mathcal{G}^{i j k l}\):
\begin{equation}
\mathbf{C}=\operatorname{cone}(\{\mathcal{G}\})
\end{equation}
By enumerating all possible \(m\) allowed by the symmetry of the theory, the \(\mathbf{C}\) cone can be constructed.

While it is possible to directly proceed in the space of \(\mathcal{M}^{i j k l}\), it is often more convenient to map \(\mathcal{M}\) to the Wilson coefficient space, to facilitate a comparison with experimental measurements, and results from global fits, etc. Doing so requires an expression of \(\mathcal{M}^{i j k l}\) as a function of operator coefficients. At the tree level, this expression is linear. One can write
\begin{equation}
\mathcal{M}^{i j k l}\left(C_{1}, C_{2}, \ldots\right)=\sum_{a} C_{a} \mathcal{M}_{a}^{i j k l}
\end{equation}
where \(C_{a}=C_{1}, C_{2}, \cdots\) are either dim-8 Wilson coefficients, or products of two dim-6 Wilson coefficients. This allows \(\mathcal{G}^{i j k l}\) to be mapped to a coefficient vector \(\vec{g}\) :
\begin{equation}
\vec{g} \equiv\left(C_{1}, C_{2}, \ldots\right), \quad \text { if } \mathcal{G}^{i j k l}=\mathcal{M}^{i j k l}\left(C_{1}, C_{2}, \ldots\right)
\end{equation}
which should be interpreted as the generator vector in the space of dim-8 coefficients. These vectors are exactly the edge vectors in our toy example. The positivity cone, when define directly by the Wilson coefficient, is simply the conical hull of all the \(\vec{g}\)'s, and we may write
\begin{equation}
\mathbf{C}=\operatorname{cone}(\{\vec{g}\})
\end{equation}
Beyond the tree-level, \(\mathcal{M}^{i j k l}\left(C_{1}, C_{2}, \ldots\right)\) can become more complicated, but a similar mapping is always possible. For the rest of the paper, we will only use a tree-level mapping, while keeping in mind that this can always be improved, once the higher-order expression of \(\mathcal{M}^{i j k l}\left(C_{1}, C_{2}, \ldots\right)\) becomes available.

\subsection{Salient cone and extremal rays}
\label{sec3.6}
An important feature of the \(\mathbf{C}\) cone is that it is always salient as predicted by the dispersion relation. A salient cone is a convex cone that does not contain a straight line. Most cones we intuitively think of are salient. Examples of non-salient cones are the entire space of \(\mathbb{R}^{n^{4}}\), its subspaces, or half spaces, etc. This feature is going to play an important role in the inverse problem. It also represents the key difference between dim-6 and dim-8 coefficient space: if we define a cone for the former in a similar way, it is not salient.

To see \(\mathbf{C}\) is salient, simply notice that all generators \(\mathcal{G}^{i j k l}\) has a strictly positive projection on the rank-4 tensor \(\delta^{i k} \delta^{j l} .\) This is easy to check with Eqs.~(\ref{eq:3.27}) and ~(\ref{eq:3.28}). Therefore all nonzero \(\mathcal{M}^{i j k l}\) in \(\mathbf{C}\) must have a positive projection on \(\delta^{i k} \delta^{j l}\), which means \(-\mathcal{M}^{i j k l} \notin \mathbf{C}\). The salient nature of \(\mathbf{C}\) can be traced back to the \(+\) sign between the two terms on the r.h.s.~of the dispersion relation. At the dim-6 level, this sign is negative. Note that \(\mathbf{C}\) being salient guarantees that \(\mathcal{M}\) is constrained in all possible directions, which is a stronger statement than simply the existence of positivity bounds. In Section 5, we will see that the salient nature of \(\mathbf{C}\) leads to very interesting physical consequences.

Once we prove \(\mathbf{C}\) is salient, the Krein-Milman theorem immediately implies that \(\mathbf{C}=\)cone\((\operatorname{ext} \mathbf{C})\), i.e. the ERs of \(\mathbf{C}\) exist, and the entire cone can be generated by positively combining these ERs. The ERs are obviously a subset of \(\{\vec{g}\}\). Let us call them \(\vec{e} \). \(\mathbf{C}\) can be written as
\begin{equation}
\mathbf{C}=\operatorname{cone}(\{\vec{e}\})
\end{equation}
In the SMEFT, when considering operators involving only one (multiplet) particle, the number of \(\mathcal{G}\) is always finite and can be enumerated using group theory. In this case, \(\{\vec{e}\}\) can be determined by directly solving the convex hull of all the \(\vec{g}\)'s.

In the toy example, we have seen how the extremality leads to uniquely determined UV particle content from an EFT on an ER. In Section 5 we will present a more detailed discussion about the role of the ERs in the inverse problem.

\subsection{Finding bounds}
\label{sec3.7}
Once \(\mathbf{C}\) is determined, we need to find the exact positivity bounds, i.e. the boundary of \(\mathbf{C}\). If the number of ERs is finite (i.e. for operators involving only one SM particle multiplet, or the ``self-quartic'' operators), finding the bounds of \(\mathbf{C}\) from all \(\{\vec{e}\}\) is a vertex enumeration (VE) problem. A VE is a classical problem which asks how to determine the vertices of a polytope by knowing its facets. This is equivalent to its own reverse: the determination of the facets from the vertices. This problem can be efficiently solved by existing algorithms  \cite{Avis,lrs}. In Section 4, we will derive bounds for a number of SM and non-SM examples, by first finding all \(\vec{e}\) and then performing a VE. This approach is referred to as the extremal positivity approach, as the bounds are found by first determining the ERs.

The same approach can be applied to cases where more than one SM particle species are involved, but the number of generators and ERs may become infinity. Taking a \(Z^{\prime}\) that couples to SM \(u_{R} \bar{u}_{R}\) and \(d_{R} \bar{d}_{R}\) as an example. The \(Z^{\prime} u_{R} \bar{u}_{R}\) and \(Z^{\prime} d_{R} \bar{d}_{R}\) couplings are individually fixed by the \(\mathrm{SU}(3)\) symmetry of the SM, but their relative coupling strength remains a free real number, and will enter the corresponding generator \(\mathcal{G}\). In general, a generator in this case is a quadratic function of several free parameters, \(r_{1}, r_{2}, \cdots\), and we have
\begin{equation}
\mathbf{C}=\operatorname{cone}\left(\left\{\vec{g}\left(r_{1}, r_{2}, \ldots\right)\right\}\right)
\end{equation}
Cones like this will have a curved boundary.

If there are not many \(r\) parameters, it is possible to derive exact expressions for the curved boundary. Examples will be given in Sections 4.1.3, 4.2.1, 4.3.2, and 4.3.6. For more complicated cases, a possible solution is to turn the problem into a programming. Consider a given coefficient vector \(\vec{C}=\left(C_{1}, C_{2}, \cdots\right)\), we want to know if it satisfies all bounds, or equivalently if \(\vec{C} \in \mathbf{C}=\) cone \(\left(\left\{\vec{g}\left(r_{1}, r_{2}, \cdots\right)\right\}\right) .\) This is equivalent to asking if one can find a hyperplane \(H\) that separates \(\vec{C}\) and \(\mathbf{C}\). Let \(\vec{n}\) be the normal vector of \(H .\) A separation is achieved if
\begin{align}
&\vec{n} \cdot \vec{g}\left(r_{1}, r_{2}, \ldots\right) \geq 0 \quad \forall r_{i} \\
&\vec{n} \cdot \vec{C}<0
\end{align}
This allows us to search for a separating hyperplane by the following programming:
\begin{align}
&\min \quad \vec{n} \cdot \vec{C} \\
&\text { subject to }\quad \vec{n} \cdot \vec{g}\left(r_{1}, r_{2}, \ldots\right) \geq 0 \quad \forall r_{i}
\end{align}
Since \(\vec{g}\) is a quadratic function of \(r_{i}\), this is essentially a polynomial matrix programming, and can be turned into a semi-definite programming (SDP) \cite{Simmons-Duffin:2015qma}. If the minimum is found to be negative, we know that \(\vec{C}\) is not contained in \(\mathbf{C}\).

Alternatively, we may directly formulate the problem as a SDP, by realizing that the dual cone of \(\mathbf{C}\) is a spectra-hedron, see the approach proposed in Ref.~\cite{Li:2021cjv}. This SDP is set up in a way independent of specific EFTs, and can be conveniently applied to a wide range of theories. However, though numerically efficient, this approach is formulated without specifying the generators, and so its connection to the UV theories is lost. Since the main purpose here is to address the inverse problem, we will not use the SDP approach in this work. Nevertheless, for complicated problems, this should be regarded as a backup option to numerically compute the exact boundary.

\section{The extremal positivity bounds}
\label{sec4}

In this section, we will illustrate the extremal positivity approach \cite{Zhang:2020jyn} with a series of examples. These examples are chosen to cover different aspects of this approach. They include scalar, vector, and fermion operators; cases with and without continuous symmetries; \(\mathrm{P}\) and CP violation operators; toy EFTs and SMEFT examples. Then in Section 4.4 we will present the collection of full positivity bounds for all SM parity-conserving self-quartic operators.

\subsection{Scalar}
\label{sec4.1}
We start with a simple EFT of two real scalars, \(\phi_{1}\) and \(\phi_{2}\) that is restricted with various discrete or continuous symmetries. First, let us define the operators. There are six independent ones at dim-8, which we simply denote by \(O_{1}, O_{2}, \cdots, O_{6}\) :
\begin{align}
&O_{i j k l} \equiv\left(\partial_{\mu} \phi_{i} \partial^{\mu} \phi_{j}\right)\left(\partial_{\nu} \phi_{k} \partial^{\nu} \phi_{l}\right) \\
&O_{1}=O_{1111}, \quad O_{2}=O_{1122}, \quad O_{3}=O_{2222}, \\
&O_{4}=O_{1212}, \quad O_{5}=O_{1112}, \quad O_{6}=O_{1222},
\end{align}
The \(\mathcal{M}^{i j k l}\) matrix can be straightforwardly computed at the tree level, in terms of the corresponding operator coefficients \(C_{1}, C_{2}, \cdots, C_{6}\)
\begin{flalign}
	\nonumber\raisebox{-8pt}{$\mathcal{M}^{ijkl}=$ }
\begin{tabular}{ r|c|c|c|c| }
	\multicolumn{1}{r}{}
	 &  \multicolumn{1}{c}{$\phi_1 \phi_1$}
	 & \multicolumn{1}{c}{$\phi_2 \phi_2$}
     &  \multicolumn{1}{c}{$\phi_1 \phi_2$}
	 & \multicolumn{1}{c}{$\phi_2 \phi_1$}  \\
	\cline{2-5}
	$\phi_1 \phi_1$ & $4C_1$ & $\bar{C}_2$&$C_5$&$C_5$ \\
	\cline{2-5}
	$\phi_2 \phi_2$ & $\bar{C}_2$&$4C_3$&$C_6$& $C_6$\\
	\cline{2-5}
    $\phi_1 \phi_2$ & $C_5$ & $C_6$&$C_4$&$\bar{C}_2$ \\
	\cline{2-5}
	$\phi_2 \phi_1$ & $C_5$&$C_6$&$\bar{C}_2$&$C_4$\\
	\cline{2-5}
\end{tabular}
\end{flalign}
where \(\bar{C}_{2} \equiv C_{2}+\frac{1}{2} C_{4}\). Here the rows correspond to \((i, j)=\left(\phi_{1}, \phi_{1}\right),\left(\phi_{2}, \phi_{2}\right),\left(\phi_{1}, \phi_{2}\right),\left(\phi_{2}, \phi_{1}\right)\), respectively, while the columns correspond to \((k, l)\) values in a similar way, as labeled explicitly above and to the left of the matrix. For the rest of the paper, we will always write \(\mathcal{M}\) or \(\mathcal{G}\) in this form. We will often omit the \(i j\) and \(k l\) labels, if they have been shown already.

\subsubsection{Two scalars with $SO(2)$}
\label{sec4.1.1}
In our first example, consider the case in which two scalars are connected by an \(\mathrm{SO}(2)\) symmetry. They are equivalent to a complex scalar \(\phi \equiv \phi_{1}+i \phi_{2}\) which carries some U(1) charge. One may write two independent operators in terms of a complex scalar at dim-8
\begin{align}
&\mathcal{L}=\sum_{i=1,2} \frac{C_{i}^{\prime}}{\Lambda^{4}} O_{i}^{\prime} \\
&O_{1}^{\prime}=\left|\partial_{\mu} \phi \partial^{\mu} \phi\right|^{2}, \quad O_{2}^{\prime}=\left|\partial_{\mu} \phi^{\dagger} \partial^{\mu} \phi\right|^{2}
\end{align}
The coefficients \(C_{1}, \ldots, C_{6}\) can be written in terms of the coefficients of the above two operators,
\begin{equation}
C_{1}=C_{3}=C_{1}^{\prime}+C_{2}^{\prime}, \quad C_{2}=-2\left(C_{1}^{\prime}-C_{2}^{\prime}\right), \quad C_{4}=4 C_{1}^{\prime}, \quad C_{5}=C_{6}=0
\end{equation}
The amplitude is
\begin{flalign}
	\nonumber\raisebox{-3pt}{$\mathcal{M}^{ijkl}=$ }
\begin{tabular}{|c|c|c|c|}
	\hline $4(C_{1}^{\prime}+C_{2}^{\prime})$&$2C_{2}^{\prime}$&0&0\\
	\hline $2C_{2}^{\prime}$&$4(C_{1}^{\prime}+C_{2}^{\prime})$&0&0\\
	\hline 0&0&$4C_{1}^{\prime}$&$2C_{2}^{\prime}$\\
    \hline 0&0&$2C_{2}^{\prime}$&$4C_{1}^{\prime}$\\
	\hline
\end{tabular}
\end{flalign}
In this section we will work with \(C_{1}^{\prime}\) and \(C_{2}^{\prime}\).

To construct the generators of the allowed parameter space, we make use of the \(\mathrm{SO}(2)\) symmetry. The incoming particles are charged under the \(\mathbf{2}\) irreps, and we have \(\mathbf{2} \otimes \mathbf{2}=\) \(\mathbf{1}_{S} \oplus \mathbf{1}_{A} \oplus \mathbf{2} .\) The subscripts \(_S\) and \(_{A}\) indicate the exchange symmetry under \(\phi_{1} \leftrightarrow \phi_{2}\). The intermediate states can be classified as living in \(\mathbf{1}_{S}, \mathbf{1}_{A}\), and \(\mathbf{2}\) irreps. The corresponding \(m\) matrices are simply the Clebsch-Gordan (CG) coefficients:
\begin{flalign}
	\raisebox{0pt}{$m_{\textbf{1}_\mathrm{\textbf{S}}}=$ }
\begin{tabular}{ r|c|c| }
	\multicolumn{1}{r}{}
	 &  \multicolumn{1}{c}{$\phi_1$}
	 & \multicolumn{1}{c}{$\phi_2$} \\
	\cline{2-3}
	$\phi_1$ & 1 & 0 \\
	\cline{2-3}
	$\phi_2$ & 0 & 1\\
	\cline{2-3}
\end{tabular}
,\quad
\raisebox{0pt}{$m_{\textbf{1}_\mathrm{\textbf{A}}}=$ }
\begin{tabular}{ r|c|c| }
	\multicolumn{1}{r}{}
	 &  \multicolumn{1}{c}{$\phi_1$}
	 & \multicolumn{1}{c}{$\phi_2$} \\
	\cline{2-3}
	$\phi_1$ & 0 & 1 \\
	\cline{2-3}
	$\phi_2$ & $-1$ & 0\\
	\cline{2-3}
\end{tabular}
\end{flalign}
\begin{align}
m_\textbf{2}^{\alpha}=\left(
\begin{tabular}{ r|c|c| }
	\multicolumn{1}{r}{}
	 &  \multicolumn{1}{c}{$\phi_1$}
	 & \multicolumn{1}{c}{$\phi_2$} \\
	\cline{2-3}
	$\phi_1$ & 1 & 0 \\
	\cline{2-3}
	$\phi_2$ & 0 & $-1$\\
	\cline{2-3}
\end{tabular}
,\quad
\begin{tabular}{ r|c|c| }
	\multicolumn{1}{r}{}
	 &  \multicolumn{1}{c}{$\phi_1$}
	 & \multicolumn{1}{c}{$\phi_2$} \\
	\cline{2-3}
	$\phi_1$ & 0 & 1 \\
	\cline{2-3}
	$\phi_2$ & 1 & 0\\
	\cline{2-3}
\end{tabular}\right)
\end{align}
where the two rows/columns correspond to \(i, j\) being \(\phi_{1}\) and \(\phi_{2}\) respectively. In the following we are going to omit the \(i, j\) labels in the \(m\) matrices.

\begin{figure}[h]
	\begin{center}
		\includegraphics[width=.5\linewidth]{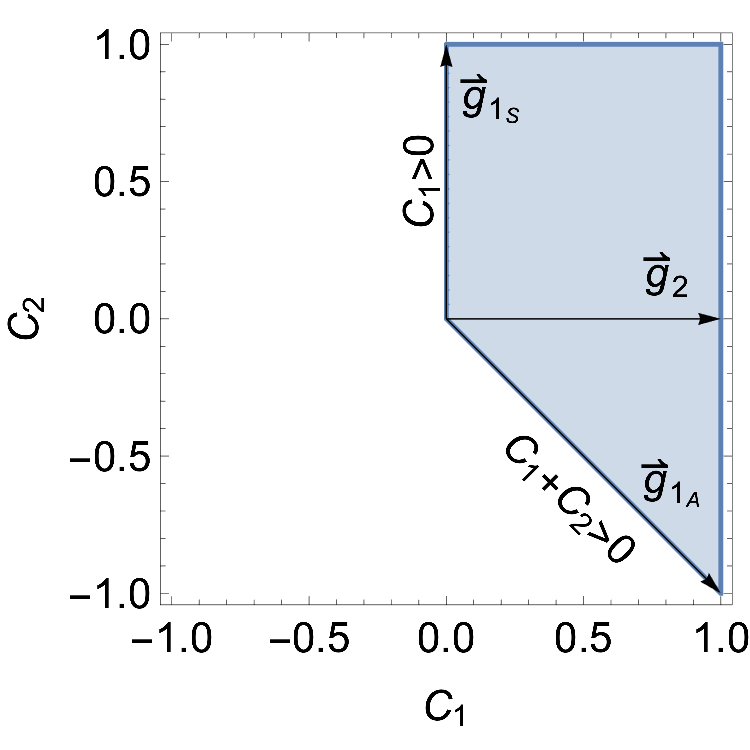}
	\end{center}
	\caption{Positivity cone for an EFT with 2-scalars under \(\mathrm{SO}(2)\). The arrows represent the generator vectors.}
	\label{fig3}
\end{figure}

The generators can be computed using Eq.~(\ref{eq:3.27}). In this simple case, they are the \(\mathrm{SO}(2)\) projector operators with the 2nd and the 4th indices symmetrized. We have
\begin{align}
\nonumber G_{\textbf{1}_{S}}^{ijkl}&=
\begin{tabular}{|c|c|c|c|}
	\hline $2$&$1$&0&0\\
	\hline $1$&$2$&0&0\\
	\hline 0&0&$0$&$1$\\
    \hline 0&0&$1$&$0$\\
	\hline
\end{tabular}
\propto P_{\textbf{1}_{S}}^{i(j|k|l)},\quad G_{\textbf{1}_{A}}^{ijkl}=
\begin{tabular}{|c|c|c|c|}
	\hline $0$&$-1$&0&0\\
	\hline $-1$&$0$&0&0\\
	\hline 0&0&$2$&$-1$\\
    \hline 0&0&$-1$&$2$\\
	\hline
\end{tabular}
\propto P_{\textbf{1}_{A}}^{i(j|k|l)},\\
\nonumber G_{\textbf{2}}^{ijkl}&=
\begin{tabular}{|c|c|c|c|}
	\hline $1$&$0$&0&0\\
	\hline $0$&$1$&0&0\\
	\hline 0&0&$1$&$0$\\
    \hline 0&0&$0$&$1$\\
	\hline
\end{tabular}\propto P_{\textbf{2}}^{i(j|k|l)}
\end{align}
where \(P_{\textbf{r}}^{i j k l}\) are the \(\mathrm{SO}(2)\) projector operators, see Appendix A, Eq.~(\ref{eq:A.1}). Comparing with the \(\mathcal{M}^{i j k l}\), we can express the generators in terms of coefficients \(\vec{g}=\left(C_{1}^{\prime}, C_{2}^{\prime}\right)\). This gives
\begin{equation}
\vec{g}_{\textbf{1} S}=(0,1), \quad \vec{g}_{\textbf{1} A}=(1,-1), \quad \vec{g}_{\textbf{2}}=(1,0).\label{eq:4.9}
\end{equation}
A vertex enumeration gives the bounds:
\begin{equation}
C_{1}^{\prime} \geq 0, \quad C_{1}^{\prime}+C_{2}^{\prime} \geq 0
\end{equation}
The same result can also be obtained from elastic channels, \(\phi_{1} \phi_{1} \rightarrow \phi_{1} \phi_{1}\) and \(\phi_{1} \phi_{2} \rightarrow \phi_{1} \phi_{2}\). The allowed positivity cone is shown in Figure \ref{fig3}.

This simple example illustrates how symmetries of the EFT can be used to enumerate \(m\), which in turn determines all \(\vec{g}\)'s. More generally, if the intermediate state \(X\) in the dispersion relation lives in a irrep \(\mathbf{r}\), then the Wigner-Eckart theorem dictates that \(M(i j \rightarrow\) \(\left.X^{\alpha}\right)\) can be written as \(\langle X|\mathcal{M}| \mathbf{r}\rangle C_{i, j}^{\mathbf{r}, \alpha}\), where \(\alpha\) labels the states of \(\mathbf{r}\) and \(C_{i, j}^{\mathbf{r}, \alpha}\) is the CG coefficients for the direct sum decomposition of \(\mathbf{r}_{i} \otimes \mathbf{r}_{j}\), with \(\mathbf{r}_{i}\left(\mathbf{r}_{j}\right)\) the irrep of \(i(j)\). Since we define the generator up to normalization, we simply need \(m^{i j} \propto C_{i, j}^{\mathbf{r}, \alpha}\), and therefore in a self-conjugate basis, we have

\begin{equation}
\mathcal{G}^{i j k l}=P_{\mathbf{r}}^{i(j|k| l)}
\end{equation}
In practice, one can simply find all \(\vec{g}\) vectors from the projectors and construct \(\mathbf{C}\), without using \(m\) and \(\mathcal{G}\). We have nevertheless presented the explicit forms of \(m\) and \(\mathcal{G}\), to facilitate comparisons with different examples in the next few sections.

We have already found all positivity bounds in this simple EFT. However, to allow a discussion of the inverse problem, we want to understand how the generators are related to the UV completions. For now we will only do this at the tree-level. As we have seen in Section 2, if the generator is an ER, the only possible UV completion is the one-particle UV completions.

Below we show the UV particles in the 3 different irreps:
\begin{equation}
\begin{array}{cccccc}
 \text { State} & \text {Spin}  &\text {Charge}  & \text { Interaction } & \mathrm{ER} & \vec{c} \\
\hline S_{1} & 0 & 0 & g_{1} S_{1} M_{1} \phi^{\dagger} \phi & \text{\Checkmark} & (0,2) \\
V & 1 & 0 & g_{V} V^{\mu} \phi^{\dagger} \stackrel{\leftrightarrow}{D}_{\mu} \phi & \text{\Checkmark}  & (2,-2) \\
S_{2} & 0 & 2 & g_{2} M_{2} S_{2}^{\dagger} \phi^{2} & \text{\XSolidBrush} & (4,0)\label{eq:4.12}
\end{array}
\end{equation}
Here the three states in this table correspond to \(\mathbf{1}_{S}, \mathbf{1}_{A}\), and \(\mathbf{2}\) irreps, respectively. The ``Charge'' columns shows the charge of the heavy state in the unit of the charge of \(\phi\); the ``Interaction'' columns shows how this state interacts with the light states \(\phi\). This interaction generates the corresponding \(m^{i j}\) matrices. The \(g\) and \(M\) are couplings and masses. When integrating out the heavy state, the resulting coefficient vector \(\vec{C}=\frac{g^{2}}{M^{4}} \vec{c}\), and \(\vec{c}\) is shown in the last column. These \(\vec{c}\) vectors are indeed proportional to the corresponding generators \(\vec{g}\) in Eq.~(\ref{eq:4.9}). A check mark in the ``ER'' column indicates that the corresponding generator is extremal in the positivity cone. In this paper, we will frequently use tables of this format to present the mapping between generators and UV particles.

The \(\mathbf{2}\) irrep is not extremal, while the other two irreps are. Later, when discussing the inverse problem, we will see that the consequence of this is that a theory with a \(S_{1}\) or a \(V\) particle can be uniquely confirmed with low energy measurements up to dim-8, while those with a \(S_{2}\) particle cannot be (as it could also be explained by combining \(S_{1}\) and \(V\) particles: \(\vec{g}_{\textbf{2}}\) stays between \(\vec{g}_{\textbf{1} S}\) and \(\left.\vec{g}_{\textbf{1} A} .\right)\) Also note that the exchange symmetry of each irrep determines the spin of the state: with Bose symmetry and no additional wave function, the antisymmetric coupling to scalars can only be realized by a spin-1 state.

\subsubsection{Two scalars with discrete symmetries}
\label{sec4.1.2}
Let us now relax the symmetry constraint by replacing the \(\mathrm{SO}(2)\) by a pair of discrete symmetries: \(\phi_{1} \rightarrow-\phi_{1}\) and \(\phi_{1} \leftrightarrow \phi_{2}\). This leaves 3 independent coefficients, which we take to be \(\vec{C}=\left(C_{1}, C_{2}, C_{4}\right)\). The other coefficients are \(C_{3}=C_{1}\) and \(C_{5}=C_{6}=0\). This is exactly the toy example we have considered in Section 2. Now instead of using elastic scattering, we will work out the bounds from the generator point of view.

An intermediate state can be now classified by its parity under both symmetries, \(\phi_{1} \rightarrow\) \(-\phi_{1}\) and \(\phi_{1} \leftrightarrow \phi_{2}\). These symmetries completely fix the \(m\) matrices up to normalization:
\[
m_{++}=\begin{array}{|c|c|}
\hline 1 & 0 \\
\hline 0 & 1 \\
\hline
\end{array}, \quad m_{+-}=\begin{array}{|l|c|}
\hline 1 & 0 \\
\hline 0 & -1 \\
\hline
\end{array}, \quad m_{-+}=\begin{array}{|c|c|}
\hline 0 & 1 \\
\hline 1 & 0 \\
\hline
\end{array}, \quad m_{--}=\begin{array}{|c|c|}
\hline 0 & 1 \\
\hline-1 & 0 \\
\hline
\end{array}
\]
Comparing with the previous example, we are simply disconnecting the two components in the \(\mathbf{2}\) representation and treat them as independent generators. Again, using Eq.~(\ref{eq:3.27}), we find four generators:
\[
G_{+\pm}^{i j k l}=\begin{array}{|c|c|c|c|}
\hline 2 & \pm 1 & 0 & 0 \\
\hline \pm 1 & 2 & 0 & 0 \\
\hline 0 & 0 & 0 & \pm 1 \\
\hline 0 & 0 & \pm 1 & 0 \\
\hline
\end{array} \quad G_{-\pm}^{i j k l}=\begin{array}{|c|c|c|c|}
\hline 0 & \pm 1 & 0 & 0 \\
\hline \pm 1 & 0 & 0 & 0 \\
\hline 0 & 0 & 2 & \pm 1 \\
\hline 0 & 0 & \pm 1 & 2 \\
\hline
\end{array}
\]
The corresponding vectors in the Wilson coefficient space are
\begin{equation}
\vec{g}_{++}=(1,2,0), \quad \vec{g}_{+-}=(1,-2,0), \quad \vec{g}_{-+}=(0,0,4), \quad \vec{g}_{--}=(0,-4,4)
\end{equation}
All four vectors are extremal. A VE directly gives the same bounds as Eqs.~(\ref{eq:2.6})-(\ref{eq:2.9}).

Finally, all 4 generators can be mapped to UV completions. The result is listed below, in completely analogy to the states shown in the previous example, Eq.~(\ref{eq:4.12}):

\begin{tabular}{c|cccccc}
Particle & Spin & Parities \(\left(\phi_{1} \rightarrow-\phi_{1}, \phi_{1} \leftrightarrow \phi_{2}\right)\) & Interaction & ER & \(\vec{c}\) &  \\
\hline
\(S_{1}\) & 0 & \(+, \quad+\) & \(g_{1} M_{1}\left(\phi_{1}^{2}+\phi_{2}^{2}\right) S_{1}\) & \text{\Checkmark} & \(2 \times(1,2,0)\) &  \\
\(S_{2}\) & 0 & \(+, \quad-\) & \(g_{2} M_{2}\left(\phi_{1}^{2}-\phi_{2}^{2}\right) S_{2}\) & \text{\Checkmark} & \(2 \times(1,-2,0)\) &  \\
\(S_{3}\) & 0 & \(-, \quad+\) & \(g_{3} M_{3} \phi_{1} \phi_{2} S_{3}\) & \text{\Checkmark} & \(2 \times(0,0,1)\) &  \\
\(V_{4}\) & 1 & \(-, \quad-\) & \(g_{4}(\phi_{1} \stackrel{\leftrightarrow}{D}_{\mu} \phi_{2}) V_{4}^{\mu}\) & \text{\Checkmark} & \(2 \times(0,-1,1)\) \\
\end{tabular}

Again all \(\vec{c}\) s are proportional to the generators. Most information in this table has been already presented in Section 2 .

\subsubsection{Two scalars with continuous ERs}
\label{sec4.1.3}
Let us continue to relax the symmetries. This time we keep only the \(Z_{2}\) symmetry \(\phi_{1} \rightarrow\) \(-\phi_{1} .\) Now all four coefficients \(\vec{C}=\left(C_{1}, C_{2}, C_{3}, C_{4}\right)\) are independent, while the other two vanish: \(C_{5}=C_{6}=0 .\) A new feature in this example is that the allowed parameter space is no longer polyhedral.

The intermediate state can have either \(+\) or \(-\) parity under \(\phi_{1} \rightarrow-\phi_{1}\). In addition, recall that another requirement on \(m\) is that it is either symmetric or anti-symmetric. This gives three possible \(m\)'s:
\[
m_{+}=\begin{array}{|c|c|}
\hline x & 0 \\
\hline 0 & y \\
\hline
\end{array},\quad m_{-S}=\begin{array}{|l|l|}
\hline 0 & 1 \\
\hline 1 & 0 \\
\hline
\end{array}, \quad m_{-A}=\begin{array}{|c|c|}
\hline 0 & 1 \\
\hline-1 & 0 \\
\hline
\end{array}
\]
Comparing with the previous example, we are essentially mixing \(m_{++}\) and \(m_{+-}\), which simply means that transitions between states with different parities under the discarded symmetry \(\phi_{1} \leftrightarrow \phi_{2}\) are now allowed. The generators are
\[
G_{+}^{i j k l}=\begin{array}{|c|c|c|c|}
\hline 2 x^{2} & x y & 0 & 0 \\
\hline x y & 2 y^{2} & 0 & 0 \\
\hline 0 & 0 & 0 & x y \\
\hline 0 & 0 & x y & 0 \\
\hline
\end{array}, \quad G_{-S}^{i j k l}=\begin{array}{|c|c|c|c|}
\hline 0 & 1 & 0 & 0 \\
\hline 1 & 0 & 0 & 0 \\
\hline 0 & 0 & 2 & 1 \\
\hline 0 & 0 & 1 & 2 \\
\hline
\end{array} \quad G_{-A}^{i j k l}=\begin{array}{|c|c|c|c|}
\hline 0 & -1 & 0 & 0 \\
\hline-1 & 0 & 0 & 0 \\
\hline 0 & 0 & 2 & -1 \\
\hline 0 & 0 & -1 & 2 \\
\hline
\end{array}
\]
Again by comparing with \(\mathcal{M}^{i j k l}\) we find the generator vectors:
\begin{equation}
\label{eq4.14addf}
\vec{g}_{+}(x, y)=\left(x^{2}, 2 x y, y^{2}, 0\right), \quad \vec{g}_{-S}=(0,0,0,4), \quad \vec{g}_{-A}=(0,-4,0,4)
\end{equation}
The first vector is a quadratic function of \(x, y \in \mathbb{R}\). Since the normalization does not matter, we may also write \(\vec{g}_{+}(r)=\left(1,2 r, r^{2}, 0\right), r \in \mathbb{R}\). There is an infinite number of generators. As a result, the allowed parameter space is not polyhedral anymore, but it has a curved boundary consisting of all \(\vec{g}_{+}(r)\) vectors.

\begin{figure}[h]
	\begin{center}
		\includegraphics[width=.45\linewidth]{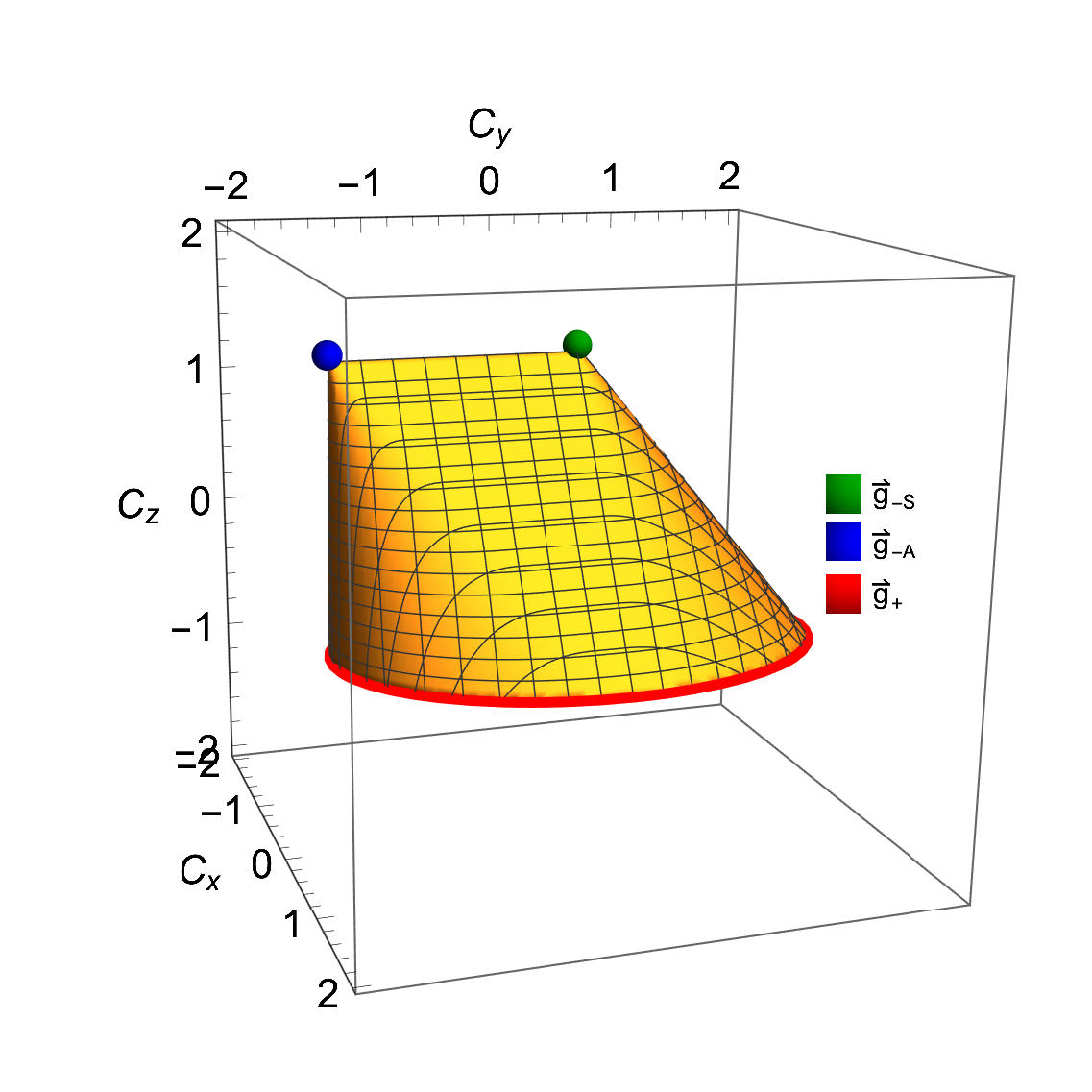}
        \includegraphics[width=.45\linewidth]{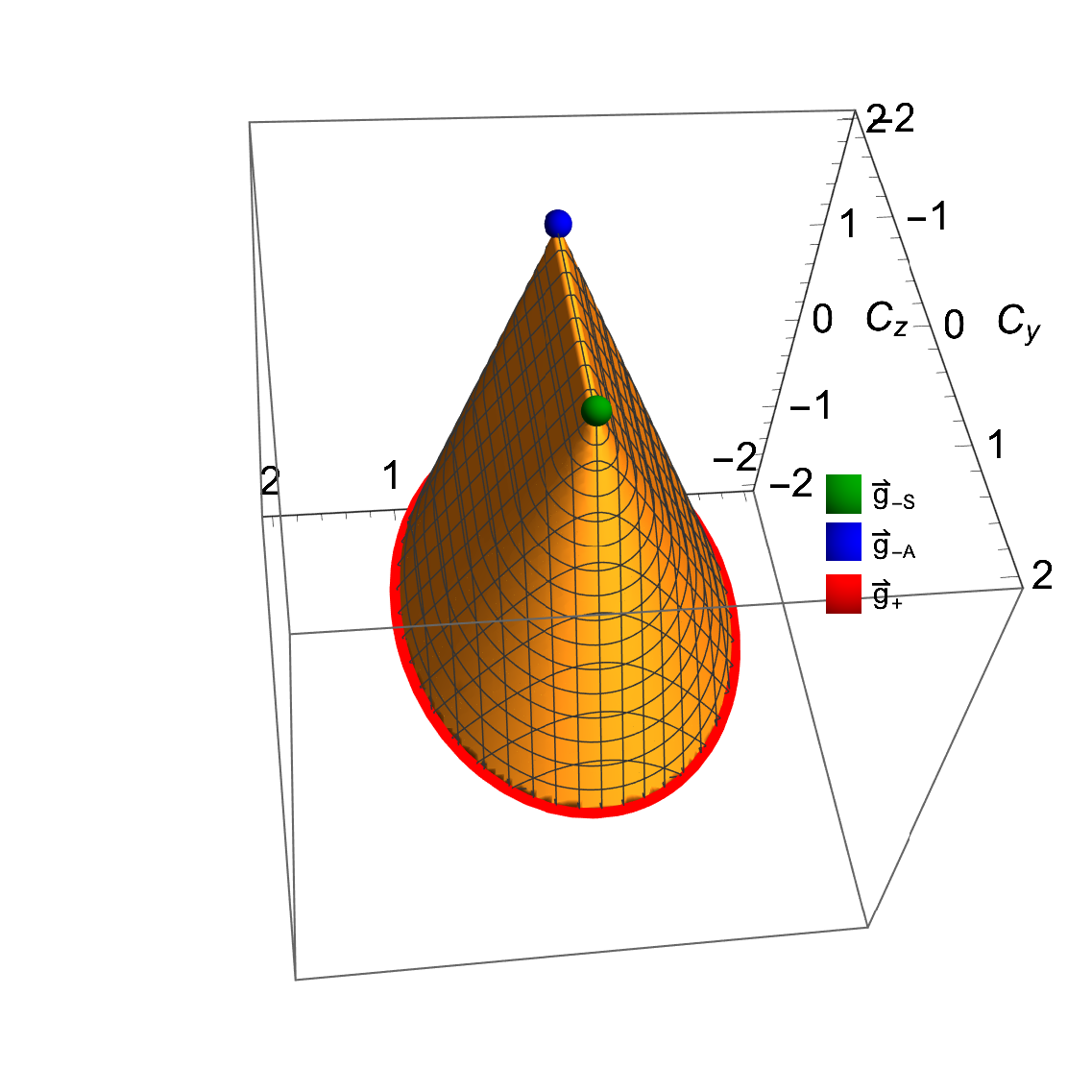}
	\end{center}
	\caption{Positivity cone of an EFT with 2 scalars and the \(Z_{2}\) symmetry. We show a cross section of the cone, in two different angles. The axes are \(C_{x}=\frac{\sqrt{3}\left(C_{1}-C_{3}\right)}{\sqrt{2}\left(C_{1}+C_{3}+C_{4}\right)}, C_{y}=\frac{\sqrt{3} C_{2}}{C_{1}+C_{3}+C_{4}}\), and \(C_{z}=-\frac{C_{1}+C_{3}-2 C_{4}}{\sqrt{C_{1}+C_{3}+C_{4}}} .\) Generators are labeled by the green and blue dots, and the red circle.}
	\label{fig4}
\end{figure}

Deriving the boundary essentially requires a continuous VE. In this simple case, one may think of \(\vec{g}_{+}(r)\) as an infinite number of ERs. A linear bound should be spanned by 3 independent ERs. At least one of them needs to be \(\vec{g}_{+}(r)\) at some \(r=r_{0}\). For the bound to be valid, a second one should be taken as its neighboring, \(\vec{g}_{+}(r)\) at \(r=r_{0}+\delta r\) or equivalently \(g_{+}^{\prime}\left(r_{0}\right) \equiv d g_{+}(r) /\left.d r\right|_{r=r_{0}}\). A third one can be either \(\vec{g}_{-S}\), or \(\vec{g}_{-A}\), or some \(\vec{g}_{+}(r)\) at some other point \(r \neq r_{0}\). These gives the normal vectors of three kinds of bounds:
\begin{align}
&\vec{n}_{1}=\epsilon_{\alpha \beta \gamma \delta}\left[g_{+}\left(r_{0}\right)\right]^{\beta}\left[g_{+}^{\prime}\left(r_{0}\right)\right]^{\gamma} g_{-S}^{\delta} \propto\left(r_{0}^{2},-r_{0}, 1,0\right) \\
&\vec{n}_{2}=\epsilon_{\alpha \beta \gamma \delta}\left[g_{+}\left(r_{0}\right)\right]^{\beta}\left[g_{+}^{\prime}\left(r_{0}\right)\right]^{\gamma} g_{-A}^{\delta} \propto\left(r_{0}^{2},-r_{0}, 1,-r_{0}\right) \\
&\vec{n}_{3} \propto(0,0,0,1)
\end{align}
The bounds are valid only if \(\vec{n}_{1,2,3} \cdot \vec{g}_{i} \geq 0\) for all \(\vec{g}_i\) in Eq.~(\ref{eq4.14addf}). These requires \(r_{0} \geq 0\) for \(\vec{n}_{1}\) and \(r_{0} \leq 0\) for \(\vec{n}_{2}\). Thus the bounds of the parameter space can be written as
\begin{equation}
C_{4} \geq 0, \quad C_{1}-r C_{2}+r^{2} C_{3} \geq 0, \quad C_{1}+r\left(C_{2}+C_{4}\right)+r^{2} C_{3} \geq 0, \quad \forall r \geq 0
\end{equation}
The above is equivalent to the following inequalities by removing \(r\):
\begin{align}
&C_{1} \geq 0, \quad C_{3} \geq 0, \quad C_{4} \geq 0 \\
&2 \sqrt{C_{1} C_{3}} \geq C_{2}, \quad 2 \sqrt{C_{1} C_{3}} \geq-\left(C_{2}+C_{4}\right)
\end{align}
These bounds are shown in Figure \ref{fig4}. Generators are also shown in the same plot: the green and blue dots represent \(\vec{g}_{-S, A}\), while \(\vec{g}_{+}(r)\) goes around the red circle as \(r\) changes. The same result has also been derived with an alternative approach \cite{Li:2021cjv}.\\

Finally, we map the ERs to heavy particles in tree-level UV completions. There are three possibilities.
\begin{equation}
\begin{array}{c|ccccc}
\text { Particle } & \text { spin } & \operatorname{Parity}\left(\phi_{1} \rightarrow-\phi_{1}\right) & \text { Interaction } & \mathrm{ER} & \vec{c} \\
\hline S_{1} & 0 & + & g_{1} M_{1}\left(x \phi_{1}^{2}+y \phi_{2}^{2}\right) S_{1} & \text{\Checkmark} & 2 \times\left(x^{2}, 2 x y, y^{2}, 0\right) \\
S_{3} & 0 & - & g_{3} M_{3} \phi_{1} \phi_{2} S_{3} & \text{\Checkmark} & 2 \times(0,0,0,1) \\
V_{4} & 1 & - & g_{4}\left(\phi_{1} \stackrel{\leftrightarrow}{D}_{\mu} \phi_{2}\right) V_{4}^{\mu} & \text{\Checkmark} & 2 \times(0,-1,0,1)\label{eq:4.21}
\end{array}
\end{equation}
We see that the free \(x, y\) parameters in the coupling of the \(S_{1}\) state are the reason of the \(x, y\) dependence in \(m_{+}\).

As one last comment, if we consider the most general case without any discrete symmetry, \(C_{5,6}\) will be allowed. Depending on the total spin of the intermediate state being even or odd, the \(m\) matrices are either symmetric or anti-symmetric. There are only two generators:
\[
m_{S}=\begin{array}{|c|c|}
\hline x & z \\
\hline z & y \\
\hline
\end{array}, \quad m_{A}=\begin{array}{|c|c|}
\hline 0 & 1 \\
\hline-1 & 0 \\
\hline
\end{array}
\]
but the first has essentially two real degrees of freedom. The generators are
\[
\begin{aligned}
G_{S}^{i j k l}&=
\begin{array}{|c|c|c|c|}
\hline 2 x^{2} & x y+z^{2} & 2 x z & 2 x z \\
\hline x y+z^{2} & 2 y^{2} & 2 y z & 2 y z \\
\hline 2 x z & 2 y z & 2 z^{2} & x y+z^{2} \\
\hline 2 x z & 2 y z & x y+z^{2} & 2 z^{2} \\
\hline
\end{array}, \quad G_{A}^{i j k l}=\begin{array}{|c|c|c|c|}
\hline 0 & -1 & 0 & 0 \\
\hline-1 & 0 & 0 & 0 \\
\hline 0 & 0 & 2 & -1 \\
\hline 0 & 0 & -1 & 2 \\
\hline
\end{array}
\end{aligned}
\]

\begin{equation}
\vec{g}_{S}(x, y, z)=\left(x^{2}, 2 x y, y^{2}, 4 z^{2}, 4 x z, 4 y z\right), \quad \vec{g}_{A}=(0,-1,0,1,0,0)
\end{equation}
The corresponding VE is difficult to calculate. The complete bounds, however, can be analytically obtained by using the alternative approach described in Ref.~\cite{Li:2021cjv}.

\subsubsection{Particle enumeration for scalars}
\label{sec4.1.4}
So far, we have been using a tree-level mapping between \(\mathcal{M}^{i j k l}\) and the Wilson coefficient space. In this case, there is an easier way to get the bounds. One simply enumerates all heavy scalars and vectors, as for example those in Eq.~(\ref{eq:4.21}), and use a tree-level matching to find all generator vectors \(\vec{g}\), directly in the space of coefficients. This is based on the observation that all the generators for the scalar EFTs can be interpreted as the tree-level exchange of either a heavy scalar \(S\) or a heavy vector \(V\), which couple to the light scalar fields via two kinds of couplings
\begin{equation}
\mathcal{L} \supset g_{i j k} M S_{k} \phi_{i} \phi_{j}+h_{i j k} V_{k}^{\mu} \phi_{i} \stackrel{\leftrightarrow}{D}_{\mu} \phi_{j}
\end{equation}
The two terms give rise to the symmetric \(\mathbf{m}^{i j} \propto g_{i j k}\) and the antisymmetric \(\mathbf{m}^{i j} \propto h_{i j k}\), respectively. Other symmetries of the theory manifest as further restrictions on \(g_{i j k}\) and \(h_{i j k} .\) Essentially, this means that one only needs to enumerate all scalars and vectors with the above couplings, and then compute the \(\mathcal{M}^{i j k l}\) from their exchanges, to obtain all generators. Other UV completions, such as loop-level completions or higher-spin states, will not give any independent generators. This is related to that we focus on forward scattering processes: the total spin of the intermediate state only affects whether \(m\) changes sign under \(i \leftrightarrow j\), and so we only need to consider the spin-0 and the spin-1 cases to cover both possibilities.

This ``particle enumeration'' approach is more convenient than directly constructing \(\mathcal{M}^{i j k l}\) from symmetries: once the UV particles are known, calculating \(\mathcal{M}^{i j k l}\) and mapping it to the coefficient space is simply a tree-level EFT matching, which can be equivalently and more conveniently carried out by solving equations of motion for the heavy fields. This is exactly what we did when presenting the UV completions of all generators, e.g. in Eq.~(\ref{eq:4.21}), but the idea here is to conversely use Eq.~(\ref{eq:4.21}) to find the bounds. Though the matching is tree level, the resulting bounds do apply to loop-level or even strongly coupled UV theories.

It is worth pointing out the limitations of this simplified approach. First, a tree-level mapping between \(\mathcal{M}^{i j k l}\) and coefficients is assumed. If higher-order effects in the EFT is not negligible, one would have to first use a tree-level mapping to convert the bounds on coefficients to bounds on \(\mathcal{M}^{i j k l}\) (as only the tree-level exchange of heavy particles corresponds exactly to the generators), and then convert the bounds back to the coefficient space by using a higher-order calculation of \(\mathcal{M}^{i j k l}\) (as bounds on \(\mathcal{M}\) is independent of the perturbative order of the mapping). In this case, this particle enumeration approach is not necessarily simpler than directly constructing the generators. In addition, we will see that this approach does not apply to vectors. It does apply to fermions, provided that some dim-5 effective coupling is taken into account.

\subsubsection{SM Higgs boson}
\label{sec4.1.5}
As a last example for scalar EFT, let us consider the SM Higgs boson. This has been worked out in Ref.~\cite{Zhang:2020jyn} in a self-conjugate basis. Here we work out the same bounds using the particle enumeration method. There are in total 6 types of particles that can generate 4-Higgs amplitude through a tree-level exchange. They are listed below:
\begin{equation}
\begin{tabular}{c|cccccc}
Particle & Spin & Charge/irrep & Interaction & \(\mathrm{ER}\) & \(\vec{c}\) & \(\vec{c}^{(6)}\) \\
\hline
\(\mathcal{B}_{1}\) & 1 & \(1_{1}\) & \(g \mathcal{B}_{1}^{\mu \dagger}(H^{T} \epsilon \stackrel{\leftrightarrow}{D}_{\mu} H)+\) $h.c.$ & \text{\Checkmark} & \(8(1,0,-1)\) & \(2(-1,2)\) \\
\(\Xi_{1}\) & 0 & \(3_{1}\) & \(g M \Xi_{1}^{I \dagger}\left(H^{T} \epsilon \tau^{I} H\right)+\) $h.c.$ & \text{\XSolidBrush} & \(8(0,1,0)\) & \(2(1,2)\) \\
\(\mathcal{S}\) & 0 & \(1_{0}(S)\) & \(g M \mathcal{S}\left(H^{\dagger} H\right)\) & \text{\Checkmark} & \(2(0,0,1)\) & \(-\frac{1}{2}(1,0)\) \\
\(\mathcal{B}\) & 1 & \(1_{0}(A)\) & \(g \mathcal{B}^{\mu}(H^{\dagger} \stackrel{\leftrightarrow}{D}_{\mu} H)\) & \text{\Checkmark} & \(2(-1,1,0)\) & \(-\frac{1}{2}(1,4)\) \\
\(\Xi_{0}\) & 0 & \(3_{0}(S)\) & \(g M \Xi_{0}^{I}\left(H^{\dagger} \tau^{I} H\right)\) & \text{\XSolidBrush} & \(2(2,0,-1)\) & \(\frac{1}{2}(1,-4)\) \\
\(\mathcal{W}\) & 1 & \(3_{0}(A)\) & \(g \mathcal{W}^{\mu I}(H^{\dagger} \tau^{I} \stackrel{\leftrightarrow}{D}_{\mu} H)\) & \text{\XSolidBrush} & \(2(1,1,-2)\) & \(-\frac{3}{2}(1,0)\) \\
\end{tabular}
\end{equation}
where the coefficients in \(\vec{c}\) are of the following three operators:
\begin{align}
Q_{H^{4}}^{(1)} &=\left(D_{\mu} H^{\dagger} D_{\nu} H\right)\left(D^{\nu} H^{\dagger} D^{\mu} H\right) \\
Q_{H^{4}}^{(2)} &=\left(D_{\mu} H^{\dagger} D_{\nu} H\right)\left(D^{\mu} H^{\dagger} D^{\nu} H\right) \\
Q_{H^{4}}^{(3)} &=\left(D^{\mu} H^{\dagger} D_{\mu} H\right)\left(D^{\nu} H^{\dagger} D_{\nu} H\right)
\end{align}
In the last column, we also give the resulting Wilson coefficients at dim-6. They are the coefficients of the Warsaw basis operators
\begin{align}
&Q_{\varphi \square}=\partial_{\mu}\left(H^{\dagger} H\right) \partial^{\mu}\left(H^{\dagger} H\right) \\
&Q_{\varphi D}=\left(H^{\dagger} D_{\mu} H\right)\left(D^{\mu} H^{\dagger} H\right)
\end{align}
and are normalized such that when each state is integrated out, the corresponding dim-6 coefficient vector is \(\vec{C}^{(6)}=g^{2} / M^{2} \vec{c}^{(6)}\). These are mainly for a discussion in Section 5.2.2.

\begin{figure}[h]
	\begin{center}
		\includegraphics[width=.45\linewidth]{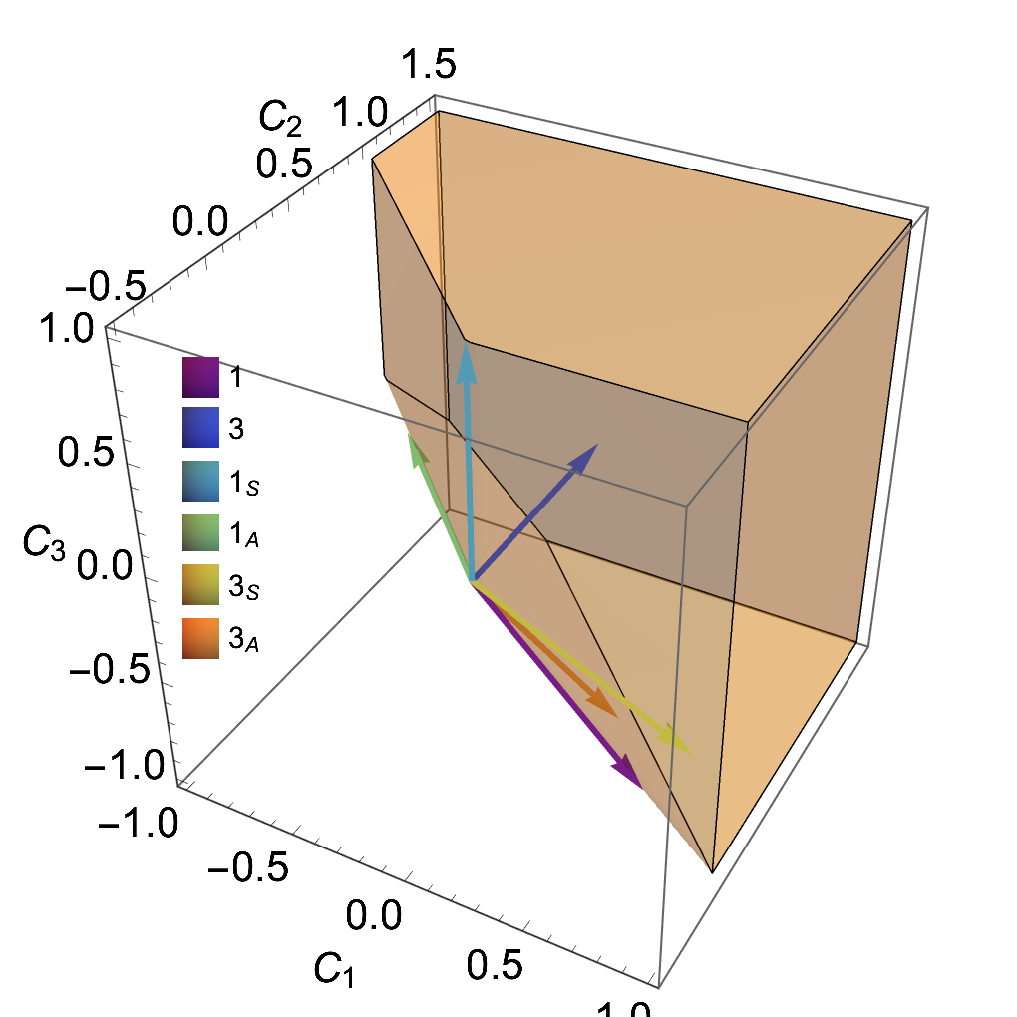}
        \includegraphics[width=.45\linewidth]{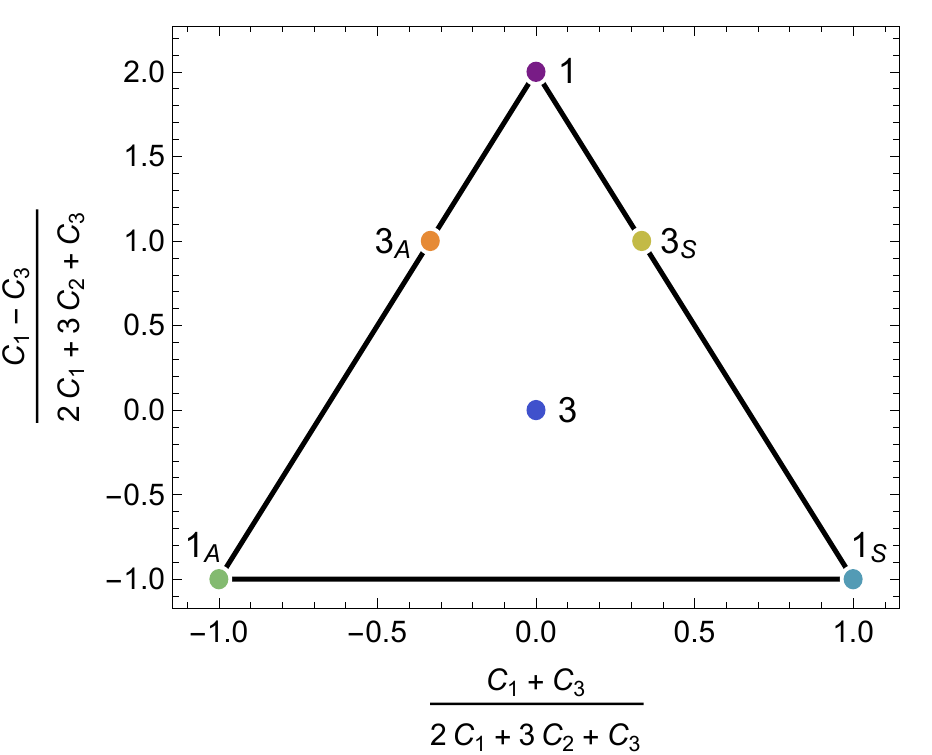}
	\end{center}
	\caption{The positivity cone for 4-Higgs operators, with the corresponding generators. Colors represent different irreps. They are only labeled with \(\mathrm{SU}(2)\) irrep \((1,3)\) and the exchange symmetry (S,A). The cone is shown in the left plot, while the right plot shows a slice of the cone. The latter can be thought of as intersecting the cone with a hyperplane \(2 C_{1}+3 C_{2}+C_{3}=1\).}
	\label{fig5}
\end{figure}

The 6 generator vectors in the dim-8 coefficient space can be read out:
\begin{equation}
\begin{array}{lll}
\vec{g}_{\textbf{1}}=(1,0,-1) & \vec{g}_{\textbf{1} S}=(0,0,2) & \vec{g}_{\textbf{1} A}=(-2,2,0) \\
\vec{g}_{\textbf{3}}=(0,1,0) & \vec{g}_{\textbf{3} S}=(4,0,-2) & \vec{g}_{\textbf{3}A}=(2,2,-4)\label{eq:4.30}
\end{array}
\end{equation}
Among the 6 generators, the three SU(2) singlets are extremal. The positivity region is thus a triangular cone, whose bounds are simply
\begin{equation}
C_{2} \geq 0, \quad C_{1}+C_{2} \geq 0, \quad C_{1}+C_{2}+C_{3} \geq 0
\end{equation}
The cone and its generators are shown in Figure \ref{fig5} with a 2-dimensional slice.

Finally, for completeness, we also present the extremal positivity approach to this problem, but using complex fields. The treatment of symmetry group projectors is similar to what we will use for the fermions in Section 4.3. A difference is that we will follow Eq.~(\ref{eq:3.28}) but without imposing the \(i \leftrightarrow j, j \leftrightarrow l\) symmetry, as this is automatic from the SU(2) irreps.

The particle indices \(i, j, k, l\) run through \(H^{1}, H^{2}, H_{1}^{\dagger}, H_{2}^{\dagger}\), where the \(^{1,2}\) and \(_{1,2}\) are the \(\mathrm{SU}(2)\) indices. The amplitude can be written as
{\footnotesize
\begin{flalign}
	\nonumber &\raisebox{0pt}{$\mathcal{M}^{ijkl}=$ }\\&\nonumber
\begin{tabular}{ r|c|c|c|c| }
	\multicolumn{1}{r}{}
	 &  \multicolumn{1}{c}{$H_c H_d$}
	 & \multicolumn{1}{c}{$H^{\dagger c}H^{\dagger d}$}
     &  \multicolumn{1}{c}{$H_c H^{\dagger d}$}
	 & \multicolumn{1}{c}{$H^{\dagger c} H_d$}  \\
	\cline{2-5}
	$H^aH^b $&$ \mathcal{M}(H H \rightarrow H H)^{a b}_{\ \ cd}$ &  & & \\
	\cline{2-5}
	$H^{\dagger }_a H^{\dagger }_b $&  &$\mathcal{M}(H^{\dagger} H^{\dagger} \rightarrow H^{\dagger} H^{\dagger})_{a b}^{\ \ c d}$& & \\
	\cline{2-5}
    $H^a H^{\dagger }_b $ &   &  &$\mathcal{M}(H H^{\dagger} \rightarrow H H^{\dagger})^{a\ \ d}_{\ bc} $&$ \mathcal{M}(H H^{\dagger} \rightarrow H^{\dagger} H)^{a\ c}_{\ b\ d}$ \\
	\cline{2-5}
	$H^{\dagger }_a  H^b  $&  & &$\mathcal{M}(H^{\dagger} H \rightarrow H H^{\dagger})^{\ b\ d}_{a\ c} $&$\mathcal{M}(H^{\dagger} H \rightarrow H^{\dagger} H)^{\ bc}_{a\ \ d}$\\
	\cline{2-5}
\end{tabular}
\end{flalign}
}
where each matrix elements carry four \(\mathrm{SU}(2)\) indices, labeled by \(a, b, c, d\). An explicity calculations of these terms give the following results
\begin{align}
&\mathcal{M}(H H \rightarrow H H)^{a b}_{\ \ c d}=\frac{1}{2}\left[\left(C_{2}+C_{3}\right) \delta_{d}^{a} \delta_{c}^{b}+\left(C_{1}+C_{2}\right) \delta_{c}^{a} \delta_{d}^{b}\right] \label{eq:4.32}\\
&\mathcal{M}\left(H^{\dagger} H^{\dagger} \rightarrow H^{\dagger} H^{\dagger}\right)_{a b}^{\ \ c d}=\frac{1}{2}\left[\left(C_{2}+C_{3}\right) \delta_{a}^{d} \delta_{b}^{c}+\left(C_{1}+C_{2}\right) \delta_{a}^{c} \delta_{b}^{d}\right] \\
&\mathcal{M}\left(H H^{\dagger} \rightarrow H H^{\dagger}\right)^{a\ \ d}_{\ bc}=\frac{1}{2}\left[\left(C_{1}+C_{2}\right) \delta_{c}^{a} \delta_{b}^{d}+\left(C_{2}+C_{3}\right) \delta_{b}^{a} \delta_{c}^{d}\right] \\
&\mathcal{M}\left(H^{\dagger} H \rightarrow H^{\dagger} H\right)_{a\ \ d}^{\ b c}=\frac{1}{2}\left[\left(C_{1}+C_{2}\right) \delta_{a}^{c} \delta_{d}^{b}+\left(C_{2}+C_{3}\right) \delta_{a}^{b} \delta_{d}^{c}\right] \\
&\mathcal{M}\left(H H^{\dagger} \rightarrow H^{\dagger} H\right)_{\ b\ d}^{a\ c}=\frac{1}{2}\left(C_{1}+C_{3}\right)\left(\delta_{d}^{a} \delta_{b}^{c}+\delta_{b}^{a} \delta_{d}^{c}\right) \\
&\mathcal{M}\left(H^{\dagger} H \rightarrow H H^{\dagger}\right)_{a\ c}^{\ b\ d}=\frac{1}{2}\left(C_{1}+C_{3}\right)\left(\delta_{a}^{d} \delta_{c}^{b}+\delta_{a}^{b} \delta_{c}^{d}\right)\label{eq:4.37}
\end{align}
where \(C_{1}, C_{2}, C_{3}\) are the coefficients of the operators \(Q_{H^{4}}^{(1)}, Q_{H^{4}}^{(2)}\) and \(Q_{H^{4}}^{(3)}\).

Now we need to enumerate the generators. An intermediate state that couples to two Higgs fields must live in the following irreps:
\begin{equation}
\mathbf{1}_{1}, \mathbf{3}_{1}, \mathbf{1}_{0 S}, \mathbf{1}_{0 A}, \mathbf{3}_{0 S}, \mathbf{3}_{0 A}
\end{equation}
The first two couple to \(H H\), while the rest couple to \(H^{\dagger} H .\) The \(S, A\) denotes the exchange symmetry between \(H \leftrightarrow H^{\dagger}\), as determined by the spin of the intermediate state. To write down the \(m\) matrices, note that the hypercharge symmetry determines which entry can be nonzero, while the \(\mathrm{SU}(2)\) symmetry determines the exact CG coefficient that appear in that entry. For example, the \(m\) matrices for the first two irreps are
\begin{flalign}
	\nonumber \raisebox{-5pt}{$m_{\textbf{1}_1}=$ }
\begin{tabular}{ r|c|c| }
	\multicolumn{1}{r}{}
	 &  \multicolumn{1}{c}{$H^b$}
	 & \multicolumn{1}{c}{$H_b^{\dagger}$} \\
	\cline{2-3}
	$H^a$ & $\epsilon^{ab}$ & 0 \\
	\cline{2-3}
	$H_a^{\dagger}$ & 0 & 0\\
	\cline{2-3}
\end{tabular}
,\quad
\raisebox{-5pt}{$m_{\textbf{3}_1}^I=$ }
\begin{tabular}{ r|c|c| }
	\multicolumn{1}{r}{}
	 &  \multicolumn{1}{c}{$H^b$}
	 & \multicolumn{1}{c}{$H_b^{\dagger}$} \\
	\cline{2-3}
	$H^a$ & $\left[\epsilon \tau^I\right]^{ab}$ & 0 \\
	\cline{2-3}
	$H_a^{\dagger}$ & 0 & 0\\
	\cline{2-3}
\end{tabular}
\end{flalign}
The generators can be computed following Eq.~(\ref{eq:3.28})
\begin{flalign}
	\nonumber\raisebox{0pt}{$\mathcal{G}_{\mathbf{1}_{1}, \mathbf{3}_{1}}^{i j k l}=$}&
\begin{tabular}{ r|c|c|c|c| }
	\multicolumn{1}{r}{}
	 &  \multicolumn{1}{c}{$H_c H_d$}
	 & \multicolumn{1}{c}{$H^{\dagger c}H^{\dagger d}$}
     &  \multicolumn{1}{c}{$H_c H^{\dagger d}$}
	 & \multicolumn{1}{c}{$H^{\dagger c} H_d$}  \\
	\cline{2-5}
	$H^aH^b $& ${P_{\mathbf{1}, \mathbf{3}}}^{ab}_{\ \ cd}$ &  & & \\
	\cline{2-5}
	$H^{\dagger a} H^{\dagger b} $&  & & & \\
	\cline{2-5}
    $H^a H^{\dagger b} $ &   &  & &  \\
	\cline{2-5}
	$H^{\dagger a}  H^b  $&  & & & \\
	\cline{2-5}
\end{tabular}
+(j \leftrightarrow l)+(i \leftrightarrow k)+(i \leftrightarrow k, j \leftrightarrow l)\\ \nonumber
=&
\begin{tabular}{ r|c|c|c|c| }
	\multicolumn{1}{r}{}
	 &  \multicolumn{1}{c}{$H_c H_d$}
	 & \multicolumn{1}{c}{$H^{\dagger c}H^{\dagger d}$}
     &  \multicolumn{1}{c}{$H_c H^{\dagger d}$}
	 & \multicolumn{1}{c}{$H^{\dagger c} H_d$}  \\
	\cline{2-5}
	$H^aH^b $& ${P_{\mathbf{1}, \mathbf{3}}}^{ab}_{\ \ cd}$ &  & & \\
	\cline{2-5}
	$H^{\dagger a} H^{\dagger b} $&  &${P_{\mathbf{1}, \mathbf{3}}}^{cd}_{\ \ ab}$ & & \\
	\cline{2-5}
    $H^a H^{\dagger b} $ &   &  &${P_{\mathbf{1}, \mathbf{3}}}^{ad}_{\ \ cb}$ &  \\
	\cline{2-5}
	$H^{\dagger a}  H^b  $&  & & & ${P_{\mathbf{1}, \mathbf{3}}}^{cb}_{\ \ ad}$\\
	\cline{2-5}
\end{tabular}
\end{flalign}
Note that we did not write the contributions of the charge conjugates of these states, as they are taken into account by the \((i \leftrightarrow k)\) symmetrization. Expressions for the projectors can be found in Appendix A, Eq.~(\ref{eq:A.2}). Next, the \(m\) matrices for the last four irreps are
\begin{flalign}
	\nonumber \raisebox{-5pt}{$m_{\textbf{1}_{0[S,A]}}=$ }
\begin{tabular}{ r|c|c| }
	\multicolumn{1}{r}{}
	 &  \multicolumn{1}{c}{$H^b$}
	 & \multicolumn{1}{c}{$H_b^{\dagger}$} \\
	\cline{2-3}
	$H^a$ & 0 & $\delta^a_b$ \\
	\cline{2-3}
	$H_a^{\dagger}$ & $[\pm]\delta^b_a$ & 0\\
	\cline{2-3}
\end{tabular}
,\quad
\raisebox{-5pt}{$m_{\textbf{3}_{0[S,A]}}^I=$ }
\begin{tabular}{ r|c|c| }
	\multicolumn{1}{r}{}
	 &  \multicolumn{1}{c}{$H^b$}
	 & \multicolumn{1}{c}{$H_b^{\dagger}$} \\
	\cline{2-3}
	$H^a$ & 0 & $\tau^{I^a}_{\ \ b}$ \\
	\cline{2-3}
	$H_a^{\dagger}$ & $[\pm]\tau^{I^b}_{\ \ a}$ & 0\\
	\cline{2-3}
\end{tabular}
\end{flalign}
Here the \(\pm\) corresponds to \(S\) and \(A\) irreps. The generators are
{\footnotesize
\begin{flalign}
	\nonumber \raisebox{0pt}{$\mathcal{G}_{\mathbf{1}_{0[S, A]}, \mathbf{3}_{0[S, A]}}^{i j k l}=$}&
\begin{tabular}{ r|c|c|c|c| }
	\multicolumn{1}{r}{}
	 &  \multicolumn{1}{c}{$H_c H_d$}
	 & \multicolumn{1}{c}{$H^{\dagger c}H^{\dagger d}$}
     &  \multicolumn{1}{c}{$H_c H^{\dagger d}$}
	 & \multicolumn{1}{c}{$H^{\dagger c} H_d$}  \\
	\cline{2-5}
	$H^aH^b $& &  & & \\
	\cline{2-5}
	$H^{\dagger a} H^{\dagger b} $&  & & & \\
	\cline{2-5}
    $H^a H^{\dagger b} $ &   &  & ${P_{\mathbf{1}, \mathbf{3}}}^{a\ \ d}_{\ bc}$  &  $\pm{P_{\mathbf{1}, \mathbf{3}}}^{a\ \ c}_{\ bd}$  \\
	\cline{2-5}
	$H^{\dagger a}  H^b  $&  & &  $\pm{P_{\mathbf{1}, \mathbf{3}}}^{b\ \ d}_{\ ac}$ &  ${P_{\mathbf{1}, \mathbf{3}}}^{b\ \ c}_{\ ad}$  \\
	\cline{2-5}
\end{tabular}
+(j \leftrightarrow l)+(i \leftrightarrow k)+(i \leftrightarrow k, j \leftrightarrow l)\\
=& 2 \times
\begin{tabular}{ r|c|c|c|c| }
	\multicolumn{1}{r}{}
	 &  \multicolumn{1}{c}{$H_c H_d$}
	 & \multicolumn{1}{c}{$H^{\dagger c}H^{\dagger d}$}
     &  \multicolumn{1}{c}{$H_c H^{\dagger d}$}
	 & \multicolumn{1}{c}{$H^{\dagger c} H_d$}  \\
	\cline{2-5}
	$H^aH^b $& ${P_{\mathbf{1}, \mathbf{3}}}^{a\ \ b}_{\ dc}$ &  & & \\
	\cline{2-5}
	$H^{\dagger a} H^{\dagger b} $&  &${P_{\mathbf{1}, \mathbf{3}}}^{c\ \ d}_{\ ba}$ & & \\
	\cline{2-5}
    $H^a H^{\dagger b} $ &   &  &${P_{\mathbf{1}, \mathbf{3}}}^{a\ \ d}_{\ bc}$ & $\pm{P_{\mathbf{1}, \mathbf{3}}}^{a\ \ c}_{\ bd}\pm{P_{\mathbf{1}, \mathbf{3}}}^{a\ \ c}_{\ db}$ \\
	\cline{2-5}
	$H^{\dagger a}  H^b  $&  & & $\pm{P_{\mathbf{1}, \mathbf{3}}}^{b\ \ d}_{\ ac}\pm{P_{\mathbf{1}, \mathbf{3}}}^{b\ \ d}_{\ ca}$ & ${P_{\mathbf{1}, \mathbf{3}}}^{c\ \ b}_{\ da}$\\
	\cline{2-5}
\end{tabular}\label{eq:4.39}
\end{flalign}
}
Expressions for the projectors can be found in Appendix A, Eq.~(\ref{eq:A.3}). Collecting all generators and comparing with the full amplitude, we obtain the same generator vectors as in Eq.~(\ref{eq:4.30}). This confirms that the particle enumeration approach derives the correct bounds that do apply to all order.

\subsection{Vector}
\label{sec4.2}
In this section, we apply the extremal positivity approach to vector bosons. The main difference w.r.t.\ the scalar case is that one needs to take into account two polarization modes. The simplest example is the hypercharge gauge boson, whose polarization in both \(x\) and \(y\) directions are denoted by \(B_{x}\) and \(B_{y}\). As we have argued, they are connected by the rotational symmetry around the beam direction (the \(z\)-direction). Therefore the problem seems identical to that of the two-scalar case discussed already in Section 4.1.1.

The relevant operators are the following:
\begin{align}
O_{1} &=\left(B_{\mu \nu} B^{\mu \nu}\right)\left(B_{\rho \sigma} B^{\rho \sigma}\right) \\
O_{2} &=(B_{\mu \nu} \tilde{B}^{\mu \nu})(B_{\rho \sigma} \tilde{B}^{\rho \sigma}) \\
O_{3} &=\left(B_{\mu \nu} B^{\mu \nu}\right)(B_{\rho \sigma} \tilde{B}^{\rho \sigma})
\end{align}
Note that \(O_{3}\) is parity-violating. Let \(C_{1}, C_{2}, C_{3}\) be their coefficients. A direct calculation of the hypercharge gauge boson scattering gives the following expression for the amplitude:
\begin{flalign}
	\nonumber \raisebox{0pt}{$\mathcal{M}^{i j k l}=8\times$}&
\begin{tabular}{ r|c|c|c|c| }
	\multicolumn{1}{r}{}
	 &  \multicolumn{1}{c}{$B_{x} B_{x}$}
	 & \multicolumn{1}{c}{$B_{y} B_{y}$}
     &  \multicolumn{1}{c}{$B_{x} B_{y}$}
	 & \multicolumn{1}{c}{$B_{y} B_{x}$}  \\
	\cline{2-5}
	$B_{x} B_{x}$& $2C_1$ & $C_1-C_2$ & $C_3$ & $-C_3$\\
	\cline{2-5}
	$B_{y} B_{y}$& $C_1-C_2$ & $2C_1$&$C_3$ & $-C_3$ \\
	\cline{2-5}
    $B_{y} B_{x}$ &  $C_3$ & $C_3$ &$2C_2$ & $C_1-C_2$ \\
	\cline{2-5}
	$B_{x} B_{y}$& $-C_3$ & $-C_3$ &$C_1-C_2$ &$2C_2$ \\
	\cline{2-5}
\end{tabular}
\end{flalign}
Consider first the parity-conserving case, and keep only \(\left(C_{1}, C_{2}\right)\). The intermediate state can be classified into three irreps: \(\mathbf{1}_{S}, \mathbf{1}_{A}\) and \(\mathbf{2}\). The generators are simply the projective operators, \(P_{\mathbf{1}_{S}}^{i(j|k| l)}, P_{\mathbf{1}_{A}}^{i(j|k| l)}\), and \(P_{\mathbf{2}}^{i(j|k| l)}\), in complete analogy with the \(\mathrm{SO}(2)\) scalar case in Section 4.1.1. Comparing with \(\mathcal{M}^{i j k l}\), this gives the following generators in the Wilson coefficient space:
\begin{equation}
\vec{g}_{\mathbf{1} S}=(1,0), \quad \vec{g}_{\mathbf{1} A}=(0,1), \quad \vec{g}_{\mathbf{2}}=(1,1)
\end{equation}
and the bounds are \(C_{1} \geq 0\) and \(C_{2} \geq 0\).

Note that the \(\mathbf{2}\) state carries a spin projection \(J_{3}=2\) in the \(z\)-direction, and therefore it cannot have a tree-level UV completion, with UV particles of spin less than \(2 .\) If UV particles carry spin equal or higher than 2, a boundary term would arise in the dispersion relation. This indicates that the theory still needs be UV completed. Without knowing this UV completion, the amplitude obtained from the \(\mathbf{2}\) generator would not match that obtained from integrating out directly this spin-2 state. Therefore \(\vec{g}_{\mathbf{2}}\) cannot have a simple tree-level interpretation, and the simplified ``particle enumeration'' method does not work for vectors. However, it is possible to generate \(\vec{g}_{\mathbf{2}}\) at the loop level, see discussions in Section 6.

\subsubsection{Parity violation}
\label{sec4.2.1}
We have so far ignored the possible mixing between \(\mathbf{1}_{S}\) and \(\mathbf{1}_{A}\) states. In the scalar case, we know this is impossible, because \(\mathbf{1}_{S}\) and \(\mathbf{1}_{A}\) corresponds to spin-even and spin-odd states, respectively. In our formalism, we have required that \(m^{i j}=\pm m^{j i}\) to implement this information. At the \(\mathcal{M}^{i j k l}\) level, this corresponds to the \((i \leftrightarrow j, k \leftrightarrow l)\) symmetry, which is a rotation of \(\pi\) around the \(y\)-axis.

However, the same requirement does not hold for vectors, if parity is violated; otherwise one cannot generate the operator \(O_{3}\). When choosing the linear polarizations as the particle basis, \((i \leftrightarrow j, k \leftrightarrow l)\) corresponds to a reflection in \(z\)-direction, and the corresponding symmetry is parity. This implies that if parity is not conserved, \(m^{i j}\) does not need to be symmetric or antisymmetric, or more specifically, a mixing between \(\mathbf{1}_{S}\) and \(\mathbf{1}_{A}\) can occur. This fact has been pointed out in Ref.~\cite{Trott:2020ebl}.

The easiest example is a scalar \(S\) that couples to hypercharge photons with the following interaction terms:
\begin{equation}
\mathcal{L} \supset \frac{g}{M} S\left(c_{\theta} B_{\mu \nu} B^{\mu \nu}+s_{\theta} B_{\mu \nu} \tilde{B}^{\mu \nu}\right)
\end{equation}
This is an effective coupling, but it is useful to illustrate the point, because no boundary term is generated in the dispersion relation. The two terms in the brackets are P-even and P-odd respectively. In P-conserving theories, they must couple to P-even and P-odd scalars separately, and the \(m^{i j}\) matrix corresponds to the CG coefficients of \(\mathbf{1}_{S}\) and \(\mathbf{1}_{A}\) respectively. In the \(\mathbf{1}_{S}\) case, the \(J^{P}\) of the diphoton system is \(0^{+} .\) In the \(\mathbf{1}_{A}\) case, the \(J^{P}\) of the diphoton system is \(0^{-}\). In theories with P-violation, angular momentum conservation does not forbid the transition between these two states. Such a transition can be generated by a mixing between P-even and P-odd scalars, and the \(m^{i j}\) matrix can in general take the following form:
\[
m_{\mathbf{1}}=\begin{array}{|c|c|}
\hline c_{\theta} & s_{\theta} \\
\hline-s_{\theta} & c_{\theta} \\
\hline
\end{array}
\]
Together with the \(\mathbf{2}\) state, two generators can be found:
\begin{equation}
\vec{g}_{\textbf{1}}(\theta)=\left(c_{\theta}^{2}, s_{\theta}^{2}, 2 s_{\theta} c_{\theta}\right), \quad \vec{g}_{\textbf{2}}=(1,1,0)
\end{equation}
Pure P-even and P-odd couplings correspond to \(\theta=0\) and \(\theta=\pi / 2 .\) In both cases the theory conserves parity, and we have \(C_{3}=0\). With a non-vanishing mixing \(\theta \in(0, \pi / 2) \cup(\pi / 2, \pi)\), \(C_{3}\) will be generated and the theory violates parity.

\begin{figure}[h]
	\begin{center}
		\includegraphics[width=.45\linewidth]{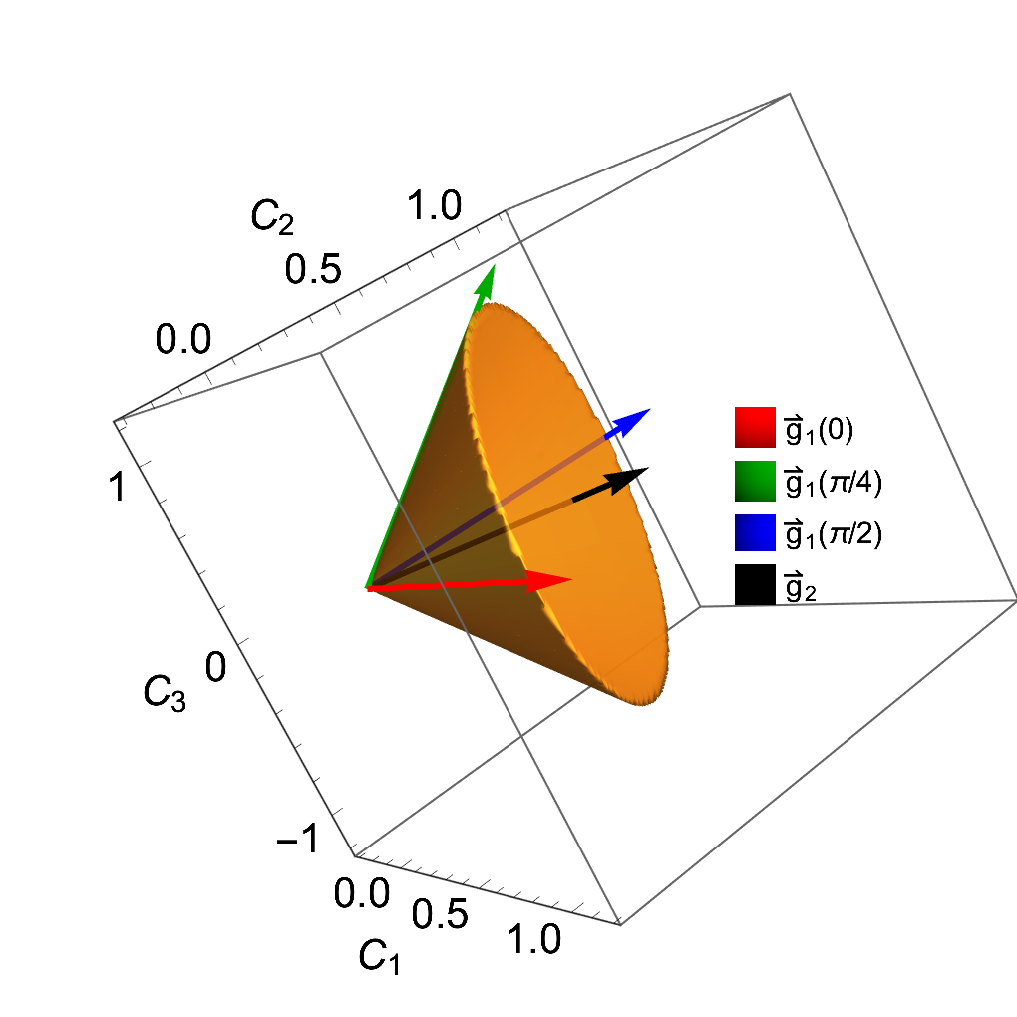}
        \includegraphics[width=.45\linewidth]{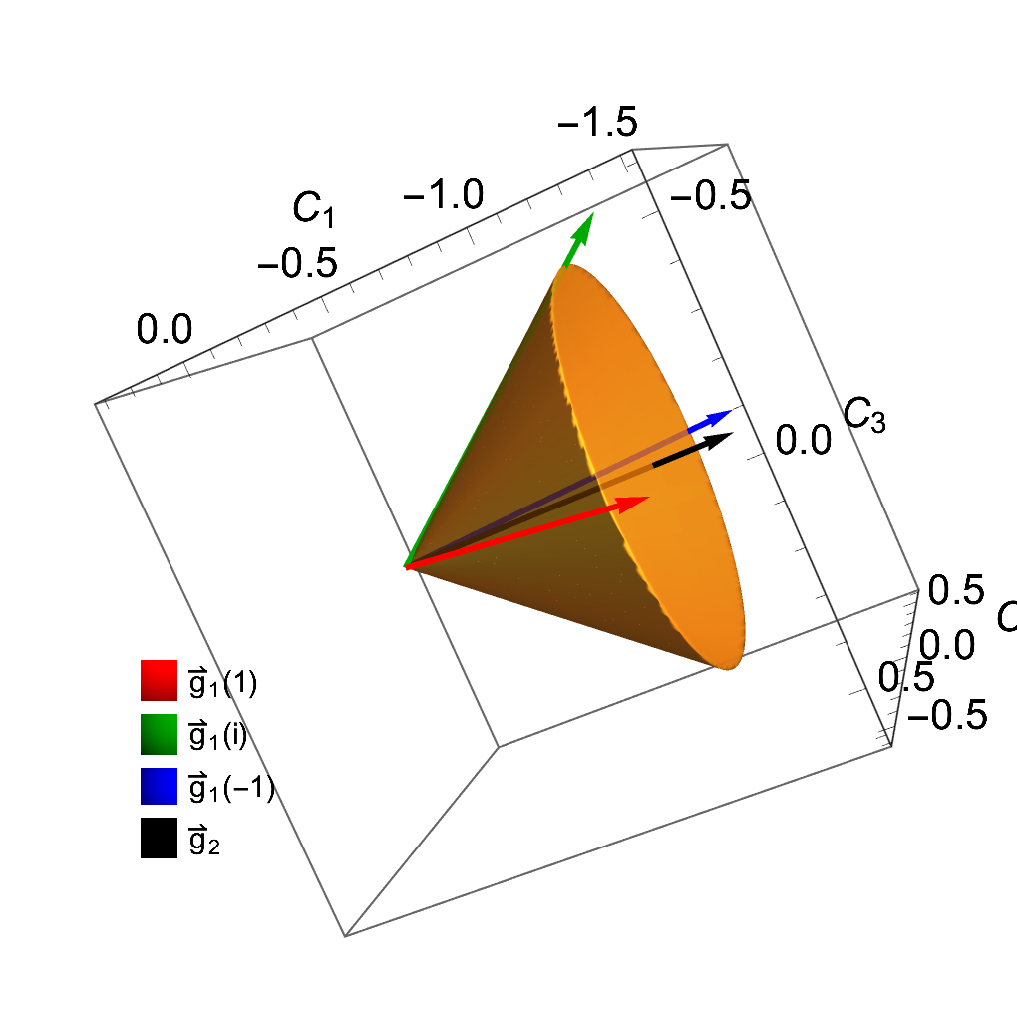}
	\end{center}
	\caption{ Left: positivity cone for hypercharge boson operators with parity violation. The cone is truncated by \(C_{1}+C_{2} \leq 1\), to make the plot easier to read. Generators are also shown, with \(\vec{g}_{1}\) taking values at \(\theta=0, \pi / 4, \pi / 2 .\) Right: positivity cone for four-fermion operators with CP violation. The cone is truncated by \(C_{1} \geq-1\), to make the plot easier to read. Generators are also shown, with \(\vec{g}_{1}\) taking values at \(\rho=1, i,-1\).}
	\label{fig6}
\end{figure}

To derive bounds, simply notice that when \(\theta\) goes from 0 to \(\pi\), the \(\vec{g}_{\textbf{1}}\) vector simply rotates around \(\vec{g}_{\textbf{2}}\) and carve out a circular cone. It is easier to see this by substituting \(C_{1}, C_{2}\) with \(\left(C_{1}^{\prime} \pm C_{2}^{\prime}\right) / 2\), so we have
\begin{equation}
\vec{g}_{\textbf{1}}(\theta)=\left(1, c_{2 \theta}, s_{2 \theta}\right), \quad \vec{g}_{\textbf{2}}=(2,0,0)
\end{equation}
in the new basis \(\left(C_{1}^{\prime}, C_{2}^{\prime}, C_{3}\right)\). The bounds of this cone are
\begin{equation}
C_{1} \geq 0, \quad C_{2} \geq 0, \quad 4 C_{1} C_{2} \geq C_{3}^{2}\label{eq:4.47}
\end{equation}
The positivity cone is shown in Figure \ref{fig6} left, together with the generators \(\vec{g}_{\textbf{1}}\) and \(\vec{g}_{\textbf{2}}\), the first taking \(\theta=0, \pi / 4, \pi / 2\). The same result has been obtained in Ref.~\cite{Yamashita:2020gtt} using a similar approach, and also in Ref.~\cite{Bi:2019phv} using the elastic scattering. The physical interpretation is that parity-violating physics is bounded by parity conserving ones from above.

Finally, the \(\vec{g}_{\textbf{1}}\) generators can be mapped to tree-level UV completions with scalars:
\begin{equation}
\begin{array}{c|ccccc}
\text { Particle } & \text { Spin } & \operatorname{Parity} & \text { Interaction } & \mathrm{ER} & \vec{c} \\
\hline S_{+} & 0 & + & \frac{g}{M} S\left(B_{\mu \nu} B^{\mu \nu}\right) & \text{\Checkmark} & \frac{1}{2}(1,0,0) \\
S_{-} & 0 & - & \frac{g}{M} S_{2}\left(B_{\mu \nu} \tilde{B}^{\mu \nu}\right) & \text{\Checkmark} & \frac{1}{2}(0,1,0) \\
S & 0 & ? & \frac{g}{M} S\left(c_{\theta} B_{\mu \nu} B^{\mu \nu}+s_{\theta} B_{\mu \nu} \tilde{B}^{\mu \nu}\right)& \text{\Checkmark} & \frac{1}{2}\left(c_{\theta}^{2}, s_{\theta}^{2}, 2 s_{\theta} c_{\theta}\right)
\end{array}
\end{equation}
The first two cases are P-even and P-odd scalars in a parity-conserving theory. The third case violates parity. The couplings are effective, so strictly speaking they cannot be viewed as UV completions. However, at the tree level, the amplitude from a heavy scalar exchange only grows as \(s\) at large energy. The dispersion relation receives no contribution from the boundary and remains valid. Therefore these scalars can be thought of as a partial UV-completion, and integrating them out would give rise to generators consistent with the dispersion relation. On the other hand, the \(\vec{g}_{\textbf{2}}\) generators comes from a spin-2 transition and is therefore possible only at the loop level, see Section 6.

The same approach can be used to derive bounds for other SM gauge bosons, including the \(W\)-boson and gluons. In fact, they have been derived in Refs.~\cite{Zhang:2020jyn, Li:2021cjv}. The treatment of the polarization is the same as hypercharge gauge bosons, while the treatment of the adjoint representation under the \(\mathrm{SU}(2)\) and \(\mathrm{SU}(3)\) is similar to previous examples with \(\mathrm{SO}(2)\) or \(\mathrm{SU}(2)\) symmetries. Since no new ingredient is required, we omit the details there, and only present the final results in Section 4.4.

\subsection{Fermions}
\label{sec4.3}
The fermion cases are slightly more complicated. Before showing concrete examples, let us first clarify a few points in the procedure. Consider \(n\) right-handed fermions, \(f_{a}\), and their corresponding anti-particles, which are left-handed, \(\bar{f}_{a}\), where \(a=1,2, \ldots, n\) labels flavor and other quantum numbers. The \(i, j, k, l\) indices run through \(2 n\) states: \(f_{1}, \bar{f}_{1}, f_{2}, \bar{f}_{2}, \ldots\) When presenting the \(m^{i j}\) matrices, we will show four distinct rows and columns: \(f_{a}, f_{b}, \bar{f}_{a}, \bar{f}_{b}\), e.g.
\begin{flalign}
	\nonumber\raisebox{0pt}{$m=$}&
\begin{tabular}{ r|c|c|c|c| }
	\multicolumn{1}{r}{}
	 &  \multicolumn{1}{c}{$f_a$}
	 & \multicolumn{1}{c}{$f_b$}
     &  \multicolumn{1}{c}{$\bar{f}_a$}
	 & \multicolumn{1}{c}{$\bar{f}_b$}  \\
	\cline{2-5}
	$f_a$&   &   &   &  \\
	\cline{2-5}
	$f_b$&   &  &  &   \\
	\cline{2-5}
    $\bar{f}_a$ &    &   &  &   \\
	\cline{2-5}
	$\bar{f}_b$&   &   &  &  \\
	\cline{2-5}
\end{tabular}
\end{flalign}
This covers all \(m^{i j}\) elements, including \(m^{a b}, m^{a \bar{b}}, m^{\bar{a} b}\) and \(m^{\bar{a} \bar{b}}\). The \(m^{a b}\) element will be shown in the \(f_{a}\) row and the \(f_{b}\) column, etc.

To construct the generators from \(m\), we shall consider an intermediate state \(X\) and the amplitude \(\mathbf{M}_{i j \rightarrow X} . X\) can be classified by its spin projection \(J_{3}\), in the forward direction. If \(J_{3}=0, f_{a} f_{b} \rightarrow X\) and \(\bar{f}_{a} \bar{f}_{b} \rightarrow X\) are allowed. If \(J_{3}=1(-1), f_{a} \bar{f}_{b}\left(\bar{f}_{a} f_{b}\right) \rightarrow X\) is allowed.

Consider the \(J_{3}=0\) case first. One may schematically write down the general form of \(m^{i j}\), for both symmetric and antisymmetric case:
\begin{flalign}
	\nonumber\raisebox{0pt}{$m_S=$}&
\begin{tabular}{ r|c|c|c|c| }
	\multicolumn{1}{r}{}
	 &  \multicolumn{1}{c}{$f_a$}
	 & \multicolumn{1}{c}{$f_b$}
     &  \multicolumn{1}{c}{$\bar{f}_a$}
	 & \multicolumn{1}{c}{$\bar{f}_b$}  \\
	\cline{2-5}
	$f_a$& $x_{aa}$  & $x_{ab}$  &   &  \\
	\cline{2-5}
	$f_b$& $x_{ab}$  & $x_{bb}$ &  &   \\
	\cline{2-5}
    $\bar{f}_a$ &    &   & $x'_{aa}$ & $x'_{ab}$  \\
	\cline{2-5}
	$\bar{f}_b$&   &   & $x'_{ab}$ & $x'_{bb}$ \\
	\cline{2-5}
\end{tabular},\quad\quad m_A=
\begin{tabular}{ r|c|c|c|c| }
	\multicolumn{1}{r}{}
	 &  \multicolumn{1}{c}{$f_a$}
	 & \multicolumn{1}{c}{$f_b$}
     &  \multicolumn{1}{c}{$\bar{f}_a$}
	 & \multicolumn{1}{c}{$\bar{f}_b$}  \\
	\cline{2-5}
	$f_a$&   & $y_{ab}$  &   &  \\
	\cline{2-5}
	$f_b$& $-y_{ab}$  &  &  &   \\
	\cline{2-5}
    $\bar{f}_a$ &    &   &  & $y'_{ab}$  \\
	\cline{2-5}
	$\bar{f}_b$&   &   & $-y'_{ab}$ &  \\
	\cline{2-5}
\end{tabular}
\end{flalign}
where \(x, y, z, x^{\prime}, y^{\prime}, z^{\prime}\) are arbitrary complex numbers. Recall that \(m\) is either symmetric or anti-symmetric as required by Eq.~(\ref{eq:3.21}). As an example of these two cases, consider a heavy scalar \(S\) and a heavy vector \(V\) that couple to two fermions through the following (effective) couplings:
\begin{equation}
\mathcal{L}_{i n t} \supset g_{a b} S^{\dagger} \bar{f}_{a}^{c} f_{b}+\frac{h_{a b}}{M} V^{\dagger \mu} \bar{f}_{a}^{c} i \overleftrightarrow{D}_{\mu} f_{b}+h . c .
\end{equation}
where \(g_{a b}, h_{a b}\) are coupling strengths. \(g_{a b}=g_{b a}\) and \(h_{a b}=-h_{b a} .\) By computing \(\mathbf{M}_{f_{a} f_{b} \rightarrow S}\) and \(\mathbf{M}_{f_{a} f_{b} \rightarrow V}\), we find that the scalar \(S\) gives rise to \(m_{S}\), and the vector \(V\) gives rise to \(m_{A}\), with
\begin{align}
x_{i j} \propto-g_{a b}, & \quad x_{a b}^{\prime} \propto g_{a b}^{*}, \\
y_{i j} \propto-h_{a b}, & \quad y_{a b}^{\prime} \propto-h_{a b}^{*}. 
\end{align}
Of course, there could be other intermediate states with higher spins, or possibly multiparticle states with higher partial waves, which could potentially also give rise to the same \(m\) matrices. A few comments are in order:

\begin{itemize}
  \item If the fermions carry other quantum numbers, such as hypercharge (as they do in the \(\mathrm{SM}), m^{i j}\) will further break into pieces, each corresponding to a certain value of the total charge of \(i\) and \(j\). For example, if \(f_{a}\) and \(f_{b}\) carry the same nonzero charge \(Q\), then \(m_{S}\) should carry charge \(2 Q\). The \(m_{S}\) matrix will split into two, one from \(S\) and the other from \(\bar{S}\) :
      \begin{flalign}
	\nonumber\raisebox{0pt}{$m_S=$}&
\begin{tabular}{ r|c|c|c|c| }
	\multicolumn{1}{r}{}
	 &  \multicolumn{1}{c}{$f_a$}
	 & \multicolumn{1}{c}{$f_b$}
     &  \multicolumn{1}{c}{$\bar{f}_a$}
	 & \multicolumn{1}{c}{$\bar{f}_b$}  \\
	\cline{2-5}
	$f_a$& $-g_{aa}$  & $-g_{ab}$  &   &  \\
	\cline{2-5}
	$f_b$& $-g_{ab}$  & $-g_{bb}$ &  &   \\
	\cline{2-5}
    $\bar{f}_a$ &    &   &   &    \\
	\cline{2-5}
	$\bar{f}_b$&   &   &   &   \\
	\cline{2-5}
\end{tabular},\quad\quad m_{\bar{S}}=
\begin{tabular}{ r|c|c|c|c| }
	\multicolumn{1}{r}{}
	 &  \multicolumn{1}{c}{$f_a$}
	 & \multicolumn{1}{c}{$f_b$}
     &  \multicolumn{1}{c}{$\bar{f}_a$}
	 & \multicolumn{1}{c}{$\bar{f}_b$}  \\
	\cline{2-5}
	$f_a$&   &    &   &  \\
	\cline{2-5}
	$f_b$&    &  &  &   \\
	\cline{2-5}
    $\bar{f}_a$ &    &   & $g^*_{aa}$ & $g^*_{ab}$  \\
	\cline{2-5}
	$\bar{f}_b$&   &   & $g^*_{ab}$ & $g^*_{bb}$ \\
	\cline{2-5}
\end{tabular}
\end{flalign}
\(-\) These two matrices actually generate the same \(\mathcal{M}^{i j k l}\). This is because
\begin{equation}
m_{\bar{S}}^{i j}=-m_{S}^{*\ \bar{i} \bar{j}}
\end{equation}
and therefore in Eq.~(\ref{eq:3.28}), the last two terms from the intermediate state \(\bar{S}\) would match the first two terms from \(S\), and vice versa. In other words, invoking the \(i \leftrightarrow k\) symmetry in Eq.~(\ref{eq:3.28}) is essentially adding the contribution from the charge conjugates of the intermediate states. The same is true also for the vector states, with \(m_{\bar{V}}^{i j}=m_{V}^{*\ \bar{i} \bar{j}}\) .
\end{itemize}

\begin{itemize}
  \item If \(\mathrm{CP}\) is conserved, we have \(g^{*}=\pm g\) and \(h^{*}=\pm h\), depending on the CP parity of \(S\) and \(V\). This suggests that \(x\) and \(y\) are either purely real or purely imaginary if CP conservation is imposed. We have \(m_{X}^{\bar{i}\bar{ j}}=\pm m_{X}^{i j}\), or \(m_{X}^{\bar{i}\bar{ j}}=\pm m_{\bar{X}}^{i j}\) if the state \(X\) carries charge. Together with Eq.~(\ref{eq:3.28}), this implies that \(\mathcal{M}^{i j k l}=\mathcal{M}^{k l i j}\) is indeed satisfied for CP-conserved amplitudes. Alternatively, one may construct the most general \(\mathcal{M}^{i j k l}\), and select its CP-conserving component, \(\left(\mathcal{M}^{i j k l}+\mathcal{M}^{k l i j}\right) / 2\).

  \item \(m^{i j}\) is either symmetric or antisymmetric, so symmetrizing \(i \leftrightarrow j, k \leftrightarrow l\) in Eq.~(\ref{eq:3.28}) has no effect.

\end{itemize}

Let us now consider the \(J_{3}=\pm 1\) states, which couple to \(f_{a} \bar{f}_{b}\) or \(\bar{f}_{a} f_{b}\). The \(m^{i j}\) for \(J_{3}=1\) and \(J_{3}=-1\) states can be respectively parametrized as

\begin{flalign}
	\nonumber\raisebox{0pt}{$m_+=$}&
\begin{tabular}{ r|c|c|c|c| }
	\multicolumn{1}{r}{}
	 &  \multicolumn{1}{c}{$f_a$}
	 & \multicolumn{1}{c}{$f_b$}
     &  \multicolumn{1}{c}{$\bar{f}_a$}
	 & \multicolumn{1}{c}{$\bar{f}_b$}  \\
	\cline{2-5}
	$f_a$&    &    &  $x_{aa}$ & $x_{ab}$ \\
	\cline{2-5}
	$f_b$&    &   & $x_{ba}$ &  $x_{bb}$ \\
	\cline{2-5}
    $\bar{f}_a$ &    &   &   &    \\
	\cline{2-5}
	$\bar{f}_b$&   &   &   &   \\
	\cline{2-5}
\end{tabular},\quad\quad m_-=\eta\times
\begin{tabular}{ r|c|c|c|c| }
	\multicolumn{1}{r}{}
	 &  \multicolumn{1}{c}{$f_a$}
	 & \multicolumn{1}{c}{$f_b$}
     &  \multicolumn{1}{c}{$\bar{f}_a$}
	 & \multicolumn{1}{c}{$\bar{f}_b$}  \\
	\cline{2-5}
	$f_a$&   &    &   &  \\
	\cline{2-5}
	$f_b$&    &  &  &   \\
	\cline{2-5}
    $\bar{f}_a$ &  $x_{aa}$  & $x_{ba}$  &   &    \\
	\cline{2-5}
	$\bar{f}_b$&  $x_{ab}$ &  $x_{aa}$ &   &   \\
	\cline{2-5}
\end{tabular}
\end{flalign}
where \(\eta\) is a pure phase. An obvious example is a heavy (complex) vector \(V\) that couples with
\begin{equation}
\mathcal{L}_{\text {int }} \supset g_{a b} V^{\dagger \mu} \bar{f}_{a} \gamma^{\mu} f_{b}+\text { h.c. }
\end{equation}
By computing \(\mathbf{M}_{f_{a} \bar{f}_{b} \rightarrow V}\) and \(\mathbf{M}_{\bar{f}_{a} f_{b} \rightarrow V}\), we find \(x_{a b} \propto g_{b a}\) and \(\eta=+1\). Its charge conjugate would give rise to a similar \(m^{i j}\), but with \(g_{a b} \rightarrow g_{b a}^{*}\). A few comments are in order:

\begin{itemize}
  \item \(V\) and its charge conjugate generate the same \(\mathcal{M}^{i j k l}\). This is because
  \begin{equation}
m_{V\left(J_{3}=1\right)}^{i j}=\eta m_{V\left(J_{3}=-1\right)}^{j i}=\eta m_{V^{\dagger}\left(J_{3}=1\right)}^{* \bar{j} \bar{ i}}=m_{V^{\dagger}\left(J_{3}=-1\right)}^{* \bar{i} \bar{j}}
\end{equation}
When writing down our master formula for the generators, we have used the crossing symmetry \(i \leftrightarrow k\) and the double exchange symmetry \(i \leftrightarrow j, k \leftrightarrow l\). The first combines \(V^{\dagger}\left(J_{3}=-1\right)\) with \(V\left(J_{3}=1\right)\), and \(V^{\dagger}\left(J_{3}=1\right)\) with \(V\left(J_{3}=-1\right)\). The second further combines the two combinations. Therefore, when constructing the generators, we only need to take into account \(m_{+}\) and without having to add \(m_{-}\) or its charge conjugate.  If \(V\) is self-conjugate, one imposes the constraint \(g_{a b}=g_{b a}^{*}\).

\end{itemize}

\begin{itemize}
  \item Similar to the \(J_{3}=0\) case, if the fermions of different flavors carry different charges, \(m^{i j}\) will split into several pieces, corresponding to the intermediate states with different charges.
\end{itemize}

Summing up the formalism, to enumerate all generators of the 4-fermion operators, one needs to consider the following three kinds of \(m\) matrices:
\begin{flalign}
	\nonumber\raisebox{0pt}{$m_S=$}&
\begin{tabular}{ r|c|c|c|c| }
	\multicolumn{1}{r}{}
	 &  \multicolumn{1}{c}{$f_a$}
	 & \multicolumn{1}{c}{$f_b$}
     &  \multicolumn{1}{c}{$\bar{f}_a$}
	 & \multicolumn{1}{c}{$\bar{f}_b$}  \\
	\cline{2-5}
	$f_a$&  $x_{aa}$  &  $x_{ab}$ &    &   \\
	\cline{2-5}
	$f_b$&  $x_{ab}$  & $x_{bb}$  &   &    \\
	\cline{2-5}
    $\bar{f}_a$ &    &   & $x'_{aa}$  & $x'_{ab}$   \\
	\cline{2-5}
	$\bar{f}_b$&   &   & $x'_{ab}$  & $x'_{bb}$  \\
	\cline{2-5}
\end{tabular},\quad\quad m_A=\eta\times
\begin{tabular}{ r|c|c|c|c| }
	\multicolumn{1}{r}{}
	 &  \multicolumn{1}{c}{$f_a$}
	 & \multicolumn{1}{c}{$f_b$}
     &  \multicolumn{1}{c}{$\bar{f}_a$}
	 & \multicolumn{1}{c}{$\bar{f}_b$}  \\
	\cline{2-5}
	$f_a$&   &  $y_{ab}$  &   &  \\
	\cline{2-5}
	$f_b$&  $-y_{ab}$  &  &  &   \\
	\cline{2-5}
    $\bar{f}_a$ &     &    &   & $y'_{ab}$   \\
	\cline{2-5}
	$\bar{f}_b$&    &    & $-y'_{ab}$  &   \\
	\cline{2-5}
\end{tabular}\\m_+=&
\begin{tabular}{ r|c|c|c|c| }
	\multicolumn{1}{r}{}
	 &  \multicolumn{1}{c}{$f_a$}
	 & \multicolumn{1}{c}{$f_b$}
     &  \multicolumn{1}{c}{$\bar{f}_a$}
	 & \multicolumn{1}{c}{$\bar{f}_b$}  \\
	\cline{2-5}
	$f_a$&    &    &  $x_{aa}$ & $x_{ab}$ \\
	\cline{2-5}
	$f_b$&    &   & $x_{ba}$ &  $x_{bb}$ \\
	\cline{2-5}
    $\bar{f}_a$ &    &   &   &    \\
	\cline{2-5}
	$\bar{f}_b$&   &   &   &   \\
	\cline{2-5}
\end{tabular}\label{eq:4.55}
\end{flalign}
If \(f_{a}\) and \(f_{b}\) have nonzero charges \(q_{a}\) and \(q_{b}\), then the \(m\) matrices will breakdown into different blocks, following in the patterns below:

\begin{flalign}
\nonumber \begin{tabular}{ r|c|c|c|c| }
	\multicolumn{1}{r}{}
	 &  \multicolumn{1}{c}{$f_a$}
	 & \multicolumn{1}{c}{$f_b$}
     &  \multicolumn{1}{c}{$\bar{f}_a$}
	 & \multicolumn{1}{c}{$\bar{f}_b$}  \\
	\cline{2-5}
	$f_a$&  \cellcolor{IndianRed}$2q_a$  & \cellcolor{LightSalmon}$q_a+q_b$   & 0  & \cellcolor{LightGreen}$q_a-q_b$ \\
	\cline{2-5}
	$f_b$&  \cellcolor{LightSalmon}  & \cellcolor{Wheat} $2q_b$ & \cellcolor{Thistle} $-q_a+q_b$ &    \\
	\cline{2-5}
    $\bar{f}_a$ &    &  \cellcolor{Thistle}  & \cellcolor[rgb]{.7,.9,.9} $-2q_a$ & \cellcolor{SteelBlue}  $-q_a-q_b$  \\
	\cline{2-5}
	$\bar{f}_b$& \cellcolor{LightCyan} &   & \cellcolor{SteelBlue}  & \cellcolor{Orchid}$-2q_b$  \\
	\cline{2-5}
\end{tabular}
\end{flalign}
The blocks with the same color correspond to the same charge \(Q(i)+Q(j)\), and are allowed to appear in the same \(\mathbf{M}_{X \rightarrow i j}\), for some intermediate state \(X\) with \(Q(X)=Q(i)+Q(j)\).

We have not taken into account any symmetries between different flavors. Most SM fermions are charged under the \(\mathrm{SU}(2)\) and/or the \(\mathrm{SU}(3)\) symmetries. The treatment of these symmetries are independent of the helicity/flavor structures that we discussed so far, and have been considered in e.g.~Section 4.1.5. If \(a, b\) are the indices of some internal gauge symmetry, one essentially replaces the arbitrary \(x_{a b}\) and \(y_{a b}\) parameters by the CG coefficients of all relevant irreps. We will illustrate this with examples.

A final remark is that in the \(\mathcal{M}^{i j k l}\) tensor, only the \(\mathcal{M}^{a b c d}, \mathcal{M}^{a b \bar{c} \bar{d}}, \mathcal{M}^{\bar{a} \bar{b} c d}\), and \(\mathcal{M}^{\bar{a} \bar{b} \bar{c} \bar{d}}\) are independent. The rest components either vanish (for example, \(\mathcal{M}^{a \bar{b} \bar{c} d}\) and \(\mathcal{M}^{\bar{a} b c \bar{d}}\) vanish by angular momentum conservation), or are fixed by \(s-u\) crossing symmetry. In other words, if we write the \(\mathcal{M}^{i j k l}\) tensor in the following form
\begin{flalign}
	\nonumber
\begin{tabular}{ r|c|c|c|c| }
	\multicolumn{1}{r}{}
	 &  \multicolumn{1}{c}{${f_{c} f_{d}}$}
	 & \multicolumn{1}{c}{${\bar{f}_{c} \bar{f}_{d}}$}
     &  \multicolumn{1}{c}{${f_{c} \bar{f}_{d}}$}
	 & \multicolumn{1}{c}{${\bar{f}_{c} f_{d}}$}  \\
	\cline{2-5}
	${f_{a} f_{b}}$&  $\mathcal{M}^{\bar{a} \bar{b} \bar{c} \bar{d}}$  &  $\mathcal{M}^{ab\bar{c} \bar{d}}$ &    &   \\
	\cline{2-5}
	${\bar{f}_{a} \bar{f}_{b}}$&  $\mathcal{M}^{\bar{a} \bar{b} cd}$  & $\mathcal{M}^{\bar{a} \bar{b} \bar{c} \bar{d}}$  &   &    \\
	\cline{2-5}
    ${f_{a} \bar{f}_{b}}$ &    &   & $\mathcal{M}^{a\bar{b}c\bar{d}}$  &     \\
	\cline{2-5}
	${\bar{f}_{a} f_{b}}$&   &   &    & $\mathcal{M}^{\bar{a} b \bar{c} d}$  \\
	\cline{2-5}
\end{tabular}
\end{flalign}
where the rows represent different \((i, j)\) pairs and the columns represent the \((k, l)\) pairs, then only the top-left quarter is independent. For the rest of this paper, we will often present a \(\mathcal{M}^{i j k l}\) in this way, with only the first two rows and columns shown explicitly, and possibly omit the row/column headings, e.g.
\begin{align}
\nonumber
\mathcal{M}^{i j k l}=\begin{array}{|l|l|}
\hline \mathcal{M}^{a b c d} & \mathcal{M}^{a b \bar{c} \bar{d}} \\
\hline \mathcal{M}^{\bar{a} \bar{d} c d} & \mathcal{M}^{\bar{a} \bar{b} \bar{c} \bar{d}} \\
\hline
\end{array}
\end{align}

\subsubsection{Single flavor fermion}
\label{sec4.3.1}
Let us first consider one fermion flavor with no internal symmetry and no charge. We will for the moment assume CP conservation. There are two relevant operators:
\begin{align}
&O_{1}=\partial_{\mu}\left(\bar{f} \gamma_{\nu} f\right) \partial^{\mu}\left(\bar{f} \gamma^{\nu} f\right), \quad O_{2}=\partial_{\mu}\left(\bar{f}^{c} f\right) \partial^{\mu}\left(\bar{f}^{c} f\right) \\
&\mathcal{L}_{\mathrm{EFT}}=\frac{C_{1}}{\Lambda^{4}} O_{1}+\left(\frac{C_{2}}{\Lambda^{4}} O_{2}+h . c .\right)
\end{align}
The operator \(O_{2}\) is not Hermitian. CP-conservation requires that \(C_{2}\) is real. The \(\mathcal{M}^{i j k l}\) can be computed straightforwardly:
\begin{equation}
\mathcal{M}^{i j k l}=-4 \times \begin{array}{|c|c|}
\hline C_{1} & 4 C_{2} \\
\hline 4 C_{2} & C_{1} \\
\hline
\end{array}\label{eq:4.58}
\end{equation}

To construct the allowed parameter space, we first write the \(J_{3}=0\) and \(J_{3}=\pm 1\) generators. There is only one flavor so \(m_{A}\) does not exist. We have
\begin{flalign}
 \raisebox{-5pt}{$m_{S}=$ }
\begin{tabular}{ r|c|c| }
	\multicolumn{1}{r}{}
	 &  \multicolumn{1}{c}{$f_a$}
	 & \multicolumn{1}{c}{$\bar{f}_a$} \\
	\cline{2-3}
	$f_a$ & $x$ &   \\
	\cline{2-3}
	$\bar{f}_a$ &   & $y$\\
	\cline{2-3}
\end{tabular},\quad
\raisebox{-5pt}{$m_{+}=$ }
\begin{tabular}{ r|c|c| }
	\multicolumn{1}{r}{}
	 &  \multicolumn{1}{c}{$f_a$}
	 & \multicolumn{1}{c}{$\bar{f}_a$} \\
	\cline{2-3}
	$f_a$ & &  $z$ \\
	\cline{2-3}
	$\bar{f}_a$ &   & \\
	\cline{2-3}
\end{tabular}\label{eq:4.59}
\end{flalign}
In the following we will omit the row/column heading \(f_{a}\) and \(\bar{f}_{a}\). From these we can write down the generators for \(\mathcal{M}^{i j k l}\), up to normalization:
\begin{equation}
\mathcal{G}_{S}=\begin{array}{|c|c|}
\hline 1 & 2 \rho \\
\hline 2 \rho^{*} & 1 \\
\hline
\end{array}, \quad\quad \mathcal{G}_{+}=\begin{array}{|l|l|}
\hline 1 & 0 \\
\hline 0 & 1 \\
\hline
\end{array}
\end{equation}
where we have defined \(\rho=\frac{2 x y^{*}}{|x|^{2}+|y|^{2}},|\rho| \in[-1,1] .|\rho|=1\) if \(|x|=|y|\), which does hold at the tree level. CP symmetry implies \(\mathcal{G}_{S}\) is symmetric (as \(\left.\mathcal{M}^{i j k l}=\mathcal{M}^{k l i j}\right)\), and so \(\rho\) is real. Comparing with Eq.~(\ref{eq:4.58}), we find the following two generator vectors in the parameter space:
\begin{equation}
\vec{g}_{1}(\rho)=-(1, \rho / 2), \quad \vec{g}_{2}=-(1,0)
\end{equation}
While \(\vec{g}_{1}\) is a function of \(\rho\), it is extremal if \(\rho=\pm 1\), and so the positivity cone is generated by \(\vec{g}_{1 \pm}=-(2, \mp 1)\) and \(\vec{g}_{2}=-(1,0)\). Its boundary is:
\begin{equation}
-C_{1} \pm 2 C_{2} \geq 0
\end{equation}

The next step is to map the generators with the simple UV completions. Consider a real CP-even heavy scalar \(S\), a real CP-odd scalar \(A\), and a real vector \(V^{2}\) \footnote{We assumed $f$ carries no charge, so $S$ and $V$ being real or complex won’t qualitatively change the result.}:
\begin{equation}
\begin{array}{cccccc}
\text { State } & \text { Spin }&\text{CP} & \text { Interaction } & \text { ER } & \vec{c} \\
\hline S & 0 & +1 & S\left(\bar{f}^{c} f+\bar{f} f^{c}\right) & \text{\Checkmark} & (-2,1) \\
A & 0 & -1 & i A\left(\bar{f}^{c} f-\bar{f} f^{c}\right) & \text{\Checkmark} & (-2,-1) \\
V & 1 & +1 & V^{\mu} \bar{f} \gamma_{\mu} f & \text{\XSolidBrush} & (-1,0)
\end{array}
\end{equation}
which correspond to \(\vec{g}_{1+}, \vec{g}_{1-}\) and \(\vec{g}_{2}\) respectively.

\subsubsection{CP-violation}
\label{sec4.3.2}
We now relax the CP-conservation requirement in the previous example. This will illustrate how CP-violation is accommodated in the present formalism, and also serves as another example with an infinite number of generators. One additional operator needs to be added:
\begin{equation}
\mathcal{L}_{\mathrm{EFT}}=\frac{C_{1}}{\Lambda^{4}} O_{1}+\frac{C_{2}}{\Lambda^{4}}\left(O_{2}+O_{2}^{\dagger}\right)+\frac{C_{3}}{\Lambda^{4}} i\left(O_{2}-O_{2}^{\dagger}\right)
\end{equation}
The new coefficient \(C_{3}\) is CP-violating. The \(\mathcal{M}^{i j k l}\) now becomes
\[
\mathcal{M}^{i j k l}=-4 \times \begin{array}{|c|c|}
\hline C_{1} & 4\left(C_{2}+i C_{3}\right) \\
\hline 4\left(C_{2}-i C_{3}\right) & C_{1} \\
\hline
\end{array}
\]

The generators \(G_{S}\) and \(G_{+}\) do not change, except that we no long require \(\rho\) to be real. By comparing with the amplitude, the generator vectors are
\begin{equation}
\vec{g}_{1}(\rho)=-(1, \Re \rho / 2, \Im \rho / 2), \quad \vec{g}_{2}=-(1,0,0)
\end{equation}
Recall that \(|\rho| \leq 1\). When \(|\rho|=1, \vec{g}_{1}(\rho)\) is extremal. Let \(\rho=e^{i \phi}\). As \(\phi\) change in \([0,2 \pi)\), \(\vec{g}_{1}\left(e^{i \phi}\right)\) circulates around \(\vec{g}_{2}\) and carves out a circular cone. The situation is very similar to vector-boson interactions with P-violation, which we have discussed in Section 4.2.1. All other \(\vec{g}_{1}(\rho)\) with \(|\rho|<1\), as well as \(\vec{g}_{2}\), are contained in this circular cone. The expressions for the bounds are
\begin{equation}
C_{1} \leq 0, \quad C_{1}^{2}-4 C_{2}^{2}-4 C_{3}^{2} \geq 0
\end{equation}
The cone is shown in Figure \ref{fig6} right. The generators are also shown, where \(\vec{g}_{1}\) takes values at \(\rho=1, i,-1\).

The tree-level UV completions can be easily found for all the ERs:
\begin{equation}
\begin{array}{cccc}
\text { State } & \text { Interaction } & \text { ER } & \vec{c} \\
\hline S  & g_{S} S\left(\bar{f}^{c} f\right)+h.c. & \text{\Checkmark} & \left(-2, \cos \left(2 \arg g_{S}\right), \sin \left(2 \arg g_{S}\right)\right) \\
V &  g_{V} V^{\mu} \bar{f} \gamma_{\mu} f & \text{\XSolidBrush} & (-1,0,0)
\end{array}
\end{equation}
Note that the \(g_{S}\) coupling is either real or imaginary in CP-conserved theories. In these cases the CPV coefficient, \(C_{3}\), vanishes. Otherwise, \(C_{3}\) will be generated by integrating out \(S\).

\subsubsection{SM leptons and \(\operatorname{SU}(2)\)}
\label{sec4.3.3}
We proceed to consider the SM fermions. There are two main differences with respect to the previous single flavor example. The first is that all SM fermions carry nonzero hypercharge. The second is that all SM fermions, except for the right-handed electron, are charged under non-abelian gauge symmetries, \(\mathrm{U}(2), \mathrm{U}(3)\), or \(\mathrm{U}(2) \times \mathrm{U}(3)\).

Let us first consider the right-handed electron \(e\). There is only one self-quartic operator, which is the \(O_{1}\) defined in the previous example, with \(f \rightarrow e \). \(O_{2}\) would violate hypercharge conservation. The \(\mathcal{M}^{i j k l}\) simply a diagonal matrix:
\[
\mathcal{M}^{i j k l}=-4 \times \begin{array}{|l|l|}
\hline C_{1} & \\
\hline & C_{1} \\
\hline
\end{array}
\]
In fact, the fact that \(M^{i j k l}\) is diagonal holds for all charged fermions, and so in the following we will omit the helicity structure and simply write \(\mathcal{M}^{i j k l}=-4\).

The \(m_{S}\) matrix in Eq.~(\ref{eq:4.59}) breaks into two pieces, as \(x\) and \(y\) components correspond to different hypercharges.
\begin{flalign}
 \nonumber \raisebox{-5pt}{$m_{S}=$ }
\begin{tabular}{ r|c|c| }
	\multicolumn{1}{r}{}
	 &  \multicolumn{1}{c}{$f_a$}
	 & \multicolumn{1}{c}{$\bar{f}_a$} \\
	\cline{2-3}
	$f_a$ & $x$ &   \\
	\cline{2-3}
	$\bar{f}_a$ &   &  \\
	\cline{2-3}
\end{tabular},\quad
\raisebox{-5pt}{$m_{\bar{S}}=$ }
\begin{tabular}{ r|c|c| }
	\multicolumn{1}{r}{}
	 &  \multicolumn{1}{c}{$f_a$}
	 & \multicolumn{1}{c}{$\bar{f}_a$} \\
	\cline{2-3}
	$f_a$ & 0 &  \\
	\cline{2-3}
	$\bar{f}_a$ &   & $y$\\
	\cline{2-3}
\end{tabular},\quad
\raisebox{-5pt}{$m_{+}=$ }
\begin{tabular}{ r|c|c| }
	\multicolumn{1}{r}{}
	 &  \multicolumn{1}{c}{$f_a$}
	 & \multicolumn{1}{c}{$\bar{f}_a$} \\
	\cline{2-3}
	$f_a$ &   &  $z$ \\
	\cline{2-3}
	$\bar{f}_a$ &   &  \\
	\cline{2-3}
\end{tabular}
\end{flalign}
We have argued that \(m_{S}\) and \(m_{\bar{S}}\) will eventually give rise to the same generator. In fact, they would both give the same \(\mathcal{G}_{S}\) but with \(\rho=0 .\) So \(\mathcal{G}_{S}\) and \(\mathcal{G}_{+}\) are both trivial identity matrices. This simply implies that the positivity bound is \(C_{1} \leq 0 .\) The one-dimensional parameter space has no further geometric features to be explored.

Now consider the left-handed lepton doublet \(l\), which lives in the fundamental rep of \(\mathrm{SU}(2)\). In this example, we will always present the full form of \(\mathcal{M}^{i j k l}\), to illustrate how internal symmetries can be taken into account in our formalism. For the rest of the paper we will again omit the helicity structure of \(\mathcal{M}^{i j k l}\).

Consider the \(J_{3}=0\) generator, from \(X \rightarrow l_{a} l_{b}\) and \(X \rightarrow \bar{l}_{a} \bar{l}_{b}\), where \(X\) is some intermediate state. While the \(m_{S}\) matrix is the same as the previous example, the fermions now carry \(\mathrm{SU}(2)\) indices \(a, b\), and need to be taken care of. The \(X\) state may exist in the \(\mathbf{1}\) and the \(\mathbf{3}\) reps of the \(\mathrm{SU}(2)\) symmetry. The \(X \rightarrow l_{a} l_{b}\left(\bar{l}_{a} \bar{l}_{b}\right)\) transitions are described by the corresponding CG coefficients. We may write
\begin{flalign}
\nonumber \raisebox{-5pt}{$m^{\alpha}_{(\textbf{r},Y=-1)}=$ }
\begin{tabular}{ r|c|c| }
	\multicolumn{1}{r}{}
	 &  \multicolumn{1}{c}{$l^b$}
	 & \multicolumn{1}{c}{$\bar{l}_b$} \\
	\cline{2-3}
	$l^a$ &  $C_{\textbf{r},\alpha}^{ab}$ &  \\
	\cline{2-3}
	$\bar{l}_a$ &   &  \\
	\cline{2-3}
\end{tabular}
\end{flalign}
where \(\alpha\) labels the states of the \(\textbf{r}\) rep. \(Y\) labels the hypercharge of \(X\). The charge conjugate \(\bar{X}\) with \(Y=1\) does not generate a independent contribution. When constructing the generators of \(\mathcal{M}^{i j k l}\), one sums over all \(\alpha\) states. This leads to
\begin{flalign}
	\nonumber\raisebox{0pt}{$\mathcal{G}_{(\mathbf{r}, Y=-1)}^{i j k l} =$}&
\begin{tabular}{ r|c|c|c|c| }
	\multicolumn{1}{r}{}
	 &  \multicolumn{1}{c}{${l_{c} l_{d}}$}
	 & \multicolumn{1}{c}{${\bar{l}^{c} \bar{l}^{d}}$}
     &  \multicolumn{1}{c}{$ {l_{c} \bar{l}^{d}}$}
	 & \multicolumn{1}{c}{${\bar{l}^{c} l_{d}}$}  \\
	\cline{2-5}
	$l^{a} l^{b}$&  $P_{\mathbf{r}}^{a b}{ }_{c d}$ &    &    &   \\
	\cline{2-5}
	$\bar{l}_{a} \bar{l}_{b}$&     & $P_{\mathbf{r}}^{c d}{ }_{a b}$  &   &    \\
	\cline{2-5}
    $l^{a} \bar{l}_{b}$ &    &   & ${P_{\mathbf{r}}}^{a d}_{\ \ c b} $  &     \\
	\cline{2-5}
	$\bar{l}_{a} l^{b}$&   &   &    & $P_{\mathbf{r}}^{c b}{ }_{a d}$  \\
	\cline{2-5}
\end{tabular}
\end{flalign}
Note that the sequence of the indices in each block is different, as they are  from crossing. Expressions for the projectors can be found in Appendix A, Eq.~(\ref{eq:A.2}).

The \(J_{3}=\pm 1\) case is similar. The intermediate state is again in 1 or \(\mathbf{3}\), but instead of \(\mathbf{2} \otimes \mathbf{2}=\mathbf{1} \oplus \mathbf{3}\) we have \(\mathbf{2} \otimes \overline{\mathbf{2}}=\mathbf{1} \oplus \mathbf{3}\), so the CG coefficients and the projectors are different. The \(m\) matrix is
\begin{flalign}
\nonumber \raisebox{-5pt}{$m^{\alpha}_{(\textbf{r},Y=0)}=$ }
\begin{tabular}{ r|c|c| }
	\multicolumn{1}{r}{}
	 &  \multicolumn{1}{c}{$l^b$}
	 & \multicolumn{1}{c}{$\bar{l}_b$} \\
	\cline{2-3}
	$l^a$ &    & ${(C_{\textbf{r},\alpha})}^{a}_{\ b}$ \\
	\cline{2-3}
	$\bar{l}_a$ &   &  \\
	\cline{2-3}
\end{tabular}
\end{flalign}
and this leads to the following generator
\begin{flalign}
	\nonumber\raisebox{0pt}{$G_{(\mathbf{r}, Y=0)} =$}&
\begin{tabular}{ r|c|c|c|c| }
	\multicolumn{1}{r}{}
	 &  \multicolumn{1}{c}{${l_{c} l_{d}}$}
	 & \multicolumn{1}{c}{${\bar{l}^{c} \bar{l}^{d}}$}
     &  \multicolumn{1}{c}{$ {l_{c} \bar{l}^{d}}$}
	 & \multicolumn{1}{c}{${\bar{l}^{c} l_{d}}$}  \\
	\cline{2-5}
	$l^{a} l^{b}$&  $P_{\mathbf{r}}^{a\ \ b}{ }_{\ dc}$ &    &    &   \\
	\cline{2-5}
	$\bar{l}_{a} \bar{l}_{b}$&     & $P_{\mathbf{r}}^{c\ \ d}{ }_{\ ba}$  &   &    \\
	\cline{2-5}
    $l^{a} \bar{l}_{b}$ &    &   & ${P_{\mathbf{r}}}^{a\ \ d}_{\ bc} $  &     \\
	\cline{2-5}
	$\bar{l}_{a} l^{b}$&   &   &    & $P_{\mathbf{r}}^{c\ \ b}{ }_{\ da}$  \\
	\cline{2-5}
\end{tabular}
\end{flalign}

We have in total 4 generators: \(\mathbf{1}_{1}, \mathbf{3}_{1}, \mathbf{1}_{0}, \mathbf{3}_{0}\), and the allowed values of \(\mathcal{M}^{i j k l}\) is positively generated by all four of them. Looking at these generators, again we see that only the first \(2 \times 2\) block is independent. In addition, the \(l l \rightarrow l l\) and \(\bar{l}\bar{ l} \rightarrow \bar{l}\bar{ l}\) entries are always equal, as \(P_{\mathbf{r}}^{a b}{ }_{c d}=P_{\mathbf{r}}^{c d}{ }_{a b}\) and \({P_{\mathbf{r}}}^{a\ \ b}_{\ dc}=P_{\mathbf{r}}{ }^{c}{ }_{b a}{ }^{d}\). Therefore, we may again omit the helicity structure and keep only the gauge structure, by writing\footnote{The \(i\) index runs through two helicities \((l, \bar{l})\) and two gauge components \(a=1,2: i \in(l, \bar{l}) \otimes(1,2)\), but if we omit the helicity structure, \(i\) should be understood as equivalent to \(a\). Same for \(j, k, l\) and \(b, c, d\).}
\begin{equation}
\mathcal{G}_{\textbf{1}_{1}}^{i j k l}=P_{\mathbf{1}}^{a b}{ }_{c d}, \quad \mathcal{G}_{\textbf{3}_{1}}^{i j k l}=P_{\mathbf{3}}^{a b}{ }_{c d}, \quad \mathcal{G}_{\textbf{1}_{0}}^{i j k l}={P_{\textbf{1}}{ }^{a\ \ b}_{\ d c}}, \quad \mathcal{G}_{\textbf{3}_{0}}^{i j k l}=P_{\mathbf{3}}{ }_{\ dc}^{a\ \ b}
\end{equation}
More generally, if \(f f \rightarrow \bar{f} \bar{f}\) is forbidden, which is true in the present example due to nonzero hypercharge, the helicity structure in \(\mathcal{M}\) is trivial, and one can always adopt the above simplification, and focus only on the \(f f \rightarrow f f\) entry.

We are now ready to construct the parameter space. There are two relevant operators:
\begin{equation}
O_{1}=\partial_{\mu}\left(\bar{l} \gamma_{\nu} l\right) \partial^{\mu}\left(\bar{l} \gamma^{\nu} l\right), \quad O_{2}=\partial_{\mu}\left(\bar{l} \gamma_{\nu} \tau^{I} l\right) \partial^{\mu}\left(\bar{l} \gamma^{\nu} \tau^{I} l\right)
\end{equation}
and the amplitude \(\mathcal{M}\) is (again omitting the helicity structure):
\begin{equation}
\mathcal{M}^{i j k l}=-4\left[\left(C_{1}-C_{2}\right) \delta_{d}^{a} \delta_{c}^{b}+2 C_{2} \delta_{c}^{a} \delta_{d}^{b}\right]
\end{equation}
This allows to map the four generators to the space of Wilson coefficients, \(\vec{g}=\left(C_{1}, C_{2}\right)\) :
\begin{equation}
\begin{array}{ll}
\vec{g}_{\textbf{1}_{1}}=(1,-1) & \vec{g}_{\textbf{1}_{0}}=(-1,0) \\
\vec{g}_{\textbf{3}_{1}}=(-3,1) & \vec{g}_{\textbf{3}_{0}}=(0,-1)
\end{array}
\end{equation}
Among them, \(\vec{g}_{\textbf{3}_{1}}\) and \(\vec{g}_{\textbf{3}_{0}}\) are not extremal. The cone is generated by \(\vec{g}_{\textbf{1}_{1}}\) and \(\vec{g}_{\textbf{1}_{0}}\), and the bounds are
\begin{equation}
C_{1}+C_{2} \leq 0, \quad C_{2} \leq 0
\end{equation}
The plot will be shown in Section 5.2.1, Figure 8, where we use the left-handed leptons as an example to discuss the inverse problem.

The last step is to map the 4 generators to simple UV completions:
\begin{equation}
\begin{tabular}{cccccc}
 { State}& Spin& Charge  & Interaction & ER & $\vec{c}$ \\
\hline $\mathcal{B}_{1}$ & 1 & $1_{1}$ & $\mathcal{B}_{1}^{\mu}\left(\bar{l}^{c} i \stackrel{\leftrightarrow}{D}_{\mu} l\right)$ & $\text{\Checkmark}$ & $\frac{1}{2}(1,-1)$ \\
$\Xi_{1}$ & 0 & $3_{1}$ & $\Xi_{1}^{I}\left(\bar{l}^{c} \tau^{I} l\right)$ & $\text{\XSolidBrush}$ & $\frac{1}{2}(-3,-1)$ \\
$\mathcal{B}$ & 1 & $1_{0}$ & $\mathcal{B}^{\mu}\left(\bar{l} \gamma_{\mu} l\right)$ & $\text{\Checkmark}$ & $\frac{1}{2}(-1,0)$ \\
$\mathcal{W}$ & 1 & $3_{0}$ & $\mathcal{W}^{I \mu}\left(\bar{l} \gamma_{\mu} \tau^{I} l\right)$ & $\text{\XSolidBrush}$ & $\frac{1}{2}(0,-1)$
\end{tabular}
\end{equation}
We will not show the explict factors of coupling \(g\) and mass \(M\) anymore in the interaction terms. Note that \(\mathcal{B}_{1}\) has a dim-5 effective couplings to the SM leptons. It is ok to think of this as a UV completion of the generator \(\vec{g}_{\textbf{1}_{1}}\), because this dim-5 coupling does not generate a boundary term in the dispersion relation, but we should keep in mind that it needs be further UV completed. This kind of couplings will also appear in the next few sections. See Section 4.3.5 for some more discussion.

\subsubsection{SM quarks, \(\mathrm{SU}(3)\) and \(\mathrm{SU}(2) \times \mathrm{SU}(3)\)}
\label{sec4.3.4}
\begin{figure}[h]
	\begin{center}
		\includegraphics[width=.5\linewidth]{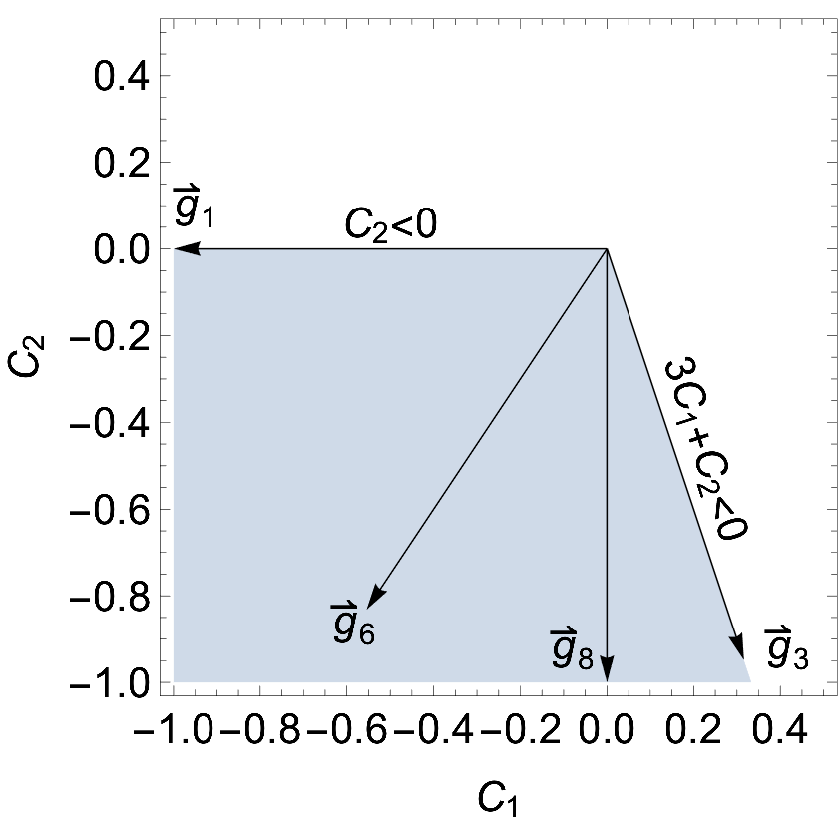}
	\end{center}
	\caption{Positivity cone for right-handed up/down-quark operators, together with the generators.}
	\label{fig7}
\end{figure}

The right-handed quarks \(u\) and \(d\) are charged under SU(3) and carry nonzero hypercharges. The situation is the same as the lepton doublet example. The only difference is the projectors. For the \(J_{3}=0\) generators, we have \(\mathbf{3} \otimes \mathbf{3}=\overline{\mathbf{3}} \oplus \mathbf{6}\), while for \(J_{3}=\pm 1\), we have \(\mathbf{3} \otimes \overline{\mathbf{3}}=\mathbf{1} \oplus \mathbf{8}\). The corresponding projectors are presented in Appendix A, Eqs.~(\ref{eq:A.4}) and ~(\ref{eq:A.5}). The rest is in completely analogy to the previous example, and we only show some key results. For the \(u\) quark, the operators are
\begin{equation}
O_{1}=\partial_{\mu}\left(\bar{u} \gamma_{\nu} u\right) \partial^{\mu}\left(\bar{u} \gamma^{\nu} u\right), \quad O_{2}=\partial_{\mu}\left(\bar{u} \gamma_{\nu} T^{A} u\right) \partial^{\mu}\left(\bar{u} \gamma^{\nu} T^{A} u\right)
\end{equation}
The amplitude is
\begin{equation}
\mathcal{M}^{i j k l}=-\frac{2}{3}\left(6 C_{1}-C_{2}\right) \delta_{d}^{a} \delta_{c}^{b}-2 C_{2} \delta_{c}^{a} \delta_{d}^{b}
\end{equation}
The generators written in terms of coefficients are
\begin{equation}
\begin{array}{ll}
\vec{g}_{\textbf{3}}=(1,-3) & \vec{g}_{\textbf{1}}=(-1,0) \\
\vec{g}_{\textbf{6}}=(-2,-3) & \vec{g}_{\textbf{8}}=(0,-1)
\end{array}
\end{equation}
The extremal ones are \(g_{\textbf{3}}\) and \(g_{\textbf{1}}\). The bounds are
\begin{equation}
3 C_{1}+C_{2} \leq 0, \quad C_{2} \leq 0
\end{equation}
And the simple UV completions are
\begin{equation}
\begin{array}{cccccc}
\text { State}& \text{Spin} &\text{Charge} & \text { Interaction } & E R & \vec{c} \\
\hline \mathcal{V}_{\frac{4}{3}} & 1 & (3,1)_{-\frac{4}{3}} & \mathcal{V}_{\frac{4}{3}}^{a \mu} \epsilon_{a b c} \bar{u}^{c b} i \stackrel{\leftrightarrow}{D}_{\mu} u^{c} & \text{\Checkmark} & \frac{2}{3}(1,-3) \\
\Omega_{4} & 0 & (6,1)_{\frac{4}{3}} & \Omega_{4}^{\dagger a b} \bar{u}^{c(a} u^{b)} & \text{\XSolidBrush} & \frac{1}{3}(-2,-3) \\
\mathcal{B} & 1 & (1,1)_{0} & \mathcal{B}^{\mu} \bar{u} \gamma_{\mu} u & \text{\Checkmark} & \frac{1}{2}(-1,0) \\
\mathcal{G} & 1 & (8,1)_{0} & \mathcal{G}^{A \mu} \bar{u} \gamma_{\mu} T^{A} u & \text{\XSolidBrush} & \frac{1}{2}(0,-1)
\end{array}
\end{equation}
and similarly for down-type quark:
\begin{equation}
\begin{array}{cccccc}
\text { State}&\text{ Spin } & \text { Charge } & \text { Interaction } & E R & \vec{c} \\
\hline \mathcal{V}_{\frac{2}{3}} & 1 & (3,1)_{\frac{2}{3}} & \mathcal{V}_{\frac{2}{3}}^{a \mu} \epsilon_{a b c} \bar{d}^{c b} i \stackrel{\leftrightarrow}{D}_{\mu} d^{c} & \text{\Checkmark} & \frac{2}{3}(1,-3) \\
\Omega_{2} & 0 & (6,1)_{-\frac{2}{3}} & \Omega_{2}^{\dagger a b} \bar{d}^{c(a} d^{b)} & \text{\XSolidBrush}& \frac{1}{3}(-2,-3) \\
\mathcal{B} & 1 & (1,1)_{0} & \mathcal{B}^{\mu} \bar{d} \gamma_{\mu} d & \text{\Checkmark} & \frac{1}{2}(-1,0) \\
\mathcal{G} & 1 & (8,1)_{0} & \mathcal{G}^{A \mu} \bar{d} \gamma_{\mu} T^{A} d & \text{\XSolidBrush} & \frac{1}{2}(0,-1)
\end{array}
\end{equation}
The positivity cone is shown in Figure  \ref{fig7}.

Lastly, the left-handed quark doublet \(q\) is charged under \(\mathrm{SU}(2) \times \mathrm{SU}(3)\). The relevant operators are
\begin{align}
&O_{1}=\partial_{\mu}\left(\bar{q} \gamma_{\nu} q\right) \partial^{\mu}\left(\bar{q} \gamma^{\nu} q\right) \\
&O_{2}=\partial_{\mu}\left(\bar{q} \gamma_{\nu} \tau^{I} q\right) \partial^{\mu}\left(\bar{q} \gamma^{\nu} \tau^{I} q\right) \\
&O_{3}=\partial_{\mu}\left(\bar{q} \gamma_{\nu} T^{A} q\right) \partial^{\mu}\left(\bar{q} \gamma^{\nu} T^{A} q\right) \\
&O_{4}=\partial_{\mu}\left(\bar{q} \gamma_{\nu} \tau^{I} T^{A} q\right) \partial^{\mu}\left(\bar{q} \gamma^{\nu} \tau^{I} T^{A} q\right)
\end{align}
The \(\mathcal{M}^{i j k l}\) in terms of Wilson coefficients are
\begin{equation}
\begin{aligned}
\mathcal{M}^{i j k l}=&-\frac{2}{3}\left[6 C_{4} \delta_{c}^{a} \delta_{d}^{b} \delta_{\gamma}^{\alpha} \delta_{\delta}^{\beta}+2\left(6 C_{2}-C_{4}\right) \delta_{c}^{a} \delta_{d}^{b} \delta_{\delta}^{\alpha} \delta_{\gamma}^{\beta}\right.\\
&\left.+3\left(C_{3}-C_{4}\right) \delta_{d}^{a} \delta_{c}^{b} \delta_{\gamma}^{\alpha} \delta_{\delta}^{\beta}+\left(6 C_{1}-6 C_{2}-C_{3}+C_{4}\right) \delta_{d}^{a} \delta_{c}^{b} \delta_{\delta}^{\alpha} \delta_{\gamma}^{\beta}\right]\label{eq:4.84}
\end{aligned}
\end{equation}
where we use \(a, b, c, d\) for the \(\mathrm{SU}(2)\) indices and \(\alpha, \beta, \gamma, \delta\) for the \(\mathrm{SU}(3)\) indices of \(i, j, k, l\) states respectively.

Now consider the generators. While the helicity structure is again trivial, the gauge group structure is determined by the direct product of \(\mathrm{SU}(2)\) and \(\mathrm{SU}(3)\) projectors:
\[
\begin{array}{ll}
J_{3}=0 & \quad J_{3}=\pm 1 \\
\mathcal{G}_{\textbf{1} \overline{\textbf{3}}}^{i j k l}=P_{\textbf{1}}^{\ a b}{ }_{c d} P_{\bar{\textbf{3}}}^{\ \alpha \beta}{ }_{\gamma \delta} & \quad \mathcal{G}_{\textbf{11}}^{i j k l}={P_{\textbf{1}}{ }^{a}{ }_{d c}{ }^{b} P_{\textbf{1}}{ }^{\alpha\ \ \beta}_{\ \delta \gamma}} \\
\mathcal{G}_{\textbf{3} \overline{\textbf{3}}}^{i j k l}=P_{\textbf{3}}{ }^{a b}{ }_{c d} P_{\overline{\textbf{3}}}{ }^{\alpha \beta}{ }_{\gamma \delta} & \quad \mathcal{G}_{\textbf{31}}^{i j k l}={P_{\textbf{3}}{ }^{a}{ }_{d c}{ }^{b} P_{\textbf{1}}{ }^{\alpha\ \ \beta}_{\ \delta \gamma}} \\
\mathcal{G}_{\textbf{16}}^{i j k l}={P_{\textbf{1}}}^{a b}{ }_{c d} P_{\textbf{6}}{ }^{\alpha \beta}{ }_{\gamma \delta} &\quad  \mathcal{G}_{\textbf{1} \mathbf{8}}^{i j k l}={P_{\textbf{1}}{ }^{a}{ }_{d c}{ }^{b} P_{\textbf{8}}{ }{ }^{\alpha\ \ \beta}_{\ \delta \gamma}} \\
\mathcal{G}_{\textbf{36}}^{i j k l}={P_{\textbf{3}}}^{a b}{ }_{c d} P_{\textbf{6}}{ }^{\alpha \beta}{ }_{\gamma \delta} & \quad \mathcal{G}_{\textbf{38}}^{i j k l}={P_{\textbf{3}}{ }^{a}{ }_{d c}{ }^{b} P_{\mathbf{8}}{ }{ }^{\alpha\ \ \beta}_{\ \delta \gamma}}
\end{array}
\]
By comparing these with Eq.~(\ref{eq:4.84}), we can write down the generators in terms of \(\vec{C}=\) \(\left(C_{1}, C_{2}, C_{3}, C_{4}\right):\)
\begin{equation}
\begin{array}{ll}
\vec{g}_{\textbf{1} \overline{\textbf{3}}}:(-1,1,3,-3) & \vec{g}_{\textbf{11}}:(-1,0,0,0) \\
\vec{g}_{\textbf{3} \overline{\textbf{3}}}:(3,1,-9,-3) & \vec{g}_{\textbf{31}}:(0,-1,0,0) \\
\vec{g}_{\textbf{16}}:(2,-2,3,-3) & \vec{g}_{\textbf{18}}:(0,0,-1,0) \\
\vec{g}_{\textbf{36}}:(-6,-2,-9,-3) & \vec{g}_{\textbf{38}}:(0,0,0,-1)
\end{array}
\end{equation}
Note that \(\vec{g}_{\textbf{36}}\) and \(\vec{g}_{\textbf{38}}\) are not extremal. The parameter space is a polyhedral cone whose edge vectors are the rest 6 generators. A VE gives the following facets:
\begin{equation}
\left(\begin{array}{llll}
3 & 3 & 1 & 1 \\
0 & 3 & 0 & 1 \\
0 & 0 & 1 & 1 \\
12 & 0 & 1 & 9 \\
1 & 0 & 0 & 1 \\
0 & 0 & 0 & 1
\end{array}\right)\left(\begin{array}{l}
C_{1} \\
C_{2} \\
C_{3} \\
C_{4}
\end{array}\right) \leq 0
\end{equation}
The plot of the cone is shown in Section 5.2.3, Figure \ref{fig15}, where we will use the left-handed quarks as an example to discuss the inverse problem. Finally, for mapping the generators to UV particles, we find
\begin{equation}
\begin{tabular}{cccccc}
State &Spin & Charge & Interaction & ER & \(\vec{c}\) \\
\hline
\(\omega_{1}\) & 0 & \((3,1)_{-\frac{1}{3}}\) & \(\omega_{1}^{a} \epsilon_{a b c} \bar{q^{c}}^{b} \epsilon q^{c}\) & $\text{\Checkmark}$ & \(\frac{1}{3}(-1,1,3,-3)\)  \\
\(\mathcal{V}_{-\frac{1}{3}}\) & 1 & \((3,3)_{-\frac{1}{3}}\) & \(\mathcal{V}_{-\frac{1}{3}}^{a I} \epsilon_{a b c} \bar{q^c}^b \epsilon \tau^{I} i \stackrel{\leftrightarrow}{D}_{\mu} q^{c}\) & $\text{\Checkmark}$ & \(\frac{1}{3}(3,1,-9,-3)\) \\
\(\mathcal{V}_{\frac{1}{3}}\) & 1 & \((6,1)_{\frac{1}{3}}\) & \(\mathcal{V}_{\frac{1}{3}}^{\dagger ab\mu}\bar{q^c}^{(a}\epsilon i \stackrel{\leftrightarrow}{D}_{\mu} q^{b)}\) & $\text{\XSolidBrush}$& \(\frac{1}{6}(2,-2,3,-3)\) \\
\(\Upsilon\) & 0 & \((6,3)_{\frac{1}{3}}\) & \(\Upsilon^{\dagger I a b} \bar{q^c}^{(a} \epsilon \tau^{I} q^{b)}\) & $\text{\Checkmark}$ & \(\frac{1}{6}(-6,-2,-9,-3)\) \\
\(\mathcal{B}\) & 1 & \((1,1)_{0}\) & \(\mathcal{B}^{\mu} \bar{q} \gamma_{\mu} q\) & $\text{\Checkmark}$ & \(\frac{1}{2}(-1,0,0,0)\)   \\
\(\mathcal{W}\) & 1 & \((1,3)_{0}\) & \(\mathcal{W}^{I \mu} \bar{q} \gamma_{\mu} \tau^{I} q\) & $\text{\Checkmark}$ & \(\frac{1}{2}(0,-1,0,0)\) \\
\(\mathcal{G}\) & 1 & \((8,1)_{0}\) & \(\mathcal{G}^{A \mu} \bar{q} \gamma_{\mu} T^{A} q\) &  $\text{\Checkmark}$ & \(\frac{1}{2}(0,0,-1,0)\) \\
\(\mathcal{H}\) & 1 & \((8,3)_{0}\) & \(\mathcal{H}^{A I \mu} \bar{q} \gamma_{\mu} T^{A} \tau^{I} q\) & $\text{\XSolidBrush}$ & \(\frac{1}{2}(0,0,0,-1)\) \\
\end{tabular}
\end{equation}

\subsubsection{Particle enumeration for fermions}
\label{sec4.3.5}
The previous example demonstrated that bounds for fermion scattering can be obtained by enumerating all UV particles that couple to two fermions, with the following three kinds of couplings:
\begin{equation}
\mathcal{L} \supset g_{i j k} S_{k} \bar{f}_{i}^{c} f_{j}+\frac{1}{M} h_{i j k} V_{k}^{\mu} \bar{f}_{i}^{c} \overleftrightarrow{D}_{\mu} f_{j}+h_{i j k}^{\prime} V_{k \mu}^{\prime} \bar{f}_{i} \gamma^{\mu} f_{j}+h . c .
\end{equation}
which correspond to the three \(m\) matrices in Eq.~(\ref{eq:4.55}), respectively. Therefore the ``particle enumeration'' method can be used for fermions, just like the scalar case. One should keep in mind that the \(h_{i j k} / M\) term in the above equation is not a real UV completion, but rather a dim-5 effective coupling. This coupling however does not generate a boundary term in the dispersion relation. The \(t\)-channel exchange of \(V\) only grows as \(s\) in the \(t \rightarrow 0\) limit. Therefore, integrating out \(V\) from the UV spectrum and then calculating \(\mathcal{M}^{i j k l}\) does give us the same generator as obtained from the dispersion relation. For this reason, the antisymmetric \(J_{3}=0\) generators can be thought of as coming from a heavy vector with a dim-5 effective coupling. Note that for massless fermions, we have
\begin{equation}
V_{k}^{\mu} \bar{f}_{i}^{c} \overleftrightarrow{D}_{\mu} f_{j}=-\frac{1}{2}\left(\bar{f}_{i}^{c} \sigma_{\mu \nu} f_{j}\right) V_{k}^{\mu \nu}
\end{equation}
so this effective coupling is simply a dipole interaction, which can be further UV completed by loops.

In summary, the generators can be constructed by enumerating all scalar, dipole, and vector couplings between two fermions. The limitation of this approach is the same as discussed in the scalar case.

\subsubsection{Two chiral fermions}
\label{sec4.3.6}
The SM fermions are chiral. In phenomenological studies, operators with both left- and right-handed fermions often need to be considered together. This increases the complexity of positivity problem. In particular, since the left-handed and right-handed fermions are charged under different groups and irreps, the resulting bounds are non-polyhedral with continuous ERs. In this section, we consider the SM leptons as an illustration.

Consider one flavor SM leptons of both left- and right-handed chirality. The number of independent operators at dim-8 is \(5 .\) We use the basis of Ref.~\cite{Li:2020gnx,Murphy:2020rsh} , which we copy here:
\begin{equation}
\begin{array}{lll}
O_{1} & =\partial^{\alpha}\left(\bar{e} \gamma^{\mu} e\right) \partial_{\alpha}\left(\bar{e} \gamma_{\mu} e\right), & O_{4}=\partial^{\alpha}\left(\bar{l} \gamma^{\mu} l\right) \partial_{\alpha}\left(\bar{l} \gamma_{\mu} l\right) \\
O_{2} & =\partial^{\alpha}\left(\bar{e} \gamma^{\mu} e\right) \partial_{\alpha}\left(\bar{l} \gamma_{\mu} l\right), & O_{5}=D^{\alpha}\left(\bar{l} \gamma^{\mu} \tau^{I} l\right) D_{\alpha}\left(\bar{l} \gamma_{\mu} \tau^{I} l\right) \\
O_{3} & =D^{\alpha}(\bar{l} e) D_{\alpha}(\bar{e} l), &
\end{array}
\end{equation}
The easiest approach is particle enumeration. The list of possible UV particles is:
\begin{equation}
\begin{tabular}{cccccc}
State& Spin& Charge & Interaction & \(\mathrm{ER}\) & \(\vec{c}\)    \\
\hline
\(\mathcal{S}_{2}\) & 0 & \(1_{2}\) & \(\mathcal{S}_{2} e^{-c} e\) & $\text{\Checkmark}$ & \((-1,0,0,0,0)\) \\
\(\varphi\) & 0 & \(2_{\frac{1}{2}}\) & \(\varphi^{\dagger} \bar{e} l\) & $\text{\Checkmark}$ & \((0,0,1,0,0)\) \\
\(\mathcal{L}\) & 1 & \(2_{\frac{1}{2}}\) & \(\mathcal{L}^{\dagger \mu} \bar{e} i \stackrel{\leftrightarrow}{D_{\mu}} l\) & $\text{\Checkmark}$ & \(\frac{1}{2}(0,0,0,-3,-1)\) \\
\(\mathcal{B}\) & 1 & \(1_{0}\) & \(\mathcal{B}_{\mu}\left(\bar{e} \gamma^{\mu} e+x \bar{l} \gamma^{\mu} l\right)\) & $\text{\Checkmark}$ & \(\frac{1}{2}\left(-1,2 x, 0,-x^{2}, 0\right)\) \\
$\Xi_{1} $ & 0 & $3_{1}$ & $\Xi_{1}^{I} \bar{l^c}\tau^{I} l $ & $\text{\XSolidBrush}$  & $\frac{1}{2}(0,0,0,-3,-1)$ \\
$\mathcal{B}$ & 1 & $1_{0}$ & $\mathcal{B}_{\mu}\left(\bar{e} \gamma^{\mu} e+{\color{red}x} \bar{l} \gamma^{\mu} l\right)$ & $\text{\Checkmark}$ & $\frac{1}{2}\left(-1,2 {\color{red}x}, 0,-{\color{red}x}^{2}, 0\right)$\\
\(\mathcal{W}\) & 1 & \(3_{0}\) & \(\mathcal{W}_{\mu}^{I} \bar{l} \gamma^{\mu} \tau^{I} l\) & $\text{\XSolidBrush}$ & \(\frac{1}{2}(0,0,0,0,-1)\) \\
\(\mathcal{L}_{3}\) & 1 & \(2_{-\frac{3}{2}}\) & \(\mathcal{L}_{3}^{\dagger} e^{c} \gamma^{\mu} l\) & $\text{\XSolidBrush}$ & \((0,-1,2,0,0)\)   \\
\end{tabular}\label{eq:4.91}
\end{equation}
Recall that the \(\vec{c}\) vectors are given by integrating out each state, and they are proportional to the generators \(\vec{g}\). Note that one of them depends on a free real parameter \(x\). A continuous VE needs to be done, similar to what we have done in Section 4.1.3. We find that the boundary is described by the following inequalities:
\begin{equation}
\begin{array}{ll}
C_{1} \leq 0, & C_{3} \geq 0 \\
C_{4}+C_{5} \leq 0, & 2 \sqrt{C_{1}\left(C_{4}+C_{5}\right)} \geq C_{2} \\
C_{5} \leq 0, & 2 \sqrt{C_{1}\left(C_{4}+C_{5}\right)} \geq-\left(C_{2}+C_{3}\right) .
\end{array}
\end{equation}
This result agrees with Ref.~\cite{Fuks:2020ujk}. The shape of this cone is similar to the example presented in Section 4.1.3, if we replace \(C_{4}+C_{5}\) here by \(C_{3}\), and \(C_{3}\) by \(C_{4}\). Therefore we do not plot the cone again, and the readers may simply refer to Figure \ref{fig4}.

For completeness, let's also present the direct construction of the generators. The amplitude \(\mathcal{M}^{i j k l}\) can be displayed as follows (note that \(\bar{e}\) and \(l\) are both left-handed):
\begin{flalign}
\nonumber \begin{tabular}{ r|c|c| }
	\multicolumn{1}{r}{}
	 &  \multicolumn{1}{c}{$\bar{e} \bar{e}, \bar{e} l, l \bar{e}, l l$}
	 & \multicolumn{1}{c}{$e e, e \bar{l}, \bar{l} e, \bar{l}\bar{ l}$} \\
	\cline{2-3}
	$\bar{e} \bar{e}, \bar{e} l, l \bar{e}, l l$ & \(\mathcal{M}_{\boxplus}\)   &   \\
	\cline{2-3}
	$e e, e \bar{l}, \bar{l} e, \bar{l} \bar{l}$ &   & \(\mathcal{M}_{\boxplus}\) \\
	\cline{2-3}
\end{tabular}
\end{flalign}
Here, the two off-diagonal entries vanish due to hypercharge conservation. The diagonal entries are denoted by \(\mathcal{M}_{\boxplus}\) and \(\mathcal{M}_{\boxplus}\). They are connected by \(s \leftrightarrow u\) crossing symmetry and are not independent. In the following, we will only show \(\mathcal{M}_{\boxplus}\) for simplicity.

First, the amplitude in terms of coefficients are given by
\begin{flalign}
	\raisebox{0pt}{$\mathcal{M}_{\boxplus}=$}&
\begin{tabular}{ r|c|c|c|c| }
	\multicolumn{1}{r}{}
	 &  \multicolumn{1}{c}{${\bar{e}\bar{e}}$}
	 & \multicolumn{1}{c}{${\bar{e} l_d}$}
     &  \multicolumn{1}{c}{$ {l_{c} \bar{e}}$}
	 & \multicolumn{1}{c}{${{l}_{c} l_{d}}$}  \\
	\cline{2-5}
	${\bar{e}\bar{e}}$&  $-4C_1$ &    &    &   \\
	\cline{2-5}
	${\bar{e} l^b}$&     & $C_3 \delta^b_d$  & $2(C_2+C_3)\times \delta^b_c$  &    \\
	\cline{2-5}
    $ {l^{a} \bar{e}}$ &    & $2(C_2+C_3)\times \delta^a_d$   & $C_3 \delta^a_c$  &     \\
	\cline{2-5}
	${{l}^{a} l^{b}}$&   &   &    & $-4\left[(C_4-C_5)\delta^a_d \delta^b_c+ 2C_5\delta^a_c\delta^b_d\right]$  \\
	\cline{2-5}
\end{tabular}\label{eq:4.93}
\end{flalign}
Here, different \(\mathrm{SU}(2)\) components of \(l\) are not shown explicitly as separate entries, but are rather indicated by the \(a, b, c, d\) gauge indices. Next, we construct the parameter space from the generators. For \(J_{3}=0\) :
\begin{flalign}
	\nonumber\raisebox{0pt}{$\left\{\colorbox{Pink}m\left[\mathbf{1}_{2}\right], \colorbox{PeachPuff}m\left[\mathbf{2}_{\frac{1}{2} S, A}\right]^{\alpha}, \colorbox{Khaki}m\left[\mathbf{1}_{-1}\right], \colorbox{Khaki}m\left[\mathbf{3}_{-1}\right]^{\alpha}\right\}=$}&
\begin{tabular}{ r|c|c|c|c| }
	\multicolumn{1}{r}{}
	 &  \multicolumn{1}{c}{$\bar{e}$}
	 & \multicolumn{1}{c}{$l^b$}
     &  \multicolumn{1}{c}{$e$}
	 & \multicolumn{1}{c}{$\bar{l}_b$}  \\
	\cline{2-5}
	$\bar{e}$& \cellcolor{Pink} 1  &  \cellcolor{PeachPuff} $\delta^b_{\alpha}$  &    &   \\
	\cline{2-5}
	$l^a$&    \cellcolor{PeachPuff}$\pm \delta^a_{\alpha}$  & \cellcolor{Khaki} $C^{ab}_{\textbf{r},\alpha}$  &   &    \\
	\cline{2-5}
    $e$ &    &   &    &     \\
	\cline{2-5}
	$\bar{l}_{a}$&   &   &    &    \\
	\cline{2-5}
\end{tabular}\\\mathcal{G}_{J_{3}=0,\boxplus}
\begin{tabular}{ r|c|c|c|c| }
	\multicolumn{1}{r}{}
	 &  \multicolumn{1}{c}{$\bar{e}\bar{e}$}
	 & \multicolumn{1}{c}{$\bar{e}l_d$}
     &  \multicolumn{1}{c}{$l_c\bar{e}$}
	 & \multicolumn{1}{c}{$l_c l_d$}  \\
	\cline{2-5}
	$\bar{e}\bar{e}$&  \cellcolor{Pink}1  &     &    &   \\
	\cline{2-5}
	$\bar{e} l^b$&      &  \cellcolor{PeachPuff} $\delta^b_{d}$  &   \cellcolor{PeachPuff}$\pm \delta^b_{c}$ &    \\
	\cline{2-5}
    $l^a \bar{e}$ &    & \cellcolor{PeachPuff}$\pm \delta^a_{d}$  &   \cellcolor{PeachPuff}$\delta^a_{c}$  &     \\
	\cline{2-5}
	${l}_{a} l^b$&   &   &    &\cellcolor{Khaki} ${P_{\textbf{r}}}^{ab}_{\ \ cd}$   \\
	\cline{2-5}
\end{tabular}\nonumber
\end{flalign}
and for \(J_{3}=-1\) :
\begin{flalign}
	\nonumber\left\{\colorbox{Aquamarine}m[\mathbf{1}_{0}],\colorbox{SkyBlue} m\left[\mathbf{2}_{-\frac{3}{2}}\right]^{\alpha}, \colorbox{LightSteelBlue}m\left[\bar{\mathbf{2}}_{\frac{3}{2}}\right]^{\alpha}, \colorbox{Plum}m\left[\mathbf{3}_{0}\right]^{\alpha}\right\}=&
\begin{tabular}{ r|c|c|c|c| }
	\multicolumn{1}{r}{}
	 &  \multicolumn{1}{c}{$\bar{e}$}
	 & \multicolumn{1}{c}{$l^b$}
     &  \multicolumn{1}{c}{$e$}
	 & \multicolumn{1}{c}{$\bar{l}_b$}  \\
	\cline{2-5}
	$\bar{e}$&    &     & \cellcolor{Aquamarine}1   &  \cellcolor{LightSteelBlue}$\delta_b^{\alpha}$ \\
	\cline{2-5}
	$l^a$&      &     & \cellcolor{SkyBlue}$\delta^a_{\alpha}$  &    \\
	\cline{2-5}
    $e$ &    &   &    &     \\
	\cline{2-5}
	$\bar{l}_{a}$&   &   &    &    \\
	\cline{2-5}
\end{tabular}\\\mathcal{G}_{J_{3}=0,\boxplus}
\begin{tabular}{ r|c|c|c|c| }
	\multicolumn{1}{r}{}
	 &  \multicolumn{1}{c}{$\bar{e}\bar{e}$}
	 & \multicolumn{1}{c}{$\bar{e}l_d$}
     &  \multicolumn{1}{c}{$l_c\bar{e}$}
	 & \multicolumn{1}{c}{$l_c l_d$}  \\
	\cline{2-5}
	$\bar{e}\bar{e}$&  \cellcolor{Aquamarine} 1  &     &   &   \\
	\cline{2-5}
	$\bar{e} l^b$&      &  \cellcolor{LightSteelBlue} $\delta^b_{d}$  &  \cellcolor{Aquamarine}$x\delta^b_{c}$ &    \\
	\cline{2-5}
    $l^a \bar{e}$ &    & \cellcolor{Aquamarine}$x\delta^a_{d}$  & \cellcolor{LightSteelBlue} $\delta^a_{c}$  &     \\
	\cline{2-5}
	${l}_{a} l^b$&   &   &    & \cellcolor{Aquamarine}$x^2 \delta^a_{d} \delta^b_{c}$   \\
      &   &   &    &\cellcolor{Plum} $2{P_{\textbf{3}}}^{a\ \ b}_{\ dc}$   \\
	\cline{2-5}
\end{tabular}\nonumber
\end{flalign}

Here, for simplicity, we use different colors to distinguish the contributions from different irreps in \(m\). The terms with different background colors in \(\mathcal{G}\) correspond to different generators. Note that \(\mathcal{G}_{J_{3}=-1, \boxplus}\) is a quadratic matrix function of a free parameter \(x\), which comes from the degeneracy between \(\bar{e} e\) and \(\bar{l} l\), both can form a singlet. This \(x\) is exactly the same \(x\) in Eq.~(\ref{eq:4.91}), appeared as the relative coupling strength between \(\mathcal{B} \bar{e} e\) and \(\mathcal{B} \bar{l} l\).

By comparing the above result with the expression in Eq.~(\ref{eq:4.93}), the same generators as shown in Eq.~(\ref{eq:4.91}) can be obtained.

\subsection{Collection of all SM self-quartic bounds}
\label{sec4.4}
We have seen how the extremal positivity approach can be used to derive the boundary of allowed EFTs in different examples. The approach proceeds by first enumerating the generators that could potentially appear on the r.h.s.~of the dispersion relation, and then finding their conical hull. The bounds can be extracted by a VE on the extremal generators. The generators can be mapped to amplitudes from a heavy particle exchange in a tree-level UV completion (except for the spin-2 ones in vector boson scatterings), which means that the resulting bounds are the tightest, as any further improvement would rule out these tree-level UV completions. For scalars and fermions, one could also use the ``particle enumeration'' approach, to directly obtain the generators in the space of coefficients, assuming a tree-level operator-amplitude mapping is used. Finally, the more symmetries the theory possesses, the easier the enumeration of generators.

Our examples complete the bounds on the SMEFT self-quartic P-conserving operators. Some of these bounds have been presented in, {\it e.g.}, Refs.~\cite{Remmen:2019cyz, Remmen:2020vts} using generalized elastic scattering. These results turn out to be complete for the Higgs boson, the hypercharge boson, the right-handed electron, up/down-quark, the left-handed lepton, but are incomplete for the \(W\)-boson, the gluon and the quark doublets. Bounds on \(W\)-boson operators were first completed in Ref.~\cite{Zhang:2020jyn}, while those on gluon operators were first completed in Ref.~\cite{Li:2021cjv}. Bounds on quark doublets have been studied in Ref.~\cite{Trott:2020ebl}, where the results are complete but have not been matched to a standard operator basis. Section 4.3.4 completes this matching. The completeness of elastic bounds on the other SM operators have been confirmed in Refs. \cite{Zhang:2020jyn, Li:2021cjv, Yamashita:2020gtt, Fuks:2020ujk, Trott:2020ebl} and in this work.

For completeness, here we give the exact bounds on all parity-conserving SM quartic operators. We follow the basis of Ref.~\cite{Li:2020gnx,Murphy:2020rsh}. The basis operators are given in the table below
\begin{equation}
\begin{array}{||l||l||}
\hline \text { Higgs } & \text { Gluon } \\
\hline Q_{H^{4}}^{(1)}=\left(D_{\mu} H^{\dagger} D_{\nu} H\right)\left(D^{\nu} H^{\dagger} D^{\mu} H\right) & Q_{G^{4}}^{(1)}=\left(G_{\mu \nu}^{A} G^{A \mu \nu}\right)\left(G_{\rho \sigma}^{B} G^{B \rho \sigma}\right) \\
\hline Q_{H^{4}}^{(2)}=\left(D_{\mu} H^{\dagger} D_{\nu} H\right)\left(D^{\mu} H^{\dagger} D^{\nu} H\right) & Q_{G^{4}}^{(2)}=\left(G_{\mu \nu}^{A} \tilde{G}^{A \mu \nu}\right)\left(G_{\rho \sigma}^{B} \tilde{G}^{B \rho \sigma}\right) \\
\hline Q_{H^{4}}^{(3)}=\left(D^{\mu} H^{\dagger} D_{\mu} H\right)\left(D^{\nu} H^{\dagger} D_{\nu} H\right) & Q_{G^{4}}^{(3)}=\left(G_{\mu \nu}^{A} G^{B \mu \nu}\right)\left(G_{\rho \sigma}^{A} G^{B \rho \sigma}\right) \\
\hline & Q_{G^{4}}^{(4)}=\left(G_{\mu \nu}^{A} \tilde{G}^{B \mu \nu}\right)\left(G_{\rho \sigma}^{A} \tilde{G}^{B \rho \sigma}\right) \\
\hline W \text { -boson } & Q_{G^{4}}^{(7)}=d^{A B E} d^{C D E}\left(G_{\mu \nu}^{A} G^{B \mu \nu}\right)\left(G_{\rho \sigma}^{C} G^{D \rho \sigma}\right) \\
\hline Q_{W^{4}}^{(1)}=\left(W_{\mu \nu}^{I} W^{I \mu \nu}\right)\left(W_{\rho \sigma}^{J} W^{J \rho \sigma}\right) & Q_{G^{4}}^{(8)}=d^{A B E} d^{C D E}\left(G_{\mu \nu}^{A} \tilde{G}^{B \mu \nu}\right)\left(G_{\rho \sigma}^{C} \tilde{G}^{D \rho \sigma}\right) \\
\hline Q_{W^{4}}^{(2)}=\left(W_{\mu \nu}^{I} \tilde{W}^{I \mu \nu}\right)\left(W_{\rho \sigma}^{J} \tilde{W}^{J \rho \sigma}\right) & O_{G}=f^{A B C} G_{\mu}^{A \nu} G_{\nu}^{B \rho} G_{\rho}^{C \mu} \\
\hline Q_{W^{4}}^{(3)}=\left(W_{\mu \nu}^{I} W^{J \mu \nu}\right)\left(W_{\rho \sigma}^{I} W^{J \rho \sigma}\right) & \\
\hline Q_{W^{4}}^{(4)}=\left(W_{\mu \nu}^{I} \tilde{W}^{J \mu \nu}\right)\left(W_{\rho \sigma}^{I} \tilde{W}^{J \rho \sigma}\right) & B \text { -boson } \\
\hline O_{W}=\epsilon^{I J K} W_{\mu}^{I \nu} W_{\nu}^{J \rho} W_{\rho}^{K \mu} & Q_{B^{4}}^{(1)}=\left(B_{\mu \nu} B^{\mu \nu}\right)\left(B_{\rho \sigma} B^{\rho \sigma}\right) \\
\hline & Q_{B^{4}}^{(2)}=\left(B_{\mu \nu} \tilde{B}^{\mu \nu}\right)\left(B_{\rho \sigma} \tilde{B}^{\rho \sigma}\right) \\
\hline \text { Left-handed lepton } & \\
\hline Q_{l^{4} D^{2}}^{(1)}=D^{\nu}\left(\bar{l} \gamma^{\mu} l\right) D_{\nu}\left(\bar{l} \gamma_{\mu} l\right) & \text { Right-handed lepton } \\
\hline Q_{l^{4} D^{2}}^{(2)}=\left(\bar{l} \gamma^{\mu} \overleftrightarrow{D}^{\nu} l\right)\left(\bar{l} \gamma_{\mu} \overleftrightarrow{D}_{\nu} l\right) & Q_{e^{4} D^{2}}=D^{\nu}\left(\bar{e} \gamma^{\mu} e\right) D_{\nu}\left(\bar{e} \gamma_{\mu} e\right) \\
\hline &\\
\hline \text { Left-handed quark } & \text { Right-handed quark } \\
\hline Q_{q^{4} D^{2}}^{(1)}=D^{\nu}\left(\bar{q} \gamma^{\mu} q\right) D_{\nu}\left(\bar{q} \gamma_{\mu} q\right) & Q_{u^{4} D^{2}}^{(1)}=D^{\nu}\left(\bar{u} \gamma^{\mu} u\right) D_{\nu}\left(\bar{u} \gamma_{\mu} u\right) \\
\hline Q_{q^{4} D^{2}}^{(2)}=\left(\bar{q} \gamma^{\mu} \overleftrightarrow{D}^{\nu} q\right)\left(\bar{q} \gamma_{\mu} \overleftrightarrow{D}_{\nu} q\right) & Q_{u^{4} D^{2}}^{(2)}=\left(\bar{u} \gamma^{\mu} \overleftrightarrow{D}^{\nu} u\right)\left(\bar{u} \gamma_{\mu} \overleftrightarrow{D}_{\nu} u\right)\\
\hline Q_{q^{4} D^{2}}^{(3)}=D^{\nu}\left(\bar{q} \gamma^{\mu} \tau^{I} q\right) D_{\nu}\left(\bar{q} \gamma_{\mu} \tau^{I} q\right) & Q_{d^{4} D^{2}}^{(1)}=D^{\nu}\left(\bar{d} \gamma^{\mu} d\right) D_{\nu}\left(\bar{d} \gamma_{\mu} d\right) \\
\hline Q_{q^{4} D^{2}}^{(4)}=\left(\bar{q} \gamma^{\mu} \overleftrightarrow{D}^{I \nu} q\right)\left(\bar{q} \gamma_{\mu} \overleftrightarrow{D}_{\nu}^{I} q\right) & Q_{d^{4} D^{2}}^{(2)}=\left(\bar{d} \gamma^{\mu}{\overleftrightarrow{D}^{\nu}} d\right)\left(\bar{d} \gamma_{\mu} \overleftrightarrow{D}_{\nu} d\right)\\
\hline
\end{array}\nonumber
\end{equation}
Here, the \(Q\) operators are dim-8 operators from Ref.~\cite{Li:2020gnx,Murphy:2020rsh}, while the \(O\) operators are dim-6 operators of the Warsaw basis~\cite{Grzadkowski:2010es}. Note that most basis operators are different from the ones used in the previous examples, but a conversion is straightforward.

We present the bounds in the form \(\vec{x} \cdot \vec{c} \geq 0\), where \(\vec{c}\) is a vector of coefficients of the same type, and we will show explicitly the \(\vec{x}\) vectors which represent the bounds. For the Higgs boson:
\begin{align}
&\vec{c}=\left[C_{H^{4}}^{(1)}, C_{H^{4}}^{(2)}, C_{H^{4}}^{(3)}\right] \\
&\vec{x}: \quad[1,0,0], \quad[1,1,0], \quad[1,1,1]
\end{align}

\(B\)-boson:
\begin{align}
&\vec{c}=\left[C_{B^{4}}^{(1)}, C_{B^{4}}^{(2)}\right] \\
&\vec{x}: \quad[1,0],\quad[0,1]
\end{align}

\(W\)-boson:
\begin{align}
&\vec{c}=\left[C_{W^{4}}^{(1)}, C_{W^{4}}^{(2)}, C_{W^{3}}^{(1)}, C_{W^{4}}^{(4)}, c_{W}^{2}\right] \\
\nonumber &\vec{x}: \quad[0,0,0,1,0], \quad[2,0,1,0,0], \quad[3,1,1,1,0], \\
&\quad\quad[0,0,4,0,-9], \quad[2,6,2,2,-9], \quad[0,8,0,4,-9]
\end{align}

Gluon:
\begin{align}
&\quad\vec{c}=\left[C_{G^{4}}^{(1)}, C_{G^{4}}^{(2)}, C_{G^{4}}^{(3)}, C_{G^{4}}^{(4)}, C_{G^{4}}^{(7)}, C_{G^{4}}^{(8)}, c_{G}^{2}\right]\\&
\begin{array}{l}
\vec{x}:\\
{\small
\begin{array}{llll}
{[0,0,0,1,0,0,0]} & {[0,0,6,3,7,2,0]} & {[24,0,12,21,15,14,0]} & {[0,0,96,24,64,40,-81]} \\
{[0,0,1,1,1,0,0]} & {[8,6,1,6,0,2,0]} & {[24,32,24,4,8,0,-27]} & {[40,32,80,4,0,0,-189]} \\
{[2,0,1,0,0,0,0]} & {[0,6,3,12,5,0,0]} & {[48,36,21,27,25,0,0]} & {[0,0,24,120,40,104,-81]} \\
{[0,2,0,1,0,0,0]} & {[8,6,1,12,0,0,0]} & {[32,40,4,80,0,0,-27]} & {[0,0,120,24,104,40,-81]} \\
{[0,0,3,0,2,0,0]} & {[0,6,6,9,10,4,0]} & {[0,48,0,48,0,40,-81]} & {[96,0,144,24,64,40,-81]} \\
{[0,0,0,3,0,2,0]} & {[0,12,0,14,0,0,-9]} & {[24,0,36,24,16,40,-81]} & {[48,0,96,24,0,40,-243]} \\
{[1,1,2,2,0,0,0]} & {[0,0,8,8,0,8,-27]} & {[0,0,48,24,32,40,-81]} & {[0,192,168,96,112,120,-405]} \\
{[6,0,3,0,2,0,0]} & {[12,0,14,0,0,0,-27]} & {[0,0,24,48,16,56,-81]} & {[168,480,168,156,56,160,-729]} \\
{[4,2,2,1,2,0,0]} & {[6,8,12,1,0,0,-27]} & {[88,32,56,4,40,0,-27]} & {[264,384,156,168,16,200,-729]} \\
{[0,0,4,0,0,0,-9]} & {[8,16,4,8,0,8,-27]} & {[96,42,27,84,25,0,0]} & {[288,384,216,168,0,200,-891]} \\
{[6,0,6,0,5,0,0]} & {[0,24,0,12,0,8,-27]} & {[96,66,42,39,50,4,0]} & {[480,384,480,168,160,200,-729]} \\
{[0,0,3,6,5,4,0]} & {[8,22,1,14,0,10,-27]} & {[120,42,39,42,40,14,0]} & {[336,768,672,216,0,200,-2187]}
\end{array}
}
\end{array}
\end{align}

Left-handed leptons, or right-handed quarks:
\begin{align}
&\vec{c}=\left[C_{l^{4} D^{2}}^{(1)}, C_{l^{4} D^{2}}^{(2)}\right], \text { or } \vec{c}=\left[C_{u^{4} D^{2}}^{(1)}, C_{u^{4} D^{2}}^{(2)}\right], \text { or } \vec{c}=\left[C_{d^{4} D^{2}}^{(1)}, C_{d^{4} D^{2}}^{(2)}\right] \\
&\vec{x}: \quad-[1,3], \quad-[0,1]
\end{align}

Left-handed quarks:
\begin{align}
\vec{c}= & {\left[C_{q^{4} D^{2}}^{(1)}, C_{q^{4} D^{2}}^{(2)}, C_{q^{4} D^{2}}^{(3)}, C_{q^{4} D^{2}}^{(4)}\right]}  \\
\vec{x}: & -[0,1,0,-1],\quad -[0,1,0,1], \quad -[0,1,1,0], \\
& -[1,3,0,0], \quad -[1,3,1,3],\quad  -[3,10,0,-3].
\end{align}
This completes the bounds for the SM self-quartic P-conserving dim-8 operators.

Finally, let us comment on the non-extremal generators. In many examples we have seen generators that are not an ER. This often happens to the largest irreps of a given problem. Examples are the \(\mathbf{2}\) in Section 4.1.1 and Section 4.2, the \(\mathbf{3}_{1}, \mathbf{3}_{0 S}\) and \(\mathbf{3}_{0 A}\) in Section 4.1.5, the \(J_{3}=1\) state of Section 4.3.1, the \(\mathbf{3}_{0}\) and \(\mathbf{3}_{1}\) in Section 4.3.3, the \((\mathbf{6}, \mathbf{1})\) and \((\mathbf{8}, \mathbf{1})\) in Section 4.3.4 for right-handed quarks, and the \((\mathbf{6}, \mathbf{3})\) and \((\mathbf{8}, \mathbf{3})\) for left-handed quarks. This may not be surprising, because for intermediate states that are large multiplet, the generator should be interpreted as \(\mathbf{m}^{i j} \cdot \mathbf{m}^{* k l}+\mathbf{m}^{i l} \cdot \mathbf{m}^{* k j}\), which is a positive sum of multiple elements. While this does not necessarily mean that the corresponding generator is not extremal, because the space itself is also restricted by the same symmetry, it seems that for the highest irrep(s) this always happen. The physical consequence is that for UV particles which are large multiplets under the SM gauge group, it will be difficult to uniquely confirm its existence, as the same contribution to \(\mathcal{M}^{i j k l}\) could also come from a combination of several other particles of different types.

\section{The inverse problem}
\label{sec5}
Having Known the accurate boundary of the UV-completable SMEFTs, we are now ready to discuss the inverse problem in SMEFT from the positivity point of view. We will assume that we will be able to measure the Wilson coefficients at dim-6 and dim-8, to some reasonable accuracy level, but not beyond. Based on the dim-8 information only, we are going to ask: how can we determine the UV models and how large is the degeneracy in this determination. By degeneracy, we mean that a given SMEFT can be matched to many different UV setups, which implies arbitrariness in reconstructing the UV theory.

It is certainly not possible to explicitly cover all possible UV theories, as we do not know all possible BSM theories. It is therefore easier if we start from tree-level UV completions, by which we mean the UV theories in which the dominant SMEFT coefficients come from integrating out heavy particles at the tree-level. We will discuss loop-level UV completions in Section 6.

Suppose the SM is extended by a number of heavy particles, \(\left\{X_{\alpha, i}\right\}\), where \(\alpha\) labels the particle type, defined by the quantum numbers including the irrep under the SM groups. \(X_{\alpha, i}\) is the \(i\) th particle of type \(\alpha\). For a given \(\alpha\), the quantum numbers determine how each \(X_{\alpha, i}\) couples to the SM particles, but up to an arbitrary normalization. Schematically, we may write their interactions to the SM particles:
\begin{equation}
\mathcal{L}_{\text {int }}=\sum_{\alpha, i} X_{\alpha i} g_{\alpha i}\left(\kappa_{\alpha}^{H} J_{\alpha}^{H}+\kappa_{\alpha}^{q} J_{\alpha}^{q}+\kappa_{\alpha}^{u} J_{\alpha}^{u}+\kappa_{\alpha}^{d} J_{\alpha}^{d}+\kappa_{\alpha}^{l} J_{\alpha}^{l}+\kappa_{\alpha}^{e} J_{\alpha}^{e}+\ldots\right)
\end{equation}
where we omit any gauge and/or Lorentz indices of \(X\), as they depend on the irrep of the particle type \(\alpha\). The \(J_{\alpha}^{i}\) are currents of SM particles, whose exact forms are fixed by \(\alpha\) (e.g. if \(X_{\alpha}\) is a scalar singlet or triplet, \(J^{H}\) is either a singlet \(H^{\dagger} H\) or a triplet current \(H^{\dagger} \tau^{I} H\), depending on the irrep of \(X_{\alpha}\)). The ...\ represents possible higher dimensional effective terms. \(g_{\alpha i}\) is an overall coupling. \(\kappa\)'s are pure numbers and cannot be fixed by \(\alpha .\) Note that we assume the \(\left(\kappa_{\alpha}^{H}, \kappa_{\alpha}^{q}, \ldots\right)\) factors are the same for all particles of the same type \(\alpha\), {\it i.e.}, they do not the label $i$. If this were not true, one can simply define an additional type \(\alpha^{\prime}\) for the states with a different \(\kappa\) value. In other words, \(\alpha\) not only fixes the irrep of \(X\), but also fixes the relative coupling strengths \(\left(\kappa_{\alpha}^{H}, \kappa_{\alpha}^{q}, \ldots\right)\)

To determine the UV model, in principle we need the following two types of information:

\begin{itemize}
  \item``UV particle spectrum'': including the masses \(M_{\alpha i}\), particle widths, possible lineshape of UV states, and the total coupling strengths \(g_{\alpha_{i}}\).

  \item ``Interaction type'': including \(\alpha\) which determines all the currents \(J\)'s, and their relative sizes \ \(\left(\kappa_{\alpha}^{H}, \kappa_{\alpha}^{q}, \ldots\right)\), but up to an overall normalization.
 \end{itemize}

The main focus of this work is the second type of information, and this is what we meant by a weaker version of the inverse problem in Section 2. As we will show, this problem is closely connected to positivity bounds at dim-8 \cite{Zhang:2020jyn}. Actually, the first type information is also related to positivity in a similar way, see discussion in Ref.~\cite{Arkani-Hamed:2020blm}. This however requires a knowledge of SMEFT coefficients beyond dim-8, and is therefore phenomenologically less interesting.

Assuming a tree-level UV completion, the goal of this section is to demonstrate how and to what extend we can determine the interaction type information of each possible UV particle, only by knowing the dim-8 coefficients. To this end, we will investigate the distribution of the degeneracy in the dim-8 SMEFT space. Similar to Section 2, we are interested in specific regions in the space, where the degeneracy are reduced or vanishing, and the interaction information of the UV origin can be determined. In this section, we aim at a generic and systematic framework for the extraction of such information.

\subsection{The PSD matrix cone}
\label{sec5.1}
Section 2 already demonstrates that extremality often leads to uniqueness in the determination of the UV theory. The toy model, however, is a simple one with only 4 extremal rays. In this section, we give some more insights to this fact, for cases where an infinite number of generators (or continuous generators) exist, by first neglecting the \(u\)-channel contributions in the dispersion relation. In this case the amplitude space is a PSD matrix cone. This helps to understand the underlying mechanism of the reverse engineering of UV models from extremality.

Recall that for tree-level UV completions, the two terms on the r.h.s.~of the dispersion relation represent the \(s\)- and the \(u\)-channel \(X_{\alpha}\) exchanges. The corresponding \(m\) matrix from a given \(X_{\alpha}\) is simply \(m^{i j} \propto\left\langle X_{\alpha}\left|J_{\alpha}\right| i j\right\rangle\), and so the \(\kappa\) factors enter as free parameters in the \(m\) matrices. The total amplitude is a positively weighted sum of generators (assuming self-conjugate particle states are used):
\begin{equation}
\mathcal{M}^{i j k l}=\sum_{\alpha} w_{\alpha} m_{\alpha}^{i(j} m_{\alpha}^{|k| l)}\label{eq:5.2}
\end{equation}
where the generator \(\mathcal{G}_{\alpha}^{i j k l}=m_{\alpha}^{i(j} m_{\alpha}^{|k| l)}\) can be normalized such that the weight
\begin{equation}
w_{\alpha}=\sum_{i} \frac{g_{\alpha i}^{2}}{M_{\alpha i}^{4}} \geq 0
\end{equation}
which represents the sum of contributions from all heavy states of type \(\alpha\). Our question is: given the measured value of \(\mathcal{M}^{i j k l}\), how can we determine the actual size of each \(w_{\alpha}\) for each particle type \(\alpha\). Note that there can be an infinite number of \(\alpha\), each having its own \(m_{\alpha}\) matrix, given that \(\kappa\) is allowed to take arbitrary real values.

Solving Eq.~(\ref{eq:5.2}) for \(w_{\alpha}\) would normally appear to be impossible as there is no algorithm that allows one to go through a summation and determine the terms inside on the r.h.s..~However, there are exceptions. In this section, as a first step, we assume that the \(u\)-channel exchange does not contribute to the amplitudes we are interested in, and simply write \(\mathcal{G}^{i j k l}=m_{\alpha}^{i j} m_{\alpha}^{k l} .\) This would allow us to view \(i j\) as one index and \(k l\) as another index, and rewrite
\begin{equation}
\mathcal{M}^{i j}=\sum_{\alpha} w_{\alpha} m_{\alpha}^{i} m_{\alpha}^{j}\label{eq:5.4}
\end{equation}
where \(i\) and \(j\) now label the incoming and the outgoing two-particle states. For continuous \(\alpha\), the summation becomes an integration. If there are no other constraints on \(m, \mathcal{M}\) is a positive linear combination of all rank-1 PSD matrices \(m_{\alpha}^{i} m_{\alpha}^{j}\), and so the positivity cone \(\mathbf{C}\) is the cone of all PSD matrices.

When can we say something about \(w_{a}\), or even uniquely solve for it? Obviously, it depends on the matrix \(\mathcal{M}^{i j}\). An obvious case is that \(\mathcal{M}^{i j}\) is rank-1. Since all \(w_{\alpha} \geq 0\), this simply means that only one \(m_{\alpha}\) on the r.h.s.~can be nonzero, otherwise the r.h.s.~will have a rank higher than one. The corresponding \(m_{\alpha}\) is simply the only non-zero eigenvector of \(\mathcal{M}^{i j}\), which in turn determines the SM current \(J_{\alpha}\) together with all the \(\kappa\) factors and \(w_{\alpha}\). This means that if \(\mathcal{M}^{i j}\) is generated by integrating out a single heavy particle, a precise low-energy measurement would leave no degeneracy in the inverse problem, in the sense that the UV origin of \(\mathcal{M}^{i j}\) can be uniquely determined.

The above fact is nothing but an application of extremality in the PSD matrix cone. The ERs of the latter are simply the rank-1 matrices, and so they cannot be written as a sum of different elements. This interpretation would eventually help us to generalize the same picture to more realistic cases, where the \(u\)-channel terms are added and symmetry conditions are imposed.

There is more information to be dug out from \(\mathcal{M}^{i j}\). The above example indicates that the rank of \(\mathcal{M}^{i j}\) plays an important role. Naively, we would expect lower degeneracy from lower rank \(\mathcal{M}^{i j}\) matrices. For example, assuming only the first \(k \times k\) block of \(\mathcal{M}^{i j}\) is nonzero, we can immediately infer that the \(m_{\alpha}\)'s on the r.h.s.~have to live in the subspace spanned by the first \(k\) components in \(m^{i}\) (or the state \(X_{\alpha}\) can only couple to the first \(k\) SM currents). More precisely, if \(\mathcal{M}\) is a rank-$r$ \(n \times n\) matrix, it has an \(n-r\) dimensional null space, spanned by \(n-r\) basis vectors \(b_{\beta}^{i}, \beta=1,2, \cdots, n-r .\) We have
\begin{equation}
\mathcal{M}^{i j} b_{\beta}^{i} b_{\beta}^{j}=\sum_{\alpha} w_{\alpha}\left(m_{\alpha}^{i} b_{\beta}^{i}\right)^{2}=0 \quad \forall \beta
\end{equation}
Since \(w_{\alpha} \geq 0\), this gives \(n-r\) constraints on each \(m_{\alpha}^{i}\) with a non-vanishing \(w_{\alpha}: m_{\alpha}^{i} b_{\beta}^{i}=0\) or \(m_{\alpha}\) is in the row space of \(\mathcal{M}\). Alternatively, if some \(m_{\alpha}\) does not satisfy this condition, the corresponding state \(X_{\alpha}\) can be excluded from the UV spectrum. If symmetries further reduce the number of non-vanishing \(w_{\alpha}\) to \(r(r+1) / 2\), then each \(w_{\alpha}\) can be uniquely determined.

This fact also has a geometric explanation related to extremality. A rank-$r$ PSD matrix is on a \(k\)-face of the PSD matrix cone, whose dimension is \(k=r(r+1) / 2\), and is defined by all rank-\(r\) matrices that have the same null space. By the definition of faces, if we split a PSD matrix on the face to be a sum of other PSD matrices, the latter must both live on the same face. Therefore only the \(m_{\alpha}^{i} m_{\alpha}^{j}\) that are on the same face are allowed on the r.h.s.~of Eq.~(\ref{eq:5.4}), which means \(m_{\alpha}\) is orthogonal to the null space. If there are only \(k\) such \(m_{\alpha}\)'s allowed by the theory, then \(w_{\alpha}\)'s are uniquely determined.\,\footnote{The symmetry relations are linear and thus they cut the PSD cone into a spectrahedron. The face structure of the latter is similar to that of the PSD cone: \(\mathcal{M}\) is contained by a face which is the set of all matrices that have the same null space \cite{Li:2021cjv}.} Note that an \(\mathrm{ER}\) is a one-dimensional face, and so this is simply the generalization of the extremality argument to larger dimensions.

A special but physically important case, is that if \(r=0\), or simply \(\mathcal{M}=0\), then all \(w_{\alpha}\) must vanish. This is obvious by taking a trace in Eq.~(\ref{eq:5.4}). The geometric interpretation is that the origin is an extreme point in the PSD matrix cone. It however has a very important physical implication: a null measurement of dim-8 coefficients would uniquely confirm the SM and universally exclude all BSM states, without using any specific assumptions about BSM. We will see that this continues to be true at dim-8 after the \(s-u\) crossing term is restored.

One last useful fact is that given the l.h.s.\ of Eq.~(\ref{eq:5.4}), we can always set an upper limit on each \(w_{\alpha}\) on the r.h.s.~This is obvious because the largest eigenvalue on the r.h.s.~cannot be larger than that of the l.h.s.. There are other ways to see this, for example by taking a trace of \(\mathcal{M}^{i j}\). In the next sections we will see how to derive a strict upper bound of \(w_{\alpha}\). Physically, this means that for any measured coefficient vector \(\vec{C}\), one can always constrain the mass scale of all heavy particles to some certain scale, independent of any assumptions about the UV theory.

The above discussion does not directly apply to realistic problems, because we have discarded the \(s-u\) crossing term in the dispersion relation, and because we did not take into account symmetry constraints. Taking them into account, \(\mathcal{M}^{i j k l}\) is not a PSD cone anymore, and so we cannot simply use its rank. However, we have also seen that the same conclusions can be equivalently drawn with a geometric picture, by using cones, extremalities, and \(k\) faces, etc. We already showed, in Section 3.6, that the UV-completable \(\mathcal{M}^{i j k l}\) forms a convex cone \(\mathbf{C}\) which is salient. It follows that \(\mathbf{C}\) must have ERs, faces, and that its origin is an extreme point. Therefore, thanks to this geometric picture, all the nice features that apply to PSD cones also apply to realistic problems  \cite{Zhang:2020jyn}. The only difference is that not all the generators \(m^{i(j} m^{|k| l)}\) are required to be extremal (as opposed to being rank-1 without \(j, l\) crossing and thus extremal in PSD cones). This is because of two reasons: the \(s \leftrightarrow u\) crossing terms, and the combination of the \(m^{i(j} m^{|k| l)}\) term from each state in a multiplet intermediate particle. In Section 4, we have already seen that the generators of the highest irreps are often not extremal.

In the following sections we will discuss real physics cases. It is then more convenient to directly consider the coefficient space, construct the generators, and then exploit the extremality of the boundaries. The discussion of this section serves as a guidance as to what kind of information about UV physics can be obtained, how this is done and where exactly in the parameter space this is possible.

\subsection{Polyhedral case}
\label{sec5.2}
Consider now the coefficient space, in which the generators are the \(\vec{g}\) vectors. The total dim-8 coefficient vector must be a positive combination of generator vectors:
\begin{equation}
\vec{C}=\sum_{\alpha} w_{\alpha} \vec{g}_{\alpha}\label{eq:5.6}
\end{equation}
Assuming for now that these generators do not depend on any free parameters, and we only have a finite number of \(\alpha\) so that the allowed parameter space is polyhedral. The weights \(w_{\alpha}=\sum_{i} \frac{g_{\alpha i}^{2}}{M_{\alpha i}^{4}} \geq 0\) are similar to the previous section. We now ask how to solve Eq.~(\ref{eq:5.6}) for \(w_{\alpha}\) by knowing the measured value of \(\vec{C}\)

It is important to realize that \(\mathbf{C}=\operatorname{cone}\left(\left\{\vec{g}_{\alpha}\right\}\right)\) is a salient cone, as explained in Section 3.6. This not only means that Eq.~(\ref{eq:5.6}) does not always admit a solution, but also implies that all interesting results we have discussed in the previous section for a PSD cone directly apply. They are:

If \(\vec{C}\) is on an \(\mathrm{ER}\) (or 1-face): there is a unique solution for \(w_{\alpha}\), as we have mentioned several times. In all examples of Section 4 we have explicitly pointed out which generators are extremal. The reason for this uniqueness is that ERs by definition cannot be a sum of two different \(\vec{g}_{\alpha}\) vectors, and therefore only one \(w_{\alpha}\) can be nonzero and its value can be uniquely fixed by \(\vec{C} / \vec{g}_{\alpha}\). Note that the existence of ERs is guaranteed by the Krein-Milman theorem, which requires \(\mathbf{C}\) is salient.

If \(\vec{C}\) is on a \(k\)-face \((k>1)\): by the same argument, only the \(\vec{g}_{\alpha}\) on the same face can have a nonzero \(w_{\alpha}\). Suppose there are \(l\) of them and \(l \geq k\). The solution for \(w_{\alpha}\) is determined by a system with \(k\) independent constraints and \(l\) unknowns, so the arbitrariness corresponds to \(l-k\) free parameters. If \(l=k\), i.e. if the \(k\)-face is a \(k\)-simplex with no additional \(\vec{g}\) inside, then the solution for \(w_{\alpha}\) is unique. This is true for all 2-faces in our toy example of Section 2. Again, the existence of \(k\)-faces is ensured by \(\mathbf{C}\) being salient.

If \(\vec{C}=0\) is at the origin (or 0-face): \(w_{\alpha}=0\) for all \(\alpha\) is the only possibility, because the origin is an extreme point of \(\mathbf{C}\). To see this, assume that Eq.~(\ref{eq:5.6}) with \(\vec{C}=0\) can be satisfied by some nonzero \(w_{\alpha}\)'s. Assume \(w_{1}\) is nonzero, we can write:
\begin{equation}
-\vec{g}_{1}=\sum_{i>1} \frac{w_{i}}{w_{1}} \vec{g}_{i} \in \mathbf{C}
\end{equation}
which contradicts with the fact that \(\mathbf{C}\) is salient, because \(\pm \vec{g}_{1} \in \mathbf{C}\). Therefore all \(w_{\alpha}\)'s have to vanish. This means that a null measurement of all coefficients is sufficient to rule out all BSM theories, and can be used as a confirmation of the SM itself.

If \(\vec{C}\) is in the interior: the arbitrariness of \(w_{\alpha}\) exists but is always finite. In other words, if \(\left\{w_{\alpha}\right\}\) is a solution of Eq.~(\ref{eq:5.6}) for a given \(\vec{C}\), then there is a maximum bound on each \(w_{\alpha}\) value. The maximum depends on \(\vec{C}\). To obtain this bound, consider
\begin{equation}
\vec{C}(\lambda) \equiv \vec{C}-\lambda \vec{g}_{k}=\sum_{i \neq k} w_{i} \vec{g}_{i}+\left(w_{k}-\lambda\right) \vec{g}_{k}
\end{equation}
We have \(\vec{C}(0)=\vec{C} \in \mathbf{C}\). We also know
\begin{equation}
\vec{C}(\infty) \|-\vec{g}_{k} \notin \mathbf{C}
\end{equation}
due to \(\mathbf{C}\) being salient. Therefore the following maximum value of \(\lambda\) exists such that \(\vec{C}(\lambda) \in \mathbf{C}:\)

\begin{equation}
\lambda_{M}=\max _{\vec{C}(\lambda) \in \mathbf{C}} \lambda \label{eq:5.10}
\end{equation}
Since \(\vec{C}\left(w_{k}\right) \in \mathbf{C}\) by definition, we have \(\lambda_{M} \geq w_{k}\) is an upper bound of \(w_{k}\). According to this equation, if \(\vec{C}\) is close to an ER (or a face), all \(w_{i}\) weights that do not correspond to this ER (or those correspond to an ER not on that face) will have a small upper bound, indicating the arbitrariness in determining the UV completion will be small.

These are four physically interesting implications of \(\mathbf{C}\) being salient.

One may wonder to what extent these implications depend on our tree-level assumptions. The answer is they almost do not. In Section 4 we have worked out the generator vectors without assuming tree-level UV completion. At the loop level, the main difference is that integrating out heavy particles could give rise to a new coefficient vector, which is still in \(\mathbf{C}\) and therefore will not change the fact that \(\mathbf{C}\) is salient. If we consider this as a new ``loop generator vector'', the geometric picture does not change, unless the new generator falls onto a certain face. In that case, the degeneracies on that face will be affected. In Section 6 we will give a more concrete discussion.

One may also wonder to what extent these implications apply at dim-6. The answer is, again, they do not. This fact represents a very important difference between dim-6 and dim-8 operators, and is crucial for building the motivation for phenomenological studies at the dim-8 level of SMEFT. Let us define the tree-level ``generator'' of dim-6 coefficients \(\vec{g}_{\alpha}^{(6)}\), in a similar way to dim-8, so that
\begin{equation}
\vec{C}^{(6)}=\sum_{\alpha} w_{\alpha}^{(6)} \vec{g}_{\alpha}^{(6)}, \quad w_{\alpha}^{(6)} \equiv \sum_{i} \frac{g_{\alpha i}^{2}}{M_{\alpha i}^{2}}\label{eq:5.11}
\end{equation}
where we use \(^{(6)}\) to indicate dim-6 coefficients. The problem with the dim-6 coefficient space is that in general \(\left\{\vec{g}^{(6)}\right\}\) do not form a salient cone, and cone \(\left(\left\{\vec{g}^{(6)}\right\}\right)\) normally spans the entire space. Any \(\vec{C}^{(6)}\) could admit a solution in Eq.~(\ref{eq:5.11}). In addition,

\begin{itemize}
  \item no \(\vec{C}^{(6)}\) admits an unique solution. This is because if cone \(\left(\left\{\vec{g}^{(6)}\right\}\right)\) is not salient, \(\vec{C}^{(6)}=0\) has at least one nontrivial solution:

\begin{equation}
0=\sum_{\alpha} \bar{w}_{\alpha}^{(6)} \vec{g}_{\alpha}^{(6)}, \quad \bar{w}_{\alpha}^{(6)}>0 \text { at least for some } \alpha
\end{equation}
Therefore, if \(w_{\alpha}^{(6)}\) is a solution of Eq.~(\ref{eq:5.6}) for some \(\vec{C}\), then \(w_{\alpha}^{(6)}+\lambda \bar{w}_{\alpha}^{(6)}\) is also a solution, for all \(\lambda \in \mathbb{R}^{+}\). Physically, this means that the reverse engineering from dim-6 coefficients is not only impossible, but in fact the intrinsic degeneracy is arbitrarily large.
\end{itemize}

\begin{itemize}
 \item \(\vec{C}^{(6)}=0\) is obviously not extremal in the full space. If \(\vec{C}^{(6)}=0\), the solution for \(w_{\alpha}^{(6)}\)’s can be \(\lambda \bar{w}_{\alpha}^{(6)}\) for any \(\lambda \in \mathbb{R}^{+}\). Physically, it means a null measurement of all coefficients is insufficient to rule out all BSM theories, and cannot be used as a confirmation of the SM. This is because different states will generate different \(\vec{g}^{(6)}\) vectors that could potentially cancel each other, as they are not confined in a salient cone.
\end{itemize}

\begin{itemize}
  \item The arbitrariness of \(w_{\alpha}^{(6)}\) is in general unlimited, because \(w_{\alpha}^{(6)}+\lambda \bar{w}_{\alpha}^{(6)}\) is a solution of Eq.~(\ref{eq:5.11}), and therefore an upper bound on any \(w_{\alpha}\) cannot be set. Physically, this means that exclusion limits for BSM particles do not exist, unless we first choose a concrete BSM theory.
\end{itemize}

We would like to emphasize that the above points provide an important motivation to study the dim-8 coefficients \cite{Zhang:2020jyn}. The SMEFT approach is mostly used in precision tests of the SM. If no deviation is seen, we would hope to confirm that the SM is correct, at least in the sense that new states do not couple to SM particles strongly enough to have a visible effect. This confirmation, unfortunately, cannot be done if the SMEFT is truncated at dim-6, as shown above, because \(\vec{C}^{(6)}=0\) could come from the cancellation between different UV states. The simplest example is actually already given in Eq.~(\ref{eq:1.2}). Precision test of the SM at the dim-6 level can thus never fully test the SM. In contrast, when promoted to dim-8, the SMEFT becomes capable of universally excluding the existence of all new states, \(\vec{C}=0\) would uniquely mean that no new states can exist, because the contributions from different states are confined in a salient cone which do not allow them to cancel each other. Another way to say this is that the origin is an extreme point in the dim-8 space, but it is not extremal in the dim-6 space. If all UV states are ruled out and the SM is confirmed, there is no need to go further to dim-10,12, and so on. This is why dim-8 is a special dimension in the SMEFT approach to BSM physics.

Before moving to the next sections, let us also discuss the possibility of setting lower bounds on \(w_{\alpha}\). While it is possible to set a universal upper bound on \(w_{\alpha}\), setting a lower bound is only possible under certain conditions. The trick is similar to Eq.~(\ref{eq:5.10}). Suppose we want to set a lower bound on \(w_{k}\) for some \(k\). Define
\begin{equation}
\mathbf{H}_{k}=\operatorname{cone}\left(\left\{\vec{g}_{\alpha \neq k},-\vec{g}_{k}\right\}\right)
\end{equation}
i.e. the conical hull of all generators except \(\vec{g}_{k}\) flipped. If \(\vec{C} \notin \mathbf{H}_{k}\), an lower bound on \(w_{k}\) can be set, by defining
\begin{equation}
\vec{C}(\lambda) \equiv \vec{C}-\lambda \vec{g}_{k}=\sum_{i \neq k} w_{i} \vec{g}_{i}+\left(w_{k}-\lambda\right) \vec{g}_{k}
\end{equation}
We have \(\vec{C}(0)=\vec{C} \notin \mathbf{H}_{k}\) and \(\vec{C}(\lambda) \in \mathbf{H}_{k}\) for \(\lambda \geq w_{k} .\) Therefore the following minimum value exists:
\begin{equation}
\lambda_{m}=\min _{\vec{C}(\lambda) \in \mathbf{H}_{k}} \lambda\label{eq:5.15}
\end{equation}
This sets a lower bound on \(w_{k}\), as \(\vec{C}\left(w_{k}\right) \in \mathbf{H}_{k} .\) There are, however,  important differences compared with setting upper bounds. Adding additional generators will not change the upper bound, but it may spoil the lower bound. For example, a lower bound derived under a tree-level UV assumption may not apply at the loop level, because the additional loop generators could change the shape of \(\mathbf{H}_{k}\). In contrast, \(\mathbf{C}\) is not affected, so the upper bound from Eq.~(\ref{eq:5.10}) always applies, independent of any BSM assumptions. Additional differences are:

\begin{itemize}
  \item If several particles \(X_{\alpha, i}\) exist for a given type \(\alpha\), the upper bound of \(w_{\alpha}\) not only applies to \(w_{\alpha}=\sum_{i} g_{i}^{2} / M_{i}^{4}\), but also applies to the contribution of each particle, \(g_{i}^{2} / M_{i}^{4} .\) In contrast, the lower bound only applies to a particle type \(\alpha\), but obviously cannot be applied to each particle of that type.

  \item The condition for setting lower bound is \(\vec{C} \notin \mathbf{H}_{k} .\) This does not require that \(\mathbf{C}\) is salient. It is therefore possible to set lower bounds on UV states using dim-6 operator coefficients. Applications of this type have been discussed in the literature, see e.g. Refs.~\cite{Low:2009di}.

\end{itemize}
The lower bound on \(w_{\alpha}\), if exists, indicates that UV particles of certain type have to exist in the spectrum. This is certainly a very useful information and provides guidance to future collider searches. It is however less solid as it can be changed by loop induced operators or strongly coupled UV theories. Therefore, in the following sections, we will instead focus more on the upper bounds. In Section 5.2.2, however, we will use the Higgs operators as an example to discuss lower bounds at the dim-6 level. A more comprehensive study of lower bounds at the dim-6 can be interesting, but we will defer it to future works.

\subsubsection{SM leptons }
\label{sec5.2.1}
Let us illustrate the previous section with some concrete examples. Consider first the example of Section 4.3.3: the SM lepton doublets. Recall the operators are
\begin{equation}
O_{1}=\partial_{\mu}\left(\bar{l} \gamma_{\nu} l\right) \partial^{\mu}\left(\bar{l} \gamma^{\nu} l\right), \quad O_{2}=\partial_{\mu}\left(\bar{l} \gamma_{\nu} \tau^{I} l\right) \partial^{\mu}\left(\bar{l} \gamma^{\nu} \tau^{I} l\right)
\end{equation}
and the UV states/generators are
\begin{equation}
\begin{array}{ccccccc}
\text { State}&\text{ Spin}&\text{ Charge } & \text { Interaction } & \mathrm{ER} & \vec{c} & C^{(6)} \\
\hline \mathcal{B}_{1} & 1 & 1_{1} & \mathcal{B}_{1}^{\mu}\left(\bar{l}^{c} i \stackrel{\leftrightarrow}{D}_{\mu} l\right) & \text{\Checkmark} & \frac{1}{2}(1,-1) & 0 \\
\Xi_{1} & 0 & 3_{1} & \Xi_{1}^{I}\left(\bar{l}^{c} \tau^{I} l\right) & \text{\XSolidBrush} & \frac{1}{2}(-3,-1) & 1 \\
\mathcal{B} & 1 & 1_{0} & \mathcal{B}^{\mu}\left(\bar{l} \gamma_{\mu} l\right) & \text{\Checkmark} & \frac{1}{2}(-1,0) & -\frac{1}{2} \\
\mathcal{W} & 1 & 3_{0} & \mathcal{W}^{I \mu}\left(\bar{l} \gamma_{\mu} \tau^{I} l\right) & \text{\XSolidBrush} & \frac{1}{2}(0,-1) & -\frac{1}{2}
\end{array}\label{eq:5.17}
\end{equation}
where we have added the last column for the dim-6 coefficient of the following operator:
\begin{equation}
O^{(6)}=\left(\bar{l} \gamma_{\mu} l\right)\left(\bar{l} \gamma^{\mu} l\right)
\end{equation}
We further rescale the generators, \(\vec{g} \rightarrow \vec{c}\), so that \(w=g^{2} / M^{4}\) :
\begin{equation}
\vec{g}_{\textbf{1}_{1}}=\frac{1}{2}(1,-1), \quad \vec{g}_{\textbf{3}_{1}}=\frac{1}{2}(-3,-1), \quad \vec{g}_{\textbf{1}_{0}}=\frac{1}{2}(-1,0), \quad \vec{g}_{\textbf{3}_{0}}=\frac{1}{2}(0,-1) .
\end{equation}

\begin{figure}[h]
	\begin{center}
		\includegraphics[width=.7\linewidth]{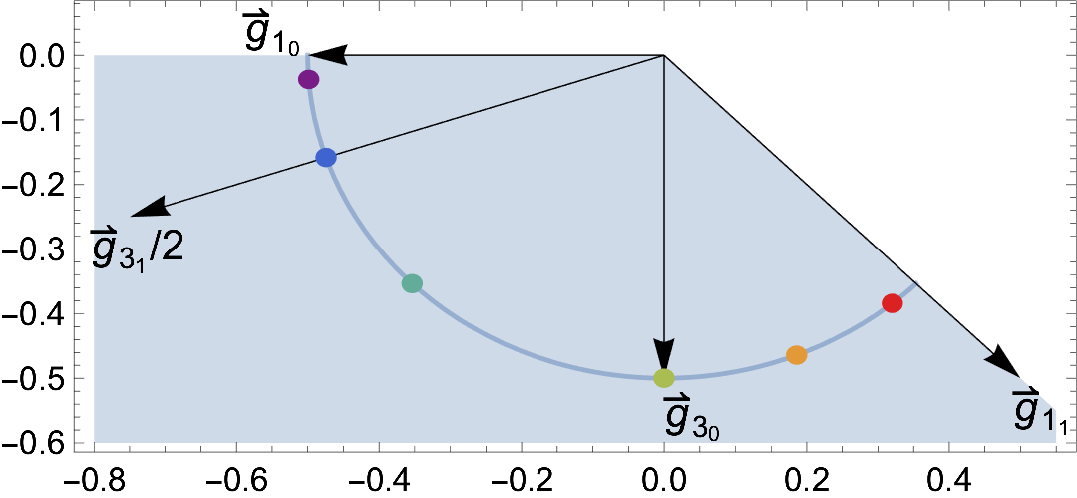}
	\end{center}
	\caption{Positivity cone for SM left-handed lepton doublets. Generators are shown in black arrows. Solution spaces for \(w_{\alpha}\) at the colored dots are further shown in Figure 9.}
	\label{fig8}
\end{figure}

This example has a two-dimensional parameter space, and therefore is somewhat trivial as most of the conclusions from the previous section can be easily obtained in many ways. It however illustrates well the physical implications of the positivity cone being salient. We first plot the generators in Figure \ref{fig8}. Obviously, \(\vec{g}_{\textbf{1}_{0}}\) and \(\vec{g}_{\textbf{1}_{1}}\) are extremal, and their positive combinations span the positivity cone. The other two generators are not extremal.

For any given value of \(\vec{C}\), the feasible solutions for \(w_{\alpha}\) forms a two-dimensional space. We choose the weights of the ERs, \(w_{\textbf{1}_{1}}\) and \(w_{\textbf{1}_{0}}\), as the two independent ones, and write \(w_{\textbf{3}_{1}}\) and \(w_{\textbf{3}_{0}}\) as
\begin{equation}
w_{\textbf{3}_{1}}=-\frac{2}{3} C_{1}+\frac{1}{3} w_{\textbf{1}_{1}}-\frac{1}{3} w_{\textbf{1}_{0}}, \quad w_{\textbf{3}_{0}}=\frac{2}{3} C_{1}-2 C_{2}-\frac{4}{3} w_{\textbf{1}_{1}}+\frac{1}{3} w_{\textbf{1}_{0}}
\end{equation}
The solution space in the \(\left(w_{\textbf{1}_{1}}, w_{\textbf{1}_{0}}\right)\) plane is confined by four inequalities:
\begin{align}
&w_{\textbf{1}_{1}} \geq 0, w_{\textbf{1}_{0}} \geq 0 \\
&w_{\textbf{3}_{1}}=-\frac{2}{3} C_{1}+\frac{1}{3} w_{\textbf{1}_{1}}-\frac{1}{3} w_{1_{0}} \geq 0, ~~ w_{\textbf{3}_{0}}=\frac{2}{3} C_{1}-2 C_{2}-\frac{4}{3} w_{\textbf{1}_{1}}+\frac{1}{3} w_{\textbf{1}_{0}} \geq 0
\end{align}
Let us now check whether the four implications of \(\mathbf{C}\) being salient, discussed in the previous section, indeed hold for this simple example.

If \(\vec{C}\) is on an ER: In fact, if the two coefficients are measured to be \(\vec{C} \propto(-1,0)\), the above inequalities uniquely fix \(w_{\textbf{3}_{1}}=w_{\textbf{1}_{1}}=w_{\textbf{3}_{0}}=0\). Similarly, \(\vec{C} \propto(1,-1)\) uniquely fixes \(w_{\textbf{1}_{0}}=w_{\textbf{3}_{1}}=w_{\textbf{3}_{0}}=0\). Physically, if the SM is extended by one vector particle of charge \(1_{1}\) or \(1_{0}\), i.e., either \(\mathcal{B}\) or \(\mathcal{B}_{1}\), then a perfect measurement of dim-8 coefficients would uniquely confirm the corresponding scenario, i.e. it not only confirms the extension of \(\mathcal{B}\) or \(\mathcal{B}_{1}\), but also rules out any alternative UV models. This does not apply to the non-extremal generators \(\vec{g}_{\textbf{3}_{1}}\) and \(\vec{g}_{\textbf{3}_{0}}\), which correspond to SU(2) triplet scalar and vector. Their contributions can also be explained by combining \(\mathcal{B}\) and \(\mathcal{B}_{1}\).

\begin{itemize}
  \item This is not possible at dim-6. The reason is that adding a scalar \(\Xi_{1}\) and a vector \(\mathcal{B}\) or \(\mathcal{W}\) with properly chosen coupling strengths would not change \(\vec{C}^{(6)}\), as their contributions come with opposite signs, see Eq.~(\ref{eq:5.17}). Therefore one can never uniquely determine the UV particle content using only dim-6 coefficients.
\end{itemize}
If \(\vec{C}\) is on a \(k\)-face \((k>1)\): in a 2-dimensional problem, the only \(k\)-face for \(k>1\) is \(\mathbf{C}\) itself.

If \(\vec{C}\) is the origin: If \(\vec{C}=(0,0)\), as a special case of the first point, all \(w_{\alpha}\) are forced to vanish. This means that if all coefficients are measured to be zero at dim-8, we can be confident that a new state cannot exist in any form to couple to a pair of lepton doublets. This serves as a strict confirmation of the SM lepton sector.

\begin{figure}[h]
	\begin{center}
		\includegraphics[width=.8\linewidth]{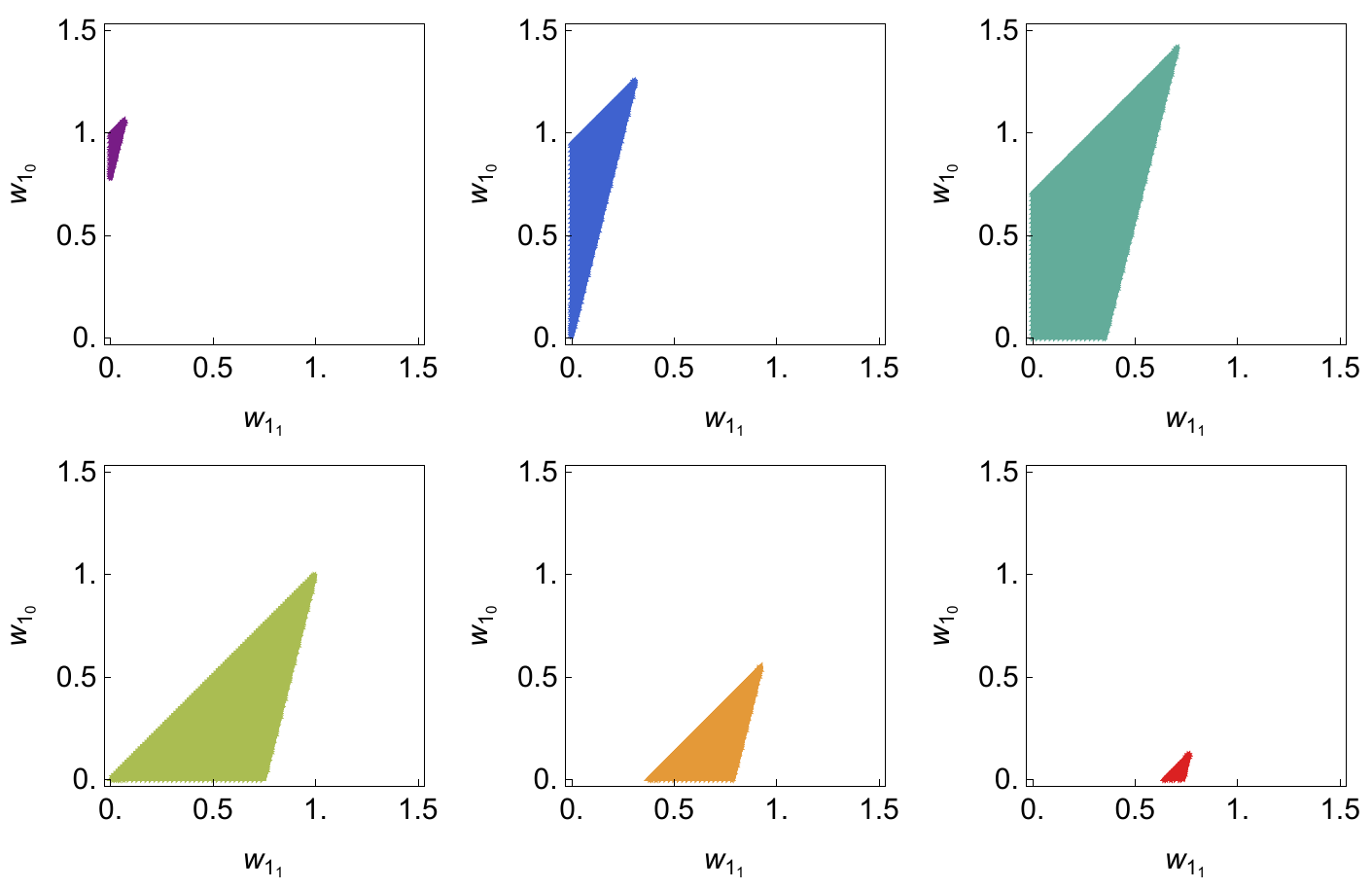}
	\end{center}
	\caption{Solution spaces for \(w_{\alpha}\), as shown in terms of \(\left(w_{\textbf{1}_{1}}, w_{\textbf{1}_{0}}\right)\), at 6 different points shown in Figure \ref{fig8}; see discussion in text.}
	\label{fig9}
\end{figure}

\begin{itemize}
  \item At dim-6, the coefficients generated by a \(\Xi_{1}\) scalar and a \(\mathcal{B}\) or \(\mathcal{W}\) vector can cancel each other. A strict confirmation of the SM is impossible.
\end{itemize}
If \(\vec{C}\) is in the interior: The allowed range of \(w_{\alpha}\) is always finite. In other words, our ignorance on the magnitude of contribution from each type of particles is limited. In Figure \ref{fig9}, we plot the allowed solution spaces for several benchmark points labeled in Figure \ref{fig8} by different colors. They are shaded with the corresponding color in Figure \ref{fig9}. We see that the solution space is always closed, finite, and becomes small as \(\vec{C}\) moves towards the edge of the cone.

\begin{figure}[h]
	\begin{center}
		\includegraphics[width=.47\linewidth]{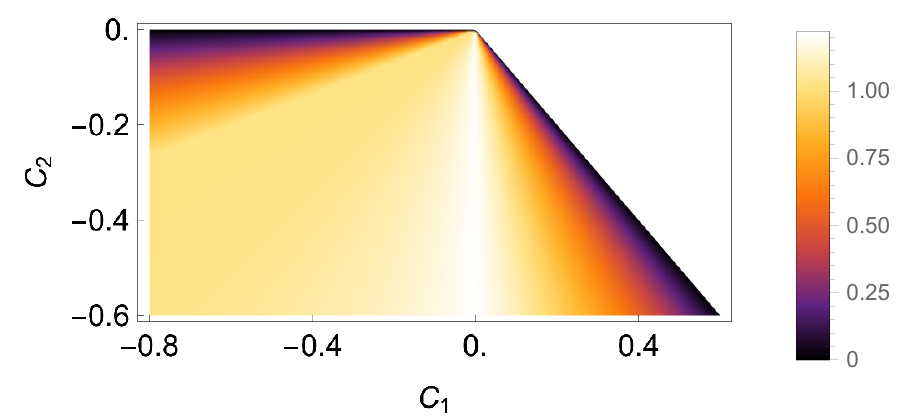}
        \includegraphics[width=.47\linewidth]{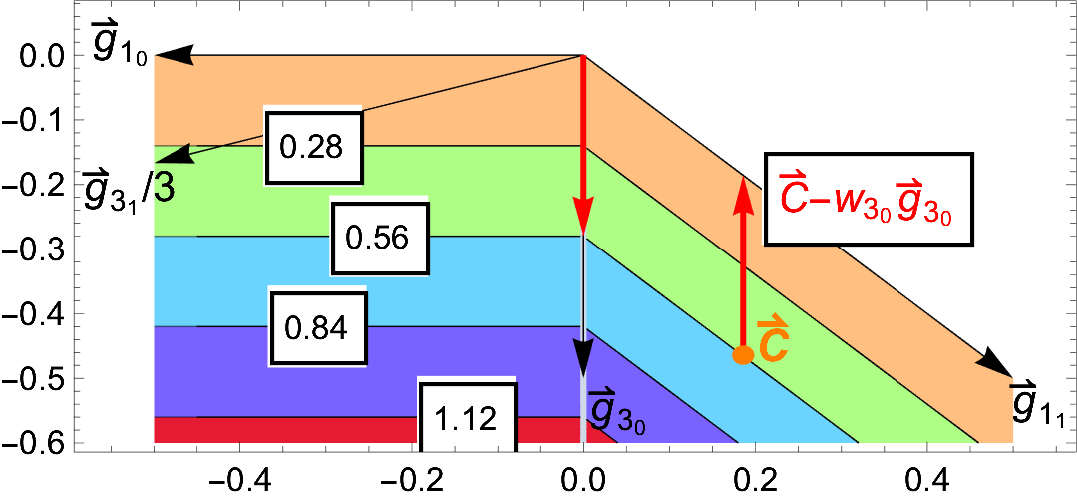}
	\end{center}
	\caption{Left: the degeneracy \(\delta\) in the positivity cone, defined in Eq.~(\ref{eq:5.24}). Right: the maximal possible value of \(w_{\textbf{3}_{0}}\) for \(\vec{C}=(0.186,-0.464)\), to illustrate Eq.~(\ref{eq:5.10}); see the discussion in the main text.}
	\label{fig10}
\end{figure}

It is more intuitive to directly plot the distribution of the degeneracy. In Section 2, we have defined the \(\Delta\) variable to quantify the degeneracy:
\begin{equation}
\Delta \equiv \max _{\vec{w}_{(1)}, \vec{w}_{(2)} \in \mathcal{W}}\left|\vec{w}_{(1)}-\vec{w}_{(2)}\right|\label{eq:5.23}
\end{equation}
which is the largest ``distance'' between two feasible solutions and where $\mathcal{W}$ is the set of feasible solutions for \(\vec{w}\). We further define a relative degeneracy \(\delta\) which is independent of the scale of the problem:
\begin{equation}
\delta \equiv \max _{\vec{w}_{(1)}, \vec{w}_{(2)} \in \mathcal{W}}\left|\vec{w}_{(1)}-\vec{w}_{(2)}\right| / \max _{\vec{w} \in \mathcal{W}}|\vec{w}|\label{eq:5.24}
\end{equation}
In Figure \ref{fig10} left, we show that \(\delta\) vanishes at the boundary of the cone, but becomes larger as it moves inside. This is a general feature which we will demonstrate with more examples.

\begin{itemize}
  \item In contrast, the dim-6 coefficient is given by \(C^{(6)}=w_{\textbf{3}_{1}}^{(6)}-w_{\textbf{1}_{0}}^{(6)} / 2-w_{\textbf{3}_{0}}^{(6)} / 2\), and with this the \(w_{\alpha}^{(6)}\) values are not bounded.
\end{itemize}
How can we obtain an upper bound on the weights \(w_{\alpha} ?\) While it is certainly possible to derive the solution space in terms of \(\left(w_{\textbf{1}_{1}}, w_{\textbf{1}_{0}}\right)\) and compute the maximum and minimum possible values of any \(w_{\alpha}\), this procedure becomes complicated as the number of \(\alpha\) increases. Furthermore, if some \(w_{\alpha}\) depends on free real parameters (e.g. the example of Section 5.3), the solution space will be infinite dimensional, and cannot be easily characterized. Fortunately, Eq.~(\ref{eq:5.10}) gives a much more convenient way to obtain an upper limit. Let us see how this works.

Suppose \(\vec{C}\) is measured to be \(\vec{C}=(0.186,-0.464)\), and we want to find the contribution of a heavy \(\mathcal{W}\) particle. What is the maximum value that \(w_{\textbf{3}_{0}}=g_{\mathcal{W}}^{2} / M_{\mathcal{W}}^{4}\) can take?\\

Obviously, \(w_{\textbf{3}_{0}}=0\) is allowed: in this case the other three generators \(\vec{g}_{\textbf{1}_{1}}\), \(\vec{g}_{\textbf{3}_{1}}\) and \(\vec{g}_{\textbf{1}_{0}}\) will positively generate the orange region and the entire space below it, which contains \(\vec{C}\). Now if we increase \(w_{\textbf{3}_{0}}\), the allowed space will be shifted down, as indicated by the red down arrow. For \(w_{\textbf{3}_{0}}=0.28\), the allowed region is the green area and downwards; for \(w_{\textbf{3}_{0}}=0.56\), it is the blue area and downwards, and so on. However, at some point the boundary of the allowed area will touch \(\vec{C}\). When this happens, \(w_{\textbf{3}_{0}}\) reaches its maximum; beyond this point, the cone generated by \(\vec{g}_{\alpha}\) cannot contain \(\vec{C}\). Conversely, one can equivalently consider moving \(\vec{C}\) above, by subtracting \(w_{\textbf{3}_{0}} \vec{g}_{\textbf{3}_{0}}\) from \(\vec{C}\), as indicated by the red up arrow. As we keep increasing \(w_{\textbf{3}_{0}}, \vec{C}-w_{\textbf{3}_{0}} \vec{g}_{\textbf{3}_{0}}\) arrives at the edge when \(w_{\textbf{3}_{0}}=0.557\), and this indicates that the maximum value is \(0.557\). This is exactly what Eq.~(\ref{eq:5.10}) means.

In Section 5.3, we will see that this same method can be used to infer an upper bound even with infinite number of generators.

\subsubsection{SM Higgs boson}
\label{sec5.2.2}
Our second example is slightly more non-trivial and also more realistic: we consider the 4-Higgs operators. Recall the dim-6 and dim-8 operators are:
\[
\begin{array}{ll}
\text { Dim-} 6 & \text { Dim-} 8 \\
Q_{\varphi \square}=\partial_{\mu}\left(H^{\dagger} H\right) \partial^{\mu}\left(H^{\dagger} H\right) ~~~~& Q_{H^{4}}^{(1)}=\left(D_{\mu} H^{\dagger} D_{\nu} H\right)\left(D^{\nu} H^{\dagger} D^{\mu} H\right) \\
Q_{\varphi D}=\left(H^{\dagger} D_{\mu} H\right)\left(D^{\mu} H^{\dagger} H\right) & Q_{H^{4}}^{(2)}=\left(D_{\mu} H^{\dagger} D_{\nu} H\right)\left(D^{\mu} H^{\dagger} D^{\nu} H\right) \\
~~~~& Q_{H^{4}}^{(3)}=\left(D^{\mu} H^{\dagger} D_{\mu} H\right)\left(D^{\nu} H^{\dagger} D_{\nu} H\right)
\end{array}
\]
and the six generators correspond to the following states:

\begin{center}
\begin{tabular}{c|cccccc}
Particle & Spin & Charge/irrep & Interaction & \(\mathrm{ER}\) & \(\vec{c}\) & \(\vec{c}^{(6)}\) \\
\hline
\(\mathcal{B}_{1}\) & 1 & \(1_{1}\) & \(g \mathcal{B}_{1}^{\mu \dagger}(H^{T} \epsilon \stackrel{\leftrightarrow}{D}_{\mu} H)+h.c.\)  & \(\text{\Checkmark}\)& \(8(1,0,-1)\) & \(2(1,2)\) \\
\(\Xi_{1}\) & 0 & \(3_{1}\) & \(g M \Xi_{1}^{I \dagger}(H^{T} \epsilon \tau^{I} H)+h.c.\)  & \(\text{\XSolidBrush}\) & \(8(0,1,0)\) & \(2(-1,2)\) \\
\(\mathcal{S}\) & 0 & \(1_{0}(S)\) & \(g M \mathcal{S}(H^{\dagger} H)\) & $\text{\Checkmark}$ & \(2(0,0,1)\) & \(\frac{1}{2}(1,0)\) \\
\(\mathcal{B}\) & 1 & \(1_{0}(A)\) & \(g \mathcal{B}^{\mu}(H^{\dagger} \stackrel{\leftrightarrow}{D}_{\mu} H)\) & $\text{\Checkmark} $& \(2(-1,1,0)\) & \(\frac{1}{2}(1,-4)\) \\
\(\Xi_{0}\) & 0 & \(3_{0}(S)\) & \(g M \Xi_{0}^{I}(H^{\dagger} \tau^{I} H)\) & $\text{\XSolidBrush}$ & \(2(2,0,-1)\) & \(-\frac{1}{2}(1,4)\) \\
\(\mathcal{W}\) & 1 & \(3_{0}(A)\) & \(g \mathcal{W}^{\mu I}(H^{\dagger} \tau^{I} \stackrel{\leftrightarrow}{D}_{\mu} H)\) & \(\text{\XSolidBrush}\) & \(2(1,1,-2)\) & \(\frac{3}{2}(1,0)\) \\
\end{tabular}
\begin{figure}[h]
	\begin{center}
		\includegraphics[width=.7\linewidth]{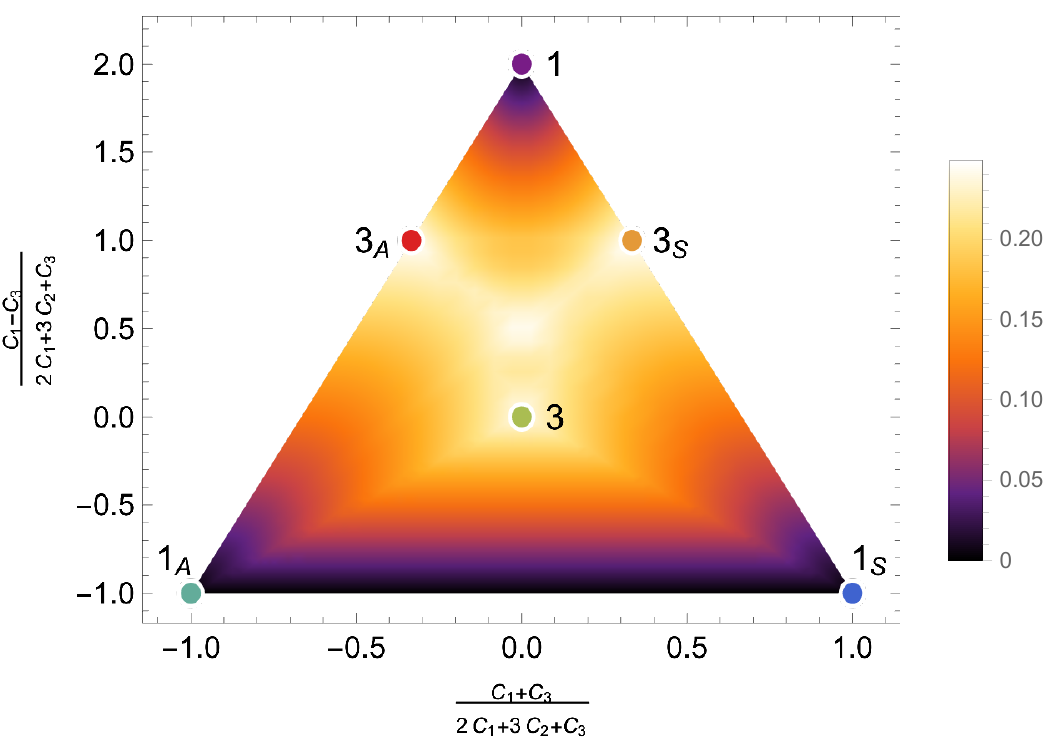}
	\end{center}
	\caption{A cross section of the positivity cone in the space of \(\vec{C}=\left(C_{1}, C_{2}, C_{3}\right)\), defined by the intersection of the hyperplane \(2 C_{1}+3 C_{2}+C_{3}=1\) with the positivity cone, and the degeneracy \(\Delta\) on this intersection, as defined in Eq.~(\ref{eq:5.23}). The color dots represent the six generators.}
	\label{fig11}
\end{figure}
\end{center}

\begin{figure}[h]
	\begin{center}
		\includegraphics[width=.9\linewidth]{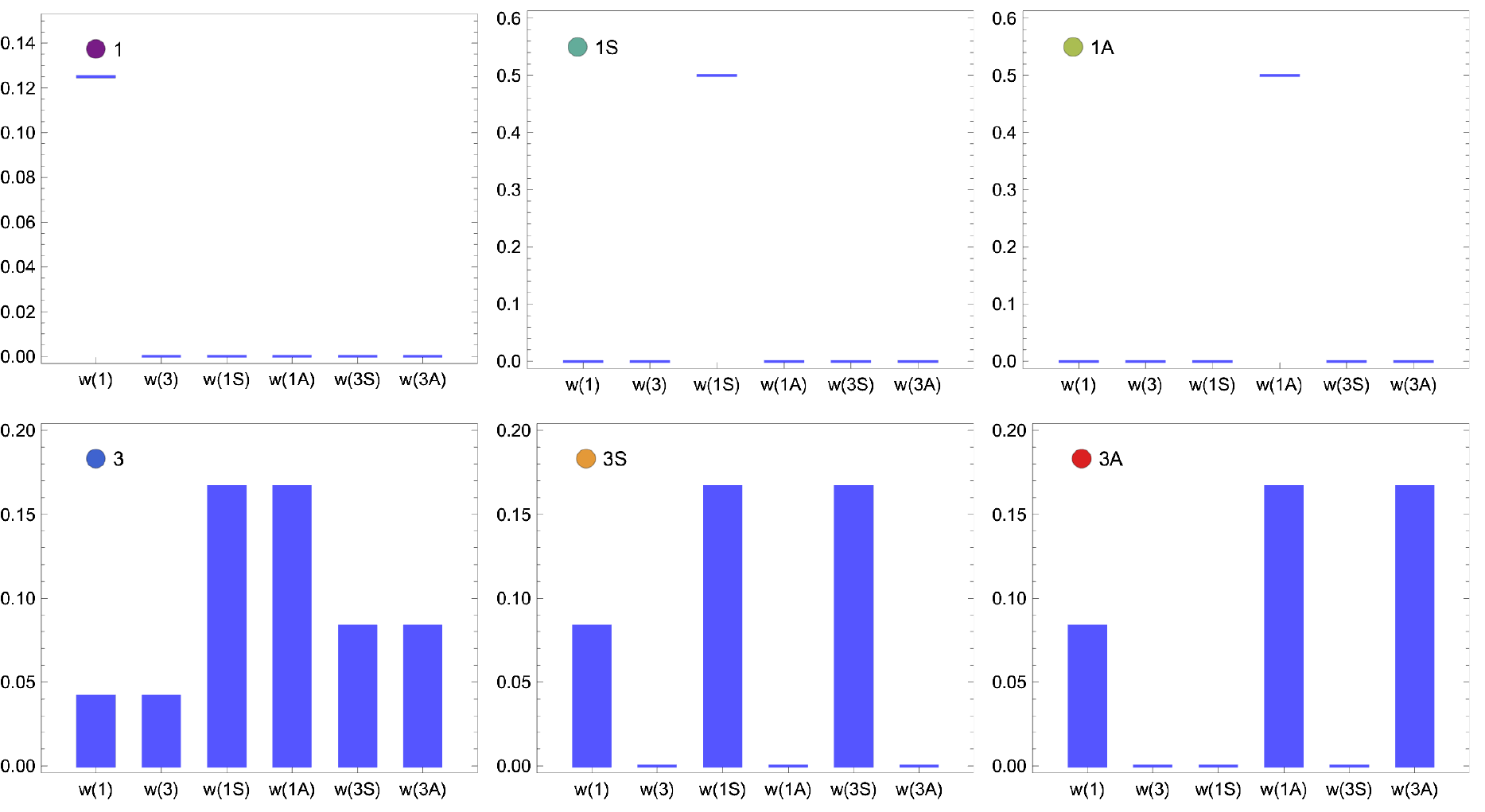}
	\end{center}
	\caption{Solutions for \(w_{\alpha}\) (denoted as \(\left.w(\alpha)\right)\) at all generators, \(\vec{g}_{\textbf{1}}, \vec{g}_{\textbf{1} S}, \vec{g}_{\textbf{1} A}, \vec{g}_{\textbf{3}}, \vec{g}_{\textbf{3} S}, \vec{g}_{\textbf{3} A}\). The blue bar indicates the allowed range of \(w_{\alpha} . \vec{C}\) is normalized such that \(2 C_{1}+3 C_{2}+C_{3}=1\).}
	\label{fig12}
\end{figure}

In Figure \ref{fig11} we show a cross section of the positivity cone, and the degeneracy for \(w_{\alpha}\), as defined in Eq.~(\ref{eq:5.23}). The black region represents where the solution for \(w_{\alpha}\) is unique and no degeneracy is present. It includes not only the three ERs: \(\mathbf{1}, \mathbf{1}_{S}\) and \(\mathbf{1}_{A}\), but also the line segment connecting \(\mathbf{1}_{S}\) and \(\mathbf{1}_{A}\), which represents a triangular face whose vertices are \(\mathbf{1}_{S}, \mathbf{1}_{A}\), and the origin. In Figure \ref{fig12}, we show the allowed values of \(w_{\alpha}\) at each generator vector, obtained by using the trick of Eqs. (5.10) and (5.15).\\

Let us check again the four physical implications of this triangular cone being salient.

If \(\vec{C}\) is on an \(\mathrm{ER}\): \(w_{\alpha}\)'s are uniquely fixed as indicated by the three plots of the first row of Figure \ref{fig12}. In contrast, for the non-extremal generators \(\mathbf{3}, \mathbf{3}_{S}\) and \(\mathbf{3}_{A}\), at least some \(w_{\alpha}\) values have some arbitrariness.

\begin{figure}[h]
	\begin{center}
		\includegraphics[width=.3\linewidth]{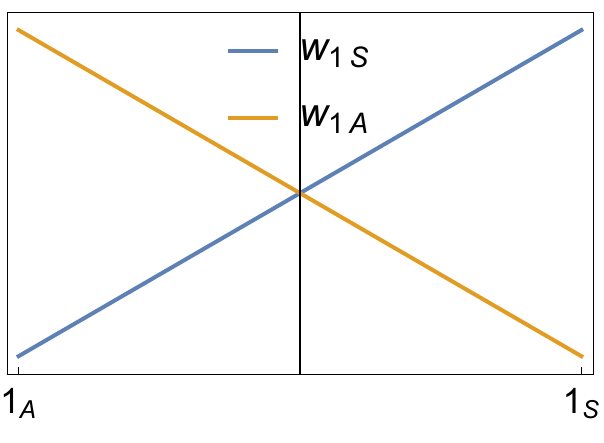}
        \includegraphics[width=.6\linewidth]{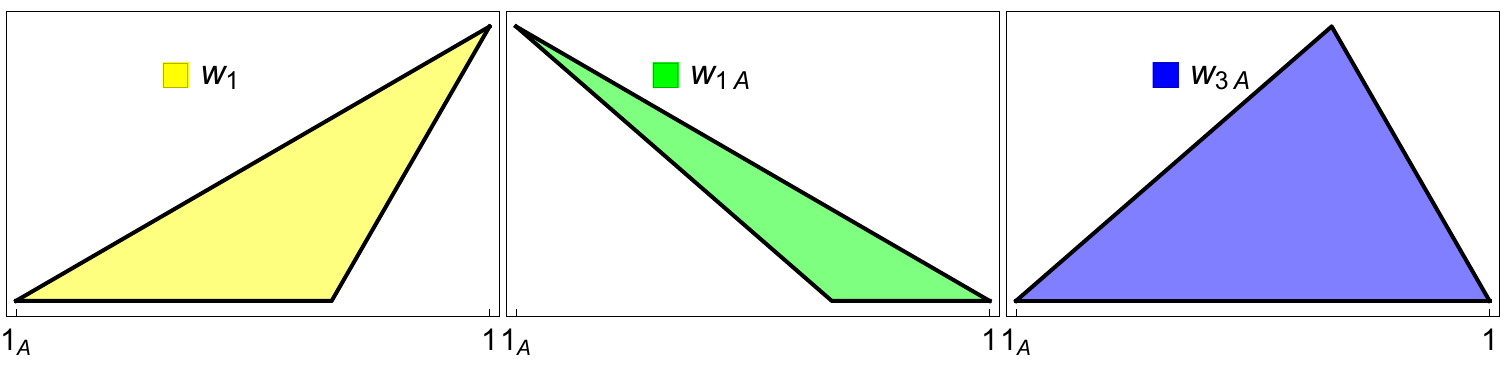}
	\end{center}
	\caption{Left: the \(w_{\alpha}\) values on the \(\mathbf{1}_{A}-\mathbf{1}_{S}\) edge. Only \(w_{\textbf{1} S}\) and \(w_{\textbf{1} A}\) are nonzero. Right: the \(w_{\alpha}\) values on the \(\textbf{1}_{A}-\textbf{1}\) edge. Only \(w_{\textbf{1}}, w_{\textbf{1} A}\) and \(w_{\textbf{3} A}\) are nonzero. The shaded regions are their allowed range. \(\vec{C}\) is always normalized such that \(2 C_{1}+3 C_{2}+C_{3}=1\).}
	\label{fig13}
\end{figure}

If \(\vec{C}\) is on a \(k\)-face \((k>1)\): for the face \(O-\mathbf{1}_{A}-\mathbf{1}_{S}\), where \(O\) is the origin, the degeneracy vanishes as it is a triangle face with no additional generator vector sitting between \(\vec{g}_{1 A}\) and \(\vec{g}_{1 S}\). In Figure \ref{fig13} left we show the \(w_{\alpha}\) values for all points on the line segment between \(\mathbf{1}_{A}\) and \(\mathbf{1}_{S}\). We see that the values of \(w_{1_{S}}\) and \(w_{1_{A}}\) are always uniquely fixed, while the other \(w_{\alpha}\)'s vanish. On the other hand, this does not hold for the other two faces, \(O-\textbf{1}-\textbf{1}_{S}\) and \(O-\textbf{1}-\textbf{1}_{A}\). The reason is that additional generator vectors, \(\vec{g}_{3_{S}}\) and \(\vec{g}_{3_{A}}\), live on these two faces respectively, so we have \(l=3\) and \(k=2\). The consequence is shown in Figure \ref{fig13} right: the \(w_{\alpha}\) values on the \(\mathbf{1}_{A}-\mathbf{1}\) edge are not unique, except for the endpoints \(\mathbf{1}_{A}\) and \(\mathbf{1}\), which are ERs. It is however worth noting that \(w_{3}, w_{1 S}\) and \(w_{3 S}\) are still fixed to be \(0 .\) This can be easily understood from extremality, as \(\mathbf{3}, \mathbf{1}_{S}, \mathbf{3}_{S}\) are not on this face. For this reason, even though \(w_{\mathbf{1}}, w_{1 A}\) and \(w_{3 A}\) cannot be uniquely fixed, the solution space is only \(l-k=1\)-dimensional.

Together with the previous point, we conclude that if the SM is extended by \(\mathcal{B}_{1}\), or by \(\mathcal{S}\), or by \(\mathcal{B}\), or by a combination of \(\mathcal{S}\) and \(\mathcal{B}\), then a perfect measurement at dim-8 would allow us to uniquely pin down the UV particle content, and exclude all alternative possibilities.

If \(\vec{C}\) is the origin: all \(w_{\alpha}\)'s are forced to vanish. The easiest way to see this in the present example is to notice that
\begin{equation}
\vec{n} \cdot \vec{g}_{\alpha}>0 \text { for all } \alpha, \quad \vec{n} \equiv(2,3,1)
\end{equation}
and therefore
\begin{equation}
0=\vec{C} \cdot \vec{n}=\sum_{\alpha} w_{\alpha} \vec{g}_{\alpha} \cdot \vec{n} \geq 0
\end{equation}
is saturated only if \(w_{\alpha}=0\) for all \(\alpha\). As a result, if the coefficients are all consistent with zero, we can exclude the existence of all BSM states that couple to the Higgs boson.

If \(\vec{C}\) is in the interior: inferring a model-independent upper limit on \(w_{\alpha}\), or equivalently a lower limit on \(M_{\alpha} / \sqrt{g_{\alpha}}\), is always possible. This is obvious from Figure \ref{fig11} which shows that the degeneracy \(\Delta\) is finite over the entire cone.

\begin{figure}[h]
	\begin{center}
		\includegraphics[width=.55\linewidth]{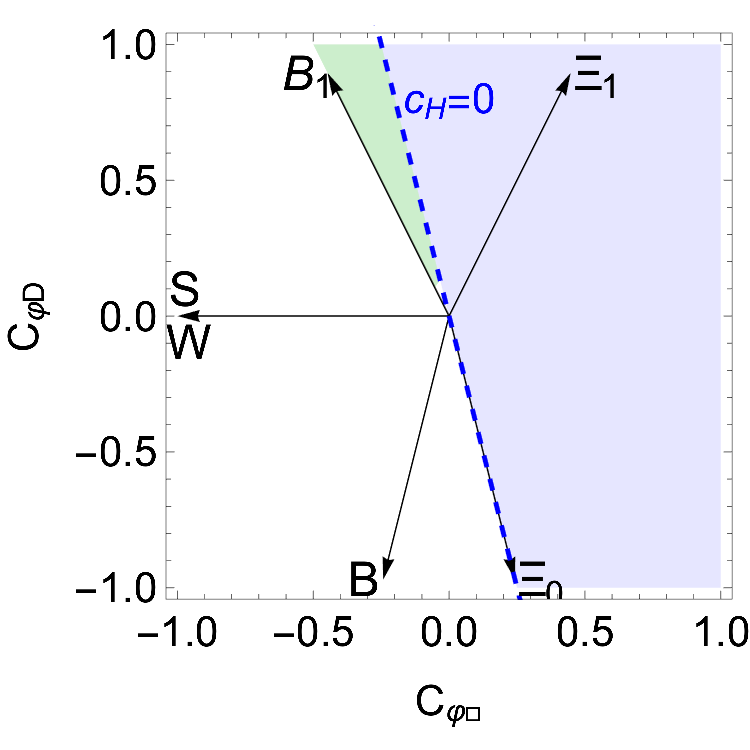}
	\end{center}
	\caption{The dim-6 space \(\left(C_{\varphi \square}, C_{\varphi D}\right)\) and its tree-level generators, normalized. The shaded region (blue and green) is outside of \(H_{\Xi_{1}} .\) In this region, lower limits on \(g_{\Xi_{1}}^{2} / M_{\Xi_{1}}^{2}\) can be set. The blue region corresponds to \(c_{H}=c_{\varphi \square}+C_{\varphi D} / 4 \geq 0\).}
	\label{fig14}
\end{figure}

Finally, we have also mentioned that it is possible to set a lower bound on certain \(w_{\alpha}\)'s, using Eq.~(\ref{eq:5.15}), even at dim-6 (where \(w_{\alpha}\) should be understood as \(\left.g_{\alpha}^{2} / M^{2}\right)\), if a tree-level UV completion is assumed. Let us see how this works. In Figure \ref{fig14} we show the dim-6 ``generators'', \(\vec{c}^{(6)}\), for the six different UV particles. We see that their conical hull is the entire 2-dimensional \(\left(C_{\varphi \square}, C_{\varphi D}\right)\) plane. The absence of a positivity cone at dim-6 is exactly what prevented us from inferring the UV models in a similar way as in the dim-8 cases. However, even though \(\mathbf{C}\) is the entire space, the \(\mathbf{H}_{k}\) set in Eq.~(\ref{eq:5.15}) may still be a proper cone, and \(\vec{C} \notin \mathbf{H}_{k}\) for some \(\vec{C}^{(6)}\)'s. In fact, in the present example, \(\mathbf{H}_{\Xi_{1}}\) is a proper cone, which can be obviously seen if we flip the arrow for \(\Xi_{1}\). The shaded region (blue and green) indicates the \(\vec{C}\)'s that are not in \(\mathbf{H}_{\Xi_{1}}\), and according to Eq.~(\ref{eq:5.15}), for these points, a lower limit on \(g_{\Xi_{1}}^{2} / M_{\Xi_{1}}^{2}\) can be set. In practice, if \(\vec{C}^{(6)}\) is found to be in the shaded area, this lower limit will guide us to search for the further evidence of the \(\Xi_{1}\) particle(s).

A similar conclusion has been given in Ref.~\cite{Low:2009di}. If \(c_{H}>0\), then a doubly charged scalar particle must exist. In the basis of 
Ref.~\cite{Low:2009di}, \(c_{H}\) is our \(C_{\varphi \square}+C_{\varphi D} / 4\), and \(c_{H}>0\) is the blue shaded region in Figure \ref{fig14}. So the conclusion of \cite{Low:2009di} is consistent with our finding from Eq.~(\ref{eq:5.15}), as the \(\Xi_{1}\) state exactly what is needed for a doubly charged scalar to exist. In general, Eq.~(\ref{eq:5.15}) is a more systematic approach: it not only shows that the SMEFTs in the green shaded region also have a doubly charged scalar, but also gives quantitatively a lower bound on \(g_{\Xi_{1}}^{2} / M_{\Xi_{1}}^{2}\).

\subsubsection{SM quarks }
\label{sec5.2.3}
As a last polyhedral example, consider the SM left-handed quark operators, where the parameter space is 4-dimensional. Recall the operators are
\begin{align}
&O_{1}=\partial_{\mu}\left(\bar{q} \gamma_{\nu} q\right) \partial^{\mu}\left(\bar{q} \gamma^{\nu} q\right) \\
&O_{2}=\partial_{\mu}\left(\bar{q} \gamma_{\nu} \tau^{I} q\right) \partial^{\mu}\left(\bar{q} \gamma^{\nu} \tau^{I} q\right) \\
&O_{3}=\partial_{\mu}\left(\bar{q} \gamma_{\nu} T^{A} q\right) \partial^{\mu}\left(\bar{q} \gamma^{\nu} T^{A} q\right) \\
&O_{4}=\partial_{\mu}\left(\bar{q} \gamma_{\nu} \tau^{I} T^{A} q\right) \partial^{\mu}\left(\bar{q} \gamma^{\nu} \tau^{I} T^{A} q\right)
\end{align}
and the UV states are

\[
\begin{array}{cccccc}
\text { State}&\text{ Spin}&\text{ Charge } & \text { Interaction } & \mathrm{ER} & \vec{c}  \\
\hline \omega_{1} & 0 & (3,1)_{-\frac{1}{3}} & \omega_{1}^{a} \epsilon_{a b c} \bar{q}^{c b} \epsilon q^{c} & \text{\Checkmark} & \frac{1}{3}(-1,1,3,-3)  \\
\mathcal{V}_{-\frac{1}{3}} & 1 & (3,3)_{-\frac{1}{3}} & \mathcal{V}_{-\frac{1}{3}}^{a I} \epsilon_{a b c} \bar{q}^{c^{b}} \epsilon \tau^{I} i \overleftrightarrow{D_{\mu}} q^{c} & \text{\Checkmark} &\frac{1}{3}(3,1,-9,-3)   \\
\mathcal{V}_{\frac{1}{3}}  & 1 & (6,1)_{\frac{1}{3}}&{ \mathcal{V}^{\dagger}}_{\frac{1}{3}}^{ab\mu}  \bar{q^c}^{(a}{\epsilon i} \overleftrightarrow{D}_{\mu} q^{b)} & \text{\Checkmark} & \frac{1}{6}(2,-2,3,-3)  \\
\Upsilon & 0 & (6,3)_{\frac{1}{3}} & \Upsilon^{\dagger I a b} \bar{q^c}^{(a}{\epsilon \tau}^{I} q^{b)} & \text{\XSolidBrush} &\frac{1}{6}(-6,-2,-9,-3) \\
\mathcal{B}& 1 & (1,1)_{0} & \mathcal{B}^{\mu} \bar{q} \gamma_{\mu} q & \text{\Checkmark} & \frac{1}{2}(-1,0,0,0)  \\
\mathcal{W} & 1 & (1,3)_{0} &\mathcal{W}^{I \mu} \bar{q} \gamma_{\mu} \tau^{I} q & \text{\Checkmark}& \frac{1}{2}(0,-1,0,0) \\
\mathcal{G} & 1 & (8,1)_{0} & \mathcal{G}^{A \mu} \bar{q} \gamma_{\mu} T^{A} q& \text{\Checkmark} & \frac{1}{2}(0,0,-1,0)\\
\mathcal{H} & 1 & (8,3)_{0} & \mathcal{H}^{A I \mu} \bar{q} \gamma_{\mu} T^{A} \tau^{I} & \text{\XSolidBrush} & \frac{1}{2}(0,0,0,-1)
\end{array}
\]

The positivity cone is described by 6 bounds:
\begin{equation}
\left(\begin{array}{llll}
3 & 3 & 1 & 1 \\
0 & 3 & 0 & 1 \\
0 & 0 & 1 & 1 \\
12 & 0 & 1 & 9 \\
1 & 0 & 0 & 1 \\
0 & 0 & 0 & 1
\end{array}\right)\left(\begin{array}{l}
C_{1} \\
C_{2} \\
C_{3} \\
C_{4}
\end{array}\right) \leq 0
\end{equation}

\begin{figure}[h]
	\begin{center}
		\includegraphics[width=.7\linewidth]{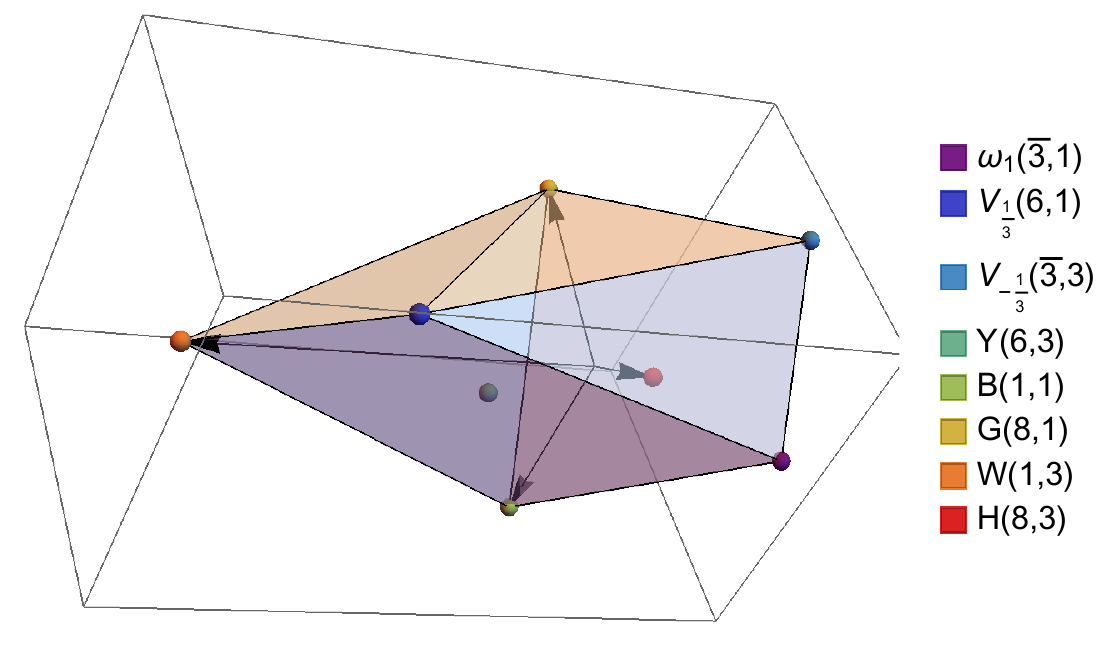}
	\end{center}
	\caption{A slice of the positive cone in the space of \(\vec{C}=\left(C_{1}, C_{2}, C_{3}, C_{4}\right)\), defined by intersecting the entire cone with a 3-dimensional hyperplane. The colored dots represent the generators. 6 of them are ERs and define the polytope cone. The black arrows are the projections of the 4 basis operators on this hyperplane.}
	\label{fig15}
\end{figure}

In Figure \ref{fig15} we show a 3-dimensional cross section of the positivity cone together with its generators. Let us focus on the SMEFTs at the \(k\)-faces, including the origin, the ERs, the 2-faces and the 3-faces. In these regions, a perfect dim-8 measurement determines the UV particle content. More concretely, we have the following conclusions:

0-face: if no BSM states are coupled to SM left-handed quark currents, a perfect measurement of dim-8 coefficients could unambiguously confirm that this is the case.

1-faces (ERs): these are the vertices in Figure \ref{fig15}: \(\vec{g}_{\overline{\textbf{3}} \textbf{1}}, \vec{g}_{\overline{\textbf{3}} \textbf{1}}, \vec{g}_{\textbf{61}}, \vec{g}_{\textbf{11}}, \vec{g}_{\textbf{13}}, \vec{g}_{\textbf{81}}\). If the SM is extended by one of the corresponding particles: \(\omega_{1}, \mathcal{V}_{\frac{1}{3}}, \mathcal{V}_{-\frac{1}{3}}, \Upsilon, \mathcal{G}, \mathcal{W}\), a perfect measurement of at dim-8 could unambiguously confirm the scenario. Note that the other two particles, \(\Upsilon\) and \(\mathcal{H}\), are not extremal. As a result, the corresponding SM extensions could be alternatively explained by combining several other UV particles.

2-faces: these are the edges in Figure \ref{fig15}. Let us use the particle names to represent the generators. Each 2-face can be defined by 2 such generators. They are:
\begin{align}
&\left(\omega_{1}, \mathcal{V}_{\frac{1}{3}}\right), \quad\left(\omega_{1}, \mathcal{V}_{-\frac{1}{3}}\right), \quad\left(\mathcal{V}_{\frac{1}{3}}, \mathcal{V}_{-\frac{1}{3}}\right), \quad\left(\mathcal{V}_{\frac{1}{3}}, \mathcal{W}\right),\quad (\mathcal{B}, \mathcal{W}) \\
&\left(\omega_{1}, \mathcal{B}\right), \quad\left(\mathcal{G}, \mathcal{V}_{\frac{1}{3}}\right), \quad\left(\mathcal{G}, \mathcal{V}_{-\frac{1}{3}}\right), \quad(\mathcal{B}, \mathcal{G}), \quad(\mathcal{W}, \mathcal{G})
\end{align}
If the SM is extended by one of these pairs of BSM particles, a perfect measurement at dim-8 could uniquely confirm the corresponding scenario. Note that the other pairs do not form 2-faces and therefore do not have this property.

3-faces: these are the facets in Figure \ref{fig15}. The following facets are a triangle in the slice shown in Figure \ref{fig15} (or a 3-simplex in the full space, \(l=k=3\)):
\begin{equation}
\left(\omega_{1}, \mathcal{V}_{\frac{1}{3}}, \mathcal{V}_{-\frac{1}{3}}\right), \quad\left(\mathcal{V}_{\frac{1}{3}}, \mathcal{V}_{-\frac{1}{3}}, \mathcal{G}\right), \quad\left(V_{\frac{1}{3}}, \mathcal{G}, \mathcal{W}\right), \quad(\mathcal{B}, \mathcal{W}, \mathcal{G})
\end{equation}
If the SM is extended by one of these four combinations of BSM particles, a perfect measurement at dim-8 could uniquely confirm the scenario.

On the other hand, there are two quadrilateral facets \((l=4, k=3)\) :
\begin{equation}
\left(\omega_{1}, \mathcal{V}_{\frac{1}{3}}, \mathcal{W}, \mathcal{B}\right), \quad\left(\omega_{1}, \mathcal{V}_{-\frac{1}{3}}, \mathcal{G}, \mathcal{B}\right)
\end{equation}
These are not 3-simplices. If the SM is extended by one of these combinations, a perfect measurement at dim-8 would still leave a degeneracy. The solution set for \(w_{\alpha}\) is however only one-dimensional \((l-k=1)\), so there is only one degree of freedom left to be fixed.

The last point can be understood by extremality, but one can also derive the same conclusions in other ways. Consider, first, an example where the SM is extended by \((\mathcal{B}, \mathcal{W}, \mathcal{G})\). We have
\begin{equation}
\vec{C}=\sum_{\alpha=(\mathcal{B}, \mathcal{W}, \mathcal{G})} w_{\alpha} \vec{g}_{\alpha}
\end{equation}
Suppose now we know \(\vec{C}\) from measurements, and want to solve
\begin{equation}
\vec{C}=\sum_{\alpha} w_{\alpha}^{\prime} \vec{g}_{\alpha}
\end{equation}
for \(w^{\prime} .\) Is the solution unique? The 3 generators \((\mathcal{B}, \mathcal{W}, \mathcal{G})\) together with the origin define a 3-dimensional facet of the positivity region, which corresponds to a positivity bound: \(C_{4} \leq 0\). Defining the normal vector to this bound
\begin{equation}
\vec{n}=(0,0,0,-1)
\end{equation}
we have
\begin{equation}
0=\vec{C} \cdot \vec{n}=\sum_{\alpha} w_{\alpha}^{\prime} \vec{g}_{\alpha} \cdot \vec{n}=\sum_{\alpha=\left(\omega_{1}, \mathcal{V}_{\pm \frac{1}{3}}, \Upsilon, \mathcal{H}\right)} w_{\alpha}^{\prime} \vec{g}_{\alpha} \cdot \vec{n} \geq 0
\end{equation}
Since \(\vec{g}_{\alpha} \cdot \vec{n}\) is strictly positive for \(\alpha=\left(\omega_{1}, \mathcal{V}_{\pm \frac{1}{3}}, \Upsilon, \mathcal{H}\right)\), the above forces \(w_{\alpha}^{\prime}=0\) for these particles. We are left with
\begin{equation}
\vec{C}=\sum_{\alpha=(\mathcal{B}, \mathcal{W}, \mathcal{G})} w_{\alpha} \vec{g}_{\alpha}=\sum_{\alpha=(\mathcal{B}, \mathcal{W}, \mathcal{G})} w_{\alpha}^{\prime} \vec{g}_{\alpha}
\end{equation}
But the three \(\vec{g}_{\alpha}\) vectors are linearly independent, therefore \(w_{\alpha}^{\prime}=w_{\alpha}\) is the unique solution, which means all SMEFTs on the \((\mathcal{B}, \mathcal{W}, \mathcal{G})\) face have no degeneracy.

Alternatively, consider the case where the SM is extended by \(\left(\omega_{1}, \mathcal{V}_{\frac{1}{3}}, \mathcal{W}, \mathcal{B}\right)\). This is also a facet that corresponds to a positivity bound: \(3 C_{2}+C_{4} \leq 0 .\) Let us now define the normal vector to this bound:
\begin{equation}
\vec{n}=(0,-3,0,-1)
\end{equation}
Similar to the previous example, using this vector one can exclude \(w_{\alpha}^{\prime}\) for all the other particles, but to determine the 4 non-vanishing \(w_{\alpha}^{\prime}\)'s, the system
\begin{equation}
\vec{C}=\sum_{\alpha=\left(\omega_{1}, \mathcal{V}_{\frac{1}{3}}, \mathcal{W}, \mathcal{B}\right)} w_{\alpha} \vec{g}_{\alpha}=\sum_{\alpha=\left(\omega_{1}, \mathcal{V}_{\frac{1}{3}}, \mathcal{W}, \mathcal{B}\right)} w_{\alpha}^{\prime} \vec{g}_{\alpha}
\end{equation}
is not sufficient, because the \(4 \ \vec{g}_{\alpha}\)'s are not linearly independent. In fact, they span a 3-dimensional subspace, and therefore the solution for \(w_{\alpha}^{\prime}\) should contain at most one free parameter.

\subsection{Non-polyhedral case}
\label{sec5.3}
When the positivity cone is non-polyhedral, the situation can be somewhat different. The number of ERs is infinite, while the number of coefficients is finite. Therefore the dimension of the solution space for \(w_{\alpha}\)'s is also infinite. Furthermore, if some ERs are continuous, setting lower bounds on a single \(w_{\alpha}\) would become impossible.\footnote{One may still set lower bounds on the total contribution of certain subsets of \(w_{\alpha}\)'s, but we will not discuss this possibility here.} It is however still possible
to use Eq.~(\ref{eq:5.10}) to infer an upper bound on \(w_{\alpha}\). In fact, the four implications of \(\mathbf{C}\) being salient are still valid. In this section we will show an example.

Consider photon operators with parity violation. Recall the operators are
\begin{align}
&O_{1}=\left(B_{\mu \nu} B^{\mu \nu}\right)\left(B_{\rho \sigma} B^{\rho \sigma}\right) \\
&O_{2}=\left(B_{\mu \nu} \tilde{B}^{\mu \nu}\right)\left(B_{\rho \sigma} \tilde{B}^{\rho \sigma}\right) \\
&O_{3}=\left(B_{\mu \nu} B^{\mu \nu}\right)\left(B_{\rho \sigma} \tilde{B}^{\rho \sigma}\right)
\end{align}
and there are two generators,
\begin{equation}
\vec{g}_{1}(\theta)=\left(c_{\theta}^{2}, s_{\theta}^{2}, 2 s_{\theta} c_{\theta}\right), \quad \vec{g}_{2}=(1,1,0)\label{eq:5.46}
\end{equation}
The first corresponds to a scalar \(S\), in a partial UV completion, that interacts with SM hypercharge photons with the following terms:
\begin{equation}
\mathcal{L} \supset \frac{g}{M} S\left(c_{\theta} B_{\mu \nu} B^{\mu \nu}+s_{\theta} B_{\mu \nu} \tilde{B}^{\mu \nu}\right)
\end{equation}
where the angle \(\theta\) represents a mixture CP-even and CP-odd interactions. The second generator corresponds to a spin-2 transition. Since it is not extremal, to large extent it does not affect our discussion. The first generator depends on a free parameter \(\theta\), and it is in this sense we say this problem involves an infinite number of ERs.
\begin{figure}[h]
	\begin{center}
		\includegraphics[width=.45\linewidth]{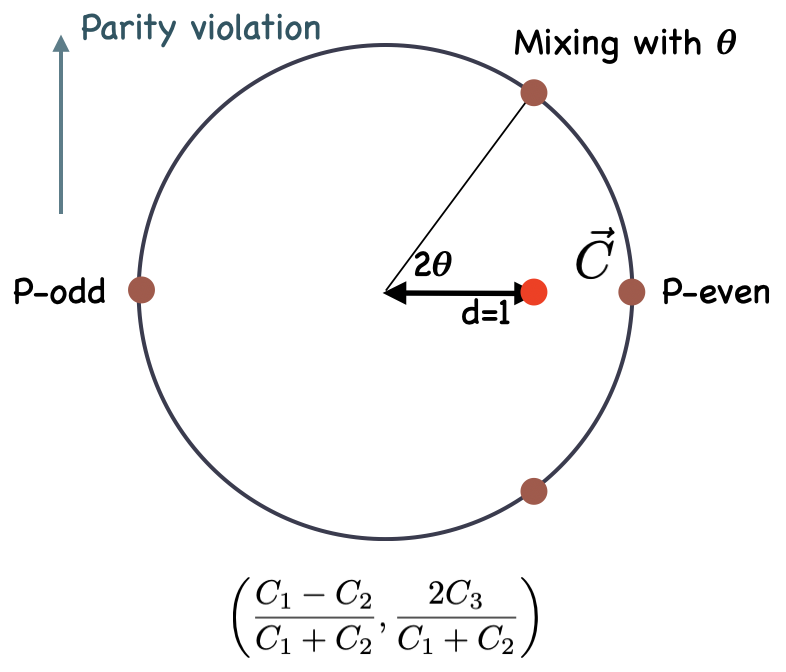}
        \includegraphics[width=.5\linewidth]{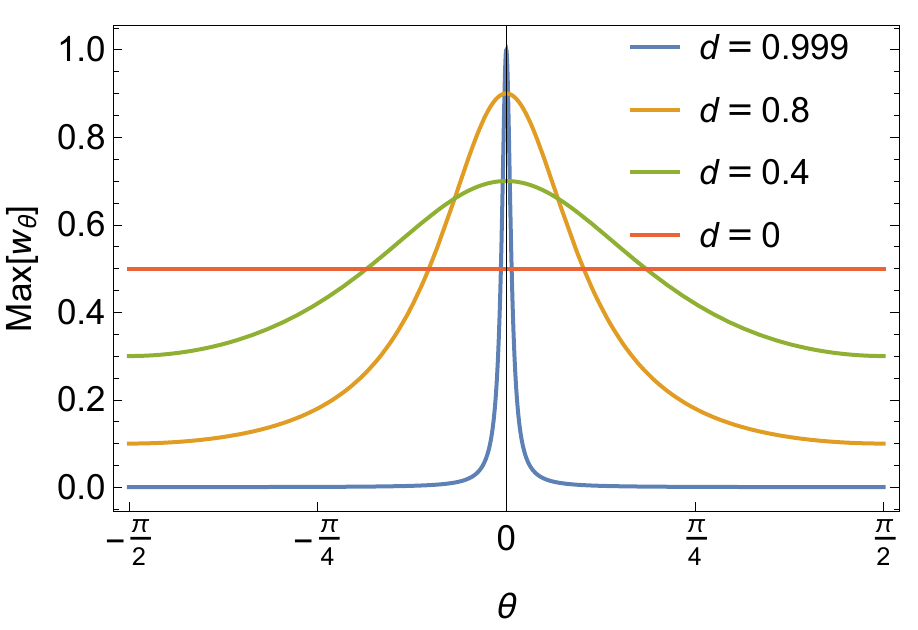}
	\end{center}
	\caption{Left: a projection of the positivity cone in the space of \(\vec{C}=\left(C_{1}, C_{2}, C_{3}\right)\) onto the hyperplane \(C_{1}+C_{2}=1\). The brown points correspond to generators with different mixing angles \(\theta\). \(\theta=0\) and \(\theta=\pi\) correspond to purely P-even and P-odd couplings, both of which are parity conserving. Right: the upper bound on \(w_{\theta}\) of different \(\theta\) values, for the point \(\vec{C}=\frac{1}{2}(1+d, 1-d, 0)\), shown as a red point on the left plot.}
	\label{fig16}
\end{figure}

The cone spanned by \(\vec{g}_{1}(\theta)\) and \(\vec{g}_{2}\) is a circular one. We show a cross section of this cone in Figure \ref{fig16} left. The brown points correspond to coefficients generated by different mixing angle \(\theta\). \(\theta=0\) and \(\theta=\pi\) correspond to purely P-even and P-odd couplings, and both are parity conserving and have a zero \(C_{3}\) value. Other \(\theta\) values imply parity violation.

 Now suppose the measured \(\vec{C}\) falls on the slice \(C_{1}+C_{2}=1\), and the distance from \(\vec{C}\) to the axis of the cone is \(d\). We are interested in the possible UV particles that generate \(\vec{C}\). In particular, suppose a set of scalars, \(S_{\theta}\), with different \(\theta\) values, interact with photons as described in Eq.~(\ref{eq:5.46}), each with some coupling \(g_{\theta}\) and mass \(M_{\theta}\), then for each \(\theta\) value, what is the maximum possible contribution, \(w_{\theta}=g^{2} / M^{4}\), allowed by the observed \(\vec{C}\) value? According to our discussion in Section 5.1 and 5.2, inferring an upper limit on \(w_{\theta}\) is always possible. In fact, Eq.~(\ref{eq:5.10}) works even with an infinite number of ERs.

Assume for the moment \(C_{3}=0\), and so \(\vec{C}=\frac{1}{2}(1+d, 1-d, 0), 0 \leq d \leq 1\), and thus the EFT has no parity violation. Following the discussion in Section 5.2, we define \(\vec{C}(\lambda) \equiv \vec{C}-\lambda \vec{g}_{\theta}=\left(\frac{1+d}{2}-\lambda c_{\theta}^{2}, \frac{1-d}{2}-\lambda s_{\theta}^{2},-2 \lambda s_{\theta} c_{\theta}\right)\). Using bounds presented in Eq.~(\ref{eq:4.47}), the condition for \(\vec{C}(\lambda) \in \mathbf{C}\) is
\begin{align}
&1+d \geq 2 \lambda c_{\theta}^{2}, \quad 1-d \geq 2 \lambda s_{\theta}^{2} \\
&2 \lambda\left(1-d c_{\theta}\right) \leq 1-d^{2}
\end{align}
the result for \(\max \left(w_{\theta}\right)\) is the largest \(\lambda\) such that all above conditions are satisfied, so
\begin{equation}
\max \left(w_{\theta}\right)=\left\{\begin{array}{ll}
1 & \text { if } d=1, \theta=0 \\
\frac{1}{2} \frac{1-d^{2}}{1-d \cos 2 \theta} & \text { otherwise }
\end{array}\right.
\end{equation}
This result is shown in Figure \ref{fig16} right for different values of \(d\). Note that the problem is symmetric under rotation \(\theta \rightarrow \theta+\delta \theta\), so if the observed \(\vec{C}\) has a nonzero \(C_{3}\), we could simply define \(d=\sqrt{\left(C_{1}-C_{2}\right)^{2}+4 C_{3}^{2}}\), and shift \(\theta\) by \(\delta \theta=\frac{1}{2} \tan ^{-1} \frac{2 C_{3}}{C_{1}-C_{2}}\), and still obtain similar results for \(\max \left(w_{\theta}\right)\)

From Figure \ref{fig16} right, we see that when \(d\) is larger, i.e., when \(\vec{C}\) is closer to the boundary, the constraints on \(w_{\theta}\) are tighter, indicating less degeneracy in finding the UV. In particular, if \(d=1\), the only nonzero \(\max \left(w_{\theta}\right)\) is when \(\theta=0\), i.e., only purely P-even scalar can exist in the UV. This is because \(\vec{C}\) is extremal when \(d=1\). For smaller \(d\)'s, \(\max \left(w_{\theta}\right)\) in general can be nonzero, but only the small \(\theta\) values are allowed to have large \(w_{\theta}\), so the degeneracy exists but is limited. As an example of this degeneracy, assuming \(C_{3}=0\), a \(d<1\) case can be explained by combining a parity-even and a parity-odd scalars:
\begin{align}
&\theta_{1}=0, \quad w_{\theta_{1}}=\frac{1+d}{2} \\
&\theta_{2}=\frac{\pi}{2}, \quad w_{\theta_{2}}=\frac{1-d}{2} \\
&\vec{C}=w_{\theta_{1}} \vec{g}_{1}\left(\theta_{1}\right)+w_{\theta_{2}} \vec{g}_{1}\left(\theta_{2}\right)=\frac{1}{2}(1+d, 1-d, 0)
\end{align}
so that the UV theory is parity-conserving. Alternatively, it could also be explained by two scalars with mixed couplings:
\begin{align}
&\theta_{3}=\frac{1}{2} \cos ^{-1} d, \quad w_{\theta_{3}}=\frac{1}{2} \\
&\theta_{4}=-\frac{1}{2} \cos ^{-1} d, \quad w_{\theta_{4}}=\frac{1}{2} \\
&\vec{C}=w_{\theta_{3}} \vec{g}_{1}\left(\theta_{3}\right)+w_{\theta_{4}} \vec{g}_{1}\left(\theta_{4}\right)=\frac{1}{2}(1+d, 1-d, 0)
\end{align}
so that the UV theory is parity-violating. The \(\vec{g}_{1}(\theta)\) for \(\theta=\theta_{1}, \theta_{2}, \theta_{3}, \theta_{4}\) are all shown in Figure \ref{fig16} left as brown dots. For all four \(\theta\) values, the corresponding weight \(w_{\theta}\) saturates the curve in Figure \ref{fig16} right.

Overall, when \(d\) is less than one, we would be able to conclude that the UV states with smaller \(|\theta|\) are expected to give the dominant contribution to \(\vec{C}\), but we would not be able to completely exclude contributions from larger \(|\theta|\)'s. The smaller the \(d\) value, the weaker the constraints on the UV couplings, and therefore larger degeneracies. This is consistent with the general picture we found for the polytope cases.

\section{Generators at one loop}
\label{sec6}
Our formalism presented in Section 4 is based on the dispersion relation and symmetry arguments, and is therefore not limited to tree-level UV completions. Integrating out a heavy particle loop, however, generates a coefficient vector, \(\vec{g}_{loop} \in \mathbf{C}\), which can be thought of as a new ``loop generator''. This new generator can be decomposed as a positive sum of the standard generators. In this section, we consider two concrete examples: photon operators and Higgs operators. The goal is to demonstrate how these generators are produced, from a dispersive point of view, and discuss their implications in the inverse problem.

To this end, we perform the loop matching following the dispersion relation, Eq.~(\ref{eq:3.9}). For one-loop UV completion, the intermediate state should run through all two-particle states. We call these two particles \(X\) and \(Y\). It is convenient to work in the helicity basis. The partial wave expansion for \(i j \rightarrow X Y\) allows to classify the intermediate states by their angular momenta:
\begin{align}
&\mathbf{M}(i j \rightarrow X Y)=\sum_{J}(2 J+1) e^{i \lambda \phi} d_{\lambda \mu}^{J}(\theta) T_{i j X Y}^{J}(s) \\
&\lambda=\lambda_{i}-\lambda_{j}, \quad \mu=\lambda_{X}-\lambda_{Y}
\end{align}
where \(\lambda_{x}\) is the helicity of particle \(x\). \( T_{i j X Y}^{J}\) is the partial wave coefficient. We shall plug this into the dispersion relation:
\begin{align}
\mathcal{M}^{i j k l}=&\frac{1}{2 \pi} \int_{(\epsilon \Lambda)^{2}}^{\infty} \frac{d s}{s^{3}} \sum_{X Y} \int \operatorname{dLIPS} \mathbf{M}(i j \rightarrow X Y) \mathbf{M}^{*}(k l \rightarrow X Y)+(j \rightarrow \bar{l}, l \rightarrow \bar{j})\label{eq:6.3} \\
\nonumber=&\frac{1}{16 \pi^{2}} \delta_{\lambda_{i}-\lambda_{j}, \lambda_{k}-\lambda_{l}} \int_{(\epsilon \Lambda)^{2}}^{\infty} \frac{d s}{s^{3}} \sum_{X Y} \sqrt{1-\frac{4 M^{2}}{s}} \sum_{J}(2 J+1) T_{i j X Y}^{J}(s) T_{k l X Y}^{* J}(s) \\
&+(j \rightarrow \bar{l}, l \rightarrow \bar{j})
\end{align}
This equation has the same form as our dispersive relation Eq.~(\ref{eq:3.9}), if we take \(T_{i j X Y}^{J}\) to be the \(m^{i j}\) matrices. The \(\delta_{\lambda_{i}-\lambda_{j}, \lambda_{k}-\lambda_{l}}\) factor represents the rotational symmetry around the forward axis. Different values of \(\lambda_{i}-\lambda_{j}\) correspond to the \(J_{3}=0, \pm\) states in Section 4.3, or the \(\mathbf{1}_{S, A}, \mathbf{2}\) states in Section 4.2. If \(\lambda_{i}-\lambda_{j}>0\), an \(i \leftrightarrow j, k \leftrightarrow l\) double exchange will add the corresponding \(\lambda_{i}-\lambda_{j}<0\) contribution. If \(\lambda_{i}-\lambda_{j}=0, T_{i j X Y}^{J}\) is either symmetric or antisymmetric under \(i \leftrightarrow j, k \leftrightarrow l\), depending on the value of \(J\). We see that a partial-wave expansion for the loop-level matching naturally reproduces various generators that have been already discussed in Section 4. Of course, loop matching can be performed without the partial wave expansion or dispersion relation, but understanding the loop generators from a dispersive point of view is helpful for a discussion of the inverse problem at the loop level.

In practice, expanding all angular momenta is not helpful, as the total angular momentum is not bounded from above, if \(i j \rightarrow X Y\) occurs in the \(t\)- or \(u\)-channel. It is also not necessary, because at dim-8 it's sufficient to study the forward scattering, and so the only relevant information is whether \(J\) is even or odd, which determines the symmetry property of \(m^{i j}\). In the following, we will work out two concrete cases, to see how various generators are produced by integrating out fermion loops. We consider fermion loops because heavy fermions do not generate purely bosonic operators at the tree level. The amplitudes are therefore loop-induced only, making the examples more relevant. Of course, the same approach can be used to understand heavy scalar/vector loops.

\subsection{Photon operators}
\label{sec6.1}
Consider integrating out a heavy fermion \(F\) of charge \(Q\) at the loop level. We follow Eq.~(\ref{eq:6.3}), but instead of expanding the angular momenta, we directly perform the phase space integration. We factor out a \(\delta_{\lambda_{i}-\lambda_{j}, \lambda_{k}-\lambda_{l}}\), and remove the azimuthal angular integration:
\begin{align}
\mathcal{M}^{i j k l} &=\frac{1}{2 \pi} \int_{(\epsilon \Lambda)^{2}}^{\infty} \frac{d s}{s^{3}} \int \mathrm{d} \operatorname{LIPS} \mathbf{M}(i j \rightarrow X Y) \mathbf{M}^{*}(k l \rightarrow X Y)+(j \rightarrow \bar{l}, l \rightarrow \bar{j}) \\
&=\frac{1}{32 \pi^{2}} \int_{(\epsilon \Lambda)^{2}}^{\infty} \frac{d s}{s^{3}} \int \mathrm{d} \cos \theta \mathbf{M}(i j \rightarrow X Y)(s, \theta) \mathbf{M}^{*}(k l \rightarrow X Y)(s, \theta)+(j \rightarrow \bar{l}, l \rightarrow \bar{j})
\end{align}
An explicit calculation of \(\mathbf{M}(\gamma \gamma \rightarrow F \bar{F})\) gives the following helicity amplitudes before the \((j \rightarrow \bar{l}, l \rightarrow \bar{j})\) crossing:
\begin{align}
\mathcal{M}^{++++} =\mathcal{M}^{----} &=\frac{7 e^{4} Q^{4}}{720 \pi^{2} M^{4}} \\
\mathcal{M}^{++--} =\mathcal{M}^{--++} &=-\frac{e^{4} Q^{4}}{240 \pi^{2} M^{4}} \\
\mathcal{M}^{+-+-} =\mathcal{M}^{-+-+} &=\frac{e^{4} Q^{4}}{180 \pi^{2} M^{4}}\label{eq:6.6}
\end{align}
Clearly, the first two lines represent a \(J_{3}=0\) transition, while the last two elements, \(\mathcal{M}^{+-+-}\) and \(\mathcal{M}^{-+-+}\), represent a \(J_{3}=2~(\)and \(J \geq 2)\) transition. After crossing, we find the following \(\mathcal{M}^{i j k l}\)'s:
\begin{flalign}
\raisebox{-8pt}{$ \mathcal{M}_{J_{3}=0}^{i j k l}=\frac{e^{4} Q^{4}}{720 \pi^{2} M^{4}} \times$}
\begin{tabular}{ r|c|c| }
	\multicolumn{1}{r}{}
	 &  \multicolumn{1}{c}{$++$}
	 & \multicolumn{1}{c}{$--$} \\
	\cline{2-3}
	$++$ & 7 & $-6$  \\
	\cline{2-3}
	$--$ &  $-6$ & 7 \\
	\cline{2-3}
\end{tabular},\quad \raisebox{-8pt}{$\mathcal{M}_{J_{3}=\pm 2}^{i j k l}=\frac{e^{4} Q^{4}}{180 \pi^{2} M^{4}} \times$}
\begin{tabular}{ r|c|c| }
	\multicolumn{1}{r}{}
	 &  \multicolumn{1}{c}{$++$}
	 & \multicolumn{1}{c}{$--$} \\
	\cline{2-3}
	$++$ & 1  &   \\
	\cline{2-3}
	$--$ &   & 1\\
	\cline{2-3}
\end{tabular}
\end{flalign}
where we only show the top-left quarter of the \(\mathcal{M}\) matrix (i.e. the \(++,--\rightarrow++,--\) entries), because the rest are either vanishing or not independent after adding the crossed term. Both of them can be converted to a coefficient vector, by using the following expression for \(\mathcal{M}\) in the helicity basis:
\begin{equation}
\mathcal{M}^{i j k l}=8 \times
\begin{array}{|c|c|}
\hline C_{1}+C_{2} & 2\left(C_{1}-C_{2}\right) \\
\hline  2\left(C_{1}-C_{2}\right) & C_{1}+C_{2} \\
\hline
\end{array}
\end{equation}
We find
\begin{equation}
\vec{g}_{J_{3}=0}=\frac{e^{4} Q^{4}}{5760 \pi^{2} M^{4}}(2,5), \quad \vec{g}_{J_{3}=\pm 2}=\frac{e^{4} Q^{4}}{2880 \pi^{2} M^{4}}(1,1)
\end{equation}
In Section 4.2, we said that the \(\vec{g}_{\textbf{2}}=(1,1)\) generator cannot be produced by a tree-level UV completion. We now see that it is naturally generated at one loop, by separating the \(J_{3}=\pm 2\) contribution. Also note that the \(J_{3}=0\) component is neither purely symmetric or anti-symmetric, but rather a positive combination of both. In total, we find
\begin{equation}
\vec{g}_{loop}=\vec{g}_{J_{3}=0}+\vec{g}_{J_{3}=\pm 2}=\frac{e^{4} Q^{4}}{6 ! \pi^{2} M^{4}} \frac{1}{8}(4,7)
\end{equation}
This result agrees with \cite{Remmen:2019cyz}.

\subsection{SM Higgs operators}
\label{sec6.2}
Let us first consider a simple case where a real scalar \(\phi\) couples to a vector-like heavy fermion \(F:\)
\begin{equation}
\mathcal{L} \supset y \bar{F} F \phi+h . c .
\end{equation}
and we want to integrate out an \(F\) loop to get the four-\(\phi\) operators at dim-8. Neglecting the \(s \leftrightarrow u\) crossing and keeping in mind that \(\phi \phi \rightarrow F \bar{F}\) has a \(t\)-channel and a \(u\)-channel diagram, the nonzero amplitudes are (we omit the \(i j k l\) indices as there is only one scalar)
\begin{align}
&\mathcal{M}=2\left(\mathcal{M}_{t t}+\mathcal{M}_{t u}+\mathcal{M}_{u t}+\mathcal{M}_{u u}\right) \\
&\mathcal{M}_{t t}=\mathcal{M}_{u u}=\frac{79 y^{4}}{20160 \pi^{2} M^{4}} \\
&\mathcal{M}_{t u}=\mathcal{M}_{u t}=-\frac{y^{4}}{10080 \pi^{2} M^{4}}
\end{align}
where \(\mathcal{M}_{t t, u u, t u, u t}\) indicate the corresponding interference contributions by keeping \(t\)- or \(u\)-channel contributions separately in \(\mathbf{M}(i j \rightarrow X Y)(s, \theta)\) and in \(\mathbf{M}^{*}(k l \rightarrow X Y)(s, \theta) .\) Separating the \(t\)- and \(u\)-channels is for later convenience. \(s \leftrightarrow u\) crossing adds an additional factor of 2 . The final result is
\begin{equation}
\mathcal{M}=\frac{11 y^{4}}{720 \pi^{2} M^{4}}
\end{equation}

Now consider a more realistic case: the SM Higgs boson. The Higgs doublet is charged under the fundamental representation of SU(2). Consider two types of vector-like heavy fermions, \(F_{1}\) in representation \(\mathbf{1}_{Y}\) and \(F_{2}\) in \(\mathbf{2}_{Y+\frac{1}{2}} .\) The following interaction terms with the Higgs boson can be written down:
\begin{equation}
\mathcal{L}=y\left(\bar{F}_{2} H\right) F_{1}+ h.c. 
\end{equation}

\begin{figure}[h]
	\centering
		\subfigure[\(H H^{\dagger}( H^{\dagger} H) \rightarrow F_{1} \bar{F}_{1}\)]{\includegraphics[width=.45\linewidth]{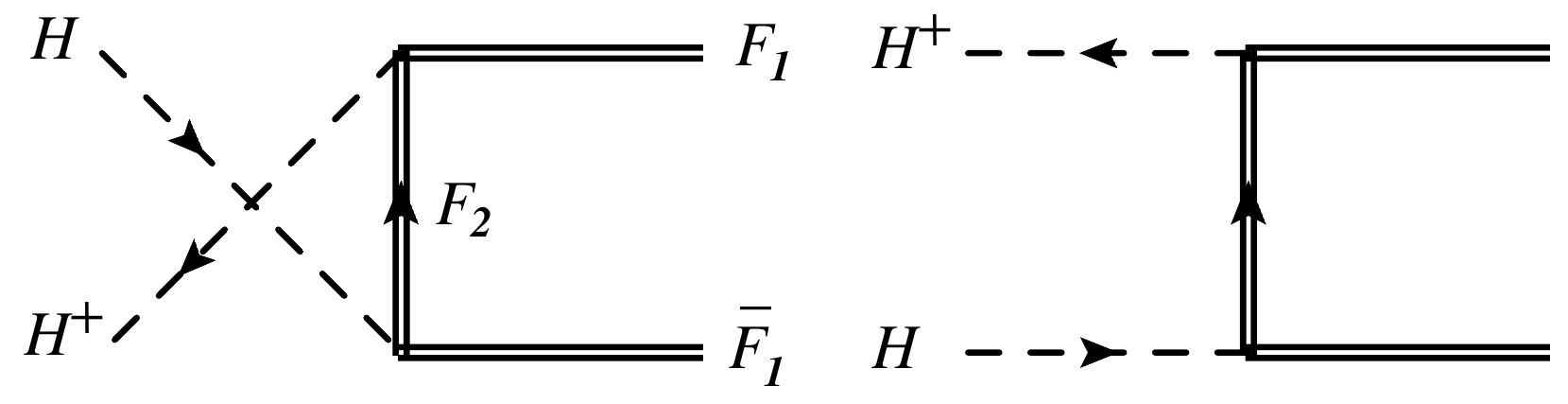}}
        \subfigure[\(H H^{\dagger}( H^{\dagger} H) \rightarrow F_{2} \bar{F}_{2}\)]{\includegraphics[width=.45\linewidth]{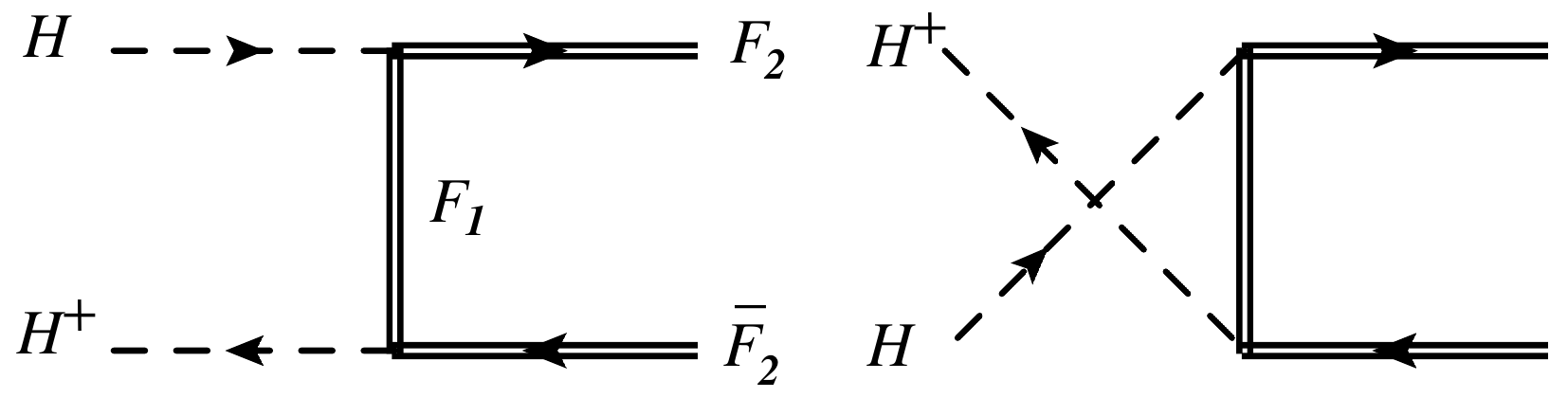}}
	\caption{Diagrams for \(H H^{\dagger}, H^{\dagger} H \rightarrow  F_{1} \bar{F}_{1}, F_{2} \bar{F}_{2}\), through either a \(t\)-channel or a \(u\)-channel diagram. The arrow represents the \(\mathrm{SU}(2)\) charge flow. Double lines represent heavy fermions.}
	\label{fig17}
\end{figure}

Compared to the previous simple example, a main difference is that \(H H^{\dagger} \rightarrow F \bar{F}\) and \(H^{\dagger} H \rightarrow F \bar{F}\) are different processes, one through \(t\)-channel and the other through \(u\)-channel, see Figure \ref{fig17}. In addition, there are two possible intermediate states: \(F_{1} \bar{F}_{1}\) and \(F_{2} \bar{F}_{2}\).

Consider first the \(F_{1} \bar{F}_{1}\) state. For simplicity we ignore the Higgs-boson self interaction. \(H H^{\dagger} \rightarrow F_{1} \bar{F}_{1}\) proceeds in \(u\)-channel, while \(H^{\dagger} H \rightarrow F_{1} \bar{F}_{1}\) proceeds in \(t\)-channel, see Figure \ref{fig17} left. Since \(F_{1}\) is an SU(2) singlet, we expect this contribution to produce the \(\mathbf{1}_{S}\) and \(\mathbf{1}_{A}\) generators. In fact,
\begin{flalign}
\mathcal{M}^{ijkl}&= 
\begin{tabular}{ r|c|c| }
	\multicolumn{1}{r}{}
	 &  \multicolumn{1}{c}{$H_cH^{\dagger d}$}
	 & \multicolumn{1}{c}{$H^{\dagger c}H_d$} \\
	\cline{2-3}
	$H^a H^{\dagger b}$ & \(\mathcal{M}_{u u} \delta_{b}^{a} \delta_{c}^{d}\)   &  \(\mathcal{M}_{u t} \delta_{b}^{a} \delta_{d}^{c}\)  \\
	\cline{2-3}
	$H^{\dagger a}H_b$ &  \(\mathcal{M}_{t u} \delta_{a}^{b} \delta_{c}^{d}\)  & \(\mathcal{M}_{tt} \delta_{a}^{b} \delta_{d}^{c}\)  \\
	\cline{2-3}
\end{tabular}+ (j\leftrightarrow l)\\
&=\begin{array}{|c|c|}
\hline \mathcal{M}_{u u} \delta_{b}^{a} \delta_{c}^{d} & \mathcal{M}_{u t}\left(\delta_{b}^{a} \delta_{d}^{c}+\delta_{d}^{a} \delta_{b}^{c}\right) \\
\hline \mathcal{M}_{t u}\left(\delta_{a}^{b} \delta_{c}^{d}+\delta_{a}^{d} \delta_{c}^{b}\right) & \mathcal{M}_{t t} \delta_{a}^{b} \delta_{d}^{c} \\
\hline
\end{array}\\
\nonumber &=\mathcal{M}^S\begin{array}{|c|c|}
\hline {P_{\mathbf{1}}}^{a\ \ d}_{\ b c} & {P_{\mathbf{1}}}^{a\ \ c}_{\ bd}+{P_{\mathbf{1}}}^{a\ \ c}_{\ db} \\
\hline {P_{\mathbf{1}}}^{b\ \ d}_{\ a c}+{P_{\mathbf{1}}}^{b\ \ d}_{\ c a} & {P_{\mathbf{1}}}^{c\ \ b}_{\ da} \\
\hline
\end{array}\\
&~~~~+\mathcal{M}^A\begin{array}{|c|c|}
\hline {P_{\mathbf{1}}}^{a\ \ d}_{\ b c} & -{P_{\mathbf{1}}}^{a\ \ c}_{\ bd}-{P_{\mathbf{1}}}^{a\ \ c}_{\ db} \\
\hline -{P_{\mathbf{1}}}^{b\ \ d}_{\ a c}-{P_{\mathbf{1}}}^{b\ \ d}_{\ c a} & {P_{\mathbf{1}}}^{c\ \ b}_{\ da} \\
\hline
\end{array}
\end{flalign}
where \(\mathcal{M}^{S} \equiv \mathcal{M}_{t t}+\mathcal{M}_{t u}\) and \(\mathcal{M}^{A} \equiv \mathcal{M}_{t t}-\mathcal{M}_{t u}\). We are only showing the bottom-right quarter of \(\mathcal{M}^{i j k l}\) in its matrix form, as the other entries are either vanishing or not independent. As expected, the \(\mathcal{M}^{S}\) and \(\mathcal{M}^{A}\) terms are proportional to the \(\mathcal{G}_{\mathbf{1}_{0 S}}\) and the \(\mathcal{G}_{\mathbf{1}_{0 A}}\) generators of Eq.~(\ref{eq:4.39}). This gives the contribution from the \(F_{1} \bar{F}_{1}\) cut:
\begin{equation}
\vec{g}_{F 1}=\frac{1}{2} \mathcal{M}^{S} \vec{g}_{\textbf{1} S}+\frac{1}{2} \mathcal{M}^{A} \vec{g}_{\textbf{1} A}
\end{equation}
where the \(\vec{g}_{\textbf{1} S, \textbf{1} A}\) vectors are defined in Eq.~(\ref{eq:4.30}). Note that both symmetric and antisymmetric components are generated. They are coming from intermediate states with different total angular momenta.

\begin{figure}[h]
	\begin{center}
        \includegraphics[width=.45\linewidth]{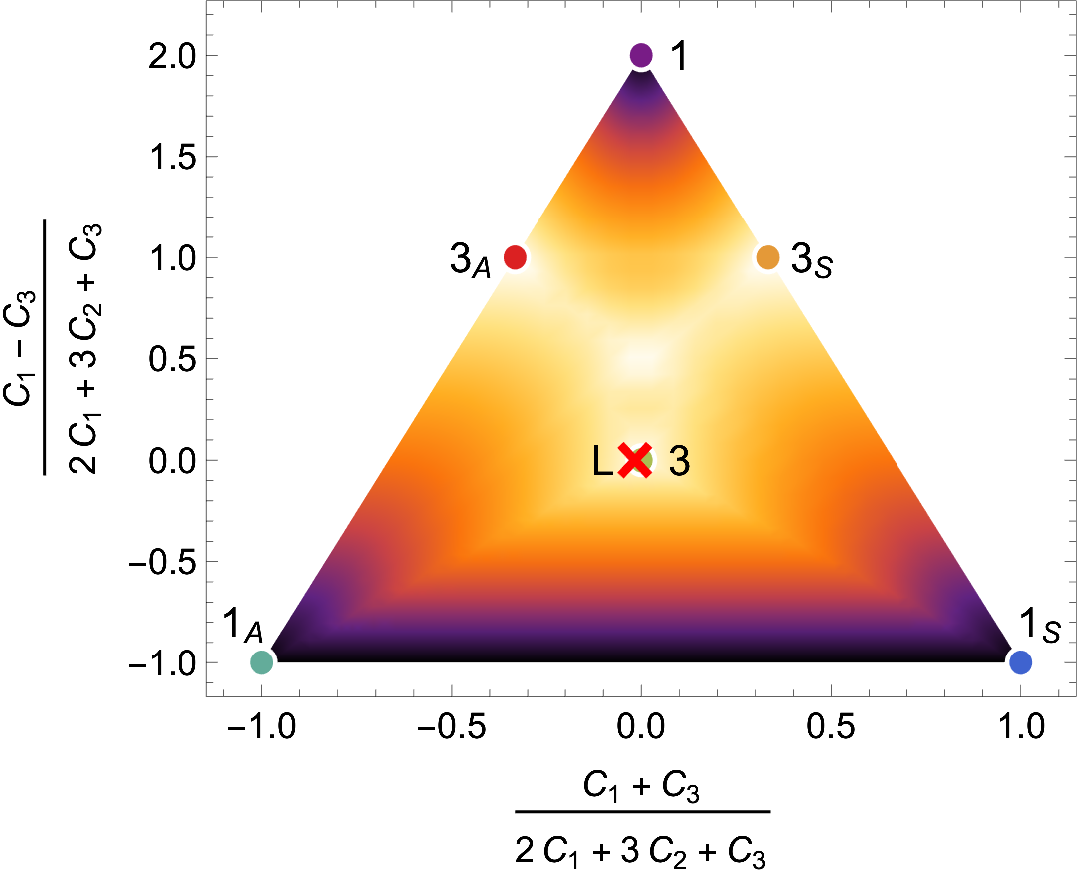}
	\end{center}
	\caption{The cross section of the positivity cone for Higgs operators, together with the 6 tree-level generators and the loop-level generator, \(\vec{g}_{loop}\), given in Eq.~(\ref{eq:6.25}). The latter is represented by a red cross labeled with ``L'', and is very close to the \(\vec{g}_{\textbf{3}}\) generator. The degeneracy is also shown, but is only computed with tree-level generators, and is the same as in Figure \ref{fig11}.}
	\label{fig18}
\end{figure}

Now consider the \(F_{2} \bar{F}_{2}\) cut, see Figure \ref{fig17} right. The only difference is the \(\mathrm{SU}(2)\) indices. For example, for \(H^{a} H_{b}^{\dagger} \rightarrow F_{2}^{X} \bar{F}_{2 Y}\), where \(X, Y\) are the \(\mathrm{SU}(2)\) indices carried by \(F_{2}\) and \(\bar{F}_{2}\), the SU(2) factor is \(\delta_{X}^{a} \delta_{b}^{Y}\). Using the \(\mathrm{SU}(2)\) projectors we can decompose this into \(\mathbf{1} \oplus \mathbf{3}\) :
\begin{equation}
\mathbf{1}: \frac{1}{2} \delta_{b}^{a} \delta_{X}^{Y}, \quad \mathbf{3}:-\frac{1}{2} \delta_{b}^{a} \delta_{X}^{Y}+\delta_{X}^{a} \delta_{b}^{Y}
\end{equation}
The rest of the calculation is the same as the \(F_{1} \bar{F}_{1}\) case. The final result is a sum of \(\mathbf{1}\) and $\textbf{3} $ generators:
\begin{equation}
\vec{g}_{F 2}=\frac{1}{4}\left(\mathcal{M}^{S} \vec{g}_{1 S}+\mathcal{M}^{A} \vec{g}_{1 A}+\mathcal{M}^{S} \vec{g}_{3 S}+\mathcal{M}^{A} \vec{g}_{3 A}\right)
\end{equation}
and the final result is
\begin{equation}
\vec{g}_{\text {loop }}=\vec{g}_{F 1}+\vec{g}_{F 2}=\frac{y^{4}}{10080 \pi^{2} M^{4}}(-2,81,-2)\label{eq:6.25}
\end{equation}
We see that by decomposing the two fermion states into a sum of different \(\mathrm{SU}(2)\) irreps, all 4 generators, \(\vec{g}_{\textbf{1} S}, \vec{g}_{\textbf{1} A}, \vec{g}_{\textbf{3} S}\) and \(\vec{g}_{\textbf{3} A}\), are generated. The final result is very close to \(\vec{g}_{\textbf{3}}\), which is not extremal; see Figure \ref{fig18}. This is partly because \(\mathcal{M}_{t t}=\mathcal{M}_{u u} \gg \mathcal{M}_{t u}=\mathcal{M}_{u t}\), and therefore \(\mathcal{M}^{S} \approx \mathcal{M}^{A}\), so the weights for symmetric and anti-symmetric ERs are almost equal.

\begin{figure}[h]
	\begin{center}
		\includegraphics[width=.25\linewidth]{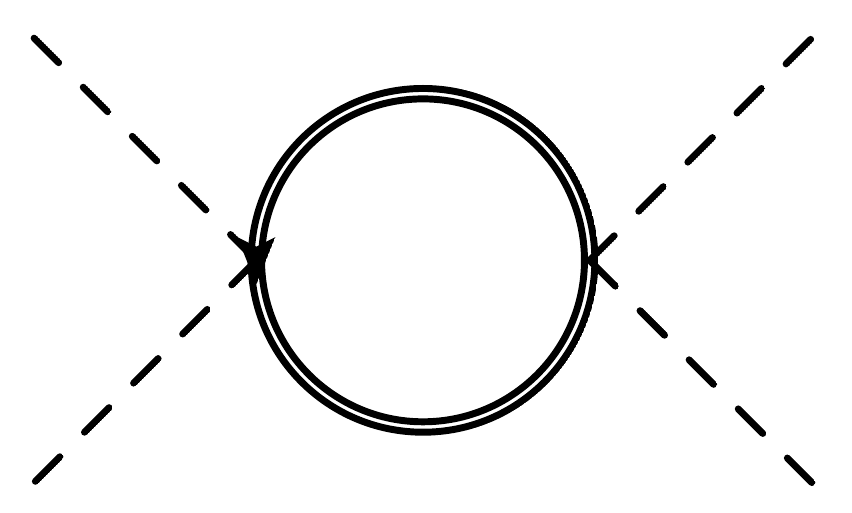}
	\end{center}
	\caption{An example diagram in which neutral heavy particle loop contributes only a s-wave component. The resulting loop generator is likely to be an ER.}
	\label{fig19}
\end{figure}

Since \(\vec{g}_{loop }\) is close to the center of the positivity cone, the degeneracy is maximized. It means that if \(F_{1}\) and \(F_{2}\) exist in the UV particle spectrum, it would be difficult to uniquely confirm this hypothesis. In general, the decomposition of two-particle intermediate states tend to generate multiple ERs, and the resulting loop generator tends to stay inside the cone. This means that all the conclusions we had for the tree-level UV completions still hold. In particular,

\begin{itemize}
  \item A SMEFT on an ER must still correspond to a one-particle extension, unless there is a loop-level generator that stays exactly on that ER. This could be possible, for example, for a singlet particle running in a loop as shown in Figure \ref{fig19}, where the loop only contributes to the \(s\)-wave component. In this case the loop generates a single ER, and it cannot be distinguished from a tree level UV completion, unless higher dimensional operator coefficients can be measured.

  \item A SMEFT on a \(k\)-face spanned by \(k\) ERs would still have a unique UV particle content, unless the loop-level generator happens to fall on the same face.

  \item The degeneracy inside the cone can be affected by the new loop generator. However, in the present example, since the loop generator almost coincides with \(\vec{g}_{\textbf{3}}\), we do not expect the degeneracy plot to change significantly, though one needs to keep in mind that distinguishing between \(\vec{g}_{\textbf{3}}\) and \(\vec{g}_{loop }\) is almost impossible.

\end{itemize}
\subsection{Light particle loops}
\label{sec6.3}
The BSM contribution to \(\mathcal{M}^{i j k l}\) may also arise from loops that involve light particles. In this case, additional complications may arise. If there are no loops with mixed heavy and light particles, the situation is easier to deal with. The light particle loops can be computed on both sides of the dispersion relation, and subtracted from both sides. The remaining contributions are from purely heavy-particle loops, and can be analyzed as in the previous section. In other words, we may define \(\mathcal{M}^{i j k l}\) to incorporate only the heavy loop contributions, and \(\mathcal{M} \in \mathbf{C}\) still holds.

If mixed heavy-light particle loops exist, the same subtraction cannot be done, because it spoils the positiveness of the r.h.s.~of the dispersion relation \cite{Bi:2019phv}. Suppose \(X\) is a two-particle state with only SM particles, then \(\mathbf{M}_{i j \rightarrow X}\) contains both SM and BSM contribution:
\begin{equation}
\mathbf{M}_{i j \rightarrow X}=\mathbf{M}_{i j \rightarrow X}^{S M}+\mathbf{M}_{i j \rightarrow X}^{B S M}
\end{equation}
an example of \(\mathbf{M}^{B S M}\) is \(i j \rightarrow X\) through some \(t\)-channel heavy propagator. If we subtract the SM part, the integrand of the dispersion relation becomes
\begin{equation}
\mathbf{M}_{i j \rightarrow X} \mathbf{M}_{k l \rightarrow X}^{*}=\mathbf{M}_{i j \rightarrow X}^{S M} \mathbf{M}_{k l \rightarrow X}^{* B S M}+\mathbf{M}_{i j \rightarrow X}^{B S M} \mathbf{M}_{k l \rightarrow X}^{* S M}+\mathbf{M}_{i j \rightarrow X}^{B S M} \mathbf{M}_{k l \rightarrow X}^{* B S M}
\end{equation}
and the positiveness of the parameter space, or the salient-ness of the resulting cone, cannot be guaranteed. To find the possible UV completions, we will have to deal with both SM particles and BSM particles.

Another complication is that the amplitude \(\mathcal{M}^{i j k l}\) can be a function of \(\epsilon \Lambda\). In the previous examples, our calculation does not depends on this scale, because the dispersive integral starts from the threshold, \(s>4 M^{2} \sim \Lambda^{2}(M\) being the loop particle mass), which is larger than \((\epsilon \Lambda)^{2}\). However, with light particle cuts, the dispersive integral from \((\epsilon \Lambda)^{2}\) to \(\Lambda^{2}\) generates additional contributions. These contributions are calculable in the EFT and are known to satisfy all positivity bounds. They will cause \(\mathcal{M}^{i j k l}\left((\epsilon \Lambda)^{2}\right)\) to flow inside \(\mathbf{C}\) when \(\epsilon \Lambda\) decreases.

In this section we consider a concrete example. We consider a vector-like \(\mathrm{SU}(2)\) singlet fermion \(F\) with hypercharge \(-\frac{1}{3}\), which interacts with the SM left-handed quark doublet \(q\) and the Higgs boson \(H\):
\begin{equation}
\mathcal{L}=y(\bar{q} H) F+h . c .
\end{equation}
We see that the heavy \(F\) exchange generates an \(H H^{\dagger} q \bar{q}\) amplitude already at the tree level, and therefore in practice its more realistic to study the \(H H^{\dagger} q \bar{q}\) operators rather than the 4-Higgs operators that are loop-induced. This is in general true when mixed loops are present. The discussion in this section is therefore mostly for the completeness of the picture.

Again we are going to assume that the SM particle masses are negligible. The \(F \bar{F}\) cut contribution is similar to the previous example. It starts above the scale \(\Lambda^{2}\), and therefore is independent of \((\epsilon \Lambda)^{2}\). We find
\begin{align}
&\mathcal{M}_{F}^{S}=\frac{484-45 \pi^{2}}{23040 \pi^{2}} \frac{y^{4}}{M^{4}}, \quad \mathcal{M}_{F}^{A}=\frac{9 \pi^{2}-68}{4608 \pi^{2}} \frac{y^{4}}{M^{4}} \\
&\vec{g}_{F}=\frac{1}{2} \mathcal{M}_{F}^{S} \vec{g}_{\textbf{1} S}+\frac{1}{2} \mathcal{M}_{F}^{A} \vec{g}_{\textbf{1} A}
\end{align}
For the \(q \bar{q}\) (SM quark) cut, for simplicity we assume that the SM Yukawa coupling vanishes. We first consider the dispersive integral above the scale \(4 M^{2}\).
\begin{equation}
\vec{g}_{q}=\frac{1}{4} \mathcal{M}_{q}^{S} \vec{g}_{\textbf{1} S}+\frac{1}{4} \mathcal{M}_{q}^{A} \vec{g}_{\textbf{1} A}+\frac{1}{4} \mathcal{M}_{q}^{S} \vec{g}_{\textbf{3} S}+\frac{1}{4} \mathcal{M}_{q}^{A} \vec{g}_{\textbf{3} A}\label{eq:6.31}
\end{equation}
where the expressions for \(\mathcal{M}_{q}^{S, A}\) are given in Appendix B, Eq.~(\ref{eq:B.2}).

Finally, the light quark loop also contributes to the integral from \((\epsilon \Lambda)^{2}\) to \(4 M^{2}\). We define \(x=(\epsilon \Lambda)^{2} / 4 M^{2}\). This additional contribution determines how \(\mathcal{M}\left((\epsilon \Lambda)^{2}\right)\) runs with \(\epsilon \Lambda\). We find
\begin{equation}
\vec{g}_{x}(x)=\frac{1}{4} \mathcal{M}_{x}^{S} \vec{g}_{\textbf{1} S}+\frac{1}{4} \mathcal{M}_{x}^{A} \vec{g}_{\textbf{1} A}+\frac{1}{4} \mathcal{M}_{x}^{S} \vec{g}_{\textbf{3}S}+\frac{1}{4} \mathcal{M}_{x}^{A} \vec{g}_{\textbf{3} A}\label{eq:6.32}
\end{equation}
Again the expressions for \(\mathcal{M}_{x}^{S, A}\) are lengthy and are presented in Appendix \(\mathrm{B}\), Eq.~(\ref{eq:B.4}). The final result is the sum
\begin{equation}
\vec{g}_{loop}(x)=\vec{g}_{F}+\vec{g}_{q}+\vec{g}_{x}(x)
\end{equation}

\begin{figure}[h]
	\begin{center}
		\includegraphics[width=.5\linewidth]{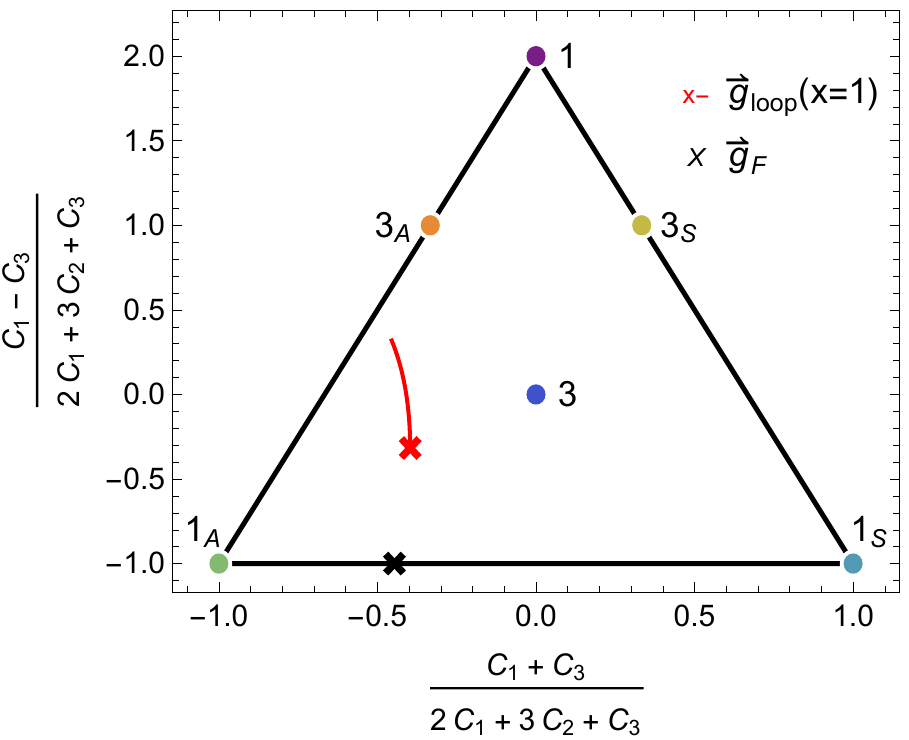}
	\end{center}
	\caption{Generators of the positivity cone for four-Higgs operators. The black cross represents \(\vec{g}_{F}\), the contribution from heavy fermion cuts. The red cross corresponds to \(\vec{g}_{loop}(1)=\vec{g}_{F}+\vec{g}_{q}\), adding the contribution of SM fermion cuts starting from \(4 M^{2}\) and up. The red curve shows \(\vec{g}_{loop}(x)\) as \(x\) decreases from 1 to 0.1, which incorporates, on top of \(\vec{g}\), the SM particle cuts from \((\epsilon \Lambda)^{2}\) to \(4 M^{2}\). See discussion in text.
}
	\label{fig20}
\end{figure}

We plot \(\vec{g}_{loop}(x)\) in Figure \ref{fig20}. The heavy cut contribution, \(\vec{g}_{F}\), is indicated by a black cross on the bottom edge. Ideally, if we were able to identify this as the only BSM contribution, we would conclude that the UV particles are \(\mathrm{SU}(2)\) singlets, as \(\vec{g}_{F}\) stays on a 2-face and excludes the existence of the rest 4 generators. Unfortunately, adding \(\vec{g}_{q}\), we see the light quark cut contribution enters. We show \(\vec{g}_{loop}(1)=\vec{g}_{F}+\vec{g}_{q}\) with a red cross mark. This corresponds to the limit \((\epsilon \Lambda)^{2} \rightarrow 4 M^{2}\), or \(x \rightarrow 1\), where the \(\mathbf{3}_{S, A}\) contributions from the SM loops are already significant, moving \(\vec{g}\) up and smearing the information from UV. Finally, adding \(\vec{g}_{x}(x)\), we see how \(\vec{g}_{loop}(x)\) runs as we decrease \((\epsilon \Lambda)^{2}\) from \(4 M^{2}\), and eventually becomes dominated by SM contributions. This is shown by the red curve. The \(\vec{g}_{loop}(x)\) flows up and slightly to the left. It flows up because the light quark cut which contributes more and more but equally to \(\mathbf{1}_{S, A}\) and \(\mathbf{3}_{S, A}\). It flows to the left because when SM dominates, \(\mathcal{M}_{x}^{A} \gg \mathcal{M}_{x}^{S}\) at low energy.

In order to obtain useful information from UV, it is clear that one should always study \(\mathcal{M}\left((\epsilon \Lambda)^{2}\right)\) with \(\epsilon \Lambda\) as large as possible, to preserve the purity of the UV information contained in \(\vec{g}_{loop}(x)\). When \(\epsilon \Lambda\) goes down, the additional contribution to \(\vec{g}_{loop }(x)\) is calculable in EFT and is known to live inside the cone, and therefore a smaller value of \(\epsilon \Lambda\) provides no additional information, but only dilutes the information from UV by adding the IR contributions.

An abuse of notation needs to be clarified here. The \(\vec{g}_{loop}(x)\) should not be directly interpreted as the vector of Wilson coefficients, \(\left(C_{1}, C_{2}, C_{3}\right)\), but rather, it is the \(\mathcal{M}^{i j k l}\) tensor, defined with \((\epsilon \Lambda)^{2}=4 x M^{2}\), expanded using the tree level amplitude-coefficient relation, given in Eqs.~(\ref{eq:4.32})-(\ref{eq:4.37}). In principle, it is possible to express \(\mathcal{M}^{i j k l}\) directly in terms of Wilson coefficients, but a loop-level calculation within SMEFT needs to be performed, which then involves dim-6 operators as well. For simplicity, we refrain from doing this and keep our discussion at the level of amplitude \(\mathcal{M}\).

It is tempting to think of \(\vec{g}_{loop }(x)\) as the RG running of coefficients, and conclude that the RG-evolved high-scale coefficient contains the most information of the UV. Strictly speaking, \(\vec{g}_{loop}(x)\) with \(x \rightarrow 1\) is not the actual coefficients as explained above, because the one-loop correction to \(\mathcal{M}\left((\epsilon \Lambda)^{2}\right)\) contains not only RG logs but also finite terms, which may come from loop diagrams with either one dim-8 insertion or two dim-6 insertions. Rather, we consider \(\mathcal{M}\left((\epsilon \Lambda)^{2}\right)\) as a physical observable, define at the energy \(\epsilon \Lambda\). Our observation is that even though \(\mathcal{M}\left((\epsilon \Lambda)^{2}\right)\) defined at any scale can be used to set bounds and study the inverse problem, choosing a larger \(\epsilon \Lambda\) is the most useful. The bounds are however set at the amplitude level. One can convert these bounds to bounds on coefficients \(\vec{C}(\mu)\) defined in any scheme and at any scale \(\mu\), but different scales are simply related by RG equations, so the running of \(\vec{C}(\mu)\) does not bring new information.

\section{Summary and discussion}
{\bf Positivity bounds} The first half of this paper (Sections 3 and 4) has been devoted to a systematic discussion of the extremal positivity approach at dim-8 \cite{Zhang:2020jyn}. The approach aims at finding the exact boundary of the UV-completable SMEFTs in the dim-8 space. Applications to various EFTs including the SMEFTs have been presented to illustrate various aspects of the approach. The main points are summarized below:

\begin{itemize}
  \item A UV-completable \(\mathcal{M}^{i j k l}\) must be a positive linear combination of the ``generators''. The latter can be constructed following Eqs.~(\ref{eq:3.27}) and (\ref{eq:3.28}), by enumerating the \(m\) matrices, and mapping them to the \(\vec{g}\) vectors in the coefficient space. Their conical hull defines the region of the UV-completable SMEFTs, which is a convex cone.

  \item A subset of \(\vec{g}\) is extremal. If their number is finite, bounds can be obtained through a vertex enumeration and are linear. If some \(\vec{g}\) depends on a free parameter, a ``continuous vertex enumeration'' may be possible and lead to curved bounds. If more free parameters are present, the problem need to be converted to a programming and solved numerically~\cite{Li:2021cjv}.

  \item All generators can be interpreted as coming from integrating out a single heavy particle from the UV spectrum. This connection provides an interface between positivity and the inverse problem. The only exception is the spin-2 generator(s) in \(V V\) scattering. They can only be interpreted as UV completion at the loop level.

  \item If a tree-level mapping between \(\mathcal{M}^{i j k l}\) and the coefficients are used, an alternative and easier way to find bounds is to enumerate all possible heavy particles that couple to scalars and fermions, through the following types of couplings:
\begin{align}
&g_{i j k} M S_{k} \phi_{i} \phi_{j}, \quad g_{i j k} V_{k}^{\mu} \phi_{i} \overleftrightarrow{D}_{\mu} \phi_{j}, \quad g_{i j k} S_{k} \bar{f}_{i}^{c} f_{j}, \quad g_{i j k} V_{k \mu} \bar{f}_{i} \gamma^{\mu} f_{j} \\
&\frac{1}{M} g_{i j k}\left(\bar{f}_{i}^{c} \sigma_{\mu \nu} f_{j}\right) V_{k}^{\mu \nu}
\end{align}
Integrating out these \(S\) and \(V\) particles directly gives all the \(\vec{g}\) vectors in the coefficient space. The bounds obtained this way apply not only to tree-level UV completions, but also to loop-level and strongly coupled ones. However, this simplified approach does not work for vector operators.
\end{itemize}

\begin{itemize}
  \item All extremal bounds for SM parity-conserving self-quartic operators from this and other works are collected in Section 4.4.
\end{itemize}
{\bf The inverse problem} The second half of this paper has been devoted to the discussion of the inverse problem for tree-level UV completions (Section 5) and for loop-level UV completions (Section 6). The main findings are presented in Section 5 and are summarized here:

\begin{itemize}
  \item Integrating out each heavy particle in the UV spectrum would give rise to generator vector \(\vec{g}\). The total coefficient vector is \(\vec{C}=\sum_{\alpha} w_{\alpha} \vec{g}_{\alpha}\), where \(\alpha\) is the particle type (specified by its spin, charge, irrep, relative couplings with different SM particles), and \(w_{\alpha}=g_{\alpha}^{2} / M_{\alpha}^{4}\). For measurements at the dim-8 level, we shall discuss a weaker version of the inverse problem: given the measured \(\vec{C}\), how to determine \(\vec{w} \equiv\left(w_{1}, w_{2}, \cdots\right)\).

  \item For each value of \(\vec{C}\), let \(\mathcal{W}\) be the set of feasible solution for \(\vec{w}\), its size \(\Delta\) is called the ``degeneracy'', which represents the amount of UV theories consistent with the measured \(\vec{C}\). \(\Delta\) represents the arbitrariness in finding a UV completion from the measured coefficients.

  \item The pattern of the degeneracy in the dim-8 space is nontrivial and is connected to the positivity bounds. The fact that the positivity cone is always a salient cone has several interesting physics implications. In particular, for SMEFTs that live on the boundary, extremality implies:
   \begin{itemize}
   \item If the SMEFT lives on the 0-face, i.e. \(\vec{C}=0\), all UV particles can be ruled out independent of any UV model assumptions. The UV completion is uniquely determined to be the SM itself. This provides a solid test of the SM.
  \item If the SMEFT lives on a 1-face, i.e. an extremal ray, the UV completion must be a ``one-particle extension'' of the SM, i.e. only one \(w_{\alpha}\) can be nonzero. Heavy particles must all belong to the same type, and their interaction type (to the SM particles) is uniquely determined.

  \item If the SMEFT lives on a \(k\)-face, \(k>1\), and there are \(l\) generators on that face, then the dimension of \(\mathcal{W}\) is \(l-k\). If \(l=k\) (e.g. restricted by symmetries of the theory), the UV particle types (and their interactions with the SM) are uniquely determined. If \(l>k\), the determination is not unique, but \(l-k\) degrees of freedom remain to be fixed.

  \item To sum up, SMEFTs that live on the boundary of the positivity cone have either vanishing or very limited degeneracy, meaning that the corresponding UV particles and their interactions to the SM can be fixed, either uniquely or at least to a large extent.

\end{itemize}

\end{itemize}

\begin{itemize}
  \item For SMEFTs that live in the interior, \(\Delta\) is always finite, and upper bounds on all \(w_{\alpha}\) (or equivalently, exclusion limits for all particles of type \(\alpha\) ) can be set, again independent of any assumptions on the UV models. This can be done with the geometric trick of Eq.~(\ref{eq:5.10}), which works even with an infinite number of \(\alpha\).

  \item All above implications require that a salient positivity cone exists. This is always true at dim-8, but not at dim-6 . Therefore all these implications are absent at dim-6.

\end{itemize}

Finally, Section 6 discusses several cases with loop-induced generators. The main conclusions are

\begin{itemize}
  \item Integrating out heavy particle loops or mixed heavy-light particle loops generates a contribution \(\vec{g}_{loop }\) to the amplitude. \(\vec{g}_{loop} \in \mathbf{C}\).
  \item \(\vec{g}_{loop }\) can be computed from two-particle cuts using the dispersion relation. A partial wave decomposition of the dispersion relation, together with the direct sum decomposition of irreps, reveals how \(\vec{g}_{loop}\) can be decomposed into a positive sum of the standard generators.

  \item In general, \(\vec{g}_{loop}\) is a sum of several different irreps, and tends to live in the interior of C. If this is the case, it does not generate additional degeneracy on the boundary of C. Many conclusions in Section 5 will still hold (for SMEFTs at the ER, on a \(k\)-face, or at the original, etc.)
  \begin{itemize}
  \item As a consequence, SMEFTs on the boundary would imply that the loop contribution can be excluded from the possible UV completion.

  \item However, the existence of \(\vec{g}_{loop }\) does increase the possible degeneracy inside the C.

  \item It is still possible that certain loops generate \(\vec{g}_{loop }\)'s on an ER or on a face. In this case, new degeneracies will be created on the boundary.
  \end{itemize}

  \item Without mixed heavy-light particle loops, one can subtract the SM loops and only focus on heavy particle loops, which are often the leading BSM effects.

  \item When mixed heavy-light particle loop exists, light particle loops contribute to the dispersion relation. \(\vec{g}_{loop }\) contains information of both BSM and SM particle states.
  \begin{itemize}
  \item As a result, \(\mathcal{M}\left((\epsilon \Lambda)^{2}\right)\) depends on the scale \(\epsilon \Lambda\) and so is \(\vec{g}_{l o o p} .\) For inverse problem, one should choose a \(\epsilon \Lambda\) as large as possible, to increase the purity of the information from UV.

  \item As \(\epsilon \Lambda\) decreases, \(\vec{g}_{loop }\) flows inside \(\mathbf{C}\) and is eventually dominated by IR contributions.

  \item In any case, the LO BSM effects would be present in other channels at the tree level, which should be studied first.
  \end{itemize}
\end{itemize}

\textbf{Discussions} The extremal positivity approach, by itself, is a powerful tool to determine the exact boundary of UV-completable EFTs, and supersede bounds from elastic scattering of two factorized and mixed states. The approach has been introduced in Refs.~\cite{Zhang:2020jyn, bellazzini_symmetries_2014}, further developed by Refs.~\cite{Yamashita:2020gtt, Fuks:2020ujk}, and systematically expanded in this work. Many SM examples have been studied and solved in these works. Several future developments can be foreseen. Cross-quartic operators can be added to the existing self-quartic bounds \cite{cross4paper}. Collider tests of these bounds can be performed at the LHC and future colliders (see, e.g. \cite{Fuks:2020ujk, Gu:2020ldn} for some initial studies in this direction). However, obtaining the full set of bounds for all SMEFT operators seems  impractical using this approach. The numerical approach proposed in Ref.~\cite{Li:2021cjv} can be a more promising alternative.

The main focus of this paper, however, is the connection between positivity and the inverse problem. We have shown that the mapping between SMEFTs and their UV completions show interesting patterns in the dim-8 space (recall that by dim-8 space we actually mean the \(s^{2}\) contributions in 2-by-2 amplitudes). If the SMEFT that describes the nature lives on or near the boundary of the positivity cone, measuring dim-8 coefficients would provide crucial information for finding the UV theory, potentially allowing us to uniquely fixes the UV particle types, their spin, charge, irrep, and couplings with the SM particles. This feature is absent at dim-6.

We consider this as an important motivation for studying SMEFT at the dim-8 level~\cite{Li:2020gnx,Murphy:2020rsh}. After all, the goal of precision measurements is not to fix the SMEFT coefficients, but rather to learn something about the UV theory. If deviations from the SM are present, we may be able to measure the dim-6 coefficients accurately, but due to the large (actually infinite) intrinsic degeneracy, an ignorance about possible UV models always remains. At dim-8, however, thanks to the existence of the positivity cone, the degeneracy of UV models is limited, and so even if the coefficients themselves are not be measured as accurate as dim-6, chances are that we can actually learn more about the UV, by exploiting the geometry information encoded in the positivity cones. This, of course, depends on where exactly the SMEFT is located in the cone, but the possibility to uniquely fix the UV particle content is already sufficiently attractive.

On the other hand, if signals of deviations from the SM are not observed, we would need confirm that the new physics cannot exist below certain scales, depending the precision level of the measurements. We have shown that this kind of conclusion is not possible at dim-6, because \(\vec{C}^{(6)}=0\) is a point with intrinsic degeneracy. Exclusion limits based on dim-6 coefficients are possible only when specific UV theories are assumed. In contrast, an experimental confirmation of \(\vec{C}^{(8)}=0\) would unambiguously rule out all types of UV states, independent of UV assumptions. This is simply because the origin is an extreme point of the positivity cone. Once we are able to exclude each heavy state individually, there is no need to further consider dim-10 operators and beyond.

In practice, investigating the dim-8 parameter space can be challenging for many reasons. One of them is that there are many dim-8 operators~\cite{Li:2020gnx,Murphy:2020rsh}, 993 counting only one flavor and 44807 counting three flavors. However, we should keep in mind that in our approach we only need to focus on operators that give rise to 2-to-2 amplitudes with an \(E^{4}\) dependence, and the number of these operators is 250 for one generation and 6076 for three. This is still a large number, but is manageable. In addition, it is always possible to focus on a subset of operators, as we have done in all the examples in Section 5. More precisely, one first picks a subset of SM particles, and take all operators that only involve these particles. All the discussions in Section 5 go through, and from the positivity cone in this subspace, one learns about how UV states interact with a subset of SM particles.

Another challenge is that it is unclear whether current and future colliders have the sensitivity to probe dim-8 effects to a desired precision level. The fact that we look for \(E^{4} / \Lambda^{4}\) rather than \(v^{4} / \Lambda^{4}\) effects is a helpful factor, as we expect to reach better sensitivity with high-mass region measurements or lepton colliders with a large beam energy. The main difficulty, however, is that dim-6 contributions are supposed to be the dominant effects, which could naively mask the \(E^{4} / \Lambda^{4}\) that we look for. A possible solution is to perform a global fit including both dim-6 and dim-8 operators. Different energy and angular dependences help to pin down dim-6 and dim-8 coefficients separately. A simple analysis of this kind has been applied to \(e^{+} e^{-}\) collision for future lepton colliders, and results do look promising \cite{Fuks:2020ujk}. For example, model-independent exclusion limits on all heavy particles can be set, using Eq.~(\ref{eq:5.10}), to a scale much larger than the collider energy. Another solution is to look for specific observables or channels where dim-6 effects are absent. This approach has been taken in Ref.~\cite{Gu:2020ldn} for the diphoton channel at future lepton colliders. While the projected precision is good, a disadvantage is that not all relevant dim-8 operators can be fixed in this way, and therefore the geometric information is incomplete. Nevertheless, we believe that more phenomenological studies in this direction can be interesting and may bring new opportunities for the bottom-up SMEFT approach.

As a final remark, in this work we have completely ignored the information from dim-6 coefficients. While the purpose is to emphasize how much we could learn just by using positivity and extremality at dim-8, in reality one should always include such information. It is likely that, under certain assumptions, the UV-completable region in the combined dim-6 and dim-8 space continues to be a salient cone, and so certain conclusions from this work may be generalized to the combined space, to develop a better understanding of the degeneracy in the space of SMEFTs. We will defer this to future works.

\section*{Acknowledgments}
We would like to thank Jiayin Gu, Xu Li, Jiang-Hao Yu, Hao Zhang and Shuang-Yong Zhou for helpful discussions and for editing this paper at the final stage. CZ is supported by IHEP under Contract No. Y7515540U1. 

\noindent {\bf Note added:} {\it The author Cen Zhang unexpectedly passed away while he was putting some final touches on this paper. He had sent out a mostly finished pdf version of this paper to most of those in the above list of acknowledgments. The present paper is recompiled from that the pdf version. To meet the publication standards, references have been added, minor physical discussions have been modified, English has been slightly improved and typos have been corrected by those in the above list.}

\appendix
\section{ Projectors}
\label{appA}

Here we give all projectors relevant for examples in Sections 4,5 and 6. 
\\ \(\textbf{SO(2):} \textbf{2} \otimes \mathbf{2}=\textbf{1}_{S} \oplus \mathbf{1}_{A} \oplus \mathbf{2}\)
\begin{equation}
\begin{aligned}
&P_{\textbf{1}_{S}}^{i j k l}=\frac{1}{2} \delta^{i j} \delta^{k l} \\
&P_{\textbf{1}_{A}}^{i j k l}=\frac{1}{2}\left(\delta^{i k} \delta^{j l}-\delta^{i l} \delta^{j k}\right) \\
&P_{\textbf{2}}^{i j k l}=\frac{1}{2}\left(\delta^{i k} \delta^{j l}+\delta^{i l} \delta^{j k}-\delta^{i j} \delta^{k l}\right)\label{eq:A.1}
\end{aligned}
\end{equation}
\(\textbf{SU(2):}\textbf{ 2} \otimes \mathbf{2}=\mathbf{1} \oplus \mathbf{3}\)
\begin{equation}
\begin{aligned}
P_{\textbf{1}}{ }^{i j}{ }_{k l} &=\frac{1}{2}\left(\delta_{k}^{i} \delta_{l}^{j}-\delta_{l}^{i} \delta_{k}^{j}\right) \\
P_{\textbf{3}}{ }^{i j}{ }_{k l} &=\frac{1}{2}\left(\delta_{k}^{i} \delta_{l}^{j}+\delta_{l}^{i} \delta_{k}^{j}\right)\label{eq:A.2}
\end{aligned}
\end{equation}
\(\textbf{SU(2):} \textbf{2} \otimes \overline{\mathbf{2}}=\mathbf{1} \oplus \mathbf{3}\)
\begin{equation}
\begin{aligned}
&{P_{\mathbf{1}}}_{j k}^{i}{ }^{l}=\frac{1}{2} \delta_{j}^{i} \delta_{k}^{l} \\
&P_{\mathbf{3}}{ }_{j k}^{i}{ }^{l}=-\frac{1}{2} \delta_{j}^{i} \delta_{k}^{l}+\delta_{k}^{i} \delta_{j}^{l}\label{eq:A.3}
\end{aligned}
\end{equation}
\(\textbf{SU(3): 3 }\otimes \textbf{3}=\overline{\textbf{3}} \oplus \mathbf{6}\)
\begin{equation}
\begin{aligned}
{P_{\overline{\textbf{3}}}}^{i j}_{\ \ kl}&=\frac{1}{2}\left(\delta_{k}^{i} \delta_{l}^{j}-\delta_{l}^{i} \delta_{k}^{j}\right) \\
{P_{\textbf{6}}}{ }^{i j}_{\ \ kl} &=\frac{1}{2}\left(\delta_{k}^{i} \delta_{l}^{j}+\delta_{l}^{i} \delta_{k}^{j}\right)\label{eq:A.4}
\end{aligned}
\end{equation}
\(\textbf{SU(3): 3 }\otimes \overline{\textbf{3}}=\textbf{1} \oplus \textbf{8}\)
\begin{equation}
\begin{aligned}
&{P_{\mathbf{1}}}_{j k}^{i}{ }^{l}=\frac{1}{3} \delta_{j}^{i} \delta_{k}^{l} \\
&{P_{\mathbf{8}}}_{j k}^{i}{ }^{l}=-\frac{1}{3} \delta_{j}^{i} \delta_{k}^{l}+\delta_{k}^{i} \delta_{j}^{l}\label{eq:A.5}
\end{aligned}
\end{equation}

\section{Some expressions for matching fermion loops}
\label{appB}

Here we list the expressions for the \(\mathcal{M}_{x, q}^{S, A}\) defined in Eqs.~(\ref{eq:6.31}) and ~(\ref{eq:6.32}):
{\small
\begin{align}
\mathcal{M}_{q}^{S}=& \frac{y^{4}\left(-2\left(\mathrm{Li}_{2}\left(-\frac{1}{4}\right)+\mathrm{Li}_{2}\left(-\frac{1}{5}\right)+1\right)-4 \log ^{2}(2)+\log (5) \log \left(\frac{36}{5}\right)\right)}{256 \pi^{2} M^{4}} \\
\mathcal{M}_{q}^{A}=& \frac{y^{4}\left(6 \mathrm{Li}_{2}\left(-\frac{1}{4}\right)+6 \mathrm{Li}_{2}\left(-\frac{1}{5}\right)-2+(\log (8)-8) \log (16)+\log (5)(20-6 \log (6)+\log (125))\right)}{768 \pi^{2} M^{4}} \label{eq:B.2}\\
\nonumber \mathcal{M}_{x}^{S}=& \frac{y^{4}}{3072 \pi^{2} M^{4} x^{3}}\bigg[24 x^{3}\left(\mathrm{Li}_{2}(-4 x-1)+\mathrm{Li}_{2}(-4 x)\right)+4\\ \nonumber
& x\left(x\left(x\left(6\left(\mathrm{Li}_{2}\left(-\frac{1}{4}\right)+\mathrm{Li}_{2}\left(-\frac{1}{5}\right)+1\right)+2 \pi^{2}+3 \log ^{2}(4)-3 \log (5) \log \left(\frac{36}{5}\right)\right)-3\right)-3\right) \\
&+3 \log (4 x+1)\left(8 x^{3} \log (4 x+2)+(3-4 x) x+1\right)\bigg] \\ \nonumber
\nonumber \mathcal{M}_{x}^{A}=& \frac{y^{4}}{3072 \pi^{2} M^{4} x^{3}}\bigg[-24 x^{3}\left(\mathrm{Li}_{2}(-4 x-1)+\mathrm{Li}_{2}(-4 x)\right)\\ \nonumber
&-4 x\left(x^{2}\left(6 \mathrm{Li}_{2}\left(-\frac{1}{4}\right)+6 \mathrm{Li}_{2}\left(-\frac{1}{5}\right)-2+2 \pi^{2}+3 \log ^{2}(4)+\log (5)(20-6 \log (6)+\log (125))\right)\right.\\
&+x+1)-64 x^{3} \log (x)+\log (4 x+1)\left(-24 x^{3} \log (4 x+2)+(4 x(16 x+3)+3) x+1\right)\bigg] \label{eq:B.4}
\end{align}
}


\bibliographystyle{JHEP}
\bibliography{cz}

\end{document}